# Building Robust Crowdsourcing Systems with Reputation-aware Decision Support Techniques

# [A Book Draft]

**YU HAN**

**Version 22-Jan-2016**

# Acknowledgements

I would like to take this opportunity to express my heart-felt gratitude towards everyone who provided me with their help without which this work would not have been possible. First and foremost, I want to thank my advisor Dr. **Chunyan Miao** for her constant support and encouragement in all aspects of my research and career. Dr. Miao always has an accurate grasp on the trend of emerging research in the field of multi-agent systems and other related areas. She taught me how to do independent research, deliver presentations, and be confident when facing different types of audience. She gave me a lot of freedom to pursue a broad range of interests and provided me with valuable opportunities to get in touch with the leading scientists in various fields. I am grateful for having the opportunity for working with such an extraordinary mentor.

I also want to give special thanks to my co-advisor Dr. **Cyril Leung** from The University of British Columbia, Vancouver, BC, Canada. He is always willing to share with me his vast experience in conducting cutting edge research. His dedicated help and insightful comments during the preparation for publishing our work have inspired me greatly.

I am deeply grateful to the **Singapore Millennium Foundation (SMF)** for providing me with the prestigious SMF Ph.D. Scholarship over the last four years. Their charitable approach towards supporting young researchers and unwavering drive towards excellence inspired me to make contributions to the state of the art with practical impact.

I want to thank my co-authors Dr. **Zhiqi Shen**, Dr. **Bo An**, Dr. **Victor R. Lesser**, Dr. **Martin G. Helander**, Dr. **Alex C. Kot**, Dr. **Tan Ah Hwee**, Dr. **Lim Ee Peng**, Dr. **Shum**

**Ping**, Dr. **Dusit Niyato**, Dr. **Boyang Li**, Dr. **Jianshu Weng**, Dr. **Yuan Miao**, Dr. **Xiaoming Li**, Dr. **Yuan Miao**, Dr. **Xuehong Tao**, Dr. **Siyuan Liu**, Dr. **Yiqiang Chen**, Dr. **Hengjie Song**, Dr. **Zhiqiang Liu**, Dr. **Xiaogang Han**, Dr. **Li Pan**, Dr **Yundong Cai**, Miss **Qiong Wu**, and Mr. **Jun Lin** who provided me with valuable suggestions and guidance. The Emerging Research Laboratory has been like a family to me. I have benefited immensely from various intellectually stimulating discussions with visiting scientists and my colleagues including Dr. **Thomas W. Calvert**, Dr. **Qiang Yang**, Dr. **Dong Huang**, Dr. **Guopeng Zhao**, and Dr. **Ailiya**.

Last but not least, I want to thank my family for their unconditional love and support from the very beginning. I want to thank my wife Dr. **Xinjia Yu** for her love and encouragement, especially for giving up everything in China and moving to Singapore with me. This thesis is dedicated to them.

# Table of Contents

# List of Figures

# List of Tables

# Abstract

Trust is a mechanism used by people to facilitate interactions in human societies where risk and uncertain are common. Over the past decade, the importance of trust management in computational intelligence research (e.g. in multi-agent systems (MASs)) has been recognized by both the industry and the academia. Computational trust models for evaluating the trustworthiness of a trustee agent based on a wide range of evidence have been proposed. Nevertheless, two important research problems remain open in this field. Firstly, how to mitigate the adverse effects of biased third-party testimonies on the accuracy of evaluating the trustworthiness of an agent? Secondly, how to make trust-aware task delegation decisions to efficiently utilize the capacities of trustee agents to achieve high social welfare?

This thesis presents the research into addressing these two problems. It first proposes a novel reinforcement learning based trust evidence aggregation model – namely the *Actor-Critic Trust* (ACT) model – to address the problem of biased testimonies. Individual truster agents can use the ACT model to dynamically learn to adjust the selection of witness agents, the weights given to each of their testimonies, as well as the weights given to the collective opinions of the witness agents and the first-hand trust evidence to produce a trustworthiness evaluation. The model operates according to observable changes in the MAS environment and has been shown to be robust against collusions among witness agents. The ACT model eliminates the need for manually tuning these weight parameters in most existing trust models and makes agents more adaptive in changing environments.

This work then goes beyond the existing trust management research framework by removing an widespread assumption implicitly adopted by existing research: that a trustee agent can process an unlimited number of interaction requests per discrete time unit without compromising its performance as perceived by the truster agents. The trust management problem is re-formalized as a multi-agent trust game based on the principles of the Congestion Game, which is solved by two trust-aware interaction decision-making approaches: 1) the *Social Welfare Optimizing approach for Reputation-aware Decision-making* (SWORD) approach, and 2) the *Distributed Request Acceptance approach for Fair utilization of Trustee agents* (DRAFT). SWORD is designed for use in MASs where a central trusted entity is available, while DRAFT is designed for individual trustee agents in fully distributed MASs. Both of these proposed approaches have been demonstrated to help an MAS achieve significantly higher social welfare than existing trust-aware interaction decision-making approaches. Theoretical analyses have shown that the social welfare produced by these two approaches can be made closer to optimal by adjusting only one key parameter. With these two approaches, the framework of research used by current multi-agent trust models can be enriched to handle more realistic operating environment conditions where the computational resources possessed by the agents are limited.

The proposed approaches can potentially be used in many application domains. In this work, we evaluated the performance of the ACT model in cognitive radio networks, which is an emerging field in wireless communication. We demonstrated its effectiveness in improving the quality of the results produced by multiple radio nodes sensing the availability of network spectrums over wide geographic locations. SWORD was evaluated under crowdsourcing conditions, while DRAFT was evaluated under open dynamic MAS conditions. The results have shown that the agents are able to make more socially beneficial interaction decisions with the proposed approaches and the social sustainability of the MASs with resource constrained trustee agents can be significantly enhanced with the help of this research.

# Chapter 1

# Introduction

For decades, trust has been one of the important foci in the study of various humanity related fields such as sociology, philosophy, and economics, etc. After the turn of the century, with a boom of computing and network technologies and the emergence of virtual online communities extending people's daily life into the cyberspace, trust research has become an important topic in computer science as well. This is especially the case in the field of open multi-agent systems (MASs) research where self-interested human-like virtual entities need to delegate tasks to each other in order to achieve their own goals. In this chapter, we provide an introduction to the original concept of trust, the research on trust in MASs, the scope of research documented in this thesis and the contributions to this area of research.

## 1.1. The Concept of Trust

Trust is a central ingredient in human relationships. The concept of "trust" was originally a topic of study in the social sciences. Our understanding of trust has mainly been derived from people's daily experience in applying it to real world situations. Many researchers have attempted at defining trust [Hardin, 2002], so far the most well accepted definition for the concept of trust was given by Gambetta [Gambetta, 1998] as:

*"the subjective probability by which an individual, A, expects that another individual, B, performs a given action which its welfare depends."*

The notion of trust defers from cryptographic security techniques in their primary aims. Cryptographic security techniques' focus on ensuring data integrity and data confidentiality through secure identification of the users through authentication and assigning access rights to them via policies is aimed to protect data or services from malicious users. Trust, on the other hand, offers a way to guard the users against malicious services or exploitation from other malicious people. The concept of trust has the following characteristics:

1) *Trust is useful only in an environment characterized by uncertainty and where the participants need to depend on each other to achieve their goals.* In a perfectly observable and predictable environment, it is not necessary to measure trustworthiness because each participant would already know the action that should be taken. In addition, if all tasks can be performed satisfactorily by individuals without the need for interactions, trust would be of no significance.
2) *Trust is context sensitive.* It is commonly agreed in the trust literature that trusting is a three-part relationship which can be expressed as "Alice trusts Bob to do X". The "to do X" part places a limit on the trust relationship based on how well the subject's capabilities suit the context in which the relationship exists. For example, Alice may trust Bob to repair her computer but she may not trust Bob to repair her car. In this case, even if Bob has good intentions towards Alice, benevolence alone does not warrant a high level of trust across all contexts. The capability of Bob in a subject matter also affects the level of trust Alice places in him.

3) *Trust is subjective.* The formation of an opinion about someone's trustworthiness depends not only on the behaviors of the subject but also on how these behaviors are perceived by the agent. The perception of the subject's behaviors often depends on some intrinsic characteristics of the agent such as a propensity to trust (how trustful the agent is towards others) and the expectation the agent has on the subject's performance. For instance, although Bob's performance in repairing computers is consistent across all customers, Malory, being a more demanding customer than Alice, may trust Bob less than Alice in terms of letting Bob repair his computer.

4) *Trust is unidirectional.* An agent's trust in a subject is based on the knowledge that it has about the subject. This knowledge may be acquired either through the agent's own observations, the recommendations from the agent's friends or other means. The subject may not necessarily know the agent and therefore may not trust the agent in this case. Even if the agent has direct observations of the subject's past behaviors, the perception of the subject on the agent's performance and benevolence may differ. Thus, an agent's trust in a subject may not be reciprocated.

5) *Trust may not be transitive.* The first time Alice meets Bob, she does not know how much to trust him. David, whom Alice trusts, comes forward and vouches for Bob. In this scenario, should Alice trust Bob? The answer can be both yes and no. This depends on the context in which Alice trusts David. David's opinion of Bob is only useful to Alice if Alice trusts David's trust assessment of others. The trust on the witness's (David's) trust assessment of the subject is an important concept in the computational trust literature. It is known as the *credibility* of the witness [Weng *et al.,* 2006]. In a system with no central trusted authority to vouch for the users, the credibility of witnesses becomes a critical facilitator for allowing an agent who has had too few interactions with a subject to form a meaningful trust opinion.

## 1.2. Computational Trust Research

In recent years, the concept of trust has been introduced into computer based systems modeled as open MASs. An MAS consists of many building blocks known as agents. According to [Wooldridge, 1997]:

> *"An agent is an encapsulated computer system situated in some environment and capable of flexible, autonomous action in that environment in order to meet its design objectives."*

With this definition, [Jennings, 2001] further specifies the desired characteristics an agent should possess as:

1) Clearly identifiable problem-solving entities with well-defined boundaries and interfaces;
2) Situated in a particular environment over which they have partial control and observability;
3) Designed to fulfill a specific role;
4) Autonomous (they have control over both their internal states and their own behaviors);
5) Capable of exhibiting flexible problem-solving behavior in pursuit of their design objectives.

An agent with autonomous reasoning and decision-making capability may act on behalf of one or more human owners in complex interactions with other agents or human beings. Agents usually live in a society of agents which is referred to as *multi-agent systems* [Wooldridge 2002]. Such a system usually possesses the following characteristics:

1) Each agent has incomplete information or capabilities for solving the problem and, thus, has a limited viewpoint;
2) There is no global control;
3) Data are decentralized;
4) Computation is asynchronous.

Although in today's world where the advance in cloud computing technologies have made centralized storage and analysis of data as well as global control in an MAS possible, these characteristics have heavily influenced agent trust research for years.

Agent trust research really took off over the last decade as more complex and large scale online transaction systems (e.g., e-commerce systems, online virtual worlds, social networking websites, etc.) emerge. Using three of the most popular annual conferences on Artificial Intelligence (AI) and MAS related topics – *the Association for the Advancement of Artificial Intelligence (AAAI) Conference*, *the International Conference on Autonomous Agents and Multi-Agent Systems (AAMAS)*, and *the International Joint Conference on Artificial Intelligence (IJCAI)* – as a gauge, we can get some rough ideas on the development of this research area.

In general, a trust agent can be conceptualized as consisting of two main modules which dictate its behavior: 1) a *trust evaluation module*, which helps the agent assess the trustworthiness of potential interaction partners; and 2) a *trust-aware decision module*, which directs the agent's selection of future interaction partners based on trust evaluations. As shown in Table 1, the majority of the research effort is currently focused in the trust evaluation sub-field. This sub-field deals with the problem of accurately evaluating the trustworthiness of potential interaction partners. The proposed solutions can be divided into four main categories, 1) *direct trust* evaluation models, which rely on past observed behaviors; 2) *indirect/reputation-based trust* evaluation models, which rely on third-party testimonies from other agents in the same environment; 3) *socio-cognitive trust* evaluation models, which rely on analyzing the social relationships among agents to estimate their trustworthiness; and 4) *organizational trust* evaluation models, which rely on the organizational affiliation or certificates issued by some trusted organizations to estimate the trustworthiness of agents. Compared with the trust evaluation sub-field, very limited research has been done in the trust-aware decision making sub-field.

Table 1. Summary of Multi-agent Trust Research Papers Published in AAMAS, IJCAI and AAAI (2002~2012)

| **Trust Evaluation** | | | | **Trust-aware Interaction Decision-making** | **Performance Assessment** |
|---|---|---|---|---|---|
| **Direct Trust** | **Indirect Trust** | **Socio-cognitive** | **Organizational** | | |
| 1. [Tran and Cohen, 2004]<br>2. [Griffiths 2005]<br>3. [Zheng *et al.*, 2006]<br>4. [Dondio and Barrett, 2007]<br>5. [Bedi *et al.*, 2007]<br>6. [Osman and Robertson, 2007]<br>7. [Reece *et al.*, 2007]<br>8. [Wang and Singh, 2007]<br>9. [Reches *et al.*, 2008]<br>10. [Teacy *et al.*, 2008]<br>11. [Pinyol and Sabater-Mir, 2009]<br>12. [Khosravifar *et al.*, 2009]<br>13. [Vogiatzis *et al.*, 2010]<br>14. [Matt *et al.*, 2010]<br>15. [Salihi-Abari and White, 2010]<br>16. [Koster *et al.*, 2011]<br>17. [Pasternack and Roth, 2011]<br>18. [Witkowski, 2011]<br>19. [Jiang *et al.*, 2012]<br>20. [Burnett and Oren, 2012] | 1. [Sen and Sajja, 2002]<br>2. [Tran, 2002]<br>3. [Yu and Singh, 2003]<br>4. [Teacy *et al.*, 2005]<br>5. [Weng *et al.*, 2006]<br>6. [Regan *et al.*, 2006]<br>7. [Wang and Singh, 2006]<br>8. [Fullam and Barber, 2007]<br>9. [Hendrix and Grosz, 2007]<br>10. [Kawamura *et al.*, 2007]<br>11. [Procaccia *et al.*, 2007]<br>12. [Hang *et al.*, 2009]<br>13. [Tang *et al.*, 2010]<br>14. [Liu *et al.*, 2011]<br>15. [Zhang *et al.*, 2011]<br>16. [Koster *et al.*, 2012]<br>17. [Piunti *et al.*, 2012]<br>18. [Liu *et al.*, 2012]<br>19. [Haghpanah and desJardins, 2012]<br>20. [Fang *et al.*, 2012]<br>21. [Serrano *et al.*, 2012] | 1. [Castelfranchi *et al.*, 2003]<br>2. [Falcone and Castelfranchi, 2004]<br>3. [Falcone *et al.*, 2004]<br>4. [Ashri *et al.*, 2005]<br>5. [Casare and Sichman, 2005]<br>6. [Massa and Avesani, 2005]<br>7. [Katz and Golbeck, 2006]<br>8. [Kuter and Golbeck, 2007]<br>9. [O'Donovan *et al.*, 2007]<br>10. [Rettinger *et al.*, 2008]<br>11. [Burnett *et al.*, 2010]<br>12. [Koster *et al.*, 2010]<br>13. [Li and Wang, 2010]<br>14. [Liu *et al.*, 2010]<br>15. [Singh, 2011]<br>16. [Noorian *et al.*, 2011]<br>17. [Parsons *et al.*, 2011]<br>18. [Liu and Datta, 2011]<br>19. [Venanzi *et al.*, 2011]<br>20. [Liu and Datta, 2012] | 1. [Kollingbaum and Norman, 2002]<br>2. [Griffiths and Luck, 2003]<br>3. [Huynh *et al.*, 2005]<br>4. [Hermoso *et al.*, 2010] | 1. [Fullam and Barber, 2006]<br>2. [Burnett *et al.*, 2011] | 1. [Fullam *et al.*, 2005]<br>2. [Kerr and Cohen, 2009] |

Apart from these two major research sub-fields, assessing the performance of proposed agent trust models is also an important sub-field in agent trust research. Although datasets concerning certain aspects of the trust evaluation problem (e.g., the *Epinions* and *Extended Epinions* datasets [Massa and Avesani, 2005]) are available, since the effectiveness of various trust models need to be assessed under different environment conditions with different types of misbehaviors, it is difficult to find suitable real world data all the time. Therefore, in the current agent trust research field, most of the existing trust models are assessed with simulation or synthetic data. One of the most popular simulation test-beds for trust models is the agent reputation and trust (ART) test-bed proposed by [Fullam *et al.*, 2005]. However, even this test-bed does not claim to be able to simulate all experimental conditions of interest. With this reason, many existing work designed their own simulation environments when assessing the performance of their proposed trust models.

In the current multi-agent trust research landscape, agents are normally considered to be *individually rational* – meaning that agents will take whatever action that is expected to maximize their own utility. Although there has been some preliminary attempt in studying the influence of inherently irrational factors (e.g., emotion) on trust agents [Jones and Pitt, 2011], irrational agents are usually not among the primary focuses of the research in MAS.

## 1.3. Scope of Research and Motivations

In systems that can be modeled as open MASs (e.g., e-commerce systems, wireless networks, distributed sensor networks), agents need to rely on services from each other to achieve their design goals. In these environments where risk and uncertainty exists when interacting with other agents, multi-agent trust is regarded as one of the most useful technologies to help agents make decisions that reduce their risk exposure in the long run. To make such decisions, agents must be able to assess the trustworthiness of potential interaction partners with a high degree of accuracy. The most reliable and relevant source of trust evidence for a truster agent is its first-hand prior interaction experience with the trustee agents. However, in large scale open MASs consisting of many agents who may dynamically leave or rejoin the MASs, individual truster agent often do not have enough direct trust evidence with a lot of trustee agents in order to make a reliable estimation of their trustworthiness.

To this end, many research works (such as those listed in Table 1 under the column labeled “Indirect Trust”) propose models that enable truster agents to obtain indirect trust evidence from each other to tap into the wisdom of the crowd. Third-party testimonies, which are supposed to be derived from an agent’s prior first-hand interaction experience with certain trustee agents, can be shared among the truster agents. They are combined with direct trust evidence, if any, to calculate the reputation of the trustee agent, which should improve the accuracy of the estimation. However, in practice, these testimonies can be biased. This may be due to the difference in personal preference among agents, or even malicious act of collusion among agents to artificially inflate or drive down the reputation of certain trustee agents. Mitigating the adverse effect of biased testimonies, therefore, becomes an important issue that need to be resolved in order to make effective trust decisions.

In some popular e-commerce systems and online rating systems, biased testimonies have become a major issue threatening the credibility, and thus the sustainability, of the business. For example, news reports concerning the rampant use of software and staged transactions to artificially boost an e-shop’s reputation in Taobao.com (which is one of China’s largest online e-commerce systems and using a reputation system similar to that used by eBay.com) have surfaced in 2010 [CIOTimes.com, 2010] and is still happening as recently as Aug 2012 [Hexun.com, 2012]. Rumors about companies in China offering services to help people build up their reputation through favorable online voting by their employees (such as http://www.hyh1688.com/) can be found through Chinese search engine Baidu.com. However, due to broken links, these rumors cannot be independently verified. Nevertheless, the problem of biased testimonies is plaguing online e-commerce systems, especially in China. One important feature of this type of collusion is that perpetrators focus on boosting their colluder's reputations rather than tarnishing the reputations of their competitors. This is due to the fact that the sheer number of competitors in a large scale online e-commerce system like Taobao.com makes the badmouthing approach prohibitively difficult and costly.

From Table 1.1, it can be observed that research on trust-aware interaction decision-making (TID) mechanisms has not attracted much attention from researchers in this field. In MASs, the goal of trust-aware interaction decision-making is to help a truster agent decide which candidate trustee agent is to be selected to perform a given task at the current time. It is a general consensus among the current multi-agent trust community that, in order to minimize a truster agent's risk exposure, it should always interact with the trustee agent with the highest reputation that it can find in the context of the given type of task. This approach is a rational choice from the perspective of a truster agent and it is adopted by the majority of the existing trust models.

However, in a system involving trust-aware interaction decision-making approaches, truster agents are not the only stakeholder affected by them. The trustee agent and the collective utility of all agents in an MAS derived through the interactions can also be affected by the interaction decisions made by the truster agents. Theoretically, TID approaches should reward trustee agents with high reputations with more tasks so that they can derive more utility through completing them.  Over time, it should help the MAS exclude untrustworthy trustee agents and sustain repeated interactions among agents over the long term. Nevertheless, a closer look at the assumptions used by existing trust models to reach this conclusion reveals that there are limitations to the applicability of this conclusion.

The perceived behavior of a trustee agent comprises of various facets, one of the most important is the timeliness in completing the given tasks [Griffiths, 2005]. The facet of a trustee agent's performance depends not only on the ability of the agent itself, but also the workload placed on it by the truster agents in the MAS. Under the existing TID approach, the more reputable it is and the wider its reputation is disseminated throughout the MAS, the more likely it will become the interaction partner of choice by a large number of truster agents for certain types of tasks. In some cases, such a development can overwhelm the trustee agent if it can only handle a limited number of tasks effectively per unit time step, and hurt its reputation. In this thesis, this phenomenon is referred to as the *reputation damage problem* (RDP).

In reality, the RDP can have deadly consequences. As recently as July 2012, multiple news reports from China about over-worked entrepreneurs managing e-shops on Taobao.com dying of fatigue related illnesses have surfaced [163.com, 2012; iFeng.com, 2012; Sina.com, 2012; Sohu.com, 2012]. The common features of such cases are 1) the victims are young (in their 20s); 2) the e-shops under their management are highly reputable on Taobao.com; 3) they receive large number of orders; 4) their e-shops are small and medium sized enterprises which require them to personally handle most business activities; and 5) they misjudged their capacity to keep handling incoming orders effectively. Although these are extreme cases and should not be entirely blamed on existing TID approaches, the assumption that the utility of a trustee agent (a human being in this case) is linearly related to the amount of business it can get appears not to be always appropriate.

Motivated by the need to the above observations, this thesis aims to achieve the following research objectives: 1) to propose a model that can effectively mitigate the adverse effect of biased testimonies which can be applied to most computational trust models; and 2) to propose TID approaches that can help truster agents make interaction decisions which not only protect their own interest, but enhance the social welfare for the entire MAS.

## 1.4. Summary of Contributions

By achieving the research objectives set out in the previous section, this thesis makes the following important contributions to the state of the art in the area of multi-agent trust research:

(i) *The ACT Model*: Although there are a large number of research works over the years focused on addressing the problem of unfair testimonies from public reports of direct trust experience by witness agents. Many of these approaches generally suffer from three main types of shortcomings: 1) *relying on assumptions about the characteristics of the witness agent population*: They are usually majority voting based and perform poorly in situations where the majority of the witness agent population are compromised; 2) *tightly coupled with specific operating environments*: They often require additional infrastructural support (e.g., payment systems, knowledge of social relationships among agents, etc.) in order to work; or 3) *involving manual tuning of parameters crucial to the performance of the model*: This reduces the adaptability of the models in the face of a dynamically changing environment and makes regular human intervention necessary. We propose a trust evidence aggregation model based on the principles of reinforcement learning called the Actor-Critic Trust (ACT) model. It enables truster agents to dynamically learn the appropriate values of a large number of parameters based on their interaction experience. Compared to existing work, the ACT model does not require additional information or infrastructure support other than the third-party testimonies received by a truster agent. Experimental results show that it outperforms related work. The ACT model was applied to solve the collaborative spectrum sensing problem in Cognitive Radio Networks (CRNs) and demonstrated effectiveness in preserving the wellbeing of the network under various attack scenarios.

(ii) *Multi-agent Trust Game*: To further study the RDP, the assumption made by most existing multi-agent trust research that trustee agents' perceived performance is not affected by the amount of workload assigned to them has to be removed. In order to do so, the multi-agent trust management problem must be reframed into a framework of thinking that is capable of taking the limitations in trustee agents' capacities into account. In this research, we redefine the TID problem as a Multi-agent Trust Game (MTG) based on the concept of Congestion Games [Monderer and Shapley, 1996]. The MTG complements existing research [Mikulski *et al.*, 2011] by providing a theoretical framework for analyzing TID approaches under more realistic conditions where trustee agents have limited resources and capabilities to handle workload and the delay experienced by truster agents is a function which is partially affected by the choices of interaction partners made collectively by them. By explicitly including these limitations into the analysis of trust in MASs, the MTG can facilitate the design of TID approaches that can produce higher system-wide social welfare.

(iii) *The SWORD Approach*: Based on the MTG, we propose the Social Welfare Optimizing Reputation-aware Decision-making (SWORD) approach. Designed for a system manager such as an e-commerce platform operator to use, the SWORD approach observes agents' real-time situations (such as their reputations, current workload, historical performance in handling assigned tasks per unit time, etc.) and help truster agents who want to delegate tasks at the current time select interaction partners. The SWORD approach is the first TID approach designed with the objective to mitigate the adverse effect of the RDP. Based on the principles of Lyapunov Drift analysis [Neely, 2010], it produces solutions to the MTG which can be proven to achieve social welfare values in an MAS close to the optimal value in the long run. Solutions are produced in polynomial time. Experiments conducted under crowdsourcing system environments have shown that the SWORD approach significantly outperforms related work in terms of promoting social equity and enhancing social welfare. The SWORD approach protects trustee agents from being overloaded with requests, makes efficient use of the overall trustee agent resources in an MAS, reduces truster agents' waiting time for delegated tasks to be completed, and increases the throughput of interactions among agents in an MAS.

(iv) *The DRAFT Approach*: in order to extend the SWORD approach to enable it to operate in fully distributed MASs, we propose the Distributed Request Acceptance approach for Fair utilization of Trustee agent services (DRAFT). Existing TID approaches are mostly designed for truster agents. The DRAFT approach bridges this gap in the state of the art by enabling trustee agents to make situation-aware decisions about whether to accept incoming task delegation requests. Through an analysis of its current situation taking into account its current reputation standing in the MAS, its current workload, and the anticipated effort level it can spend on completing tasks over the current time step, the trustee agent can adjust the degree of greediness for accepting incoming requests with the DRAFT approach. Based on the same design principle as the SWORD approach, the DRAFT approach can also be proven to produce solutions for the MTG achieving close to optimal social welfare over the long run. The solutions can also be produced in polynomial time. Experiments conducted in open dynamic MAS environments show that the DRAFT approach achieves significantly higher social welfare than related work and promotes social equity in the community.

The research proposed in this thesis can be applied to the general context of MASs complementing existing trustworthiness evaluation models and trust-aware interaction decision-making approaches.

## 1.5. Outline of the Thesis

The rest of this report is organized as follows:

- Chapter 2 provides an overview of the current state of the art in multi-agent trust research.
- In Chapter 3 proposes the ACT model and describes applications of the model in the domains of e-commerce and cognitive radio networks.

- Chapter 4 presents a detailed analysis of the adverse effects of the reputation damage problem and proposes the multi-agent trust game.
- Chapter 5 proposes the SWORD approach and analyzes its performance through theoretical proof.
- Chapter 6 applies the SWORD approach in crowdsourcing system scenarios and analyzes its performance under different conditions.
- Chapter 7 proposes the DRAFT approach, analyzes its performance through theoretical proof, and evaluates its effectiveness under e-commerce application environments.
- Chapter 8 concludes the thesis and outlines potential future research directions.

# Chapter 2

# Literature Review

*Trust* was first introduced as a measurable property of an entity in computer science in [Marsh, 1994]. Following this seminal work, a significant number of computational models focusing on various facets of trust management has been proposed in MAS research. Trustworthiness evaluation models employ probabilistic, socio-cognitive, and organizational techniques to enable truster agents to estimate the potential risk of interacting with a given trustee agent. Once the trustworthiness evaluations for a set of candidate trustee agents have been produced, trust-aware interaction decision-making approaches help the truster agent to select a trustee agent for interaction at a particular point in time. By reviewing the key advancements published in multi-agent trust research, it can be seen that most existing research effort is concentrated on improving the accuracy of trust evaluation models. These models can be further classified into four categories according to their approaches, namely:

1) *Direct trust evaluation models*,

2) *Reputation-based trust evaluation models,*

3) *Socio-cognitive trust evaluation models*, and

4) *Organizational trust evaluation models*.

In this chapter, we review a number of notable works in multi-agent trust management research and summarize the assumptions commonly used in this field.

## 2.1. Direct Trust Evaluation Models

One of the ways people establish trust between each other is through observing the outcome of past interactions between them. This evidence based approach of evaluating the trustworthiness of a potential interaction partner has been adopted by the multi-agent trust research community as one of the most widely used basis for enabling agents to establish trust. An intuitive way of modeling trust between agents is to view interaction risk as the probability of being cheated by the interaction partner. Such a probability can be established by a truster agent through looking back at the outcomes of past interactions with a trustee agent. The historical interaction outcomes serve as the direct evidence available for the truster agent to make an educated guess about a trustee agent's trustworthiness.

One of the earliest models that attempt to derive a trustworthiness value based on direct evidence is the Beta Reputation System (BRS) proposed by Jøsang and Ismail in [Jøsang and Ismail, 2002] which is inspired by the Beta Distribution. The model projects past interaction experience with a trustee agent into the future to give a measure of its trustworthiness. BRS estimates the trustworthiness of a trustee agent by calculating its reputation, which is defined as the probability expectation value of a distribution that consists of positive and negative feedbacks about the trustee agent. This expectation value is then discounted belief, disbelief and uncertainty about the truthfulness of the feedbacks (in the case of direct evidence, the truster agent can be very sure about the truthfulness since the feedbacks were produced by itself) and then discounted by a forgetting factor to allow past evidence can be gradually discarded. The resulting value is the reputation of the trustee agent.

In BRS, the outcome of an interaction is represented by a binary value (i.e., the interaction is regarded as either a complete success or a complete failure). As an extension to this model in order to handle cases where the interaction outcomes are rated on a multinomial scale, Jøsang and Haller extended their previous work by proposing the Dirichlet Reputation System (DRS) in [Jøsang and Haller, 2007]. The basic intuitions of this model are similar to that used in BRS except when modeling the outcomes of historical interactions. However, instead of rating an interaction outcome as a binary value, the outcome of an interaction can take a value of $i$ where $i = 1, \dots, k$ (e.g.. a rating of 1 to 5 where 1 represents most unsatisfactory and 5 represents the most satisfactory). With more finely grained ratings on the outcomes of past interactions, multiple ways of deriving the reputation of a trustee agent are available in DRS. It can be represented as 1) an evidence representation, 2) a density representation, 3) a multinomial probability representation, or 4) a point estimate representation. Nevertheless, the first two representations are more difficult for human interpretation than the third and fourth types of representations. In general, BRS is more widely adopted than DRS.

To gauge the performance of a trustee agent, various aspects of the quality of services provided by it should be analyzed. [Griffiths, 2005] proposed a multi-dimensional trust model that models the trustworthiness of a trustee agent along four dimensions: 1) the likelihood it can successfully produce an interaction result, 2) the likelihood of producing an interaction result within the expected budget, 3) the likelihood of completing the task within the deadline specified, and 4) the likelihood that the quality of the result meets expectation. A weighted average approach is used to compute the trustworthiness of an agent based on these dimensions where the weights are specified by individual truster agents according to their personal preferences. The advantage of the multi-dimensional trust model is that the reason for assigning an agent a trustworthiness value can be viewed in more details which may facilitate more complex interaction decision-making.

The work by Wang and Singh in [Wang and Singh, 2007] focused on another important aspect in evidence-based trust models – quantifying the uncertainty present in the trust evidence. Consider a scenario where one truster agent *A* has only interacted with a trustee agent *C* twice, and in both instances, the outcomes are successful; whereas truster agent *B* has interacted with *C* for 100 times and only 50 interactions are successful. Which set of evidence contains more uncertainty for evaluating *C*'s trustworthiness? [Wang and Singh, 2007] addressed this problem by proposing a method to calculate the uncertainty in a set of trust evidence based on the distribution of positive and negative feedbacks. Based on statistical inference, the method produces a certainty value in the range of [0, 1] where 0 represents the most uncertain and 1 represents the most certain. The method satisfies the intuition that 1) certainty is high if the amount of trust evidence is large; and 2) certainty is high if the conflicts among the feedbacks are low.

In practice, the trustworthiness of a trustee agent is often defined within certain context. This allows individual truster agents to simplify complex decision-making scenarios and focus on evidence which is the most relevant to the interaction decision that has to be made at the moment. Existing evidence-based trust models often handle context by storing past evidence according to the context they belong to. This makes the evaluated trustworthiness valid only within the stipulated context (e.g., a trustee agent's trustworthiness in repairing computers tells other agents nothing about its trustworthiness in selling T-shirts).

## 2.2. Reputation-based Trust Evaluation Models

While direct evidence is one of the most relevant source of information for a truster agent to evaluate a trustee agent, such information may not always be available. This is especially the case where a large number of agents exist in an MAS and interactions among them are sparse. Therefore, indirect evidence (third-party testimonies which are derived from direct interaction experience with a trustee agent from other agents which are called *witness agents*) may be needed to complement direct evidence with estimating a trustee agent's trustworthiness. Nevertheless, doing so exposes the truster agents to a new category of risk – the possibility of receiving biased testimonies which can negatively affect the trust-aware interaction decisions.

It is widely recognized within the research community that the importance of incorporating mechanisms to mitigate the adverse effects of biased testimonies. In this section, we discuss some recent research work on aggregating trust evidence from different sources and filtering out biased testimonies.

### 2.2.1. Trust Evidence Aggregation Approaches

Evidence-based trust models often make use of two distinct sources of information to evaluate the trustworthiness of a trustee agent: 1) *direct trust evidence*: a truster agent's personal interaction experience with a trustee agent, and 2) *indirect trust evidence*: third-party testimonies about the trustee agent. The majority of existing trust models adopts a weighted average approach when aggregating these two sources of trust evidence. Direct trust evidence is often assigned a weight of $\gamma$ ($0 \leq \gamma \leq 1$), and indirect evidence is assigned a corresponding weight of $(1-\gamma)$. Existing approaches for aggregating direct and indirect trust evidence can be divided into two broad categories: 1) *static approaches*, where the value of $\gamma$ is pre-defined; and 2) *dynamic approaches*, in which the value of $\gamma$ is continually adjusted by the truster agent.

In many papers, static $\gamma$ values for trust evidence aggregation. The majority of them tend to take a balanced approach by assigning a value of 0.5 to $\gamma$ [Weng *et al.*, 2006; Weng *et al.*, 2010; Liu *et al.*, 2011; Shen *et al.*, 2011; Yu *et al.*, 2011]. In some studies, the authors assign the value 0 [Jonker and Treur, 1999; Schillo *et al.*, 2000] or 1 [Shi *et al.*, 2005] to $\gamma$ to exclusively use only one source of trust information. Barber and Kim [Barber and Kim, 2003] have empirically shown, without considering the presence of biased testimonies, that direct trust evidence is the most useful to a truster agent over the long term while indirect trust evidence gives an accurate picture more quickly. Thus, approaches that discard one source or the other, forfeit some of the advantages provided by evidence based trust models. However, using a static value for $\gamma$ is also not always a good strategy.

Some researchers have explored adjusting the value of $\gamma$ dynamically based on different rationales. In [Mui and Mohtashemi, 2002], the value of $\gamma$ is varied according to the number of direct observations on the behavior of a trustee agent available to a truster agent. It is assumed that every truster agent starts with no prior interaction experience with a trustee agent and gradually accumulates direct trust evidence over time. Initially, the truster agent relies completely on indirect trust evidence (i.e. $\gamma = 0$) to select trustee agents for interaction. As the number of its interactions with a trustee agent $s_j$ increases, the value of $\gamma$ also increases according to the formula

$$\gamma = \begin{cases} \frac{N_j^b}{N_{min}}, \ if \ N_j^b < N_{min} \\ 1, \ otherwise \end{cases} \tag{2.1}$$

where $N_j^b$ is the total number of direct observations of $s_j$'s behavior by $c_b$, and $N_{min}$ is the minimum number of direct observations required in order to achieve a pre-determined acceptable level of error rate $\varepsilon$ and confidence level $\vartheta$. $N_{min}$ is calculated following the *Chernoff Bound Theorem*:

$$N_{min} = -\frac{1}{2\epsilon^2}\ln(\frac{1-\vartheta}{2}). \tag{2.2}$$

This approach is not concerned with filtering potentially biased third-party testimonies. Rather, its aim is to accumulate enough direct trust evidence so that a truster agent can make a statistically accurate estimation on the trustworthiness of a trustee agent without relying on indirect trust evidence. In order to achieve a high level of confidence and a low error rate, $N_{min}$ may be very high. In practice, this may mean a significant risk to the truster agent. Moreover, since the value of $\gamma$ increases to 1, this approach implicitly assumes that agent behaviors do not change with time. This may not always be true and limits the applicability of the approach under more dynamic scenarios.

In [Fullam and Barber, 2007], an approach based on the Q-learning technique [Sutton and Barto, 1998] to select an appropriate $\gamma$ value from a predetermined static set of values $\Gamma$ has been proposed. In order to select appropriate values for the set $\Gamma$, expert opinions about the underlying system characteristics are assumed to be available. Based on the reward accumulated by a truster agent under different $\gamma$ values, Q-learning selects the $\gamma$ value associated with the highest accumulated reward at each time step. This work provided the first step towards using interaction outcomes to enable the truster agent to weight the two sources of trust evidence. However, as this method uses a predetermined set of $\gamma$ values, its performance is affected by the quality of the expert opinions used to form the set of permissible $\gamma$ values.

### 2.2.2. Testimony Filtering Approaches

Over the years, many models for filtering potentially biased third-party testimonies have been proposed. However, these models are usually based on assumptions of the presence of some infrastructure support or special characteristics in the environment. In this section, some representative models in this sub-field are discussed.

The ReGreT model [Sabater and Sierra, 2002] makes use of the social relationships among the members of a community to determine the credibility of witnesses. Pre-determined fuzzy rules are used to estimate the credibility of each witness which, in turn, is used as the weight of its testimony for a trustee agent when aggregating all the testimonies. This model relies on the availability of social network information among the agents which may not be present in many systems.

In [Whitby *et al.*, 2004], unfair testimonies are assumed to exhibit certain characteristics. The proposed approach is closely coupled with the Beta Reputation System [Jøsang and Ismail, 2002] which records testimonies in the form of counts of successful and unsuccessful interactions with a trustee agent. The received testimonies are aggregated with equal weights to form a majority opinion and then, each testimony is tested to see if it is outside the $q$ quartile and (1 - $q$) quartile of the majority opinion. If so, the testimony is discarded and the majority opinion updated. This model assumes that the majority opinion is always correct. Thus, it is not effective in highly hostile environments where the majority of witnesses are malicious.

In [Weng *et al.*, 2006], it is assumed that the direct experience of the truster agent is the most reliable source of belief about the trustworthiness of a particular trustee agent, and it is used as the basis for filtering testimonies before aggregating them to form a reputation evaluation. An entropy-based approach is proposed to measure how much a testimony deviates from the current belief of the truster agent before deciding whether to incorporate it into the current belief. However, by depending on having sufficient direct interaction experience with a trustee agent, this assumption conflicts with the purpose for relying on third-party testimonies, which is to help truster agents make better interaction decisions when they lack direct trust evidence.

The model in [Liu *et al.*, 2011] supports interaction outcomes recorded in multi-dimensional forms. It applies two rounds of clustering of the received testimonies to identify testimonies which are extremely positive or extremely negative about a trustee. If neither the extremely positive opinion cluster nor the extremely negative opinion cluster forms a clear majority, they are both discarded as unfair testimonies and the remaining testimonies are used to estimate the reputation of a trustee agent. Otherwise, the majority cluster is considered as the reliable testimonies. Due to its iterative nature, the computational complexity of this method is high, with a time complexity of $O(mn^2)$ where $m$ is the number of candidate trustee agents whose reputations need to be evaluated and $n$ is the number of testimonies received for each candidate trustee agent. The method is also not robust in hostile environments where the majority of the witnesses are malicious.

## 2.3. Socio-cognitive Trust Evaluation Models

Another school of thought in multi-agent trust research emphasizes on the analysis of the intrinsic properties of the trustee agents and the external factors affecting the agents to infer their likely behavior in future interactions. This category of trust models are mainly designed to complement evidence-based trust models in situations where there is not enough evidence to draw upon when making trusting decisions.

In [Castelfranchi *et al.,* 2003], a trust decision model based on the concept of fuzzy cognitive maps (FCMs) is proposed. It constructs a generic list of internal external factors into FCMs to allow truster agents to infer if a trustee agent is worthy of interacting with. Each truster agent can determine the values to be given to the causal links between different factors so as to express their own preferences. Nevertheless, belief source variations and the variations in choosing values for the causal links can heavily affect the performance of the model and it is difficult to verify the validity of the models produced since there is a large degree of subjective preference involved.

The model proposed in [Ashri *et al.,* 2005] narrows down the scope of analysis to focus on the relationship between agents. The relationships used in their model are not social relationships but market relationships built up through interactions. The model identifies the relationships between agents (e.g., trade, dependency, competition, collaboration, tripartite, etc.) by analyzing their interactions through the perspective of an agent-based market model; these relationships are then filtered to identify which are most relevant to the analysis of agent trustworthiness; then, the relationships are interpreted to derive the trustworthiness evaluations for the agents.

The SUNNY model [Kuter and Golbeck, 2007] is the first trust inference model that computes a confidence measure based on social network information. The model maps a trust network into a Bayesian Network which is useful for probabilistic reasoning. The generated Bayesian Network is then used to produce estimates of the lower and upper bounds of confidence values for trust evaluation. The confidence values are used as heuristics to calculate the most accurate estimations of the trustworthiness of the trustee agents in the Bayesian Network.

In [Burnett *et al.,* 2010], the bootstrapping problem facing evidence-based trust model is investigated. In bootstrapping, it is assumed that neither prior interaction experience nor social relationship information is available about trustee agents who are new comers into an MAS. In this work, the underlying intuition used to design the model is that the intrinsic properties of a trustee agent can reflect its trustworthiness to some degree. The model learns a set of stereotypes based on the features in trustee agents' profiles using a decision tree based technique. New comer trustee agents are then classified into different stereotypes and stereotypical reputation values are produced for them. Nevertheless, due to the lack of suitable data, this paper did not point out which features may be useful in estimating a trustee agent's trustworthiness.

[Noorian *et al.*, 2011] enriches trust evaluation models by incorporating human dispositions such as optimism, pessimism and realism into the process of selecting whose opinions to believe in. The model proposed in this work consists of a two-layered cognitive filtering algorithm. The first layer filters out the agents whose lacks required experience or reliability using the BRS the uncertainty measure proposed in [Wang and Singh, 2007]. The second layer calculates a similarity measure for opinions received from advisor agents and the current belief by the truster agent. Combining it with the truster agent's innate disposition, the model produces credulity measures for the advisor agents and enables the truster agent to know whose opinions it should trust more.

## 2.4. Organizational Trust Evaluation Models

Another approach to maintaining trust in an MAS is to introduce organizational structure into multi-agent trust management. Such a goal can be accomplished only if there exists at least one trusted third-party in an MAS who can act as a supervising body for the transactions among other agents.

[Kollingbaum and Norman, 2002] is one of the earliest research works in this area. The proposed framework consists of three components: 1) a specific transactional organization structure made of three roles (i.e., the addressee, the counter-party and the authority), 2) a contract specification language for contract management, and 3) a set of contract templates created using the contract specification language. In order to conduct transactions, an agent needs to register with the authority, negotiate with other agents to set up the terms in the contracts, and carry out the work required by the contracts under the supervision of the authority.

The Certified Reputation (CR) model is proposed in [Huynh *et al.*, 2005]. It provides a mechanism for a trustee agent to provide truster agents with certified ratings about its past performance. It is possible to make sharing certified ratings a standard part of setting up a transaction between agents. By putting the burden of demonstrating past performance on the trustee agents, truster agents can enjoy savings on effort required to solicit third-party testimonies and filtering these testimonies. In addition, the certified ratings are provided by the trustee agent's previous interaction partners, thus making the CR model a distributed approach which is suitable for use in MASs.

In [Hermoso *et al.*, 2010], an agent coordination mechanism based on the interplay of trust and organizational roles for agents is proposed. It provides a mechanism for agents to establish which task a trustee agent is good at through multiple interactions and gradually allow the role each agent can play in an agent society to evolve and thus, dynamically changing the organizational structure by evolving an organizational taxonomy in the MAS. In subsequent interactions, the updated roles for the trustee agents act as a reference for truster agents to decide how to delegate tasks.

## 2.5. Trust-aware Interaction Decision-making Approaches

Existing trust-aware interaction decision making approaches can be broadly divided into two categories: 1) *greedy* and 2) *dynamic*. Such a classification is based on the strategy adopted by different approaches in terms of selecting trustee agents for interaction. Static approaches tend to use simple rules while dynamic approaches often attempt to assess the changing conditions in its operating environment in an effort to balance the exploitation of known trustworthy trustee agents with the exploration for potentially better alternatives.

### 2.5.1. Greedy Approach

In a typical greedy approach, a truster agent explores for trustee agents with a desired reputation standing through either some supporting infrastructure (e.g., peer recommendation, social network analysis, etc.) or random exploration. The reputation values of the candidate trustee agents are evaluated using a trust evaluation model of choice, and the one with the highest estimated reputation is selected for interaction. Such an approach is currently the most widely adopted in computational trust literature [Jøsang and Ismail, 2002; Yu and Singh, 2003; Teacy *et al.,* 2005; Weng *et al.,* 2006; Teacy *et al.,* 2008; Weng *et al.,* 2010; Liu *et al.,* 2011; Shen *et al.,* 2011]. From an individual truster agent's point of view, in order to maximize its own long term wellbeing, it is rational to always select the best possible option that can be found as often as possible.

### 2.5.2. Dynamic Approaches

Compared to static approaches, there are significantly fewer dynamic approaches in the current computational trust literature. A reinforcement learning based approach is proposed in [Teacy *et al.,* 2008]. The gain derived by a truster agent from choosing each trustee agent for interaction consists of the Q-value from Q-learning as well as the expected value of perfect information. At each time step, a truster agent chooses an action (i.e., exploration v.s. exploitation) which can maximize its gain.

In [Muñoz *et al.,* 2009], the authors measure a truster agent's knowledge degrees about trustee agents and use this metric to determine which trustee agent to select for interaction. The knowledge degree depends on the amount of direct past interaction experience with the trustee agent, third-party testimonies about that trustee agent, and the self reported trustworthiness by that trustee agent available to the truster agent. The value of the knowledge degree is normalized within the range of [0, 1], with 1 representing "completely known" and 0 representing "no direct interaction experience". In the local record of a truster agent, candidate trustee agents are organized into four different groups according to their knowledge degree values. If there are enough trustee agents with reputation values higher than a predefined threshold in the most well known group, the truster agent will only select from these trustee agents for interaction; otherwise, a number of exploration rounds will be allocated to trustee agents in groups to build up the knowledge degree about them and promote them into higher order groups.

Another dynamic approach proposed in [Hoogendoorn *et al.*, 2010] measures how much the behavior of the trustee agents has changed to determine the amount of effort a truster agent should devote to exploration. In this approach, each truster agent keeps track of the *long term trust* ($LT_i(t)$) and the *short term trust* ($ST_i(t)$) of candidate trustee agent $i$, where $ST_i(t)$ reflects the changes in $i$'s behavior faster than $LT_i(t)$. The average absolute difference between $LT_i(t)$ and $ST_i(t)$ is used to estimate the collective degree of change $C(t)$ in trustee agents' behavior. When $C(t)$ is larger than 0, an exploration extent value $E(t)$ is calculated. Together with the reputation value of each trustee agent, this value is used to derive a selection probability $RP_i(t)$ for every trustee agent. The candidate trustee agents are then selected using a Monte Carlo method based on their $RP_i(t)$ values. When $C(t) = 0$, the trustee agent with the highest reputation evaluation is always selected for interaction.

Both the greedy and the dynamic approaches eventually settle in the same strategy which is to "always select the known trustee with the highest reputation for interaction". While this strategy is rational from the perspective of an individual truster, it disregards the potential influence on the perceived performance of the trustee agents by the choices of interaction partners made by the truster agents.

Apart from decisions on the balance of exploration and exploitation, a truster agent can also decide on when to use additional mechanisms to induce the desired behavior from trustee agents following the framework proposed in [Burnett *et al.,* 2011]. In this framework, strategies a truster agent can adopt include 1) explicit incentives, 2) monitoring, and 3) reputational incentives. Based on the consideration of a wide range of factors including reputation, cost of monitoring, expected loss, and expected value of monitoring an activity, etc., a truster agent dynamically makes a choice among these strategies in addition to the decision on which trustee agent to select for an interaction.

While most trustworthiness evaluation models and trust-aware interaction decision-making approaches are designed for truster agents to use, [Fullam and Barber, 2006] proposed an interesting model that includes mechanisms to help trustee agents determine how trustworthy to be. Based on the Agent Reputation and Trust (ART) Testbed, the interdependencies, rewards and complexities of trust decisions are identified. A Q-learning based method is then used to help truster agents determine who to trust, how truthful to be in sharing reputation information and what reputations to believe in; and help trustee agents to determine how trustworthy to be.

## 2.6. Common Assumptions

In order to advance the research in multi-agent trust, many assumptions have been made. These assumptions can be classified into two categories:

1) *Fundamental assumptions*: the ones that are essential for multi-agent trust research to be carried out and are commonly accepted by researchers in this field. They include:

   a. Trustee and truster agents are self-interested;

b. An identity is associated with each trustee agent although it can be discarded by the agent;

c. Every truster agent always prefers to interact with the most trustworthy trustee agent.

2) *Simplifying assumptions*: the ones that are made to enable certain trust models to operate and not necessarily adopted by many researchers in this field. They include:

a. The outcome of an interaction between a truster agent and a trustee agent is binary (success or failure);

b. The effect of an interaction between a truster agent and a trustee agent on the wellbeing of the truster agent can be known immediately after the interaction is complete;

c. Interactions between a truster agent and a trustee agent occur in discrete time steps;

d. The majority of third-parties testimonies are reliable;

e. A truster agent's own direct interaction experience with a trustee agent is the most relevant to itself;

f. The properties of a trustee agent are useful for predicting its future behavior;

g. A truster agent needs to select only one trustee agent for each interaction;

h. A trustee agent can service an unlimited number of requests from truster agents during a time step without affecting its quality of service.

While the fundamental assumptions have stayed the same over the past decade, some of the simplifying assumptions have been relaxed. For example, Assumption 2.a was relaxed by [Jøsang and Haller, 2007], Assumption 2.c was relaxed by [Liu *et al.*, 2012] and Assumption 2.d was relaxed by [Teacy *et al.*, 2005]. One the other hand, new assumptions are sometimes added into the list. For example, Assumption 2.f was proposed by [Burnett *et al.*, 2010].

## 2.7. Evaluating the Performance of Trust Models

To evaluate the effectiveness of a proposed model in multi-agent trust research, two major evaluation methods are available: 1) simulation-based evaluation, and 2) evaluation through test datasets. Each of these methods has its own merits and has been observed to be applied either individually or in combinations by researchers in the field.

Currently, the most widely used method for evaluating a trust model is through simulations. In an effort to standardize the evaluation of trust models through simulations, the Agent Reputation and Trust (ART) testbed [Fullam *et al.*, 2005] was proposed and a series of competitions in this testbed were held in the International Conference on Autonomous Agents and Multi-Agent Systems (AAMAS). The ART testbed creates an environment where agents need to delegate task to each other to produce appraisals for virtual artworks and earn virtual profits during this process. Nevertheless, the testbed is designed mainly for evaluating models that aim to mitigate the adverse effect of biased third-party testimonies which is only one of the problems multi-agent trust research aims to address. Three ART testbed competitions were held in AAMAS from 2006 to 2008.

Currently, researchers are still creating their own simulation environments in order to produce conditions under which their trust models are designed to operate.

Another way of evaluating the performance of a trust model is by feeding data collected from real world applications into them. Depending on the specific features in a trust model, data from different types of online sources may be selected. For example, the Epinions dataset and the Extended Epinions dataset compiled by [Massa and Avesani, 2006] have been used to analyze the performance of trust models concerned with bootstrapping and collaborative recommendation [Massa and Avesani, 2007; Massa and Avesani, 2009; Li and Wang, 2010; Ray and Mahanti, 2010]; in [Kuter and Golbeck, 2007], data from the FilmTrust social network are used to analyze the performance of the proposed SUNNY socio-cognitive trust model; rating information from eBay was used in [Rettinger *et al.*, 2008]; the web spam dataset from Yahoo! was used in [Zhang *et al.*, 2011] to evaluate their model for propagating trust and distrust in the web; and data crawled from the Internet auction site Allegro were used as part of the evaluation in [Liu and Datta, 2012].

Real world data enables researchers to have a better idea of how their models would work in realistic environment conditions. However, the behavior patterns of the users in such datasets are fixed which makes it difficult for researchers to vary experimental conditions to simulate different ways the model can be attacked. In addition, many datasets are not specifically collected for the purpose of evaluating trust models. Thus, they may lack the ground truth about the user behavior and intention to facilitate more in-depth analysis of the performance of proposed trust models. In order to comprehensively evaluate a trust model, we believe a combination of these two methods should be employed. Nevertheless, it remains a difficult task collecting data from real world sources or convincing the industry to release datasets related to trust research with the concerns of privacy and trade secret protection. As shown in Figure 1, by analyzing research papers on the topic of trust management published in the *Association for the Advancement of Artificial Intelligence*

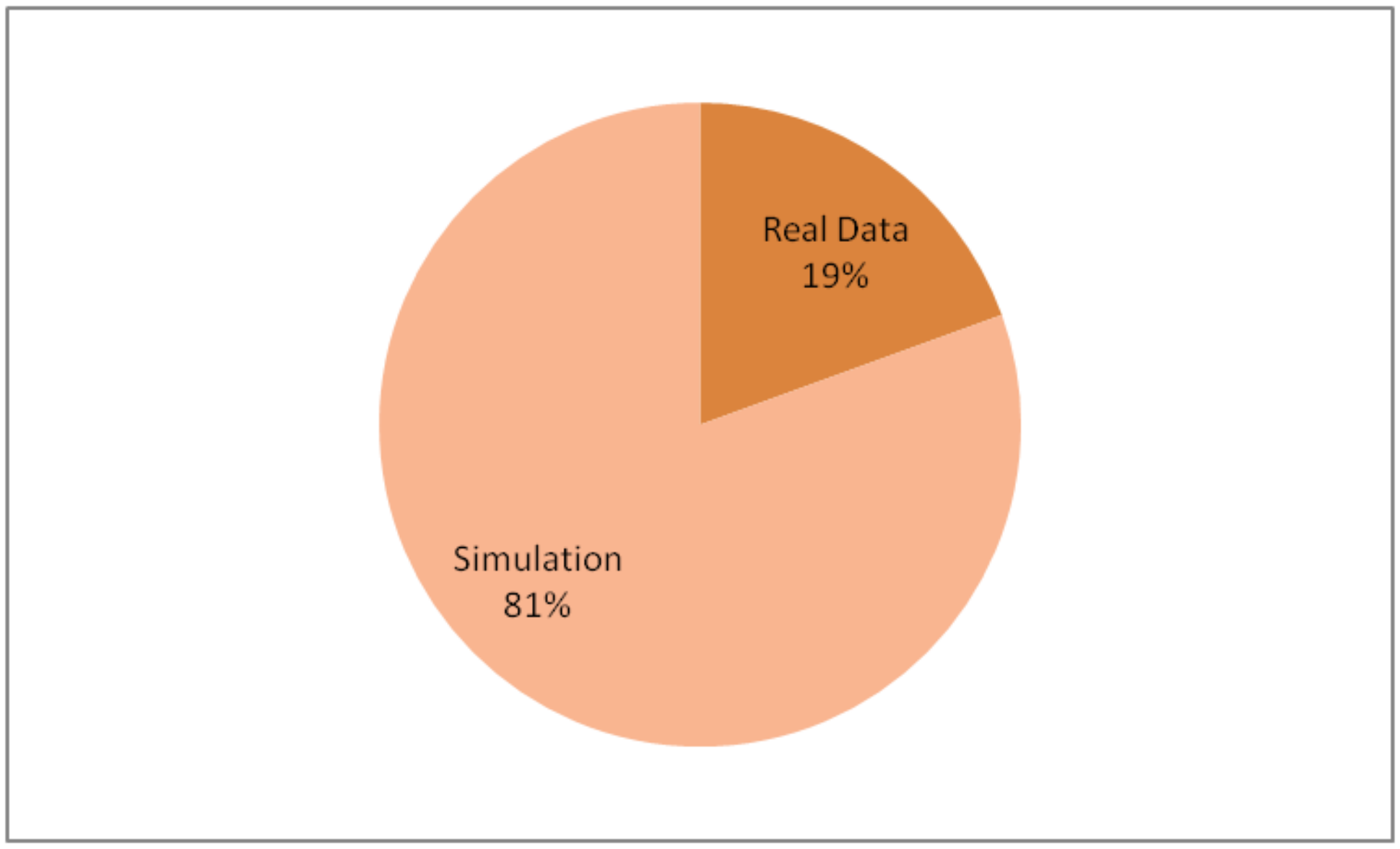

Figure 1. The percentage of research papers on trust management published in AAAI, AAMAS and IJCAI from 2002 to 2012 adopting the two methods of performance evaluation.

*(AAAI) Conference* (15 papers), the *International Conference on Autonomous Agents and Multi-Agent Systems (AAMAS)* (45 papers) and *the International Joint Conference on Artificial Intelligence (IJCAI)* (11 papers) from 2002 to 2012, it appears that over 80% of them use simulations to assess the performance of their proposed trust models while the rest using real world data as part of the evaluation.

## 2.8. Summary

In this chapter, we presented an overview of multi-agent trust research. We divided existing work into five categories according to the problems they are trying to address and reviewed selected notable works. Then we summarized a list of assumptions used by existing work. After that, we discussed the relative merits of evaluating trust models using simulation and real world data. The focus of this thesis is on the challenging problem of filtering biased testimonies in reputation-based trust evaluation modeling as well as helping agents make socially sustainable trusting decisions so as to enhance the social welfare in an MAS without sacrificing the wellbeing of individual agents. In the next chapter, we will present our work in mitigating the adverse effect of biased third-party testimonies in reputation-based trust evaluation models.

# Chapter 3

# ACT: Protecting Truster Agents against Collusion

## 3.1. Background

In open and highly dynamic distributed computing systems where users are from diverse backgrounds and may have conflicting goals, distributed social control is needed to sustain long term interactions among them. Nowadays, such systems are quite common (e.g., service oriented computing systems [Jøsang *et al.*, 2007], e-commerce systems [Noorian and Ulieru, 2010], wireless communication networks [Yu *et al.*, 2010], etc.). In such environments in which services and devices usually have limited capabilities, users often have to interact with each other in order to achieve their goals. These interactions usually involve an exchange of services, information, or goods with value. Selfish users may renege on their commitments, thereby breaching the trust placed in them by others. Therefore, trust and reputation management mechanisms are often used to minimize the negative impact of selfish users.

Generally, users in an open distributed computing system may play two types of roles [Jøsang *et al.*, 2007]:

1) *service providers (SPs),* who provide services, goods or information requested by others and do not need to rely on others to perform these services; and
2) *service consumers (SCs),* who need to rely on service providers to accomplish certain tasks.

The main objective of evidence-based trust models is to estimate the *trustworthiness* of a potential interaction partner which represents its true behavior pattern. Evidences about a service provider from the perspective of a service consumer are usually from two sources:

1) *direct trust evidence*: which are a service consumer's direct interaction experience with the service provider; and
2) *indirect trust evidence*: which are third-party testimonies about the service provider from other service providers in the system.

In practical systems, it is not possible to definitively know the trustworthiness of a service provider. Therefore, it is often estimated using trust evidences. The estimation of a service provider's trustworthiness derived from the direct trust evidence of a service consumer alone is called *direct trust,* while that derived from the indirect trust evidence is called *indirect trust*. An estimation derived from both sources of trust evidence is commonly known as the *reputation* of a service provider. In the eyes of a service consumer, other service consumers who provide it with indirect trust evidence (i.e. testimonies) about a service provider are regarded as *witnesses*. A witness's reliability in terms of providing useful testimonies is referred to as its *credibility*.

Since such systems tend to be very large in practice, service consumers often have to interact with service providers with whom they may not be very familiar (i.e. have little or no prior interaction experience with) [Fullam and Barber, 2007]. Thus, it is both necessary and advantageous to allow service consumers to act as witnesses to provide their own first-hand interaction experience as testimonies to other service consumers who lack such information.

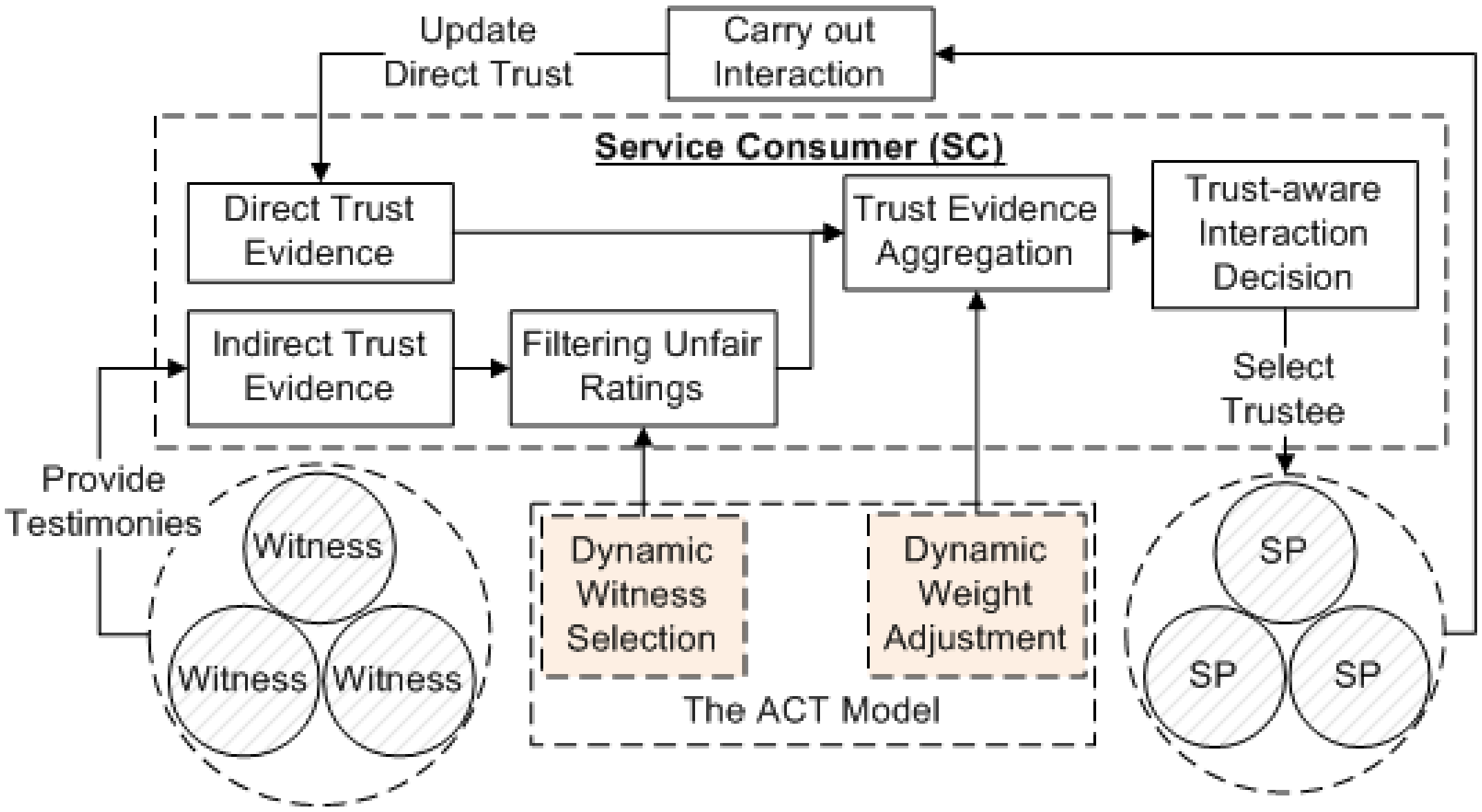

Figure 2. The general flow of trust-aware interaction decision making for evidence-based trust and reputation management models, and the contributions by the proposed ACT model.

However, such an approach is not without its perils. A witness might fabricate or hide information to promote the reputations of service providers who are related to it in some way, due to self-interest; the behavior patterns of service providers might change over time, thus rendering some service consumers' existing record of past interaction experience obsolete; service consumers may use different criteria to define the success and failure of an interaction, thereby making it difficult to use testimonies in a uniform way. Biased testimonies resulting from one or more of these factors can degrade the accuracy of trust decisions [Jøsang *et al.*, 2007]. Therefore, testimonies from witnesses need to be filtered before being used to evaluate a service consumer's reputation.

To this end, a number of evidence-based trust and reputation management (TRM) models have been proposed over the years. The general flow for a service consumer to decide which service provider to select for interaction is illustrated in Figure 2. Each service consumer continuously records its direct interaction experience with service providers over time. When a service provider's trustworthiness needs to be evaluated, the service consumer may request third-party testimonies from witnesses, depending on the service consumer's confidence on its own direct trust evidence. These testimonies are preprocessed in an attempt to filter out unfair ratings. The resulting direct and indirect trust evidences are then aggregated to form a trustworthiness evaluation for that particular service provider. At the end of this process, the service consumer decides which service provider to interact with based on their trustworthiness evaluations. Recently, there are a number of adaptive trust evaluation approaches being proposed [Moe *et al.*, 2008; Teacy *et al.*, 2008; Liu and Datta, 2012]. However, these approaches are all focused on predicting trustee performance based on direct trust evidence alone which is not the focus of this study.

Existing approaches for third-party testimony filtering and aggregation fall into three main categories:

1) filtering witness testimonies before aggregating them without recording the witnesses' credibility in terms of providing useful testimonies [Ba and Pavlou, 2002; Whitby *et al.*, 2004; Weng *et al.*, 2006; Liu *et al.*, 2011];
2) aggregating testimonies according to witnesses' credibility evaluations [Jurca and Faltings, 2003; Yu and Singh, 2003; Weng *et al.*, 2010]; and

3) incorporating incentive mechanisms into existing trust models to induce witnesses to provide fair testimonies [Miller *et al.*, 2005; Zhang *et al.*, 2008; Marsh and Briggs, 2009].

However, these approaches tend to be either based on majority voting which negatively affects their perform in scenarios where the majority of the witnesses are compromised, or tightly coupled with certain infrastructure support (e.g., payment systems, knowledge of social relationships among agents, etc.) which may be lacking in many application domains.

In this research, we address these limitations by proposing the Actor-Critic learning Trust (ACT) model based on the principles of the Actor-Critic Learning Method [Tesauro, 1995]. The ACT model enables a service consumer to dynamically make two important decisions when presented with third-party testimonies for a service provider: 1) how much weight to give to its own personal direct trust evidence and the collective opinions from witnesses, and 2) how much weight to assign to the testimonies of each witness. The reward and penalty strategy design in the ACT model enables the service consumer to base its learning process on the actual outcomes of its past interactions with service providers. As a result, its performance is less affected by the fraction of malicious witnesses in the agent population. It also does not require additional infrastructure support for its operations. The contribution of the proposed ACT model is mainly focused on the third-party testimony filtering and aggregation procedures.

## 3.2. System Model

Before discussing details of the proposed model, we introduce the notation that will be used in this chapter. At each time step $t$, a service consumer $c_i$ will interact with at most one service provider $s_j$ in our target system. For each interaction, $c_i$ chooses a service provider from among several candidates based on their estimated trustworthiness values. Whenever $c_i$ needs to assess the trustworthiness of $s_j$, it draws upon both its own direct trust evidence about $s_j$ (if there is any) as well as testimonies from a list of witnesses $W_{i,j}$ which are known by $c_i$ to have previously interacted with $s_j$. A witness $w_k$ may reply to $c_i$'s request at time step $t$ with a testimony $d_j^k(t)$. A malicious $w_k$ may distort its testimonies before sharing them with others. The service provider chosen for interaction by $c_i$ at time step $t$ is affected by the selection of witnesses as well as the weights given to the direct and indirect trust evidence by $c_i$.

For each interaction with $s_j$, $c_i$ incurs an utility cost $C$. If $s_j$ successfully completes the task assigned to it by $c_i$, $c_i$ receives an utility gain of $G$. We assume that the outcome of the interaction $O_j^i(t)$ can be observed by $c_i$ within the same time step in which the interaction occurs. We further assume that the interaction outcome is either successful $(O_j^i(t) = 1)$ or unsuccessful $(O_j^i(t) = 0)$. By comparing the recommendation $d_j^k(t)$ by each $w_k \in W_{i,j}$ about $s_j$ at time $t$ with $O_j^i(t)$, $c_i$ can learn the ranking of each $w_k$ in $W_{i,j}$. New witnesses for $s_j$ discovered by $c_i$ over time are added into $W_{i,j}$. The interaction outcome value, $O_j^i(t)$, is further compared with the recommended interaction decision value, $D_d^{i,j}(t)$, based on direct trust evidence and the value, $D_{ind}^{i,j}(t)$, based on indirect trust evidence from the testimonies of selected witnesses. Reward and penalty values are assigned to these two sources of trust evidence by $c_i$ in its local record to determine how much to rely on either source in the future.

The individually rational objective of a service consumer is to maximize its utility over its lifetime in the presence of malicious service providers and malicious witnesses. For convenience, the main symbols used in this chapter are listed in Table 2.

Table 2. Key Notations used in this Chapter

| | |
|---|---|
| $c_i$ | A service consumer. |
| $s_j$ | A service provider. |
| $w_k$ | A witness. |
| $W_{i,j}$ | A set of witnesses for $s_j$ known to $c_i$. |
| $O_j^i(t)$ | The binary outcome of an interaction between $c_i$ and $s_j$ at time *t*. |
| $test_j^k(t)$ | A testimony from $w_k$ with regard to $s_j$ at time *t*. |
| $d_j^k(t)$ | The interaction decision as suggested by $test_j^k(t)$. |
| $D_d^{i,j}(t)$ | The decision by $c_i$ on whether to interact $s_j$ with at time *t* based on direct trust evidence only. |
| $D_{ind}^{i,j}(t)$ | The decision by $c_i$ on whether to interact $s_j$ with at time *t* based on indirect trust evidence only. |
| $D_j^i(t)$ | The overall decision by $c_i$ on whether to interact $s_j$ with at time *t* based on both direct and indirect trust evidence. |
| $C$ | The cost in utility incurred by $c_i$ when engaging the service of $s_j$. |
| $G$ | The gain in utility after a successful interaction $c_i$ and $s_j$. |
| $\gamma_{i,j}(t)$ | The weight assigned to direct trust evidence about $s_j$ by $c_i$. |
| $R$ | The reward assigned to a source of trust evidence. |
| $P$ | The penalty assigned to a source of trust evidence. |
| $\tau_{i,j}^d(t)$ | The direct trust for $s_j$ by $c_i$. |
| $\tau_{i,j}^{ind}(t)$ | The indirect trust for $s_j$ by $c_i$. |
| $rp_j(t)$ | The reputation of $s_j$. |
| $\pi_{k,j}(t)$ | The credibility ranking of $w_k$ in $c_i$'s local record. |

### 3.3. The Basics of Actor-Critic Learning

Before proposing the ACT model, we briefly introduce the basic concept of the actor-critic method of reinforcement learning. The actor-critic methods are temporal difference (TD) methods that represent the decision-making policy in a separate memory structure independent of the value function. The decision-making policy is known as the *actor* and the value function is known as the *critic*. The critic criticizes the actions made by the actor. Learning is always on-policy, which means that the critic must learn about and critique whatever policy is currently being followed by the actor. The critique is in the form of a temporal difference error. The output of the critic is this TD error and it drives all learning in both the actor and the critic, as shown by Figure 3. The actor-critic method

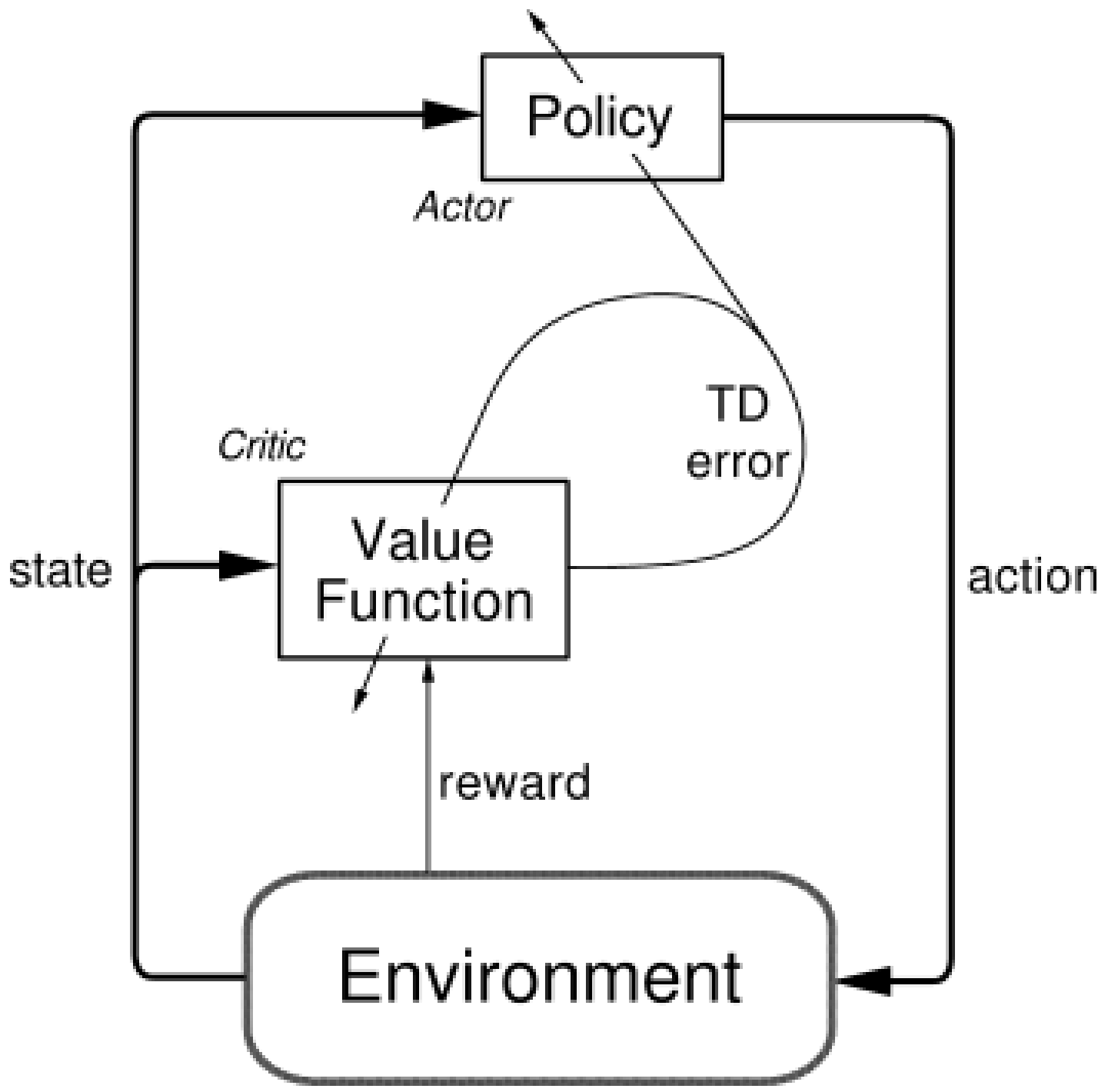

Figure 3. The general framework of the ACT model [Tesauro, 1995].

requires minimal computation when selecting an action.

## 3.4. The ACT Model

We now propose the ACT model to assist service consumers make trust-aware interaction decisions in the presence of potentially unreliable third-party testimonines. The general framework of the proposed ACT model is presented in Figure 4. Each service consumer $c_i$ keeps two local lists: 1) a *list of known witnesses*, and 2) a *list of known service providers*. Since a witness may only have interacted with a few service providers, the list of known witnesses organizes the witnesses into sub-lists indexed according to known service providers. The list of known service providers stores the direct trust evidence $c_i$ has for each known service provider and the weight assigned to the direct trust evidence $\gamma_{i,j}(t)$ in the case of that service provider. These two lists grow as $c_i$ acquires more interaction experience with these two types of system participants.

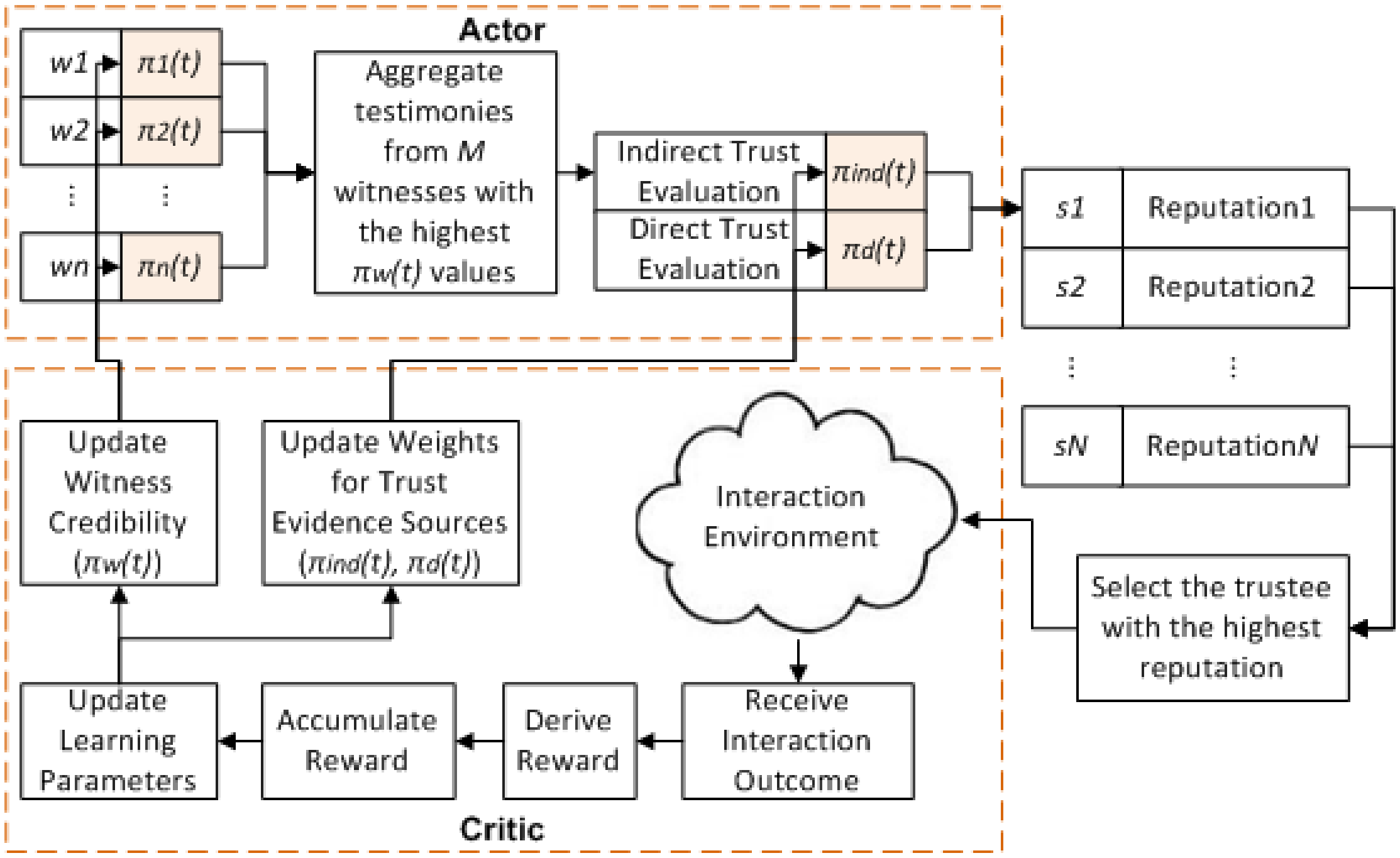

Figure 4. The general framework of the ACT model.

The ACT model is designed based on a variant of the reinforcement learning (RL) approach – the actor-critic method [Konda and Tsitsiklis, 2002]. The actor module represents the policy used to choose which witnesses' testimonies should be selected and how much weight each of them should have when aggregating them together to form the indirect trust evidence. The policy also determines how much weight should be given to the direct and indirect trust evidence in order to evaluate the service provider's trustworthiness.  The critic module represents the value function that determines whether the service provider $c_i$ is better off or worse off after each interaction with a selected service provider $s_j$. Overtime, the learning parameters of the ACT model are updated in such a way that more preference is given to witnesses and the source of trust evidence that enhance $c_i$'s wellbeing.

Although the ACT model can be used together with many possible trust evaluation models, to be specific, we assume that the popular Beta Reputation System (BRS) [Jøsang and Ismail, 2002] is used as the underlying trust evaluation method. The direct trust for $s_j$ by $c_i$ can be calcaulated using the BRS as

$$\tau_{i,j}^{d}(t) = E\left[\Pr\left(s_j\right)\right] = \frac{\alpha}{\alpha + \beta}$$

$$\alpha = N_p(t) + 1, \beta = N_n(t) + 1 \qquad (3.1)$$

It is equivalent to the expectation of the probability that $s_j$ will successfully serve a requests from $c_i$.  $N_p(t)$ and $N_n(t)$ are variables representing the total number of successful and unsuccessful interactions between $s_j$ and $c_i$ until time step *t* respectively.

### 3.4.1. Learning Witness Credibility Ranking

In the critic module, the reward function for $c_i$ is

$$r_{i,j} = (\mu_j^i(t) \cdot (G - C) - (1 - \mu_j^i(t)) \cdot C). \tag{3.2}$$

$r_{i,j}$ is computed at the end of each interaction between $c_i$ and $s_j$. It reflects the level of success achieved by $c_i$ following the current interaction decision policy. $T_{i,j}$ is the number of interactions between $c_i$ and $s_j$, and the parameter $\mu_j^i(t)$ is defined as

$$\mu_j^i(t) = \begin{cases} 0, & O_j^i(t) = 0 | D_j^i(t) = 1 \\ 1, & O_j^i(t) = 1 | D_j^i(t) = 1 \end{cases}. \tag{3.3}$$

$D_j^i(t)$ denotes the overall decision by $c_i$ to interact with $s_j$ at time *t* based on both the direct and indirect trust evidence currently available. Here, we only consider the case when the decision is to interact with a service provider (i.e. $D_j^i(t) = 1$), because in order for a service consumer $c_i$ to be able to observe the actual interaction outcome with a service provider $s_j$ at the end of time step *t*, $s_j$ must be selected by $c_i$ for interaction in that time step. When $D_j^i(t) = 0$, it implies that $c_i$ deems $s_j$ untrustworthy based on its reputation value. Thus, in these cases, no interaction between them will take place at that time and no $O_j^i(t)$ value can be observed. In this study, we assume that the agents' direct trust values and indirect trust values are normalized to a range [0, 1]. A testimony $test_j^k(t)$ is simply $w_k$'s direct trust value for $s_j$ based on its own direct trust evidence upto time step *t*. Thus, its value is also within the range [0, 1].

Once the latest interaction outcome is known, a reward correction value $\theta_{k,j}$ is computed for each of the $M$ selected witnesses whose testimonies have been used to calculate the reputation of $s_j$ namely:

$$\theta_{k,j} = \frac{1}{T_{k,j}} \sum_{t=1}^{T_{k,j}} [d_j^k(t) \cdot (1 - O_j^i(t))]. \quad (3.4)$$

In (3.2), $T_{k,j}$ denotes the total number of times that $w_k$'s testimonies about $s_j$ has been used by $c_i$, and $d_j^k(t)$ represents the interaction recommendation implied by $w_k$'s testimony, $test_j^k(t)$, on $s_j$ at time step $t$ and is given by

$$d_j^k(t) = \begin{cases} 0, & if\ test_j^k(t) < Th \\ 1, & if\ test_j^k(t) \geq Th \end{cases}. \quad (3.5)$$

In (3.3), $Th \in [0,1]$ is a predefined threshold value. $\theta_{k,j}$ increases with the number of times that $w_k$ has given testimonies suggesting a service provider is trustworthy but the actual interaction outcome is unsuccessful. It is used to penalize the act of unfairly praising a service provider, which is the most common form of collusion between service providers and witnesses.

After the interaction outcome with a service provider is known, the critic process is carried out by updating the learning parameter $p_{k,j}$ for each of the $M$ witnesses whose testimonies resulted in the selection of $s_j$ by $c_i$ as follows

$$p_{k,j} \leftarrow p_{k,j} + \rho \cdot \left(r_{i,j} - \widetilde{r_{i,j}} - \delta \cdot \theta_{k,j}\right) \cdot (1 - \pi_{k,j}(t-1)). \quad (3.6)$$

The constant $\rho$ ($0 < \rho \leq 1$) denotes the learning rate. As $\rho$ increases, the learning parameter $p_{k,j}$ change more rapidly as new interaction outcomes become available. In this study, we choose a $\rho$ value close to zero to make $p_{k,j}$ vary more smoothly. The constant $\delta$ ($0 < \delta \ll 1$) represents the bias towards penalizing collusion when updating the learning parameter; its value should be significantly smaller than 1 to avoid drastic changes in the value of $p_{k,j}$.

The credibility ranking value $\pi_{k,j}(t)$ of each known $w_k$ with regard to a service provider $s_j$ is calculated using the Gibbs softmax method [2] as

$$\pi_{k,j}(t) = \frac{e^{p_{k,j}}}{\sum_{l=1}^{M} e^{p_{l,j}}}. \tag{3.7}$$

The resulting values of $\pi_{k,j}(t)$ is used to rank the witnesses known to $c_i$ to facilitate subsequent witness selections. The sum of all $\pi_{k,j}$ values always equals to 1. Thus, the credibility ranking value can be thought of as the probability of soliciting testimonies from each of the known witnesses.

After the credibility ranking values are calculated, the total accumulated reward $\widetilde{r_{i,j}}$ is updated. It is used as a reference in the process of evaluating the well-being of $c_i$ resulted from interactions with $s_j$. It is updated as

$$\widetilde{r_{i,j}} \leftarrow \varphi \cdot \widetilde{r_{i,j}} + (1 - \varphi) \cdot r_{i,j} \tag{3.8}$$

where constant $\varphi$ $(0 < \varphi \leq 1)$ determines the influence of the latest rewards in the smoothed baseline reward $\widetilde{r_{i,j}}$. When $\varphi = 1$, only the current reward is used to evaluate the credibility of each witness.

The indirect trust for $s_j$ by $c_i$ can be computed as

$$\tau_{i,j}^{ind}(t) = \frac{\sum_{k=1}^{M}[\pi_{k,j}(t) \cdot test_j^k(t)]}{\sum_{k=1}^{M}\pi_{k,j}(t)}. \tag{3.9}$$

### 3.4.2. Learning Trust Evidence Source Preference

With the values of $\tau_{i,j}^{d}(t)$ and $\tau_{i,j}^{ind}(t)$ calculated using (3.1) and (3.9), the next step is to aggregate them to compute the reputation of $s_j$. In the ACT model, for each $s_j$ known to $c_i$, two critic modules are used to learn the weights for the two sources of trust evidence and one actor module is used for estimating the trustworthiness of $s_j$. The critic module in the proposed method determines the relative merit of each source of trust evidence through reward accumulation. The learning process is similar to that presented in the last section. Since the two critic modules are essentially the same but only use different sources of trust evidence as input data, in the following, we only discuss the critic module for direct trust evidence source.

The value function of the critic module is designed as:

$$r_d = \tilde{\mu}(t) \cdot R + (1 - \tilde{\mu}(t)) \cdot P$$

$$\tilde{\mu}(t) = \begin{cases} 0, & if O_j^i(t) = 0 | D_d^{i,j}(t) = 1 \ or \ O_j^i(t) = 1 | D_d^{i,j}(t) = 0 \\ 1, & if O_j^i(t) = 0 | D_d^{i,j}(t) = 0 \ or \ O_j^i(t) = 1 | D_d^{i,j}(t) = 1 \end{cases}$$

$$D_d^{i,j}(t) = \begin{cases} 0, \ \tau_{i,j}^{d}(t) < Th \\ 1, \ \tau_{i,j}^{d}(t) \geq Th \end{cases}. \tag{3.10}$$

$r_d$ can be considered as the time averaged per interaction reward achieved by $c_i$ through relying on its direct trust evidence source about $s_j$ with the current weight value $\gamma_{i,j}$. $R$ and $P$ are predetermined constant values for reward and penalty, based on the consequences of the interaction decision. The ratio of $R$ to $P$, rather than their absolute values, is important to the learning process. A small $R{:}P$ ratio means that trust is hard for a service provider to gain, but easy to lose. The variable $\tilde{\mu}(t)$ determines whether this trust evidence source should be rewarded or penalized at time step $t$. Its value toggles between 0 and 1 according to the relationship between the interaction decision $D_d^{i,j}(t)$, which is related to the direct trustworthiness evaluation $\tau_{i,j}^d(t)$ $(0 \leq \tau_{i,j}^d(t) \leq 1)$, and the actual interaction outcome $O_j^i(t)$. As $D_d^{i,j}(t)$ is only one component of the overall interaction, it is possible that even as $D_d^{i,j}(t)$ suggests not to interact with $s_j$, the overall decision is otherwise.

Once the latest $r_d$ is calculated, it is compared with the baseline reward $\widetilde{r_d}$ accumulated by this trust evidence source to update the learning parameter $p_d$ according to

$$p_d \leftarrow p_d + \rho \cdot (r_d - \widetilde{r_d}) \cdot (1 - \pi_d(t-1)). \tag{3.11}$$

After $p_d$ is updated, $\widetilde{r_d}$ is updated to incorporate the latest reward $r_d$:

$$\widetilde{r_d} \leftarrow \varphi \cdot \widetilde{r_d} + (1 - \varphi) \cdot r_d. \tag{3.12}$$

$\widetilde{r_d}$ can be treated as a basis for comparing whether $c_i$ is better off or worse off by aggregating the direct trust evidence into the estimation for the trustworthiness of $s_j$ using the latest $\gamma_{i,j}(t)$ value.

Similarly, the learning parameter $p_{ind}$ for the indirect source of trust evidence can be obtained. When both $p_d$ and $p_{ind}$ are obtained, the learning parameters $\pi_d(t)$ and $\pi_{ind}(t)$ are updated as:

$$\pi_d(t) = \frac{e^{p_d}}{e^{p_d} + e^{p_{ind}}}$$

$$\pi_{ind}(t) = \frac{e^{p_{ind}}}{e^{p_d + e^{p_{ind}}}}. \tag{3.13}$$

$\pi_d(t)$ and $\pi_{ind}(t)$ can be treated as the probability of selecting each source of trust evidence $\pi_d(t) + \pi_{ind}(t) = 1$. In the ACT model, $\gamma_{i,j}(t) = \pi_d(t)$.

While the strategy for exploiting known witnesses with high credibility is relatively straightforward (i.e. selecting the top *M* most credible witnesses to request testimonies from), balancing it with exploration for addition witnesses requires careful design. In the ACT model, the exploration process is controlled by two parameters: 1) an exploration probability *Pr*, and 2) the magnitude of *M*. The value of *Pr* is initialized to 1 at the start of a service consumer $c_i$'s life time to enable $c_i$ to explore when the list of known witnesses is empty. The value of *Pr* is gradually decreased over time until it reaches a pre-defined minimum value, $Pr_{min}$. Testimonies returned by previously unknown witnesses are given the benefit of the doubt and included in the calculation of the service provider's reputation with weight values equal to the lowest $\pi_{k,j}(t)$ among that of the selected known witnesses. This is to ensure that $c_i$ will always have some opportunity to discover new witnesses.

A service provider $s_j$'s reputation is calculated as

$$rp_j(t) = \gamma_{i,j} \cdot \tau_{i,j}^{d}(t) + (1 - \gamma_{i,j}(t)) \cdot \tau_{i,j}^{ind}(t). \tag{3.14}$$

$rp_j(t)$ represents the overall reputation of $s_j$ and is used by $c_i$ to estimate $s_j$'s trustworthiness. At each time step, $c_i$ might have more than one candidate service providers to choose from. In this study, we assume that $c_i$ always selects the service provider with the highest overall reputation for interaction.

## 3.5. Evaluation

To evaluate the performance of the ACT model, we have designed a test-bed which allows the well-being of service consumers adopting different approaches to be gauged. Through extensive simulations, it has been shown that the ACT model significantly outperforms existing approaches in terms of the reduction in normalized average utility loss and, in the case of colluding witnesses, the reduction in their collusion power.

### 3.5.1. Design of the Simulation Test-bed

The test-bed simulates a scenario where a number of service consumers need the services offered by service providers. A service consumer incurs a cost of $C$ in order to utilize the service of a service provider. If the service provider acts honestly, i.e. satisfies the service consumer's request, the service consumer gains an amount of utility of $G$ after the interaction; otherwise, it gains zero utility. Therefore, the maximum average utility gain a service consumer can achieve is $G - C$, corresponding to all its interactions with service providers being successful; the minimum of this value is $-C$, if all its interactions are unsuccessful.

The main purpose of this test-bed is to investigate the effectiveness of the proposed ACT model in mitigating the adverse effects of unfair testimonies relative to existing approaches. Since it is impractical to investigate the proposed model for a large number of possible service provider population configurations, we focus on a highly hostile service provider population consisting of

- 10% honest service providers (which renege randomly with a probability of 10%);
- 10% Type I dishonest service providers (which renege randomly with an initial probability of 40%);
- 40% Type II dishonest service providers (which renege randomly with an initial probability of 60%); and
- 40% Type III dishonest service providers (which renege randomly with an initial probability of 80%).

Except for the honest service provider group, the behavior patterns of all other groups changes gradually during the simulation. A service provider’s behavior can change according to three different profiles: 1) increasing reneging probability, 2) decreasing reneging probability, or 3) unchanging reneging probability. The magnitude of each change is randomly chosen from the interval [0, 0.01]. Each dishonest service provider chooses one of the three profiles in each interaction with equal probability (i.e. 1/3). If the proposed ACT model can achieve good performance in this hostile environment, it should also perform well under more benign environments. The test-bed environment consists of 100 service providers with different behavior patterns. During each round of simulation, each service consumer attempts to solve a total of 200 problems. The service consumers select service providers for interaction based on their reputation. The outcome of the interaction is assumed to be binary, namely *successful* or *unsuccessful*, depending on whether the service provider provides the requested service.

There are 100 witnesses who accumulate direct trust evidence about the service providers and respond to service consumers requesting testimonies. When a request for testimony is received by a witness it will return a testimony to the requester if it has prior interaction experience with the particular service provider in the request; otherwise, it will decline the request. Two categories of malicious testimony sharing strategies are studied: 1) *random lying*, and 2) *collusive lying*.

In the case of random lying, a malicious witness does not collude with any other service provider. It either positively distorts a testimony (*ballot-stuffing*) or negatively distorts a testimony (*badmouthing*) following a preset lying probability. In the case of collusive lying, a number of service providers collude with lying witnesses to inflate their reputation in the eyes of service consumers (*ballot-stuffing*). The colluding witnesses do not give unfair testimonies about service providers who are outside the collusion ring. In both random lying and collusive lying cases, the distortions are implemented as offset values added to or subtracted from the original testimony. Two types of unfair testimonies are supported in the test-bed:

1) *Moderately Unfair Testimonies (MUT)*: the magnitude of the offset is randomly chosen in the range [0.1, 0.4];
2) *Highly Unfair Testimonies (HUT)*: the magnitude of the offset is randomly chosen in the range [0.8, 1.0].

Table 3. Parameter Values used in the Experiments

| **Parameter** | **Value** |
|---|---|
| $Th$ | 0.5 |
| $\varphi$ | 0.6 |
| $\delta$ | 0.05 |
| $\rho$ | 0.4 |
| $M$ | 10 |
| $G$ | 5 |
| $C$ | 1 |
| $R$ | 1 |
| $P$ | -10 |
| $Pr_{min}$ | 0.1 |

The values of the distorted testimonies are always kept within the range [0, 1] by hard-limiting to 1 (or 0) if the distorted testimonies after adding (or subtracting) exceeds 1 (or falls below 0). In the proposed ACT model, we use BRS as the trust evaluation model in this study. The values selected for the parameters in the ACT model are listed in Table 3.

### 3.5.2. Evaluation Metrics

Two evaluation metrics from [Weng *et al.* 2010] are adopted to facilitate comparisons with state-of-the-art methods: 1) *Normalized Average Utility Loss (NAUL)*, and 2) *Collusion Power*. The normalized average utility gain $\sigma$ ($0 \leq \sigma \leq 1$) measures the average per time step utility gain for each service consumer over its lifetime. It is calculated as

$$\sigma = \frac{\frac{1}{T \cdot N}\sum_{t=1}^{T}\sum_{i=1}^{N} g_i(t) - g_{min}}{g_{max} - g_{min}}. \quad (3.15)$$

In (3.15), $T$ is the total number of times a service consumer $c_i$ has interacted with the service providers, $N$ is the number of service consumers adopting the same approach as $c_i$'s in the test-bed and $g_{max} = G - C, g_{min} = -C$. $g_{i,t}$ is the actual utility gain of each $c_i$ after each interaction at time step $t$. If the interaction is successful, $g_i(t) = g_{max}$;

otherwise, $g_i(t) = g_{min}$. The normalized average utility loss is then $(1 - \sigma)$. A lower value corresponds to a better performance.

The Collusion Power is a measure of the effectiveness of different strategies in the face of collusion. It is defined as [Weng *et al.*, 2010]

$$collusion\ power = \frac{\sum_{c_i \in A_{nc}} \#try(c_i)}{|A_{nc}| * N_m} \quad (3.16)$$

where $A_{nc}$ denotes the set of non-colluding service consumers, $|A_{nc}|$ is the total number of service consumers in set $A_{nc}$, $c_i$ is a service consumer in this set, and $\#try(c_i)$ is the number of times $c_i$ interacted with any colluding service provider (i.e. considered a colluding service provider as a potential interaction partner) during the simulation. In essence, the collusion power represents the percentage of all tasks delegated to any of the colluding service providers. The lower the collusion power, the more effective a strategy is against collusion.

### 3.5.3. Experiment Setup

For each experiment, the composition of the common witness population is altered to simulate different scenarios. In the following sections, *Hon* denotes a population consisting entirely of honest common witnesses. *BMn* denotes a population consisting of *n*% badmouthing witnesses and (100-*n*)% honest witnesses. *BSn* denotes a population consisting of *n*% ballot-stuffing witnesses and (100-*n*)% honest witnesses. The malicious witness populations consist of half giving out MUTs and half giving out HUTs.

The experiments conducted in this study include two parts: 1) verifying the effectiveness of the adaptive trust evidence aggregation module of the ACT model (labeled as *ACT'*), and 2) verifying the effectiveness of the ACT model as a whole (labeled as *ACT*). In Part 1 of the study, five groups of service consumers are used for comparison. They are:

1) Group $\gamma = 0$: service consumers who completely rely on indirect trust evidence;
2) Group $\gamma = 0.5$: service consumers who rely on a balanced mix of direct and indirect trust evidence;
3) Group $\gamma = 1$: service consumers who completely rely on direct trust evidence;
4) Group M2002: service consumers who use the method described in [Mui and Mohtashemi, 2002] to set the $\gamma$ value;
5) Group F&B2007: service consumers who use the method described in [Fullam and Barber, 2007] to set the $\gamma$ value.

The group of service consumers equipped with the proposed method is labeled as Group *ACT'*. Each group consists of 10 agents. All competing groups only request for testimonies from the common witness group.

In Part 2 of this study, we compare the performance of the ACT model against:

1) Group *W2010*: service consumers which employ an existing state-of-the-art method [Weng *et al.*, 2010];
2) Group *YS2003*: service consumers which employ a classic method [Yu and Singh, 2003];
3) Group *NoCred*: service consumers which employ a baseline method which adopts BRS as the trust evaluation model and blindly aggregating any testimonies they receive;

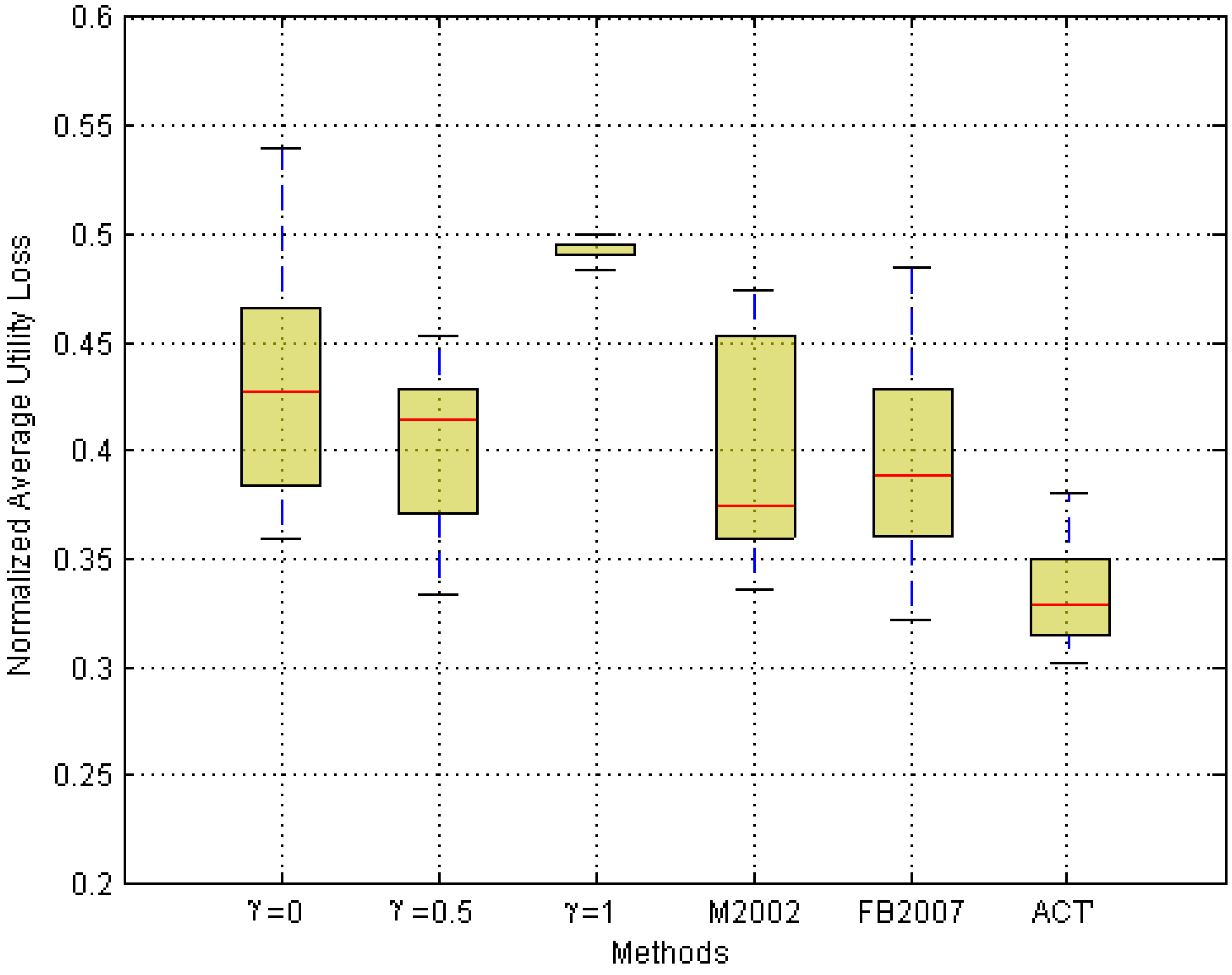

Figure 5. Ranges of variation of NAUL by service consumer groups under non-collusive conditions from Part 1 of this study.

4) Group *BRS2002*: service consumers who only rely on their direct interaction experience to evaluate a service provider's trustworthiness using BRS [Jøsang and Ismail, 2002].

The group of service consumers equipped with the ACT method is labeled as Group *ACT*. Each group also consists of 10 agents. All groups only request for testimonies from the common witness group same as in Part 1 of this study.

## 3.6. Analysis of Results

### 3.6.1. The Effect of Adaptive $\gamma$ Values

Part 1 of this study is conducted assuming non-collusive common witnesses. The common witness population composition is altered from *BM80* to *Hon* and then to *BS80* to test the performance of service consumers employing different testimony aggregation methods. The results are summarized in Figure 5. It can be observed that Group $\gamma = 1$ achieves the highest NAUL values as they need more exploration to identify trustworthy service providers. Its performance is not affected by the changes in the common witness population composition. Completely relying on indirect trust evidence is also not a good strategy as the performance of Group $\gamma = 0$ is heavily affected by the presence of unreliable witnesses of both BM and BS types. However, the saving in exploration from completely relying on third party testimonies allows Group $\gamma = 0$ to achieve lower NAUL values than Group $\gamma = 1$. Nevertheless, the advantage drops with number of misbehaving witnesses as shown in Figure 6. The performance of the Group $\gamma = 0.5$ is the best among the three groups using static $\gamma$ values. Group *F&B2007*'s performance is similar to that of Group *M2002*. As *F&B2007* tries to learn which static strategy ($\gamma = 0, 0.5, or\ 1$) is the best under different conditions, its performance more or less tracks that of Group $\gamma = 0.5$ in our experiments. Group *ACT'* outperforms all other methods under all testing conditions by an average of 20.79% in terms of the reduction in NAUL. A detailed breakdown of the comparisons is shown in Table 4.

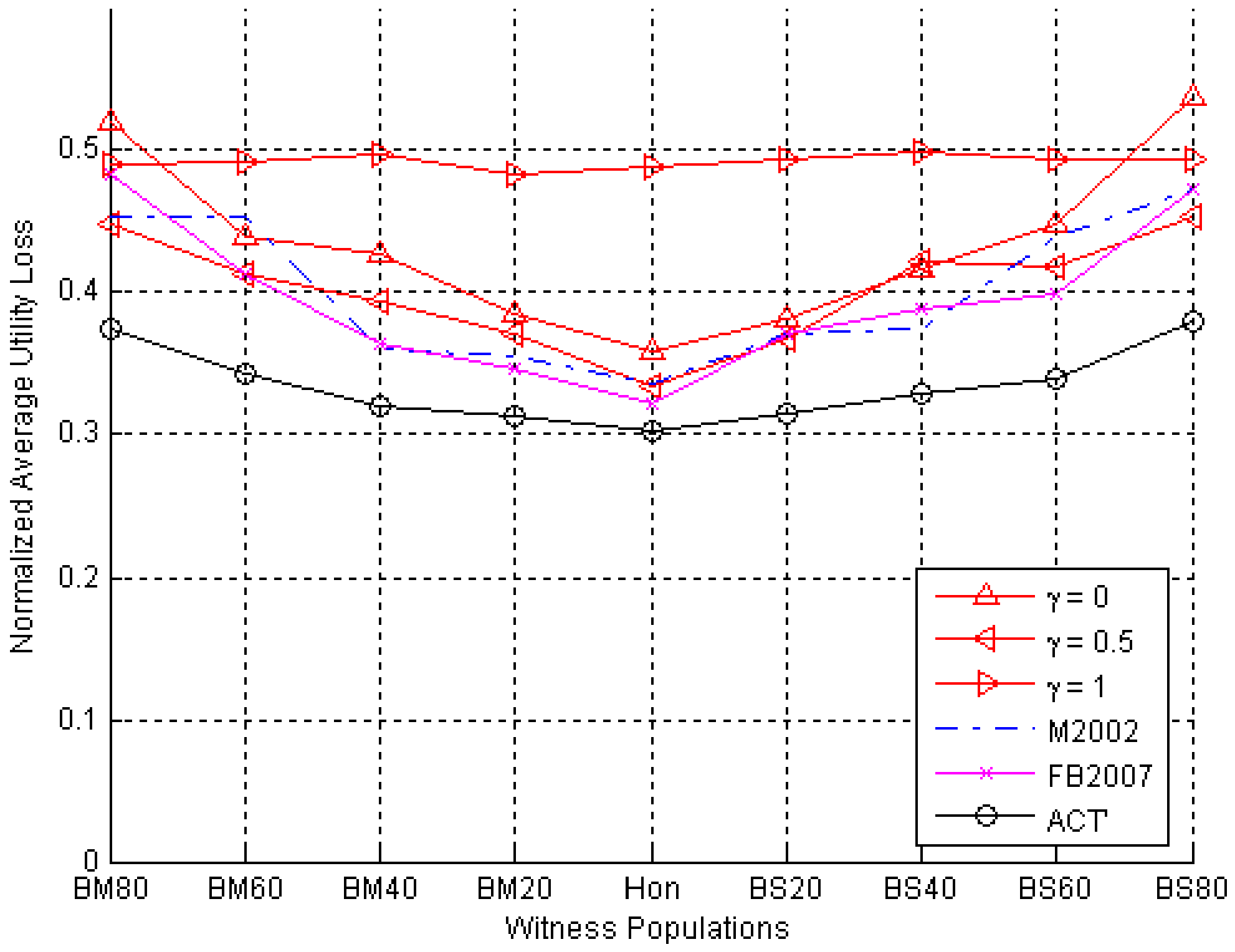

Figure 6. Performance of various service consumer groups under different non-collusive common witness populations from Part 1 of this study.

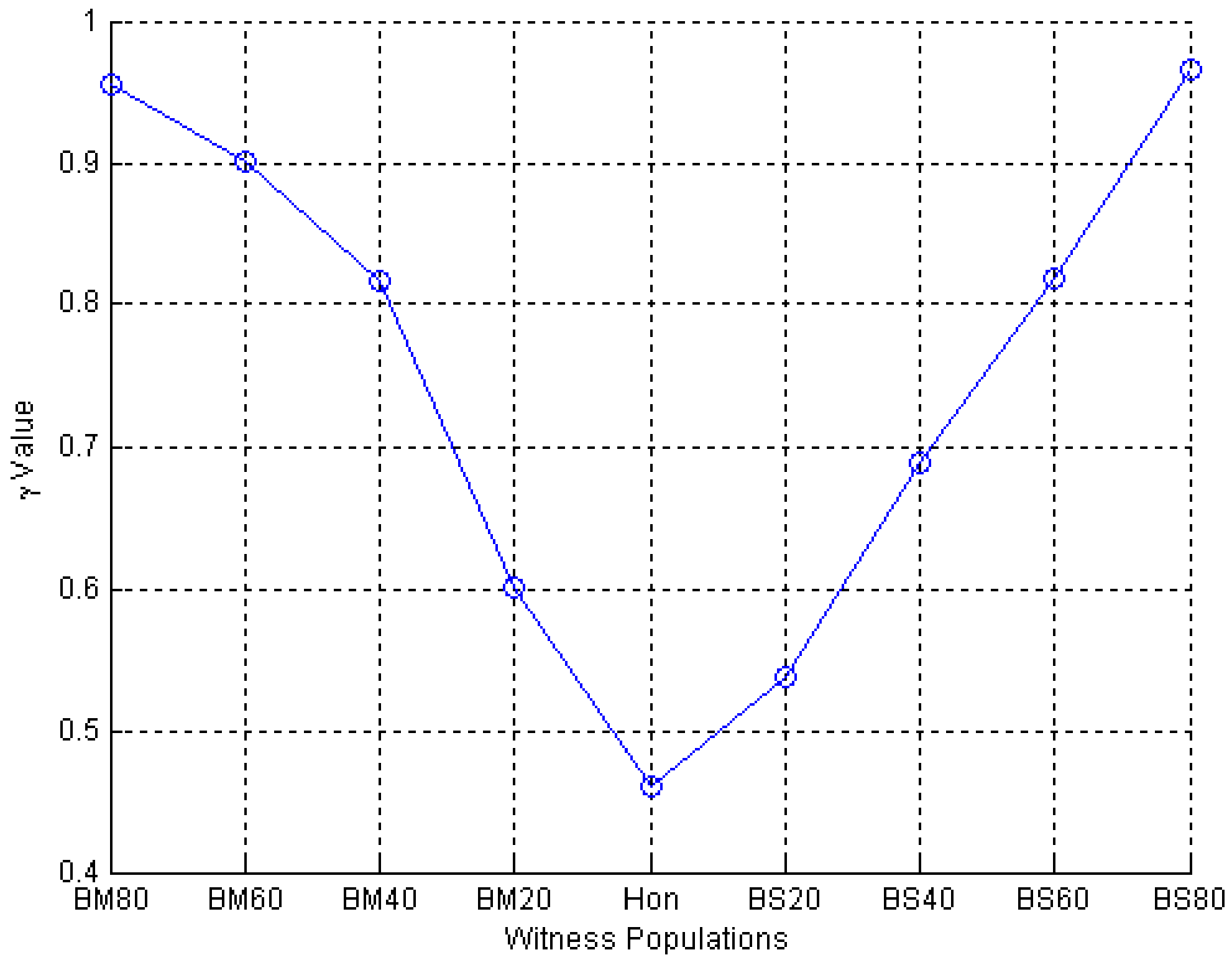

Figure 7. The variation of the $\boldsymbol{\gamma}$ value from the record of a service consumer in Group ACT' with respect to an honest service provider under different non-collusive common witness populations from Part 1 of this study.

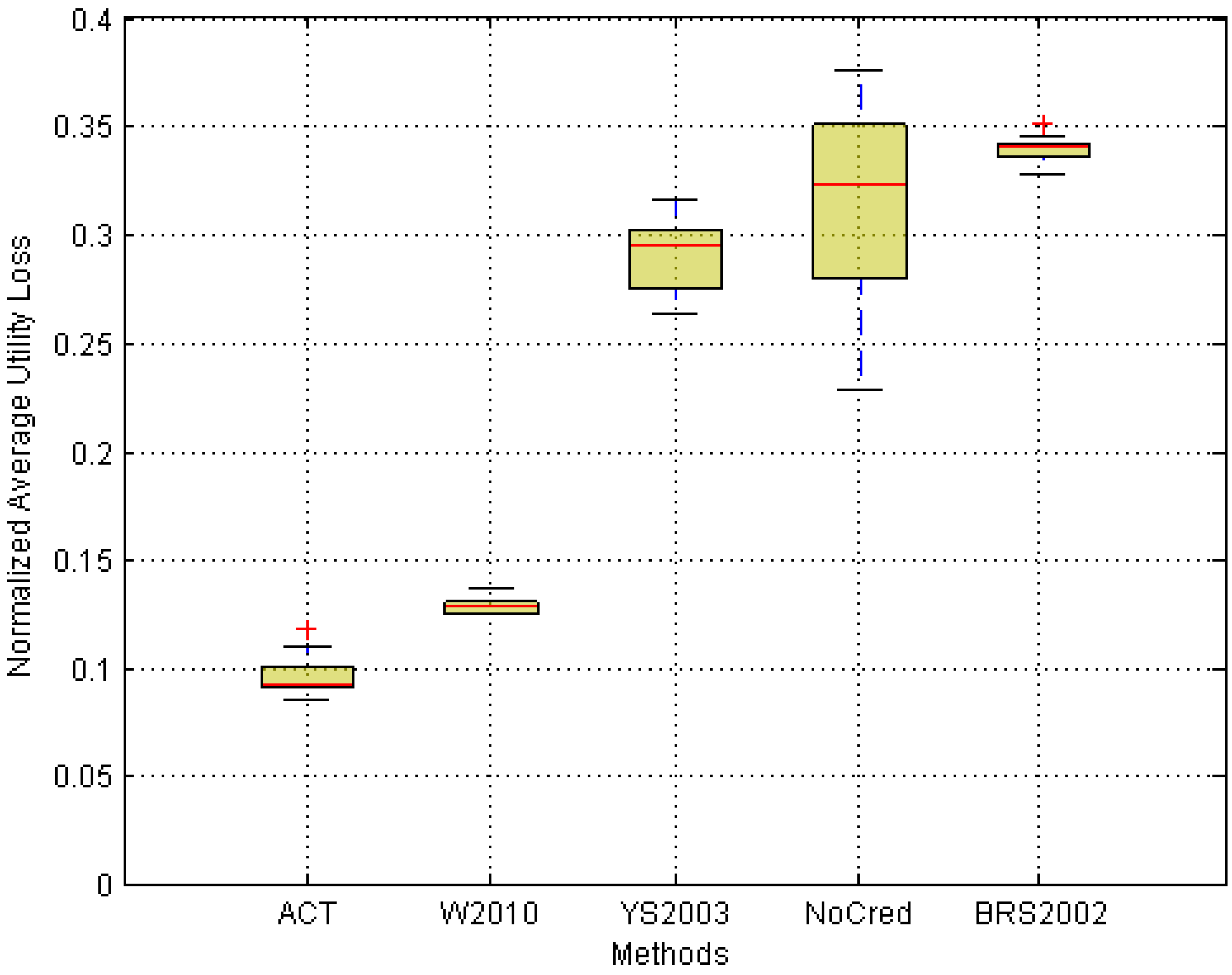

Figure 8. Ranges of variation of NAUL by service consumer groups under non-collusive conditions from Part 2 of this study.

The performance achieved by the proposed *ACT'* service consumers can be attributed to their ability to adapt the values of $\gamma$ for each service provider as the environment conditions change in a continuous manner. Figure 7 shows a snap-shot of the $\gamma$ value from a service consumer in Group *ACT'* with respect to an honest service provider in its local record. It can be seen that as the witness population becomes increasingly hostile, the reliance on third-party testimonies is reduced to mitigate their negative influence on the service consumer's interaction decisions.

Table 4. Improvement of Group ACT' over other groups.

| **Methods** | | **Improvement** |
|---|---|---|
| **Static** | $\gamma$ **= 0** | 23.00% |

| Methods | | Improvement |
|---|---|---|
| **Methods** | **$\gamma$ = 0.5** | 16.82% |
| | **$\gamma$ = 1** | 31.99% |
| **Dynamic Methods** | **M2002** | 16.73% |
| | **F&B2007** | 15.41% |
| **Average** | | **20.79%** |

### 3.6.2. Performance of ACT under non-Collusive Lying

In Part 2 of this study, the performance of the complete ACT model is investigated. The distributions of the NAUL achieved by all five models in this study are shown in Figure 8. It can be seen that Group *NoCred* is outperformed by all other groups as it blindly includes all testimonies into the reputation evaluation process. Group *ACT* has achieved

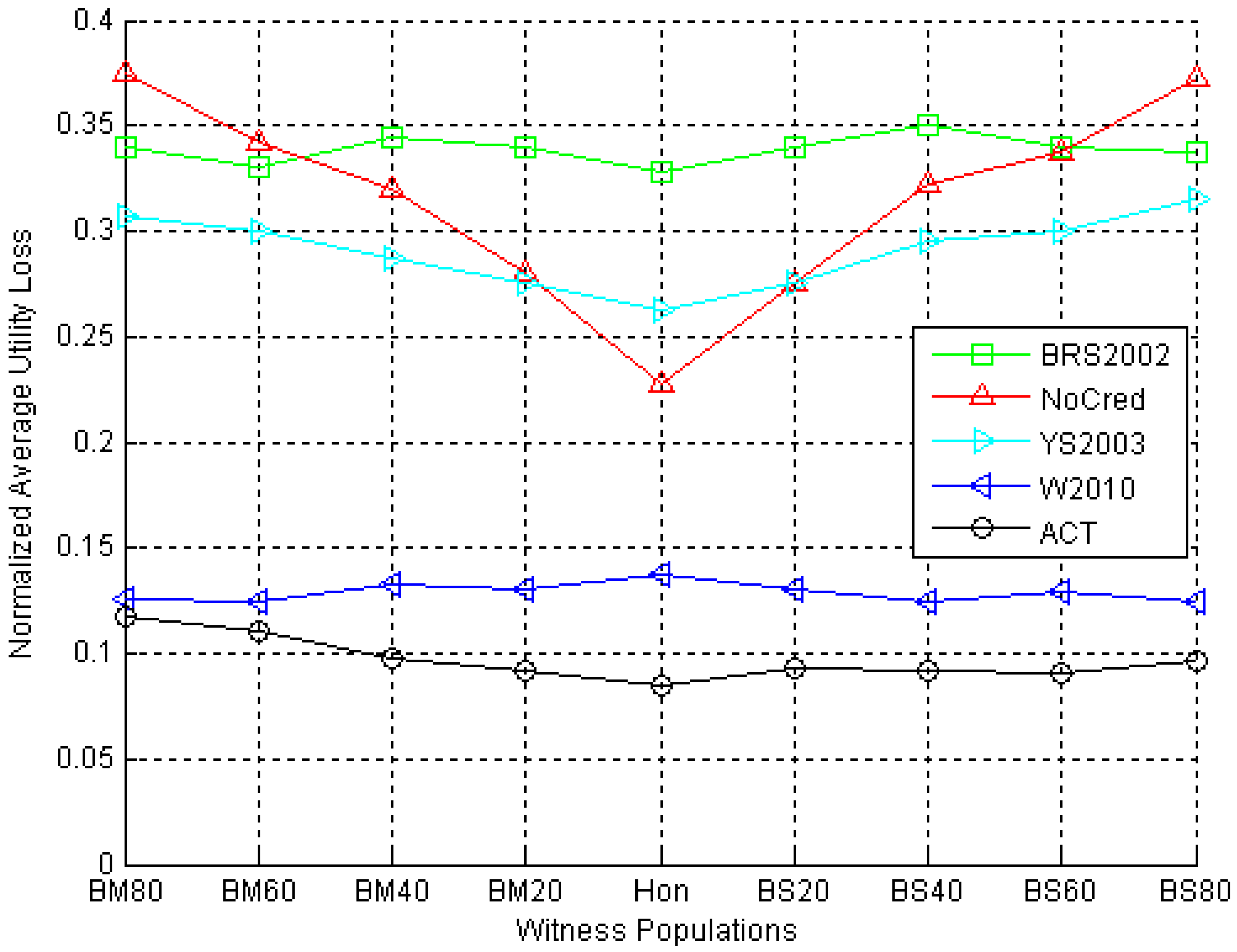

Figure 9. Performance of various service consumer groups under different non-collusive common witness populations from Part 2 of this study.

significantly lower level of NAUL than existing models.

As shown in Figure 9, when the percentage of malicious witnesses increases, the performance of Group *BRS2002* is relatively stable as it does not take into account testimonies from witnesses when making trustworthiness evaluations. However, the NAUL of Group *NoCred* deteriorates significantly. The performance of groups *W2010* and *ACT* are relatively consistant across different witness population configurations. The consistent performance achieved by the ACT model is due to that fact that it uses the interaction outcomes with the service providers rather than the majority opinion of the witnesses to update the credibility ranking of known witnesses, as well as its ability to adjust its preference of the two trust evidence sources dynamically.

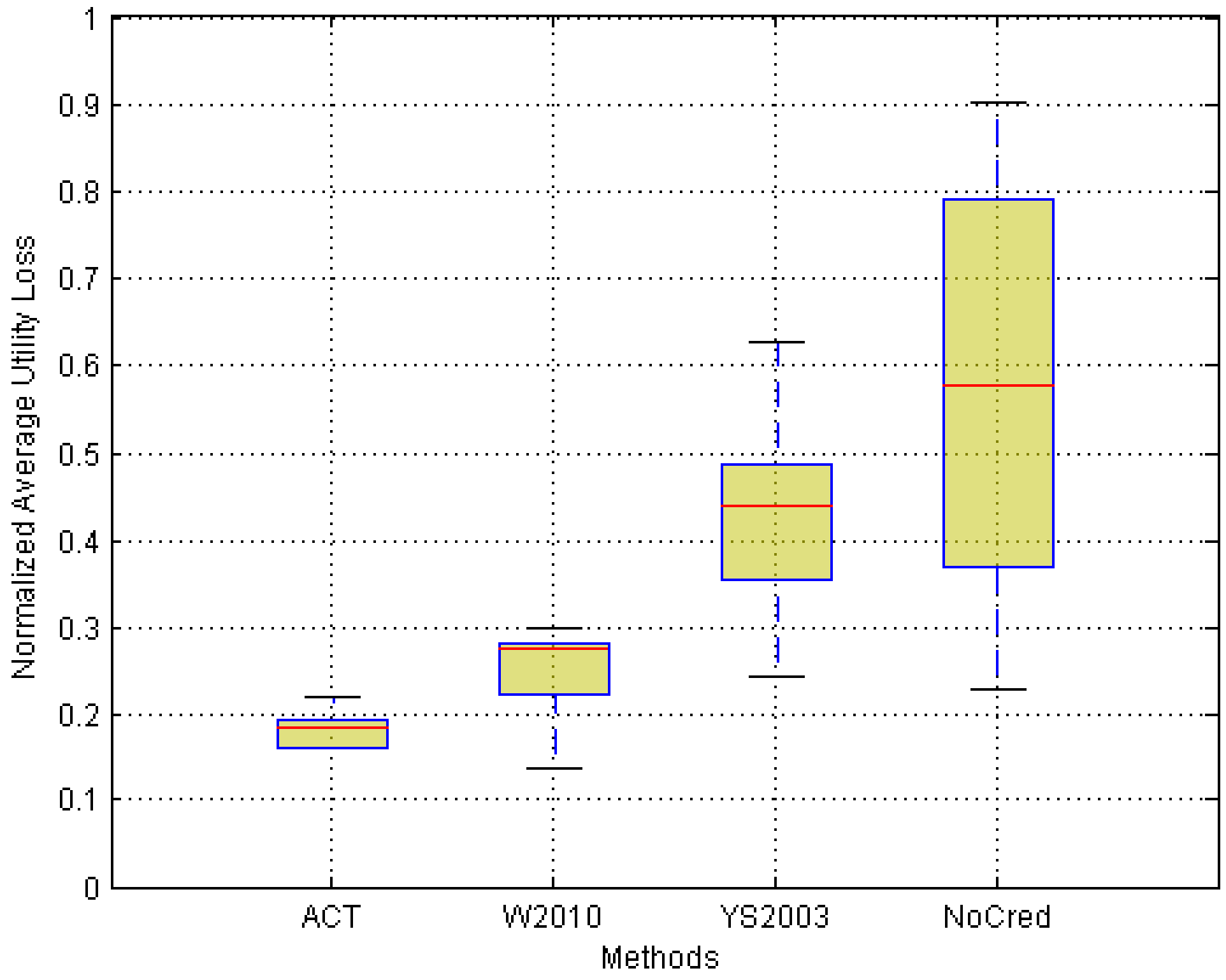

Figure 10. Ranges of variation of NAUL by service consumer groups under collusive conditions from Part 2 of this study.

As can be seen from Table 5, overall, Group *ACT* outperforms all other groups in terms of reduction in NAUL by significant margins. The advantage is more significant under ballot-stuffing conditions due to the addition of the reward correction value $\theta_{k,j}$ in (3.6) that penalizes positively biased testimonies.

Table 5. Improvement of the ACT model over other Models

| Trust Models | Improvement | | |
|---|---|---|---|
| | **Badmouthing** | **Ballot-stuffing** | **Overall** |
| W2010 | 18.44% | 29.91% | 25.16% |
| YS2003 | 64.18% | 70.20% | 66.98% |
| NoCred | 68.19% | 72.99% | 69.70% |
| BRS2002 | 69.07% | 74.18% | 71.66% |

### 3.6.3. Performance of ACT under Collusive Lying

In our test-bed, the collusive witnesses always form collusion rings with Type III

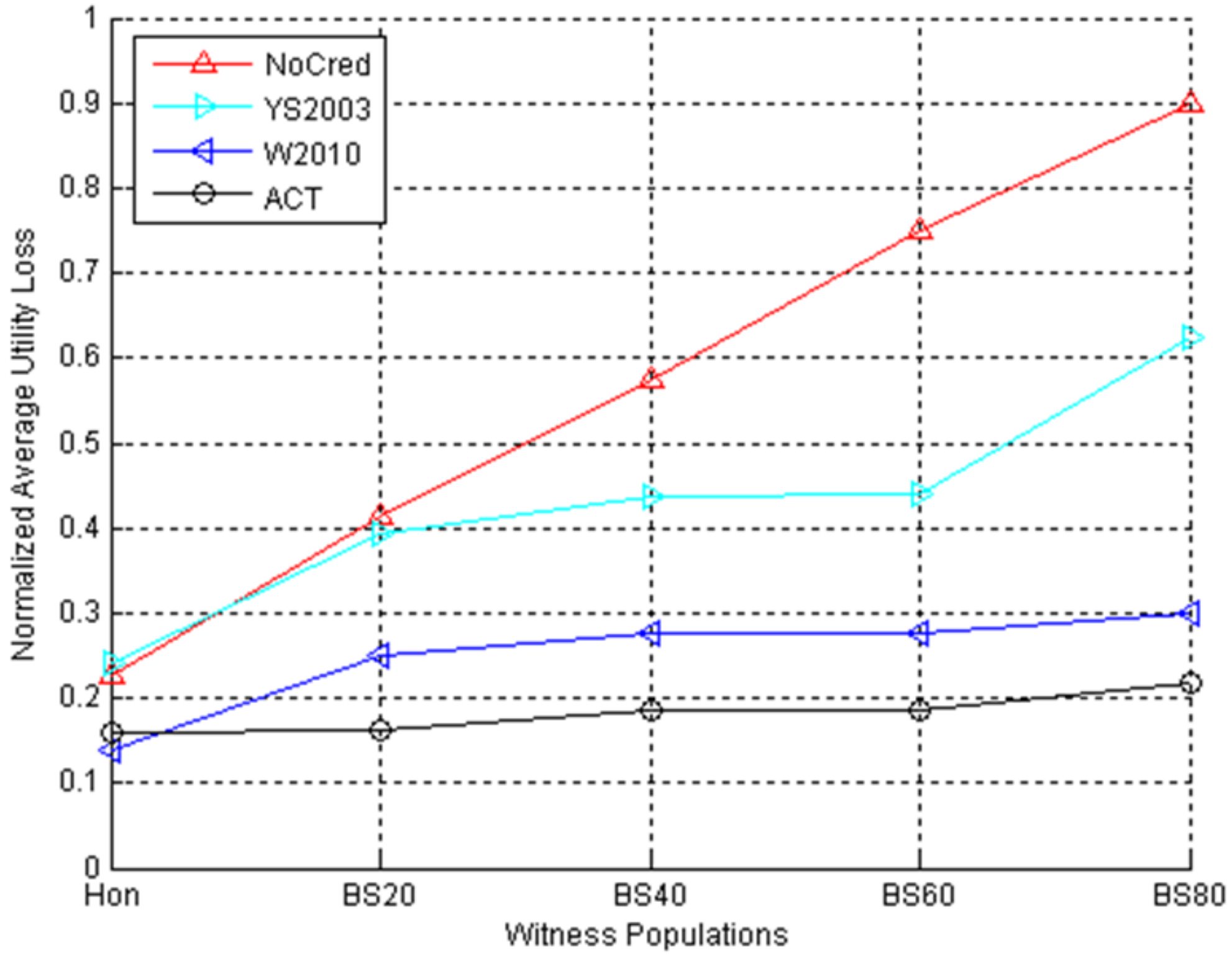

Figure 11. Performance of various service consumer groups under different collusive common witness populations from Part 2 of this study.

malicious service providers to try to promote their reputation. The proportion of collusive witnesses in the total common witness population is varied from *Hon* to *BS80*. From Figure 10, it can be seen that the presence of colluding witnesses tricks the *NoCred* group into interacting more often with collusive service providers than other groups. In addition, by comparing Figure 10 with Figure 8, we find that the negative impact of collusion is more powerful than that of non-collusive random lying. The most adversely affected group is still the *NoCred* group. The highest NAUL of this group is about 0.4 under *BS80* without collusion. However, under *BS80* with collusion, this value increases to around 0.9 (as shown in Figure 11). This is due to the fact that colluding witnesses do not give unfair testimonies about non-colluding service providers, so that their testimonies are considered accurate in these cases. Thus, they are essentially strategically building up their credibility with the service consumers in order to mislead them into interacting with collusive service providers later. The performance of all the models studied in our test-bed deteriorated under the influence of collusion as shown in Table 6. Although Group *ACT* and Group *W2010* managed to maintain the witness agents' collusion power at relatively low levels compared to other groups as illustrated in Figure 12, their performances in terms of NAUL still deteriorated under collusion. It is observed, from Table 7, that the ACT model significantly outperforms all other approaches in terms of mitigating the adverse effect of collusion. The outperformance in terms of reduction in collusion power is the most significant when the majority of the witness population consists of collusive witnesses, as can be seen from Figure 13.

Table 6. Performance Deterioration of various Models due to Collusion

| **Trust Models** | **Average NAUL** | |
|---|---|---|
| | **Non-collusive** | **Collusive** |
| ACT | 0.0890 | 0.1825 |

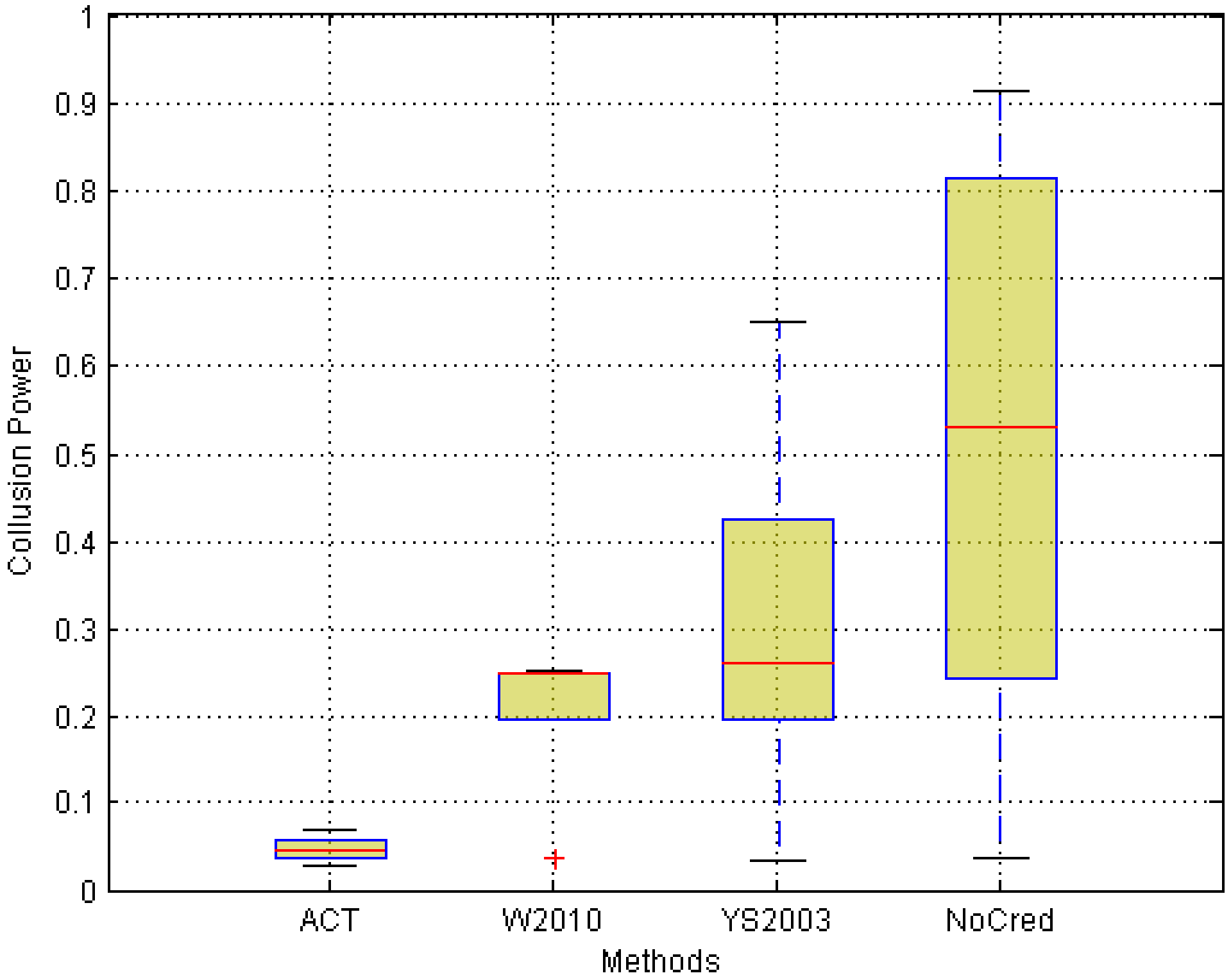

Figure 12. Ranges of variation of Collusion Power by service consumer groups under collusive conditions from Part 2 of this study.

| | | |
|---|---|---|
| W2010 | 0.1283 | 0.2479 |
| YS2003 | 0.2895 | 0.4145 |
| NoCred | 0.3070 | 0.5737 |

Table 7. Improvement of the ACT Model over other Models

| **Trust Models** | **Improvement** | |
|---|---|---|
| | **Collusion Power** | **NAUL** |
| W2010 | 77.60% | 26.37% |
| YS2003 | 85.94% | 55.97% |
| NoCred | 90.90% | 68.18% |

## 3.7. Sensitivity Analysis

Several reasons contribute to the ACT model's superior performance over *W2010* and *YS2003*:

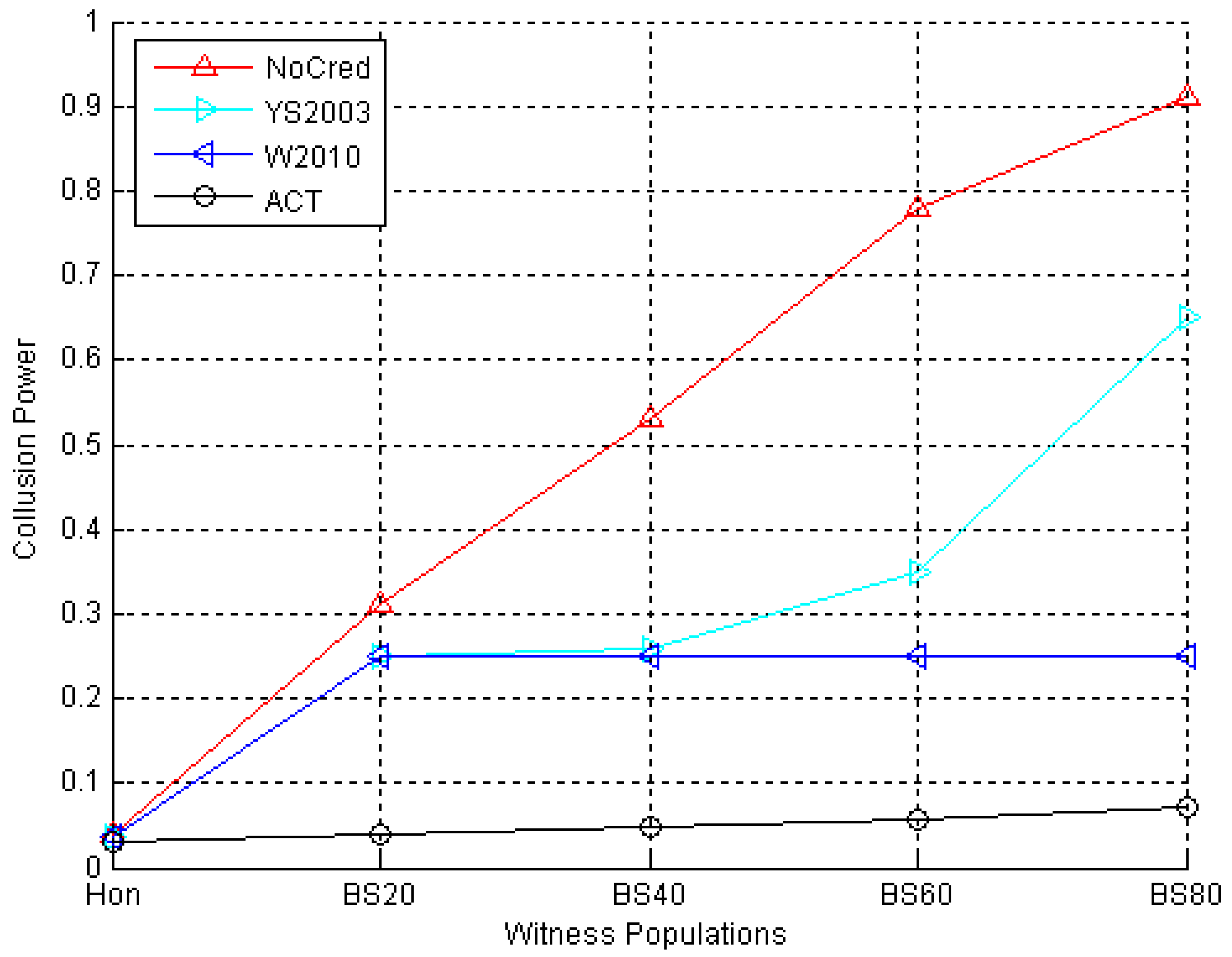

Figure 13. Performance of various service consumer groups under different collusive common witness populations from Part 2 of this study.

1) *YS2003* uses the number of past interactions between a service consumer $c_i$ and the service provider of interest $s_j$ to determine whether third-party testimonies are required. If the number of past interactions between $c_i$ and $s_j$ exceeds a predefined threshold, $c_i$ will not ask for testimonies when estimating $s_j$'s trustworthiness. However, since the behavior of the witnesses are changing in the experiments, $c_i$'s direct trust evidence may become outdated. This increases $c_i$'s risk exposure in the long run.

2) *W2010* applies an adaptive strategy in aggregating third-party testimonies. However, it also uses a service consumer $c_i$ 's own evaluation of a service provider $s_j$ 's trustworthiness as a baseline to determine which testimonies are potentially unfair. It proposed a measure of uncertainty induced by additional testimonies. If a new testimony contradicts $c_i$'s current belief about the trustworthiness of $s_j$, it would be regarded as increasing $c_i$'s uncertainty and discarded. While this approach is more dynamic than *YS2003*, it still suffers from the effect of changing service provider behavior to some degree.

3) In contrast, the ACT model always seeks testimonies from witnesses when estimating a service provider's reputation. By learning the weights assigned to different witnesses' testimonies based on the outcomes after each interaction, the ACT model dynamically decides which witnesses to keep in the top *M* list for each service provider based on their contributions to the well-being of the service consumer. Even in the face of highly hostile witness populations, the ACT model still can maintain a relatively good performance by relying more on the direct trust evidence source. This mechanism also helps the service consumers when the behavior of a service provider changes. If this change is reflected first in the testimonies, the service consumer can increase the weight given to the indirect trust evidence source to reduce the need for trial and error; if this change is detected first by the service consumer itself, it can increase the weight given to the direct trust evidence source to reduce its chance of being misled by outdated opinions from others.

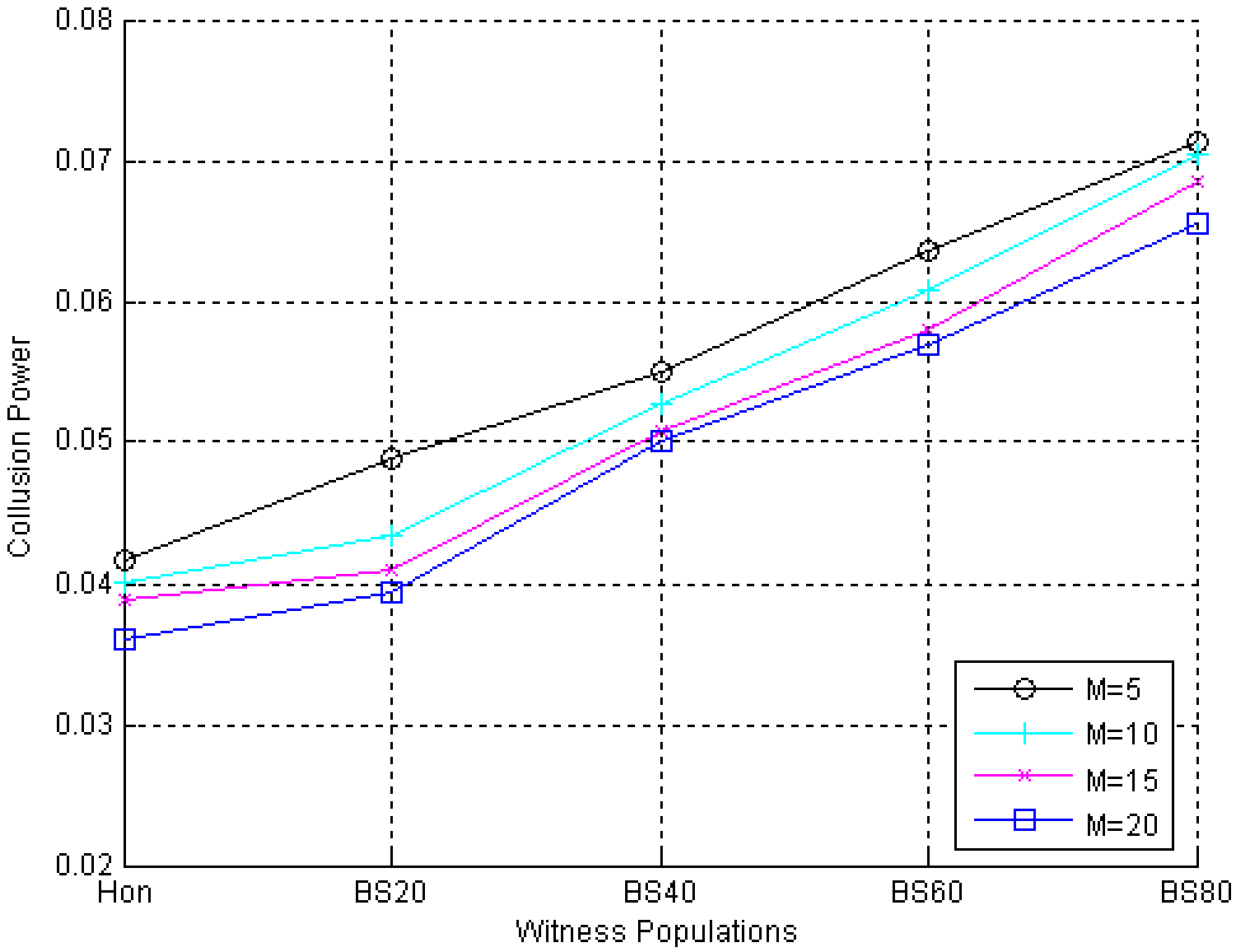

Figure 14. The influence of the parameter *M* on the performance of the proposed approach under different witness population compositions.

To study the influence of *M* on the proposed ACT model, we alter the value of *M* and re-run the experiments. The value of *M* is varied to be equivalent to between 5% to 100% of the common witness population. The experiments are re-run only for the cases where collusion exists since collusive testimonies are more powerful in affecting the credibility models.

From Figure 14, it can be seen that generally, collusion power increases with the fraction of colluding witness agents. However, the value of collusion power is maintained at a relatively low level by the ACT model. This trend is true for the different values of *M*.

By taking an average of the collusion power over different witness population configurations under different *M* values, the effect of different choices of *M* on the average collusion power and the time taken for the ACT model to estimate the reputation of a service provider is plotted in Figure 15. *T(M)* represents the time taken for the ACT model to estimate the reputation of a service provider measured in multiples of the time taken when *M*=5% of the common witness population (i.e. T(3.3)=1). As *M* increases, initially, the performance of the ACT model is improved (reduced average collusion power). However, increasing *M* beyond about 15% yields no further reduction in collusion power. Nevertheless, *T(M)* consistently increases with *M*.

There is a trade-off in the selection of a value for *M*. It is expected that the effectiveness of the ACT model improves with *M*. However, the value of *M* also determines the storage capacity required at each individual service consumer as well as the time taken to estimate the reputation of a service provider. From Figure 15, a good choice of *M* appears to be in the range of [5%, 15%] which achieves a balance between effectiveness and computation time.

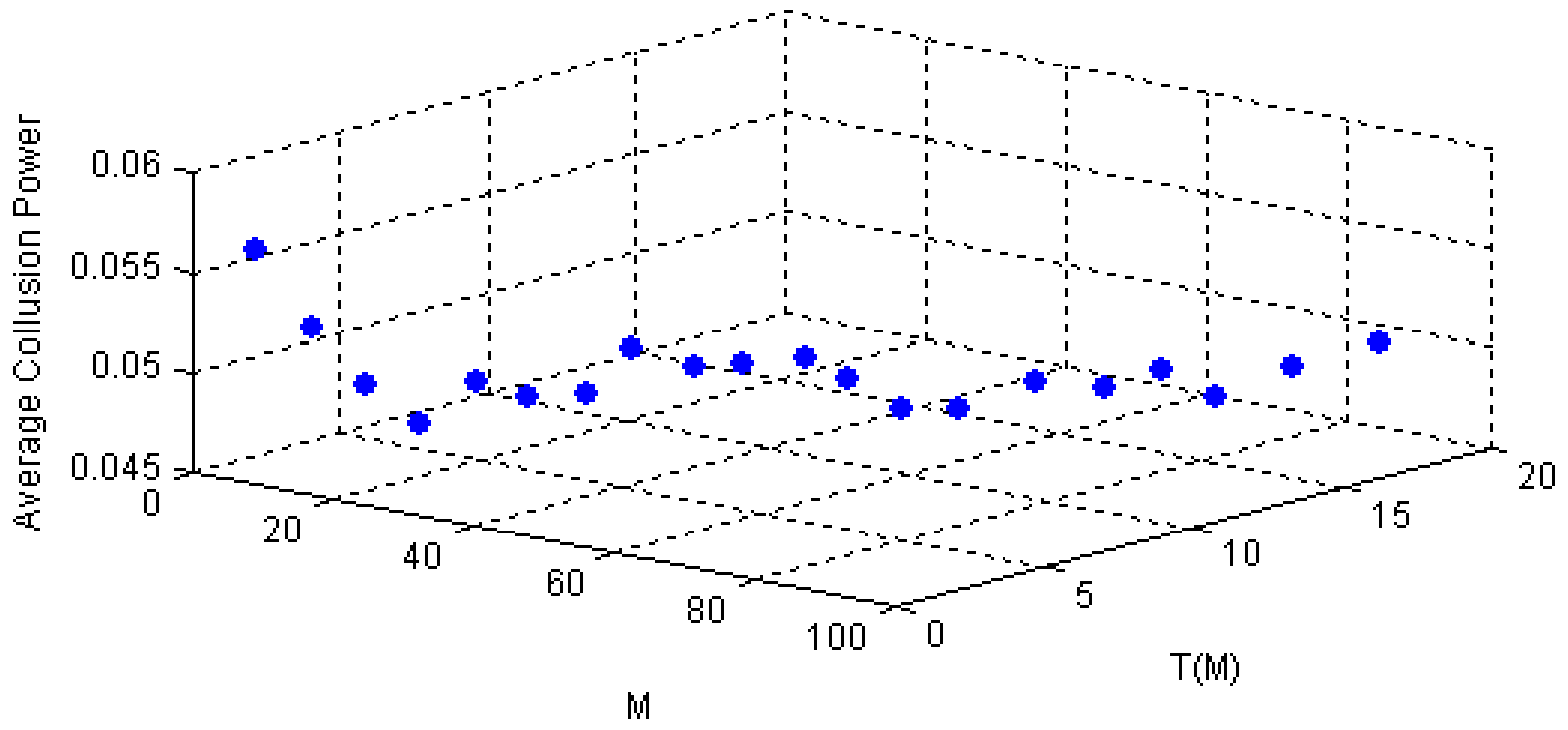

Figure 15. The trade-off between reduction in average collusion power and time required for computation under different choices of value for parameter *M*.

## 3.8. Application in Cognitive Radio Networks

In this section, we report the study of the proposed ACT model under an emerging wireless network application scenario. As the demand for wireless communications grows, so does the importance for efficient utilization of the scarce radio spectrum resource. The Federal Communications Commission (FCC) has reported that most of the licensed spectrum is currently under-utilized [Marcus *et al.*, 2002]. Cognitive radio (CR) is a novel approach for improving the utilization by making it possible for a group of secondary (unlicensed) users (SUs) to access spectrum bands which are not being used by the primary (licensed) users (PU) in some geographical location [Haykin, 2002]. A CR system is an intelligent wireless communication system capable of learning from its radio environment and dynamically adjusting its transmission characteristics accordingly.

Dynamic spectrum allocation (DSA) is one of the central ideas in the cognitive radio network (CRN) paradigm. Efficient DSA requires the SU to be able to accurately determine when a PU spectrum band is idle. In a CRN, the sensing accuracy is affected by a number of factors such as terrain features, the types of the sensing devices, etc. [Ilisei, 2006]. Sensing accuracy can be improved by collecting measurements from a number of sensing devices located in a wider geographic area rather than relying on one or two dedicated devices [Mishra *et al.*, 2006; Chen *et al.*, 2008]. SU devices may possess different sensing capabilities or may purposely choose to misbehave in order to maximize their own utility gains. Although efforts have been directed at making CRNs more robust against traditional security attacks [Clancy & Goergen, 2008], few attempts have been made to deal with legitimate users who misbehave. Authors in [Jøsang et al, 2007] coined the term "soft security threat" to describe the aforementioned situation and suggested the use of trust management to mitigate these threats. A SU's actions in the PU spectrum usage sensing process can impact the collective decision making process either positively or negatively depending on both its intentions and its capabilities. Therefore, we study the effectiveness of incorporating trust management into the CR system architecture in improving the robustness of the distributed PU spectrum sensing process.

### 3.8.1. System Architecture

The CRNs can be deployed in various kinds of network architectures such as centralized, Ad-hoc and mesh architectures [Chen *et al.*, 2008] as shown in Figure 16. In this study, we adopt an infrastructure-based CRN with centralized base station in cellular networks which is an extension of the one in [Lee and Akyildiz, 2008] as the underlying habitat on top of which a trust ecosystem is to be overlaid. We assume that PUs coexist with SUs in some geographical area and PUs are controlled by a PU base station (PUBS). In this CRN, SUs are distributed in the coverage area of an SU base station (SUBS) and SUs within the transmission range of the SUBS can only communicate with each other through the SUBS. The SUs are not able to communicate with either PUs or the PUBS but communication between SUBS and PUBS is possible through backbone networks. During the sensing process, SUs sense the PU spectrum individually and report the results to SUBS. By integrating sensing results reported by SUs and its own sensing result, SUBS determines the availability of PU spectrum band and allocates resources to SUs in its range. While a PU experiences undue interference, it informs its PUBS which then sends a complaint message immediately to SUBSs within range. This complaint message from PUBS to SUBS will cause a temporary shutdown of SUBS operation and may cause a great utility loss for SUs. In this study, we assume that PUs, PUBS and SUBS are trustworthy entities in the CRN.

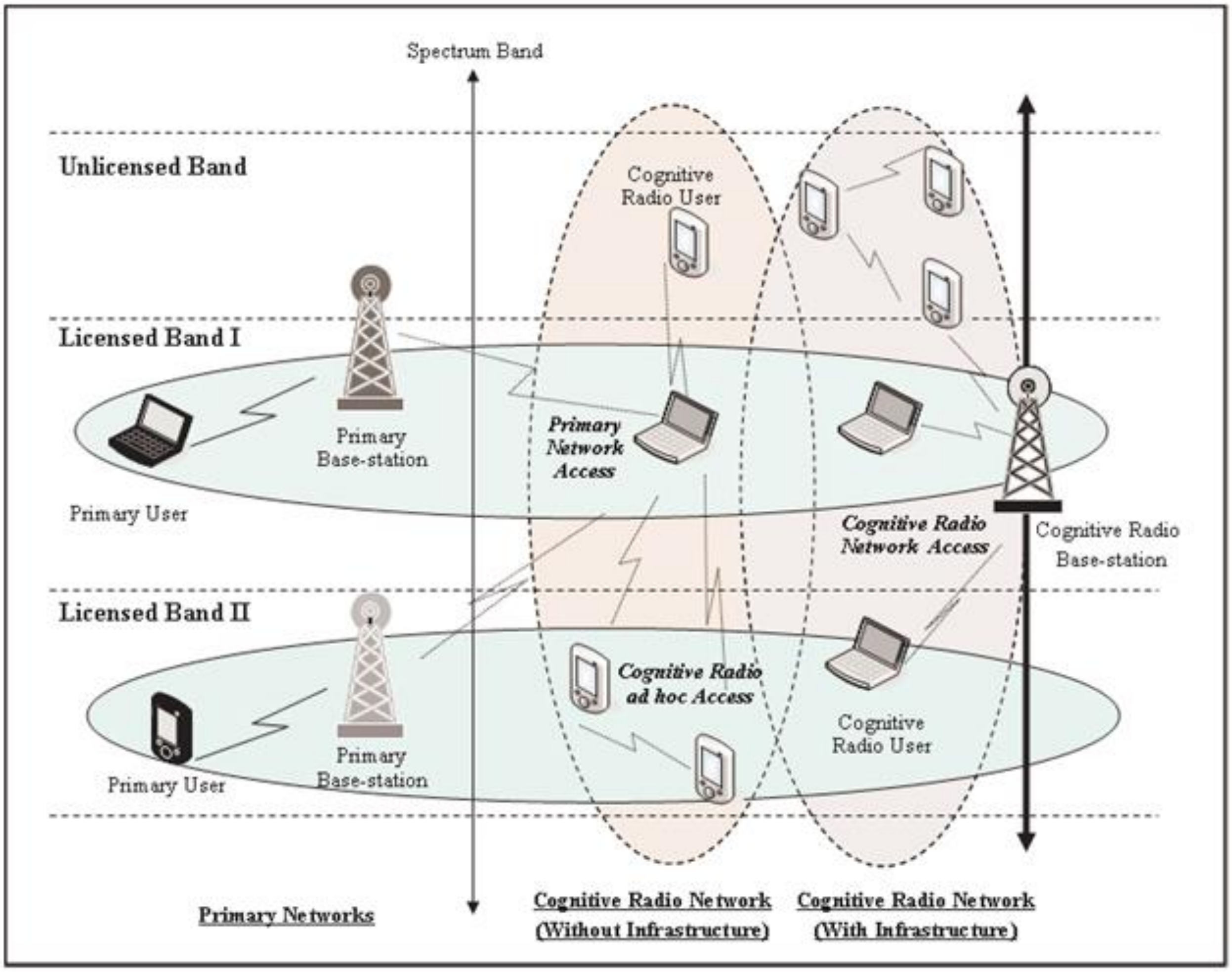

Figure 16. An overview of CRN Architectures [Akyildiz *et al.*, 2008]

Under the CRN architecture mentioned above, the decision on whether a PU is using its assigned spectrum band or not depends on two major sources of information: 1) the sensing result from the SUBS, and 2) the collective sensing results obtained from the SUs managed by this SUBS. This mode of operation enables the SUBS to contribute to the overall sensing result while still harnessing the power of distributed sensing by participating SUs which could enhance the overall sensing result accuracy provided they act in an honest way. However, as the sensing results can be significantly influenced by the relative locations between the PU transmitter and the SUBS as well as between the PU transmitter and the SUs, sometimes, the SUBS or the SUs may not be confident about the sensing results they have obtained. Therefore, when SUs report their sensing results to the SUBS and when the results are aggregated, the confidence level, $\mu$, of each result is taken into account. The confidence level of the sensing result relies on a range of factors. For simplicity, in our subsequent experiments, we assume it to be related only to the channel gains between the transmitters and the sensors. Three types of sensing results can be obtained when trying to establish whether PU is using its allocated spectrum band: 1) the PU spectrum band is in active use (denoted by numeric value 1); 2) the PU spectrum band is not in active use (denoted by numeric value -1), and 3) the usage situation of the PU spectrum band is not clear or abstain from the sensing operation (denoted by numeric value 0). An honest SU can choose to return any one of these three types of results to the SUBS on the basis of its sensing algorithm output and its own confidence level on the result. Therefore, the aggregated sensing result for a particular PU spectrum band *p* can be calculated with the formula:

$$R_p = \theta \Gamma_{\mathrm{BS}} + (1-\theta) \frac{\sum_{\mathrm{i}=1}^{\mathrm{M}} \tau_{\mathrm{ip}} \Gamma_{\mathrm{ip}}}{\sum_{\mathrm{i}=1}^{\mathrm{M}} \tau_{\mathrm{ip}}} \quad (3.17)$$

where $R_p$ is the overall sensing result for PU spectrum band *p*; $\mu$ is the confidence level of the SUBS; *ΓBS* is the sensing result provided by the SUBS; *τip* is the trustworthiness of the *i*th SU in the context of PU spectrum band *p*; *Γip* is the sensing result for PU spectrum band $p$ provided by $i$th SU; $M$ is the number of SU whose trustworthiness with respect to PU spectrum band $p$ is above a predefined threshold of $\eta$.

In the case when the variance in trustworthiness of each SU in the context of a primary user $p$ is not considered, $\eta$ is set to 0 and all $\tau ip$ are set to 1. Thus, the second term in the above formula becomes a simple average of all the sensing results obtained from the SUs. It is effectively a weighted sum of the SUBS sensing result and a majority voting from all the SUs who choose to participate in the distributed sensing operation. The final decision $D_p$ is made based on the sign of $R_p$:

$$D_p = \begin{cases} -1, R_p < 0 \\ 0, R_p = 0 \\ 1, R_p > 0 \end{cases} . \tag{3.18}$$

In a CRN, two types of errors in the decision to utilize the PU spectrum can occur: 1) when the PU spectrum band is idle but the final decision of the SUBS is not to use the spectrum band; and 2) when the PU spectrum band is actively used but the final decision of the SUBS suggests otherwise. We will refer to these two category errors as false alarm error, $\varepsilon_1$, and miss detection error, $\varepsilon_2$, respectively. The false alarm error, $\varepsilon_1$, tends to reduce the efficiency of the system by reducing the system throughput while the miss detection error, $\varepsilon_2$, tends to cause interference to the normal usage of the PU spectrum by the primary user which may result in a forced shutdown of the SUs sharing the spectrum. In view of these characteristics, $\varepsilon_2$ is considered a more severe error than $\varepsilon_1$.

### 3.8.2 Trust Aware Collaborative Spectrum Sensing Model

As a spectrum band can only have two states: 1) occupied by a PU, or 2) vacant, the sensing result from each SU which chooses not be abstain can only be either right or wrong. The behavior of a SU can be generally divided into three categories: honest, neutral and dishonest. Although there can be many forms of misbehaviors in a particular system, their common characteristic is the tendency to deviate the final decision from the truth. The evaluation of the behavior of a SU in one sensing operation is in the form of [$\alpha$, $\beta$]. For example, if a SU is regarded as having provided a right sensing result in this turn, a value of [1, 0] can be added to the storage of its historic behaviors. A window size $N$ of past behaviors are recorded. This provides a way for the trust and reputation score of a SU to vary with deviations in its behaviors from the past but also dampens the rate of change of these scores to prevent them from being overly affected by the latest behavior. It is commonly acknowledged in computational trust and reputation model literatures that trust is a context-dependent concept. For instance, an agent who behaves honestly when dealing with a large and well established organization may be malicious towards an individual person in order to maximize its utility gain. Therefore, to better account for the possible dichotomy in a secondary user's sensing behaviors towards different PU spectrum bands, we divide the trustworthiness score of each SU into contexts based on the PU's characteristics (e.g. geographical location, general spectrum usage rate etc.). Therefore, a history of behaviors and the corresponding forgetting factor of each SU for each context is stored in the trust and reputation database. The context dependent trustworthiness score can be computed as: $\tau_{ic} = \frac{\sum_{j=0}^{N-1} \rho_{ic}^{N-1-j} \alpha_{jic}}{\sum_{j=0}^{N-1} \rho_{ic}^{N-1-j} (\alpha_{jic} + \beta_{jic})}$, where $\tau_{ic}$ is the trustworthiness score of $i$th SU in the context $c$; $N$ is the window size of store past behavior history; $\rho_{ic}$ $(0 \leq \rho_{ic} \leq 1)$ is the forget factor for the $i$th SU in the context $c$; $\alpha_{jic}$ is the $j$th positive behavior score of $i$th SU in the context $c$; $\beta_{jic}$ is the $j$th negative behavior score of $i$th SU in the context $c$.

Since in our system, $\varepsilon_2$ is regarded as having more serious consequences than $\varepsilon_1$, it is desirable to drastically reduce the trust and reputation score of a SU when its sensing result could contribute to the complaint from the PU. In our trust and reputation model, two techniques are employed to accomplish this goal. The first one is the adaptive forget factor technique. The context-dependent forget factor is adapted as: $\rho_{ic} = \begin{cases} \rho_1, if\ i\ contributed\ to\ \varepsilon_2 \\ \rho_2, if\ the\ latest\ \tau_{ic} \geq \eta \end{cases}$, where $0 \leq (\rho_1, \rho_2) \leq 1$ and $\rho_1 \geq \rho_2$. A larger forgetting factor translates into more weight assigned to past behaviors. So when new ratings come in, the past negative ratings exert more dampening effect to prevent the trustworthiness score to vary too much.

However, this technique alone is not adequate in combating $\varepsilon_2$ causing behaviours since it magnifies both past negative and positive behaviours. Therefore, we employ a second technique which assigns larger weight to negative behaviours causing $\varepsilon_2$. In the case when the sensing result of a SU is considered to have caused a complaint from a PU, a record of [0, $N$] is entered into its behaviour history. This technique can dramatically reduce the trustworthiness of a dishonest SU when a complaint from a PUBS is received by the SUBS. The $\varepsilon_2$ errors result in definitive feedbacks from the PUBS in the form of complaints to the SUBS. In this case, those SUs who reported that the PU spectrum band was idle will be subject to both the aforementioned punitive measures. However, in the situation when the overall decision suggests the PU spectrum band is in active use, there is no definitive feedback which can be used to guide the subsequent reward and punishment decisions. Therefore, in these cases, our trust model adopts a conservative approach which assigns a unit positive rating of [1, 0] to those who agrees with the overall decision $D_p$ and a unit negative rating [0, 1] to those who disagree with the overall decision $D_p$. Abstaining SUs will always get a rating of [1, 1] regardless of $D_p$. However, in the situation where $D_p$ is 0 (i.e. not sure), no rewards or punishments will be bestowed on any SUs.

The reputation of a SU which represents its overall probability of behaving honestly across different contexts is calculated using the proposed ACT model.

### 3.8.3. Evaluating the ACT Model under Various Attack Scenarios

In this section, we investigate the robustness and efficiency of trust management using computer simulations. We considered a system with 8 PU spectrum bands and a total of 100 secondary users. The confidence level of SUBS and SU, $\mu$, was outcomes of independent, identically distributed (i.i.d.) Gaussian random variables (rvs) with mean equal to 0.5. A confidence level below 0.25 was considered to be inadequate to justify the sensing result which led to the reporting of a not-sure message (0 in our case). The trustworthiness threshold was set to $\eta$ = 0.65. Only SUs with trustworthiness scores exceeding this threshold have their sensing results used by the SUBS in the decision making process. The forgetting factors was set to $\rho_1$ = 1 and $\rho_2$ = 0.9. Each experiment consisted of 10,000 iterations for each PU spectrum band.

To investigate the effectiveness of trust management against various attacks, we introduced a performance evaluation metric, TUL, to describe the total system utility loss due to the attacks. The TUL can be expressed as:

$$TUL = W_1\varepsilon_1 + W_2\varepsilon_2 \tag{3.19}$$

where $W_1$ and $W_2$ denote the weight factors for false alarm error rate and miss detection error rate respectively. In the proposed CRN, since a misdetection is regarded as more damaging to the system than false alarm, $W_2$ is assigned a larger value than $W_1$ in the following simulations. We denote the attack rate by $\sigma$. To show that trust management is robust and efficient, we consider a number of attack scenarios.

### A. *Fabrication Attack*

In the fabrication attack, a malicious SU deliberately reports inverted sensing results to SUBS all the time. This kind of misbehavior aims to cause deterioration to the overall performance of the network since it will either prevent other secondary users from accessing network resources or introduce excessive interference to PU spectrum bands.

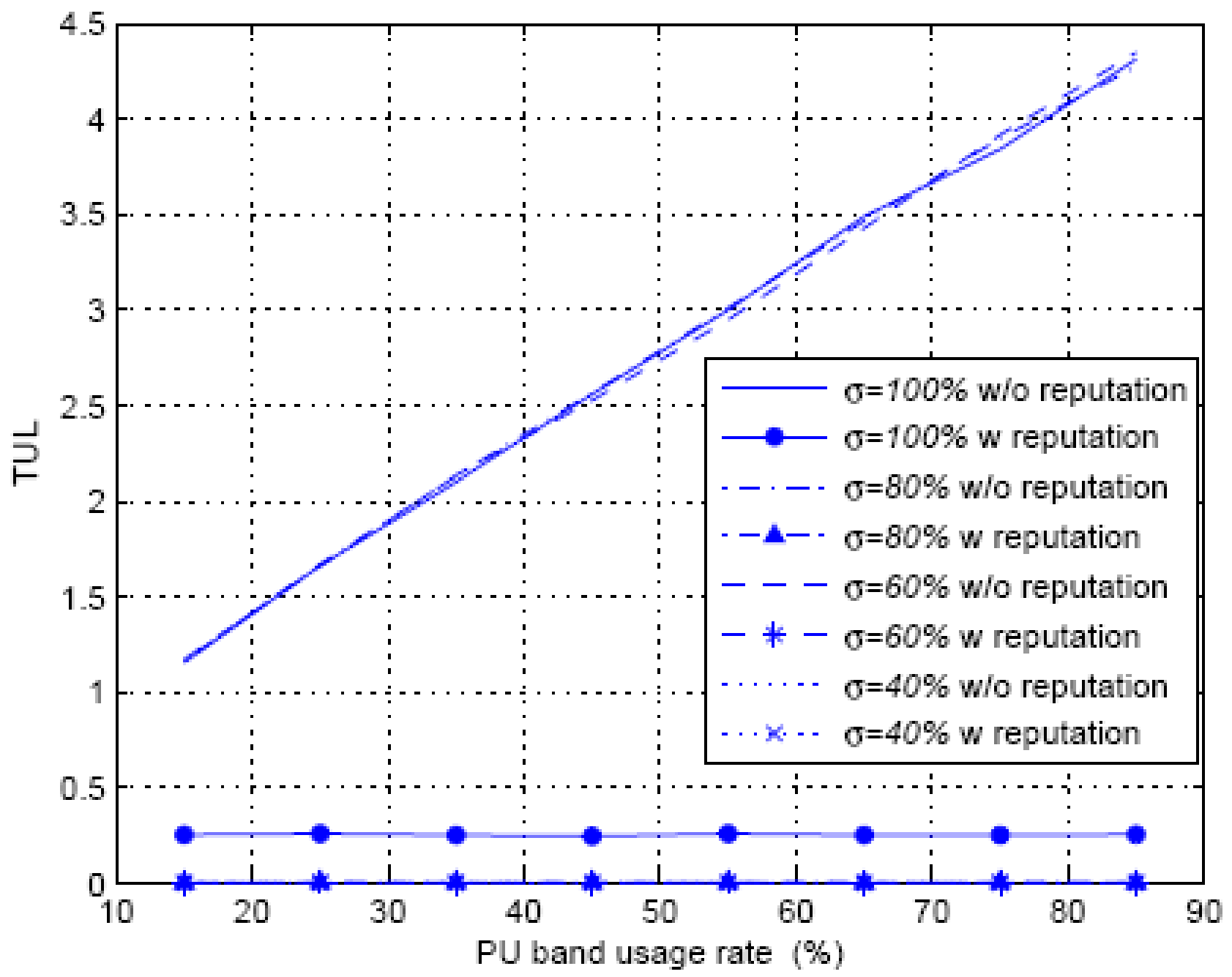

Figure 17. Total utility loss, TUL, of CRN under the fabrication attack.

Figure 17 shows the total utility loss under the fabrication attack. As $\sigma$ increases above 40%, trust management can reduce the total utility loss significantly. Without trust management, the decision depends on the majority vote from the SUs. If the majority of the SUs present fabricated sensing results, the probability of miss detection error increases with increasing PU spectrum usage rate. Thus, when the PUs are using their designated spectrum bands more actively, the fabrication attack will cause more miss detection errors in the long run which, in turn, result in dramatic increases in the TUL. The simulation results with $\sigma < 40\%$ are not presented here since both CRN with and without trust management can handle this case well. The cumulative rates of false alarm error and misdetection error with the fabrication attack are shown in Figure 18. It can be seen that as the number of iterations increases, both false alarm rate and miss detection rate increase if trust management is not used. However, with trust management in place, both cumulative rates remain close to 0. It can be concluded that the trust management model is robust under the fabrication attack.

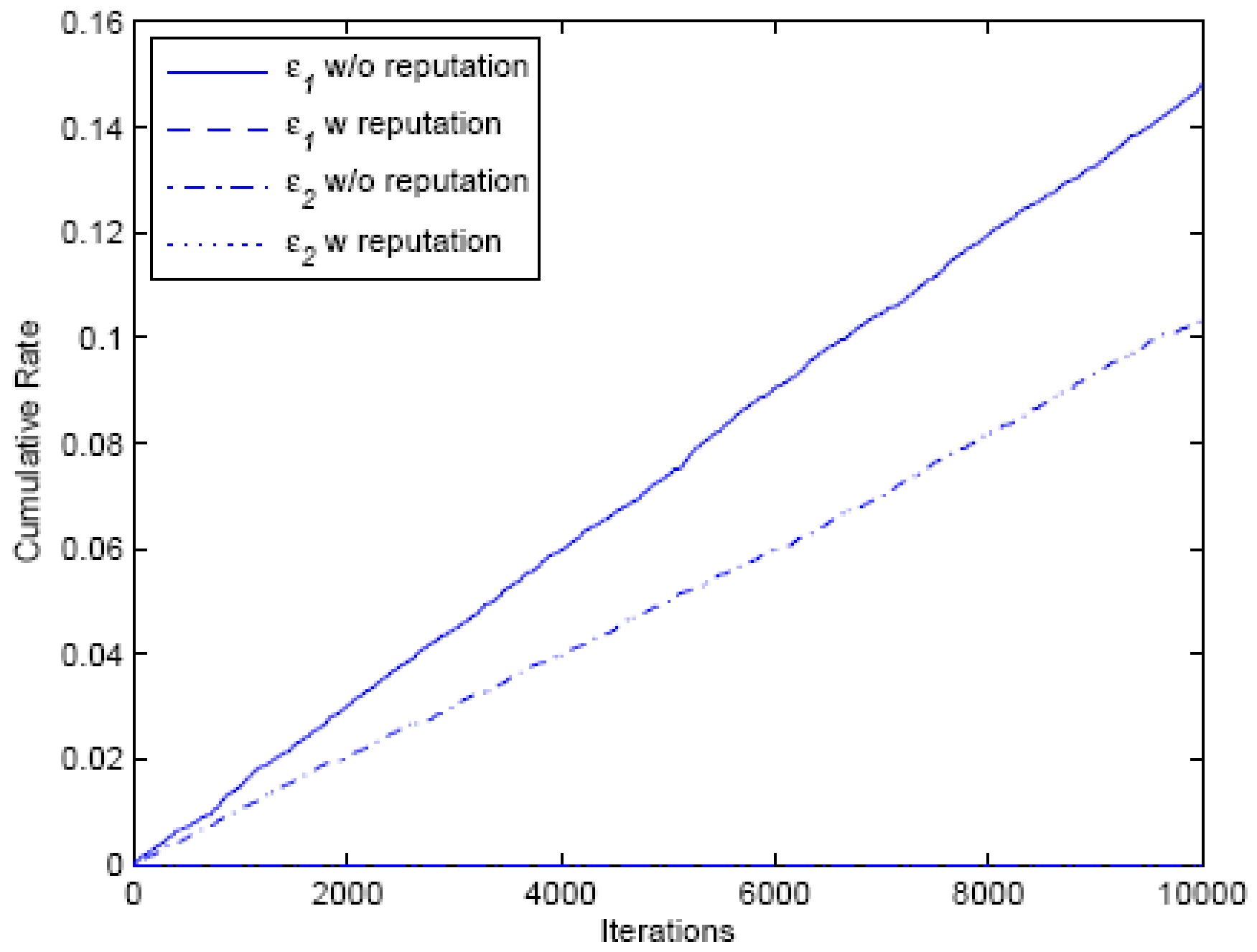

Figure 19. Cumulative false alarm rate, $\varepsilon_1$, and misdetection rate, $\varepsilon_2$, with the fabrication attack rate, $\sigma$ =50% for PU band usage rate at 45%.

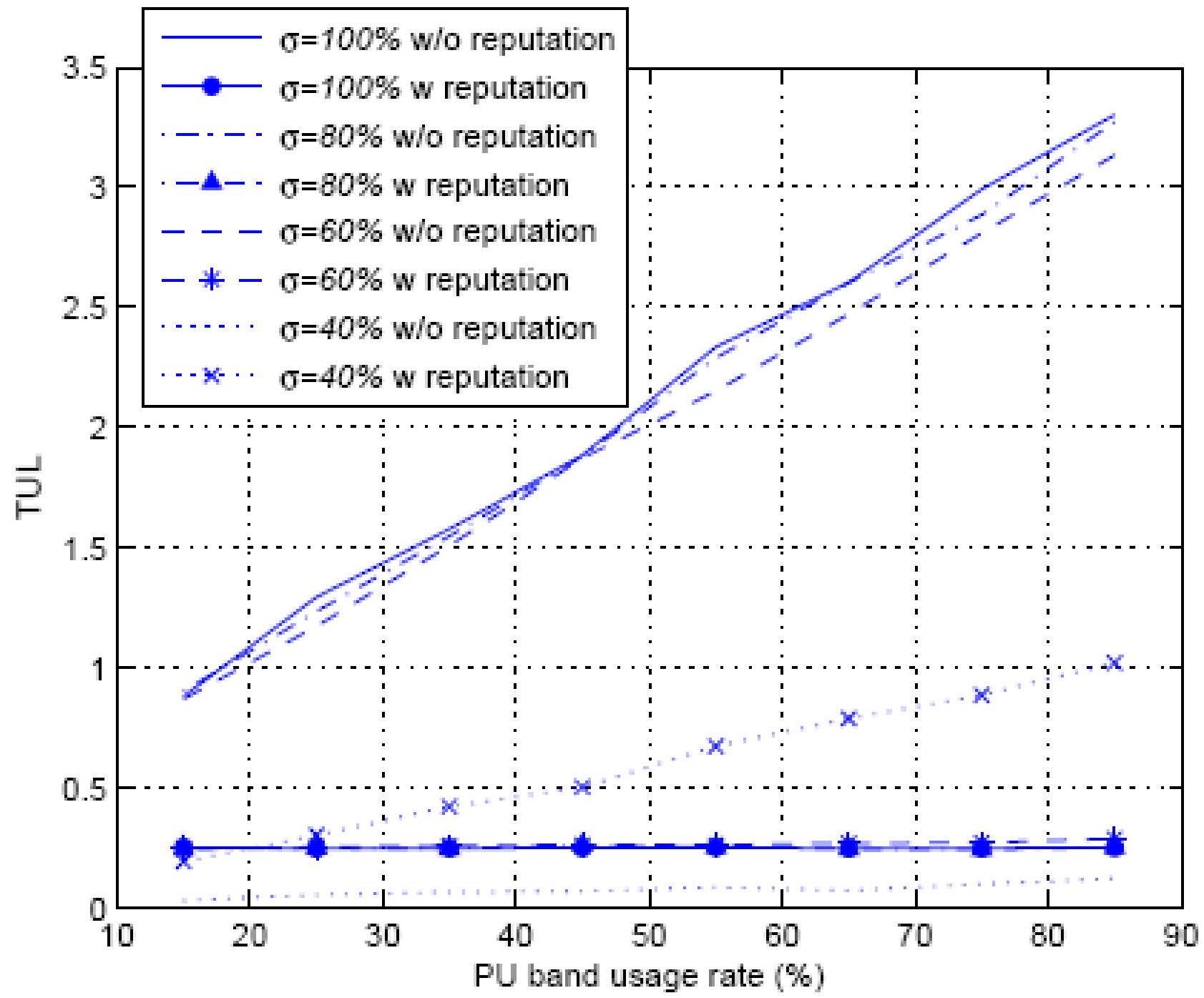

Figure 18. Total utility loss, TUL, of the CRN under the on-off attack.

## B. *On-off Attack*

The on-off attack refers to malicious SUs behaving honestly or dishonestly alternatively to avoid punishments while sending wrong sensing results on purpose. Unlike the fabrication attack in which the behaviors of the attackers are static, the on-off attack follows a dynamic pattern which makes the misbehaviors harder to detect.

Figure 19 shows the TUL as a function of PU band usage under high rate of the on-off attack. The results with an attack rate of below 40% are not shown since the TUL in these cases approaches 0. It can be seen that trust management reduces TUL greatly under heavy on-off attacks. However, we notice that at $\sigma$ = 40% the TUL with trust management is slightly higher than the one without trust management. Figure 20 shows that both false alarm and miss detection rates with trust management are higher than without trust management. This is because, without trust management, when the confidence level of SUBS for the sensing result falls below the threshold which is set at 0.25 and if there is 40% on-off attack rate, i.e. SUs who behave well with probability of 0.6, the majority voting in this situation is more likely to help the SUBS make a correct sensing decision. But with trust management, after filtering out most of the on-off attackers, the sensing result just depends on the SUBS and those whose trustworthiness scores are higher than $\eta$ which is set at 0.65, that is, SUBS has a higher probability to make a false alarm or miss detection error decision. This situation happens only when $(1-\sigma) \approx \eta$.

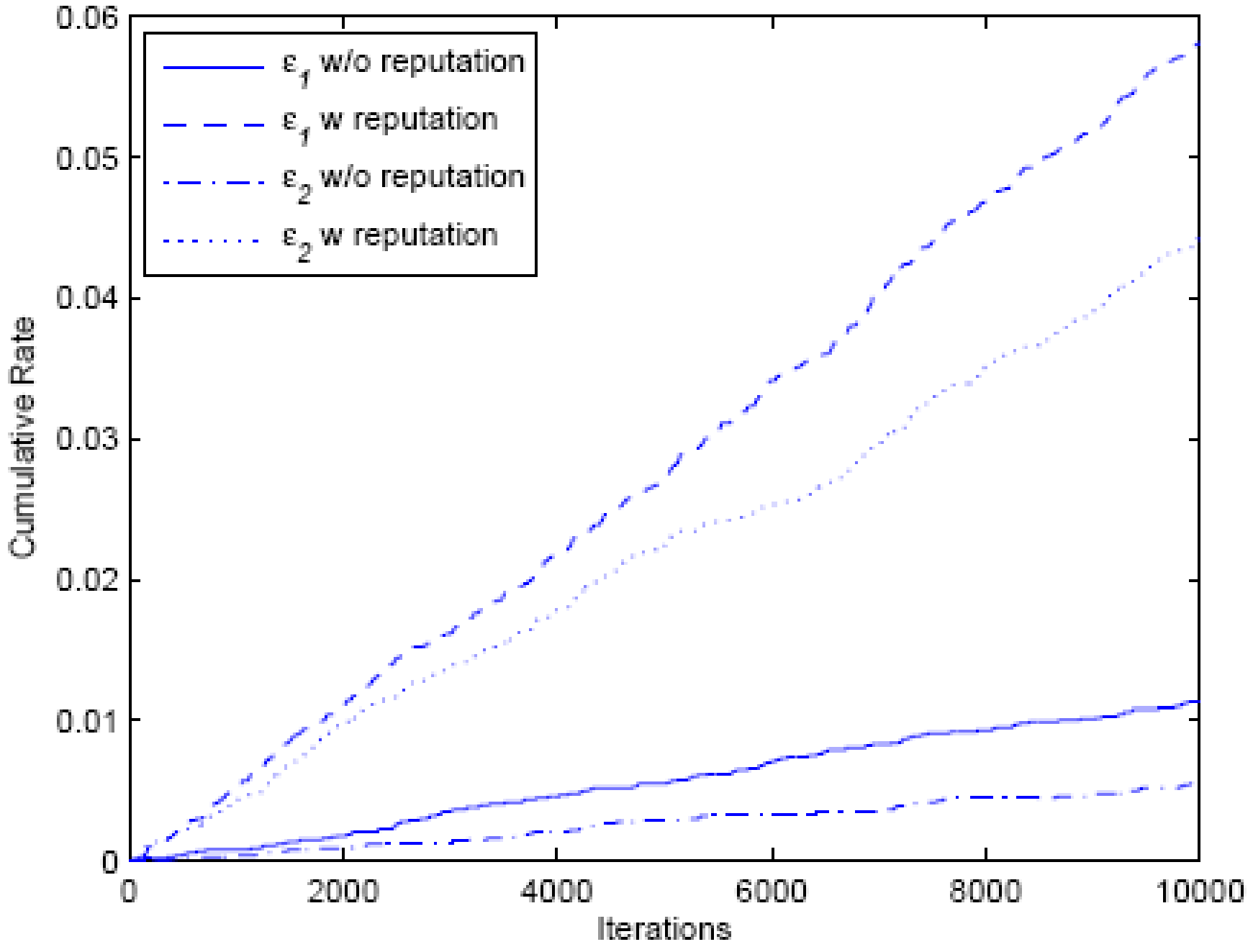

Figure 20. Cumulative false alarm rate, $\varepsilon_1$, and misdetection rate, $\varepsilon_2$, with the on-off attack rate, $\sigma$ =50% for PU band usage rate at 45%.

### C. *Denial of Service Attack*

The denial of service attack prevents SUs from utilizing the PU spectrum band. The attackers generate sensing results that will mislead the SUBS to make a false alarm error. This means the SUs will lose the opportunity to utilize the PU spectrum band which is actually not occupied by primary users. If the attacks are successfully conducted, the system performance will degrade sharply.

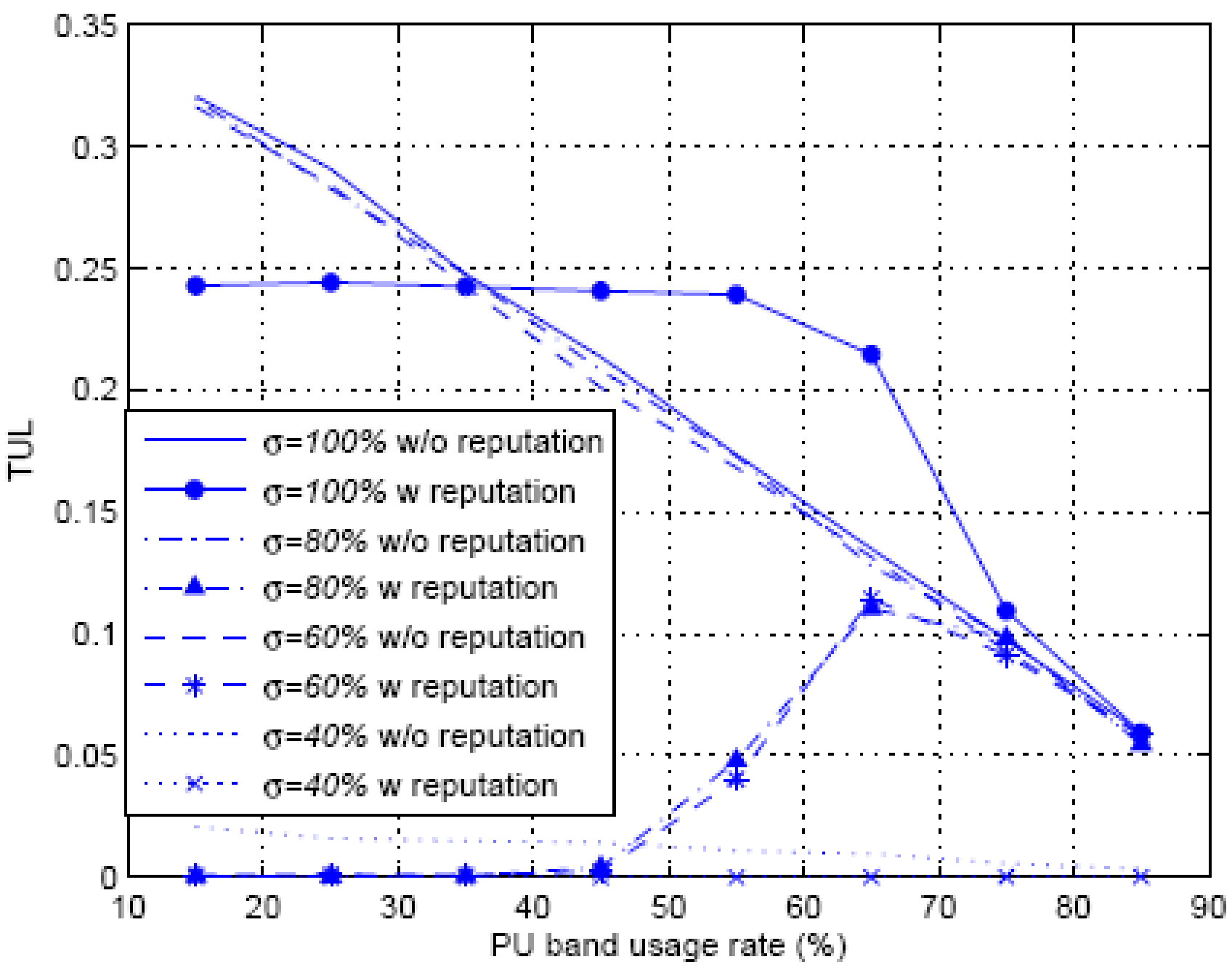

Figure 21. Total utility loss, TUL, of CRN under the denial of service attack.

Figure 21 shows the total utility loss under the denial of service attack. The TUL results with an attack rate below 40% are not presented because they are close to 0. It can be seen that trust management generally works well under this kind of attack. However, it can be seen that, when the attack rate is 100% and PU band usage is low, the TUL of trust management is higher than the one without trust management. Since with a population consisting entirely of denial of service attackers, the sensing decisions of SUBS depend solely on its own sensing results while the PU band usage rate is below $\eta$. Since there is no misdetection error under this kind of attack, the false alarm error rate is close to the SUBS confidence level threshold which is predefined.

### *D. Resource Hungry Attack*

In resource hungry attacks, malicious SUs always report to SUBS that the PU spectrum band is not in use. By doing so, they hope that the SUBS will make a miss detection error and consequently allocate resources to them. This kind of misbehavior, if successful, will introduce undue interference to the PU using the same spectrum band. This could lead to serious consequences such as the SUBS being shut down for a certain period of time which will impair the whole system significantly.

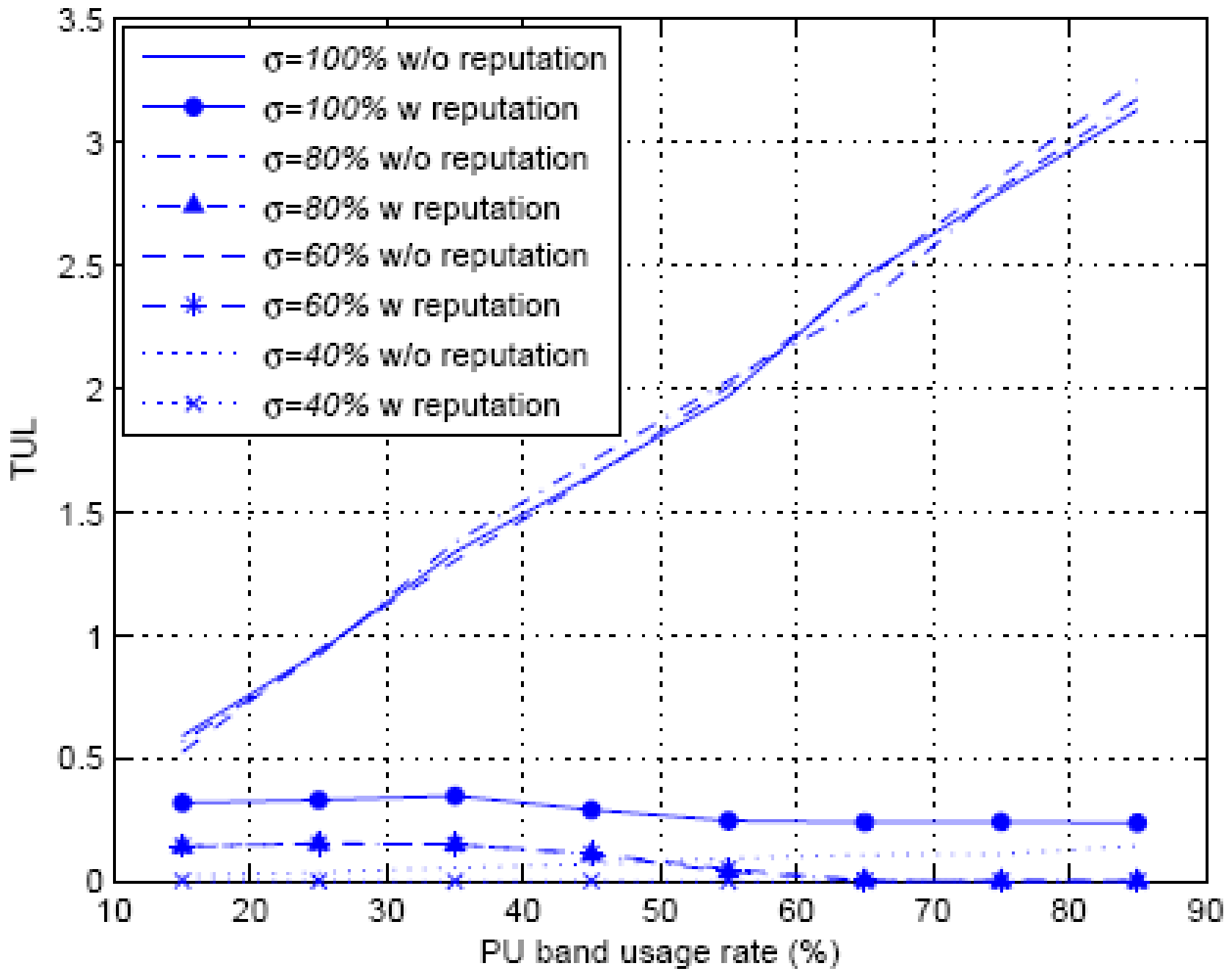

Figure 22. Total utility loss, TUL, of CRN under the resource hungry attack.

Figure 22 shows total utility loss, TUL, under the resource hungry attack. The results of TUL with the attack rate below 40% are not presented because they are close to 0. It can be seen that trust management significantly reduces the TUL compared to the one without trust management. Without trust management, TUL increases as PU band usage rate increases since more misdetection error occur and this will cause a large utility loss since the misdetection error is regarded as a serious offense in this CRN.

## 3.9. Summary

Biased third-party testimonies can distort the interaction decisions made by agents relying on trust models and cause havoc in an open MAS. In this chapter, we proposed a reinforcement learning based testimony aggregation model – the ACT model to address this problem. We go beyond the state of the art by providing an adaptive method for a truster agent to fine tune important parameters such as the selection of witness agents, the weight given to each of their testimonies as well as the weights given to the collective recommendation based on third-party testimonies and the direct trust evidence. By eliminating the need for human operators to help the agent fine tune these parameters, the ACT model enables truster agents to operate with more autonomy and adaptability when facing risk and uncertainty in an MAS. The ACT model is verified with extensive simulations under general multi-agent interactions as well as in the application domain of cognitive radio networks. The performance of the proposed model is upheld under various attack scenarios. In the next chapter, we will take a new perspective toward trust-aware interaction decision-making by removing one of the most widely used implicit assumptions in existing trust models.

# Chapter 4

# The Multi-agent Trust Game – A Preliminary

In Chapter 3, we have analyzed how the trust evaluation produced by a truster may be influenced by the behavior of other trusters in an MAS, and the proposed ACT model to address these problems. However, even with accurate trust information, trusters face an additional challenge. How should a truster delegate tasks to trustees given their reputation in such a way which not only maximizes its long term utility, but also does not negatively affect other trusters in the MAS? In this chapter, we formalize the trust-aware task delegation problem in MASs with resource constrained trustee agents into a trust game similar in spirit to the *Congestion Game*. We then analyze the unintended negative consequences produced by existing approaches in this environment. This chapter defines an important problem few researchers in this area have realized. Solutions for this problem will be proposed in Chapter 5 to 7.

## 4.1. The Unintended Consequence of Self-interested Trust-aware Decision-making

From Chapters 1 to 3, it is obvious that the most widely accepted method of studying the topic of trust management in MASs is through simulations which emulate the conditions of the target environments. Existing works often take an individual truster agent's perspective when evaluating the effectiveness of their proposed trust models. This gives rise to a variety of individual-performance-centric evaluation metrics. Two of the most widely adopted such metrics are the average accuracy rate of the estimated trustworthiness [Falcone *et al.,* 2004; Teacy *et al.,* 2005; Weng *et al.,* 2006; Reece *et al.,* 2007; Hang *et al.,* 2009; Matt *et al.,* 2010; Vogiatzis *et al.,* 2010; Liu *et al.,* 2011; Liu and Datta, 2011; Jiang *et al.,* 2012; Koster *et al.,* 2012; Piunti *et al.,* 2012] and the average utility achieved by individual truster agents [Dash *et al.,* 2004; Tran and Cohen, 2004; Huynh *et al.,* 2005; Zheng *et al.,* 2006; Fullam and Barber, 2007; Reches *et al.,* 2008; Teacy *et al.,* 2008; Kerr and Cohen, 2009; Liu *et al.,* 2010; Burnett *et al.,* 2010; Burnett *et al.,* 2011; Burnett and Oren, 2012; Haghpanah and desJardins, 2012].

To achieve high individual welfare, most existing trust models adopt a greedy decision-making approach after evaluating the reputations of trustee agents – a truster agent should interact with the most trustworthy trustee agent known to it as often as possible. This doctrine is consistent with the assumption that, in an open MAS, individual agents are self-interested and do not have a common goal to work towards. Although several types of methods for exploration for potentially better alternative trustee agents have been proposed [Weng *et al.,* 2010; Wierzbicki and Nielek, 2011], in general, this doctrine is upheld in these methods. Thus, the behavior exhibited by truster agents using most existing trust models to make interaction decisions tend to be - to concentrate most of their interactions with a few known trustworthy trustee agents.

In many existing trust models, such a general approach seems to yield good performance results. In order for this to happen, one implicit assumption is made by existing studies:

> [**The Unlimited Processing Capacity (UPC) Assumption**]: *trustee agents have unlimited request processing capacities compared to the number of incoming requests from truster agents at each time step*.

In MASs entirely consisting of software agents offering digital services, such an assumption is generally valid. Thus, from the perspective of a trustee agent, the more business it can attract from truster agents, the better off it will become. However, in MASs where the services being exchanged are performed by human beings and involving physical processes (e.g., e-commerce systems, crowdsourcing systems, etc.) which can be referred to as *human-agent collectives*, such an assumption stops being valid.

Evidence-based trust models depend on the feedback provided by truster agents in order to function. Such feedbacks are often regarded as a way to reflect the subjective belief in the quality of the result received by the truster agent from an interaction with a trustee agent. The quality of an interaction is judged by the commonly used quality-of-service (QoS) metrics suitable for the context of the interaction. It is often made up of two main aspects 1) metrics related to the correctness of the interaction results, and 2) timeliness of completion of the interaction. For example, when a truster agent *a* wants to send a message to another agent using the messaging service provided by trustee agent *b*, only if the message is successfully received by the recipient agent within the expected time will *a* consider the interaction with *b* to be successful.

Under the UPC assumption, the satisfaction of the correctness and timeliness requirements depends only on the intrinsic characteristics of trustee agents. The collective choice of interaction partners made by a population of truster agents has no effect on the observed performance of the trustee agents. However, when the UPC assumption is removed (i.e., trustee agents have limited process capacities for incoming requests), the timeliness aspect starts to be influenced by two factors 1) the process capacity (or effort level) of the trustee agent which is innate to the trustee agent, and 2) the current workload of the trustee agent which is exogenous to the trustee agent. In this situation, the trustee agent receiving a good feedback not only depends on its own trustworthiness, but also depends on the collective interaction decisions made by the truster agents. Here, we assume that truster agents do not purposely distort their feedbacks.

Interaction outcome evaluation is simple with the UPC assumption. Since results can always be assumed to be received on time, a truster agent just needs to produce a rating based on the correctness of the results. Such a rating can be binary (i.e., success/failure) [Jøsang and Ismail, 2002] or multi-nominal (i.e., on a scale of 1 to *n*) [Jøsang and Haller, 2007]. Nevertheless, existing literatures are generally vague about how a rating based on a received interaction result and a truster's own preferences can be derived. This is mainly because of the difficulty of designing a generic model to judge the correctness of a received result relative as this may be manifested in different ways for different domains of application. For example, in an e-commerce system, receiving a parcel containing the purchased item with expected quality may be consider a successful interaction result while, in a crowdsourcing system, receiving a file containing a piece of properly transcribed audio may be considered a successful interaction result. These ratings can be relatively easily produced by human beings, but are difficult for software agents to determine.

With the removal of the UPC assumption, the timeliness aspect of an interaction result needs to be explicitly taken into account when evaluating the outcome of that interaction. Keeping track of the completion time v.s. deadlines of a large number of interactions is a task that is more tractable for a software agent to perform and help lighten the cognitive load for its user. In this situation, the rating produced by the user based on the quality of the interaction result needs to be discounted by the timeliness of its reception in order to derive a feedback for future evaluation of the trustee agent's reputation.

Intuitively, if no result is received when the predetermined hard deadline has passed, the interaction should be rated as a failure. For example, agent $a$ depends on agent $b$ to provide it with a component in order to build a product and deliver to agent *c* by a certain date *t*, and *a* needs at least *n* days to assemble the product once the component from *b* is received. If *b* fails to make the delivery by day (*t* – *n*), there is no way for *a* to serve *c* on time. In this case, *a* will consider its interaction with *b* as a failure regardless of whether *b* delivers the component on the next day. In this case, the timeliness discount factor can be a binary function as:

$$f_{td}(T_{end}) = \begin{cases} 1, & T_{end} \leq T_{dl} \\ 0, & otherwise \end{cases} \quad (4.1)$$

where $T_{end}$ is the time at which the interaction ended and its result is received by the truster agent, and $T_{dl}$ is the stipulated deadline.

To distinguish the performance of different trustees further, the timeliness discount factor can be made into a smoother function with respect to the difference between $T_{end}$ and $T_{dl}$. The closer $T_{end}$ is to the time the interaction started ($T_{start}$), the closer $f_{td}(T_{end})$ should be to 1; the closer $T_{end}$ is to $T_{dl}$, the closer $f_{td}(T_{end})$ should be to 0. A simple example of such a function may be

$$f_{td}(T_{end}) = 1 - \frac{T_{end} - T_{start}}{T_{dl} - T_{start}}. \tag{4.2}$$

Incorporating the concept of timeliness discount into one of the most widely used trust evaluation models – the BRS [Jøsang and Ismail, 2002] by as:

$$\tau_{i,j} = \frac{\alpha}{\alpha + \beta}$$

$$\alpha = \sum_{k=1}^{N_p} p_k + 1, \beta = \sum_{k=1}^{N_p} n_k + 1$$

$$p_k = \begin{cases} p_k, & if\ f_{td}\left(T_{end}^k\right) > 0 \\ 0\ and\ n_k = 1, & if f_{td}\left(T_{end}^k\right) = 0 \end{cases}$$

(4.3)

where $\tau_{i,j}$ is the trustworthiness evaluation of trustee agent *j* from the perspective of truster agent *i*, $T_{end}^k$is the completion time of the *k*th interaction between *i* and *j*, $p_k$ equals to 1 if the *k*th interaction is successful and 0 otherwise. Here, we use Equation (4.1) as to calculate the timeliness discount. This extended version of BRS is referred to as *BRS2012*.

Although not explicitly designed to cope with the removal of the UPC assumption, existing trust models can operate under the revised condition. To understand the dynamics of existing trust models when the UPC assumption is removed, we design a simple experiment as follows. A simulated MAS consists of 200 trustee agents and 1,000 truster agents. At each time step of the simulation, a truster agent needs to engage the service of a trustee agent in order to achieve its goal. Truster agents employs a variation of *BRS2012* in which for a randomized 15% of the time, a truster agent will explore for potentially better alternative trustee agents by randomly selecting a trustee agent for interaction. The rest of the time, the truster agent always selects the known trustee agent with the highest trustworthiness value for interaction. The trustee agent population consists of 50% of agents who produce correct results 90% (*Hon*) of the time on average and 50% of agents who produce results 10% (*Mal*) of the time on average. Throughout the simulation, the behavior patterns of the trustee agent do not change. A trustee agent can serve at most 10 interaction requests in its request queue. A uniform deadline of 3 time steps is used for all interaction requests. Each simulation is run for 500 time steps and the reputation of each trustee agent (i.e., the average trustworthiness evaluation calculated from the local trust evidence of all truster agents who has interaction experience with that trustee agent) are recorded at each time step.

If the trustee agents are aware of the deadline requirements of the requests when the requests are accepted, they can periodically clean up their request queues to get rid of the pending requests whose deadlines have passed and inform the requesting truster agents of this decision. We call this operation *clean sweep*. Assuming the clean sweep operation is not used (i.e., the trustee agents keeps working on requests in their queues regardless of whether their deadlines have passed), the changes of five agents belonging to the *Hon* group of trustee agents are shown in Figure 23. The changes in trustee agents' reputation values as evaluated by *BRS2012* are as follows:

1) *Reputation Building Phase*: During this phase, the agent's (for example Agent 2's) reputation starts from a low or neutral level. At this stage, not many truster agents want to interact with this Agent 2. However, due to random exploration by some truster agents, Agent 2 can get some requests. Since its reputation is relatively low compared to those of other trustee agents, the workload of Agent 2 is likely to be within a level which it can easily handle. Since Agent 2 belongs to the *Hon* group of trustee agents, the quality of its service is high on average. Gradually, its reputation is built up due to the positive feedbacks received from satisfied truster agents.

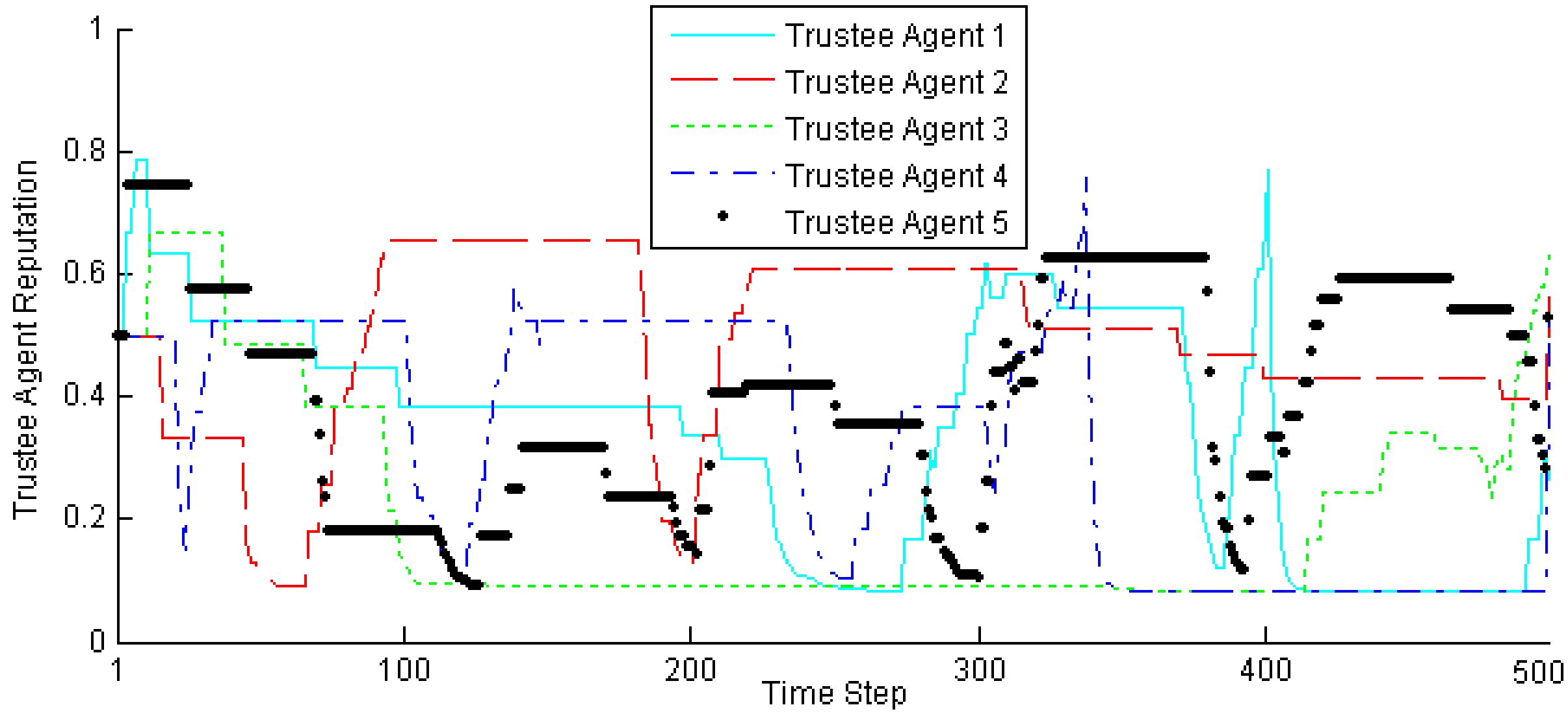

Figure 24. Changes in reputation values of five trustee agents from the *Hon* group under *BRS2012* without *clean sweep*.

2) *Reputation Damage Phase*: As Agent 2 builds up its reputation, it is known to an increasing number of truster agents. More truster agents start to request its services. Gradually, the workload of Agent 2 increases past its processing capacity which results in longer delays for some requests. As more requests fail to be served within their stipulated deadlines, negative feedbacks from disgruntled truster agents start to damage Agent 2’s reputation.

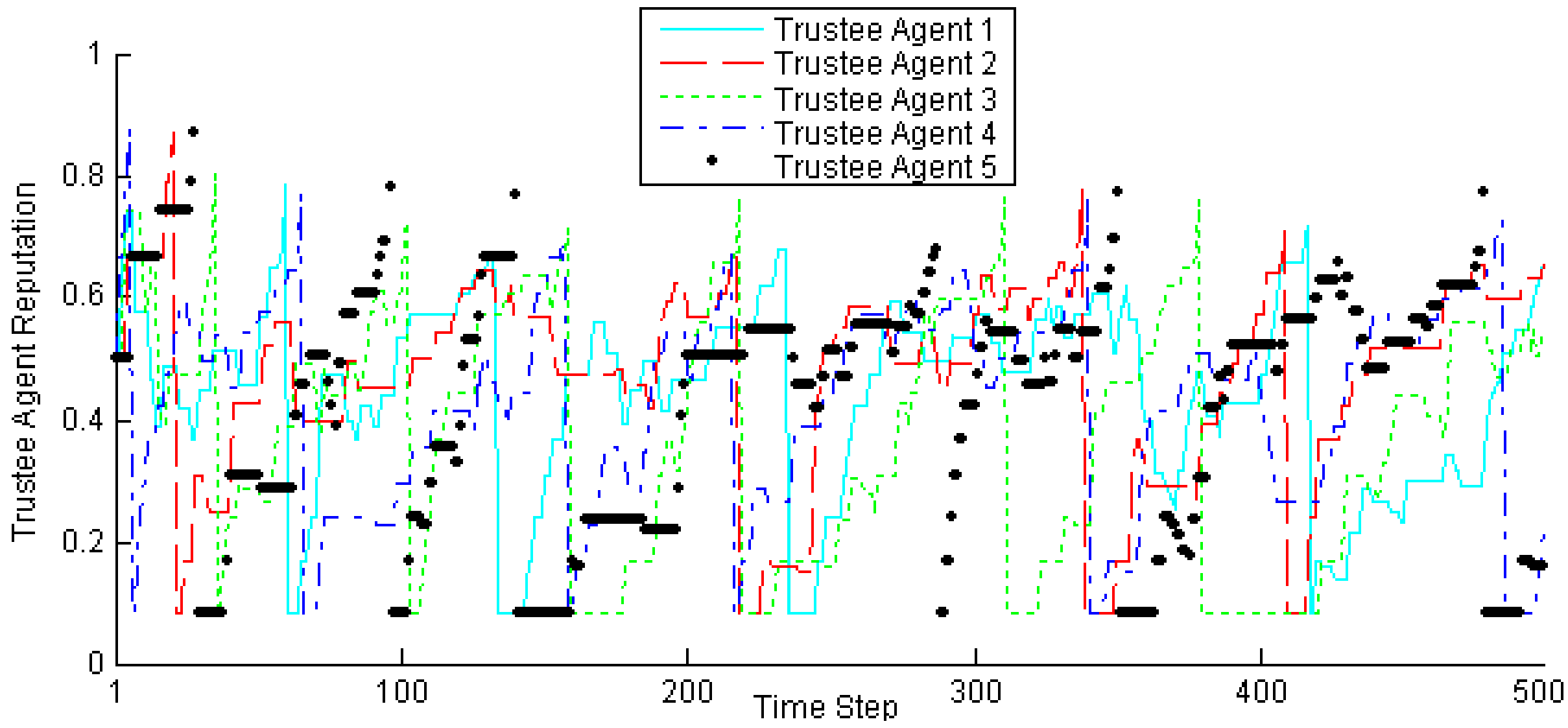

Figure 23. Changes in reputation values of five trustee agents from the *Hon* group under *BRS2012* with *clean sweep*.

From Figure 23, it can be seen that the reputation values of the trustee agents alternate between these two phases. Their reputation values fluctuate around an average of 0.5272 which is 41.42% lower than their actual trustworthiness which is 0.9.

Figure 24 shows the changes in trustee agents’ reputation values, as evaluated by *BRS2012*, with the use of clean sweep operations by the trustee agents. The two phases can still be observed although the lengths of their cycles have become visibly shorter than in the case of no clean sweep operation. This is due to the fact that once a clean sweep operation is performed, the truster agents are informed of the fact that their requests cannot be served by the trustee agents. Therefore, their negative feedbacks can be issued more quickly than in the case of no clean sweep operation and have an impact on the trustee agents’ reputation values. From Figure 23, it can be seen that the reputation values of the trustee agents alternate between these two phases. Their reputation values fluctuate around an average of 0.4298 which is 52.25% lower than their actual trustworthiness which is 0.9.

However, such a drastic deviation from the ground truth is not due to the fault of the trustee agents. On the contrary, they are victims of their own success. The greedy approach employed by truster agents when using the reputation evaluations to guide their interaction decisions with the aim of maximizing their own chances of success has caused the reputation evaluation to reflect not only the behavior pattern of the trustee agents, but also the impact of the collective interaction decisions made by the truster agents. Since the interaction decision-making mechanism employed by existing trust models have not taken this factor into account, this phenomenon results in instability in the MAS and negatively affects the wellbeing of trustee agents and truster agents. In this dissertation, we refer to this phenomenon *Reputation Damage Problem (RDP)*.

Among the factors affecting the perceived trustworthiness of a trustee agent, the timeliness of task completion is one that is affected both by the ability and willingness of the trustee agent as well as the task delegation decisions made by the truster agents. If the RDP is not mitigated, the resulting reputation values will not fully reflect the behavior of the trustee agent, and become biased with the influence of the environment in which the trustee agent operates. In this case, the reputation value will lose its fairness and become less useful in guiding the decision making process of truster agents in subsequent interactions.

## 4.2. Multi-agent Trust as a Congestion Game

Congestion games are a type of games where the payoff for each player depends on the resources it selects and the number of players selecting the same resources. For example, commuting to work can be modeled as a congestion game where the time taken by an agent from its home to its workplace depends on how many other agents are taking the same public transport it chooses on each day. In a Discrete Congestion Game (DCG) [Monderer and Shapley, 1996], the following components are present:

- A base set of congestible resources $\boldsymbol{E}$;
- $n$ players;
- A finite set of strategies $\boldsymbol{S_i}$ for each player where a strategy $P \in \boldsymbol{S_i}$ is a subset of $\boldsymbol{E}$;

- For each resource $e$ and a vector of strategies $(P_1, \ldots, P_n)$, a load $x_e$ is placed on $e$;
- For each resource $e$, a delay function $d_e$ maps the number of players choosing $e$ to a delay represented by a real number;
- Given a strategy $P_1$, player $i$ experiences a delay of $\sum_{e \in P_i} d_e(x_e)$, assuming that each $d_e$ is positive and monotonically increasing.

The root cause of the RDP is similar to the situation described in the DCG. The performance perceived by a truster agent delegating tasks to a reputable trustee agent it has found is partially dependent on how many other truster agents are making the same choice as it is at the same time. A key difference in our case is that the perceived performance also depends on the behavior of the trustee agent which is uncertain in an open MAS. In order to give the RDP a concrete theoretical grounding, we formalize it as a special type of DCG where trustee agents (resources) may behave maliciously. We call this formalization the Multi-agent Trust Game (MTG).

[**Definition 4.1** (Multi-agent Trust Game)]: A multi-agent trust game is specified by an ordered quadruple $\mathcal{q} = (\boldsymbol{E}, \vec{\ell}, \boldsymbol{Cn}(t), \boldsymbol{V}(t))$ where

- $\boldsymbol{E}$ is a finite set of trustee agents with limited task processing capacity per unit time;
- $\vec{\ell}$ is a vector of latency functions expressing the delay experienced by truster agents selecting the set of trustee agents for task delegation. The $e$-th component of $\vec{\ell}$ is denoted as $\ell_e$. It is a non-decreasing function mapping from $\mathbb{R}^+$ to $\mathbb{R}^+$;
- $\boldsymbol{Cn}(t)$ is the set of possible connections between a finite set of truster agents $\boldsymbol{W}$ and trustee agents in $\boldsymbol{E}$ in the MAS when delegating their tasks to trustee agents in $\boldsymbol{E}$ at time step $t$. The connections depend on what types of tasks each truster agent wants to delegate to each trustee agent and the qualifications of the trustee agents to perform these tasks. For example, truster agent *A* may want to find a trustee agents who sells apples at time step $t$. A subset of trustee agents $\boldsymbol{C}'(t)$ are qualified for this task. At time step $t+1$, $A$ is looking to buy computers. Therefore, a different subset of trustee agents $\boldsymbol{C}''(t+1)$ are qualified for this task. The difference between the set of possible connections between truster agents and trustee agents can also be caused by agents dynamically joining or leaving an MAS;

- $\boldsymbol{V}(t)$ is the set of functions for calculating the potential cost for truster agents who choose to delegate tasks to the same trustee agent at time step *t*.

[**Definition 4.2** (Task Delegation Flow)]: A task delegation flow in the MTG is a function *f* mapping from $\boldsymbol{Cn}(t)$ to $\mathbb{R}^+$. It can be regarded as the amount of workload assigned to a trustee agent.

[**Definition 4.3** (Task Delegation Cost)]: The load on a trustee agent $e \in \boldsymbol{E}$ is

$$x_e(f) = \sum_{c(t)\in\boldsymbol{C}(t)|e\in c(t)} f(c(t)). \tag{4.4}$$

The delay on a connection $c(t) \in \boldsymbol{C}(t)$ is

$$L(c(t)) = \sum_{e\in c(t)} \ell_e\,(x_e(f)). \tag{4.5}$$

The cost of a task delegation is

$$v(f) = \sum_{c(t)\in\boldsymbol{C}(t)} \frac{f\big(c(t)\big)L(c(t))}{\tau_e(t)} = \sum_{e\in\boldsymbol{E}} \frac{x_e(f)\ell_e(x_e(f))}{\tau_e(t)} \tag{4.6}$$

where $\tau_e(t)$ is the reputation of trustee agent e in performing a given type of tasks as assessed at time step *t*.

## 4.3. Summary

In this chapter, we analyzed the adverse effect of the reputation damage problem that, left to chance, can happen under the greedy trust-aware interaction decision-making approach which is widely adopted by existing trust models in MAS. We propose a modified version of the BRS – the BRS2012 to explicitly take into account the timeliness of completion of a task delegated to a trustee agent. Under such an evaluation model, the reputation values of trustee agents who are programmed to be trustworthy fluctuate significantly during the course of interaction in a simulated MAS. We further propose a formalization of trust-aware interaction decision-making problem into a multi-agent trust game based on the concept of discrete congestion games. In the Chapters 5 to 7, we will investigate possible approaches that can provide solutions to this type of games and help agents in MASs make better trust decisions. Two different approaches will be taken to address the RDP: 1) through coordinating how truster agents make decisions (Chapter 5 and 6), and 2) through helping trustee agents to decide how to respond to uncoordinated truster agent decisions (Chapter 7).

# Chapter 5

# SWORD: A Centralized Decision Approach for Social Sustainability in Multi-Agent Systems

In Chapter 4, the problem of delegating tasks to trustees with capacity constraints in practical applications has been formulated into a multi-agent trust game based on the congestion game. It provides a framework under which trust-aware task allocation can be analyzed in a principled way to derive appropriate solutions. In this chapter, we aim to propose a task trust-aware task allocation approach – SWORD - to be used by a central trusted entity in an MAS. The objective is to compute a decision policy in real time to maximize the total expected utility of all agents in the MAS over potentially infinite horizon of interactions. We demonstrate that the SWORD approach balances the consideration of multiple, and potentially conflicting objectives in a situation-aware manner, and is capable of producing results in polynomial time. Through theoretical analysis, we prove the existence of various performance bounds that make SWORD a boundedly-optimal approach.

## 5.1. System Model and Problem Definition

In order to address the RDP defined in Chapter 4, we first enrich the system model used by existing trust research to expressly include the timeliness consideration and remove the UPC assumption. From the perspective of trust management, an MAS consists of a set of truster agents $\boldsymbol{R}$ requesting for services from a set of trustee agents $\boldsymbol{W}$ who work on these requests. For simplicity, in the following parts of this section, we will refer to truster agents as *requesters* and trustee agents as *workers*. A requester $r \in \boldsymbol{R}$ may need to request more than one services at each time step. These services can be of the same type (e.g., 5 book chapters requiring proof reading) or different types (e.g., 2 book chapters requiring proof reading and 2 videos requiring transcription). These tasks are represented by a task group $\boldsymbol{H}(t) = (h_1, h_2, \dots, h_N)$ proposed at time $t$. The requester will look for suitable workers to complete these tasks based on their reputations. As the time and skills of a worker $w \in \boldsymbol{W}$ are limited, each worker can only complete up to a maximum of $\mu_w^{max}(t)$ tasks per time step without sacrificing quality. We assume $\mu_w^{max}(t)$ is an innate characteristic of $w$ and will not vary significantly from time to time. Once $r$ receives the results, he can review whether they are up to the required standard.

The outcome of an interaction is denoted as $O_w^r(t)$. It reflects the performance of $w$ from the perspective of requesters $r$ receiving the results at time step $t$ and is assumed to be a binary value [0, 1]. If $f_{td}(t) = 1$ and $O_w^r(t) = 1$, $r$ will give $w$ a positive feedback rating; otherwise, $r$ will give $w$ a negative feedback rating. A series of such ratings reported by various requesters can be converted into an evaluation of $w$'s reputation $\tau_w(t)$ using a trust evaluation model. There are many different trust evaluation models have been proposed [Noorian and Ulieru, 2010], but they are not the focus of our study here.

Let $\boldsymbol{A}(t) = \big(A_w(t)\big), w \in \boldsymbol{W}$ be a vector of decision variables representing the number of tasks assigned to the workers in an MAS at time step $t$, and let $\boldsymbol{\mu}(t) = \big(\mu_w(t)\big), w \in \boldsymbol{W}$ be a vector representing the number of tasks the workers in an MAS completed at time step $t$. The values $A_w(t)$ and $\mu_w(t)$ are non-negative integers for all $w \in \boldsymbol{W}$. For each task allocated to a worker $w$ by a requester $r$, a utility cost $c_h(t)$ is incurred by $r$. In this study, we assume this cost to be a constant value $c$ for all tasks of the same type. If the task is completed on time and considered to be of acceptable quality by the requester $r$, a utility reward is derived by $r$. In this study, we assume this utility is the same for all tasks of the same type and represented by $g_{max}$; otherwise, 0 reward is derived.

Assuming each task requires one time step to complete. At each time step $t$, a worker $w$ may complete $\mu_w(t)$ HITs according to the following constraints:

$$\mu_w(t) \in \{0, \dots, \mu_w^{max}(t)\} \text{ for all } w \in \boldsymbol{W} \tag{5.1}$$

$$\mu_w(t) \leq Q_w(t) \text{ for all } w \in \boldsymbol{W} \tag{5.2}$$

where $Q_w(t)$ is a variable representing the total number of pending tasks in a worker $w$'s task queue at any time step $t$. Constraint (5.1) ensures that only an integer number of no more than $\mu_w^{max}(t)$ tasks can be completed by a worker $w$ at any time step. Although the value of $\mu_w^{max}(t)$ may change over time, it is generally assumed to be remaining relatively stable over time. Constraint (5.2) requires that the number of HITs completed be less than or equal to the current number of tasks in $w$'s task queue.

The objective of our study is to design a Trust-aware Interaction Decision-making (TID) approach that maximizes the *social welfare* for a given MAS which is defined as the sum of utility gain from all requesters minus the sum of the cost of allocation tasks to workers:

$$U(t) \triangleq \sum_{w:w\in\boldsymbol{W}} [g(t)\mu_w(t) - cA_w(t)]$$

$$g(t) = \begin{cases} g_{max}, & if\ f_{td}(t) = 1 \text{ and } O_w^r(t) = 1 \\ 0, & otherwise \end{cases}. \quad (5.3)$$

Since an MAS may exist for an unknown period of time, the time averaged social welfare is a better gauge into the performance of a TID approach:

$$\overline{U} \triangleq lim_{T\to\infty} \frac{1}{T} \sum_{t=0}^{T-1} U(t). \quad (5.4)$$

## 5.2. Similarities and Differences with the Constraint Optimization Problem

A Constraint Optimization Problem (COP) is usually defined as follows:

- It consists a set of *n* variables $Var = \{x_1, \dots, x_n\}$;
- Each of the variables may be assigned values from discrete and finite domains $Dom = \{D_1, \dots, D_n\}$;
- A set of cost functions $f = \{f_1, \dots, f_m\}$ where each $f_i(d_{i,1}, \dots, d_{i,j})$ is $f_i: D_{i,1} \times \dots \times D_{i,j} \to N$.

The problem is to find an assignment of values to the variables $A^* = \{d_1, \dots, d_n | d_j \in D_i\}$ which minimizes the global cost *F*, which is defined as:

$$F(A) = \sum_{i=1}^{m} f_i(A). \quad (5.5)$$

The objective of our study is set in the same framework as COP. The set of variables, in our case, is the new task requests to be assigned to each worker at every time step. Instead of minimizing a global cost function, we aim to maximize a global reward function with the cost of operation being infused in it – the social welfare of an MAS. The computational complexity of COP has been shown to be above NP-Hard. Nevertheless, there are several features in our proposed system model that make the problem even more challenging than COP.

Firstly, it is generally assumed by COP solutions that agents in an MAS are working towards a common goal which is expressed by the objective function. Thus, most solutions to multi-agent COP assumes that a logical neighbourhood among agents exists and agents are willing and able to communicate with their neighbours in order to achieve their common goal [Mailler and Lesser, 2004]. This is true for cooperative MASs. However, in open and dynamically formed MASs which are the focus of study of the trust management field, competition rather than cooperation is assumed to be the main relationship among workers and among requesters. Without adequate incentives, self-interested agents from diverse backgrounds who are not working towards a common goal of maximizing the social welfare of an MAS cannot be assumed to be willing to expend their own resources to communicate with each other.

Secondly, some variants of the COP (e.g., Synchronous Branch and Bound (SBB) and Iterative Distributed Breakout (IDB) [Hirayama and Yokoo, 1997]) require that an agent may be able to know the value assigned to variables by other agents (a.k.a. *inter-agent constraints*). Again, in the case of an open dynamic MAS where agents are self-interested and competing with each other to maximize their own utility, such information cannot be assumed to be available.

Thirdly, some variants of the COP [Dantzig and Thapa, 2003] assume the probability of variations in agents' characteristics to be known. This is not the case in our system model. The reputation of a worker depends on both its own characteristics and the collective interaction decisions made by requesters at each time step. Thus, it does not have a clearly defined probability distribution. The dynamics of the change in a worker's workload depends on its reputation. Therefore, it, too, does not have a clearly defined probability distribution.

These unique features in our system model pose new challenges to COP. It requires an approach that can translate the objective function in (5.4) into an expression that can be directly related to the strategy for request distribution that resolves the conflict between individual agents' self-interest and the social welfare of the entire MAS. As clearly defined probability distributions for key metrics in our system model are not available, the solution approach must be adjusted based on observed performance feedbacks from previously iterations. In the following sections, we first propose a centralized reputation-aware decision-making approach based on the theory of *Lyapunov Drift*. Then, in Chapter 6, we further refined this approach to enable it to operate in a distributed fashion.

## 5.3. The SWORD Approach

To achieve the objective of maximizing social welfare in a trust-aware MAS, we first analyze what considerations constitute a good request distribution plan. Since social welfare is defined as the summation of the welfare of individual agents in an MAS, intuitively, a good request distribution plan should satisfy the following two properties:

- Maximizing the average correctness of the interaction results; and

- Maximizing the total number of requests a given MAS can complete per time step.

Ideally, if the worker population in a given MAS entirely consists of highly trustworthy workers and each of them are able to process a large number of request at each time step, both properties can be satisfied in the long run by randomizing allocation of requests to workers. However, in practical systems, these two properties are often in conflict with each other.

Since highly trustworthy workers may only be a minority, a request assignment approach should achieve a balance between these two properties in order to maximize the social welfare of an MAS. We hypothesize that such balance can be achieved through satisfying the following two requirements:

- *Just Fairness*: where more trustworthy workers should receive more requests than less trustworthy workers over the long run; and
- *Distributive Fairness*: where workers whose reputation values are close to each other should receive similar number of requests over the long run.

Just fairness affects the quality of the results received which while distributive fairness affects total number of requests a given MAS can complete per time step. Essentially, the adverse effect of the RDP needs to be mediated in order to improve social welfare over existing trust management models.

The root cause of the RDP is congestion of requests at a small number of reputable workers. A solution to such a problem could be an algorithm for choosing control actions over time in reaction to the existing state each worker is in. The state consists of both the innate characteristics of the worker as well as the exogenous factors. In our study, the state of worker is represented by the 3-tuple $S_w(t) = \{\tau_w(t), \mu_w^{max}(t), Q_w(t)\}$. Inefficient control actions incur larger backlog in some workers' task queues. These backlogs can act as sufficient statistic on which to base the next control decision. This makes it possible to optimize the objective function in (7) without knowledge of the probability associated with changes in $S_w(t)$.

### 5.3.1. The Theory of *Lyapunov Drift*

The theory of *Lyapunov Drift* [Neely, 2010] provides a way to measure the backlog in a population of agents. It defines a function $L(t)$ as the sum of squares of backlog in all queues in time step *t*. This is called a *Lyapunov function* and is a scalar measure of the level of congestion. If $L(t)$ is small, then all queues are short; and if $L(t)$ is large, then at least one queue is long. Use $\Delta(t) = L(t+1) - L(t)$ to represent the difference between the Lyapunov function from one time step to the next. If control decisions are made to greedily minimize $\Delta(t)$, then the backlogs of the queues are consistently pushed towards lower levels of congestion. Intuitively, this helps improve the distributive fairness among workers with similar level of trustworthiness.

Minimizing the Lyapunov drift is only helpful in increasing the total number of requests that can be completed per time step for a given MAS by more efficiently utilizing the workers' capacities. The requirement for the average quality of the interaction results still need to be incorporated. The objective function can be mapped to an appropriate reward function $reward(t)$. Instead of greedily minimize only the Lyapunov drift, control actions are taken at every time step $t$ to greedily minimize the following *drift-minus-reward* expression:

$$a \times \Delta(t) - b \times reward(t) \tag{5.6}$$

where $a$ and $b$ are positive constant values representing the relative importance given to the two components of this expression. Maximizing the negative of $\Delta(t)$ is equivalent to minimizing $\Delta(t)$. Since both $a$ and $b$ are positive constant values, the expression can be further simplified by dividing it with $b$. It then becomes:

$$\Delta(t) - V \times reward(t). \tag{5.7}$$

Then, $V \in R^+$ is the only control parameter that needs to be chosen by the system administrator for the MAS. If $V = 0$, then only the Lyapunov drift is minimized. If $V > 0$, then the weighted reward term is included into the control actions to arrive at a smooth trade-off between queue backlog reduction and utility maximization. To the best of our knowledge, this research is the first introduction of this form of Lyapunov Drift to enhance social welfare in open and non-cooperative MASs.

### 5.3.2. Making Situation-aware Trusting Decisions

In this chapter, we reframe the objective function in (5.4) based on the principle of Lyapunov Drift and propose simple Social Welfare Optimizing approach for Reputation-aware Decision-making (SWORD). The SWORD approach will be part of the administration system in an MAS (e.g., the transaction management system in an e-commerce platform) to act as a broker between requesters and workers. By being a centralized approach, its domain of application may be limited. However, with the advance of cloud computing, scalability and resource constraints have become less critical issues for centralized modes of operation. In many large scale systems deploying reputation management in real world applications (e.g., eBay.com, Taobao.com,

tripadvisor.com, etc.), feedbacks and ratings from users are stored in a centralized manner by the system rather than in a distributed manner by individual users. For this type of systems, a centralized approach like SWORD is useful. Therefore, it is assumed that the proposed approach is able to access the state information about workers regarding their past performance in terms of serving requests, the statistical information about the maximum number of requests they can process per time step, and the number of requests currently pending in their request queues.

Since SWORD is mainly concerned with making interaction decisions based on the reputation values of the workers, it relies on existing reputation evaluation models to supply it with the reputation values. In this study, we choose *BRS2012* to perform this task. Nevertheless, many other models can be used in place of *BRS2012* as long as the reputation values they produce can be normalized to a range of [0, 1]. However, they are not the focus of this study.

The output produced by the SWORD approach is a request allocation policy that determines the control variables $\boldsymbol{A}(t)$ for all workers in a given MAS when new requests are submitted to the system by requesters. $\boldsymbol{A}(t)$ is constrained as follows:

$$A_w(t) \in \{0,1,2,\ldots,\theta_w(t)\} \text{ for all } w \in \boldsymbol{W}, \text{if } \tau_w(t) \geq Th_r \quad (5.8)$$

Where $0 \leq \tau_w(t) \leq 1$ is the estimated reputation of worker *w* at time step *t*, and $Th_r$ is the minimum reputation threshold specified by the requester *r*. $\theta_w(t)$ is the target HIT queue length for worker *w*. It is dynamically determined by the SWORD approach based on the ability of each worker as:

$$\theta_w(t) \triangleq N\mu_w^{max}(t) + Vg_{max}\tau_w^{max}(t). \quad (5.9)$$

where $\tau_w^{max}(t)$ is the maximum reputation of a worker $w$ observed over a given period of time, the values of the parameters *N* and *V* determines the relative importance given to the expected waiting time and the expected quality of the results, and need to be selected by the system administrators based on their preferences.

A *Lyapunov function* $L(\boldsymbol{Q}(t))$ is constructed to as the sum of the squares of difference between the actual queue length and the target queue length for each worker at time step *t*:

$$L(\boldsymbol{Q}(t)) \triangleq \frac{1}{2}\sum_{w:w\in\boldsymbol{W}} \omega_w(Q_w(t) - \theta_w(t))^2. \quad (5.10)$$

$\{\omega_w\}_{w\in\boldsymbol{W}}$ are a collection of positive weights to enable different queues to be treated differently. Here, we use $\omega_w = 1$ for all *w* since we do not have prior knowledge of the relative importance of the task queues at each worker in an open MAS. Define $\Delta(\boldsymbol{Q}(t))$, the *per time step conditional Lyapunov drift*, as

$$\Delta(\boldsymbol{Q}(t)) \triangleq \mathbb{E}\{L\big(\boldsymbol{Q}(t+1)\big) - L(\boldsymbol{Q}(t))|\boldsymbol{Q}(t)\} \quad (5.11)$$

which is the expected difference between the value of (5.10) at time step *t+1* and that at time step *t* given the current task queue lengths of all the workers at time step *t*. It should be minimized if enough new requests are being proposed by requesters in a given MAS to ensure workers who can serve more requests per time step are assigned more requests and the task queues will not keep growing longer. In addition to this, the potential gain (which is directly corelated to the selected workers' reputations) by requesters should also be maximized. Based on these considerations, we define a variation of the *drift-minus-reward* expression – the *drift-plus-penalty* expression - as follows whose bound needs to be minimized:

$$\Delta\big(\boldsymbol{Q}(t)\big) + V \cdot \mathbb{E}\{P(t)|\boldsymbol{Q}(t)\} \tag{5.12}$$

where $P(t)$ is a penalty function representing the risk exposure involved in allocating requests to a selection of workers. The expectations in (5.11) and (5.12) are not clearly defined because the state transition probability distributions depend partially on the control actions taken by the SWORD approach at each time step. Therefore, instead of estimating them, the SWORD approach attempts to ensure the constraints in (5.8) and the drift conditions in (5.10) are satisfied in each time step. For a worker *w*, at time step *t*, when $A_w(t)$ new requests are assigned to it, the term for potential penalty involved in (5.12) can be expressed as:

$$P(t) \triangleq \sum_{h=1}^{A_w(t)} [p_h(t) + c]. \tag{5.13}$$

Assuming a worker stores different types of requests into separate queues, with regard to one of the queues, the maximum reward for successfully serving requests of the same type is denoted by the constant $g_{max}$. $p_h(t)$ can be estimated using the reputation of the worker *w* (i.e., $\tau_w(t)$). Thus, (5.13) becomes:

$$p_h(t) \triangleq [\big(1 - \tau_w(t)\big) \cdot g_{max} + c] A_w(t). \tag{5.14}$$

In essence, $p_h(t)$ represents the potential waste of utility which should be gained from allocating $A_w(t)$ new requests to *w* at time step *t*. In our case, instead of pushing the congestion level in each worker's request queues towards 0, it is more desirable to push it towards their respective target queue sizes $\theta_w(t)$. This is to ensure the rate of utilization of the workers' capacities to be maintained at a reasonable level if there are enough incoming requests. When $A_w(t)$ new requests are assigned to worker *w* at time step *t*, the Lyapunov drift which is related to both the incoming requests and the current queue size in relation to the target queue size can be expressed as:

$$\Delta\big(Q(t)\big) \triangleq (Q_w(t) - \theta_w)A_w(t). \qquad (5.15)$$

Thus, by combining (5.14) with (5.15), the overall *reward-minus-drift* expression for all the request queues for requests of the same type at each time step *t* is:

$$\sum_{w:w\in\boldsymbol{W}}\{V\big[\big(1-\tau_w(t)\big)\cdot g_{max} + c\big]A_w(t) + (Q_w(t) - \theta_w(t))A_w(t)\}. \qquad (5.16)$$

Minimizing (5.16) at each time step is equivalent to minimizing (5.12). In this way, instead of trying to estimate the probability distribution representing what the queue drift may be given each configuration of current queues among the workers in a given MAS, the SWORD approach uses the observed current queue sizes and the potential reward to guide future request assignment decisions. Thus, it determines how many request to be assigned to which workers by choosing $\boldsymbol{A}(t) = \big(A_w(t)\big), w \in \boldsymbol{W}$ so as to minimize (5.16) which can be simplified to:

Minimize: $\sum_{w:w\in\boldsymbol{W}}\{V\big[\big(1-\tau_w(t)\big)\cdot g_{max} + c\big] + Q_w(t) - \theta_w\}A_w(t)$

Subject to: Constraint (5.8)

Since we assume that the cost for serving a request is constant, request assignment decisions for each worker can be made independently. When new requests are submitted by requesters to the SWORD broker, the desirability score ($D_w(t)$) for all workers are calculated as:

$$D_w(t) = \theta_w(t) - Q_w(t) - V\big[\big(1-\tau_w(t)\big)\cdot g_{max} + c\big]. \qquad (5.17)$$

Therefore, the SWORD approach calculates $\boldsymbol{A}(t)$ as illustrated in Algorithm 5.1. The computational complexity of the SWORD approach is $O(|\boldsymbol{W}|)$.

The SWORD approach spends a fixed $\Pr(Exp)$ percentage of time exploring for unfamiliar workers. During this process, HITs are randomly allocated to workers in the MAS. Workers with low desirability scores are either having low reputation or already working a long backlog in their request queues. The resulting queuing dynamic for the request queues of the workers in a given MAS is thus:

$$Q_w(t+1) = max[Q_w(t) - \mu_w(t), 0] + A_w(t). \tag{5.18}$$

**Algorithm 5.1** SWORD

**Input**: $S_w(t)$values for all trustee agent $w$, and the task requests in the incoming request queue at the task broker $\lambda(t)$ at time step $t$.
Re-evaluate $\theta_w(t)$ for all $w$ based on their $S_w(t)$values.
Re-evaluate $D_w(t)$ for all $w$.
Rank all $w$ in descending order of their $D_w(t)$ values.
**for** $\forall w \in \boldsymbol{W}$ **do**
    **if** $D_w(t) > 0$ **then**
        **if** $\lambda(t) \leq \mu_w^{max}(t)$ **then**
            $A_w(t) \leftarrow \lambda(t)$
        **else**
            $A_w(t) \leftarrow \mu_w^{max}(t)$
        **end-if**
        $\lambda(t) \leftarrow \lambda(t) - A_w(t)$
    **else**
        $A_w(t) \leftarrow 0$
    **end-if**
**end-for**
Return($\boldsymbol{A}(t)$)

## 5.4. Theoretical Analysis of the Performance of SWORD

As $\mu_w(t)$ is determined by worker $w$ at time step $t$ at his own discretion, it is out of the control of SWORD. Nevertheless, by controlling $A_w(t)$ alone, SWORD can still ensure an upper bound on the request queue sizes for all workers in a given MAS.

[***Lemma 5.1***]: *(Upper bound of the request queue sizes)* Under SWORD, and for arbitrary reputation variation processes that satisfies $0 \leq \tau_w(t) \leq 1$ for all $w$ and $t$, if $Q_w(t) > \theta_w$ for some particular queue at $w$ and time step $t$, then $A_w(t) = 0$ and thus, the queue cannot increase in the next time step. It follows that if $Q_w(0) \leq \theta_w + \mu_w^{max}$, then $Q_w(t) \leq \theta_w + \mu_w^{max}$ for all $t$.

*Proof*: Suppose that $Q_w(t) > \theta_w$ for a particular worker $w$'s request queue at time step $t$, then $Q_w(t) - \theta_w > 0$. Let $\boldsymbol{A}(t) = (A_1(t), \dots, A_{|\boldsymbol{W}|}(t))$ be a vector representing the request allocation decision made by SWORD that minimizes the expression:

$$\sum_{n=1}^{|\boldsymbol{W}|}\{V[(1-\tau_n(t)) \cdot g_{max}] + Q_n(t) - \theta_n\}A_n(t) + \sum_{n=1}^{|\boldsymbol{W}|} VcA_n(t) \qquad (5.19)$$

subject to (5.8).

Suppose $A_n(t) > 0$. Since $V > 0$, $g_{max} \in R^+$, $1 - \tau_n(t) \in [0,1]$, and $c \in R^+$, the term $V[(1-\tau_n(t)) \cdot g_{max}] + Q_n(t) - \theta_n$ is strictly positive. The value of expression (5.19) can be strictly reduced by choosing $A_n(t) = 0$. By choosing $A_n(t) = 0$, we can obtain a strictly smaller value for the expression (5.19) than choosing any $A_n(t) > 0$ while satisfying Constraint (5.8). Thus, $A_n(t) > 0$ contradicts the assumption that $\boldsymbol{A}(t)$ minimizes expression (5.19). Therefore, if $Q_w(t) > \theta_w$, then $A_w(t) = 0$.

Since at each time step t, the maximum $A_w(t)$ SWORD allocates to a worker $w$ is equal to $\mu_w^{max}$ and if $Q_w(t) > \theta_w$, then $A_w(t) = 0$, it can be deduced that $Q_w(t) \leq \theta_w + \mu_w^{max}$ for all $t$ if $Q_w(0) \leq \theta_w + \mu_w^{max}$. Following the definition for $\theta_w$ in (5.9), we have

$$0 \leq Q_w(t) \leq (N+1)\mu_w^{max} + V g_{max} \tau_w^{max}, \tag{5.20}$$

provided the conditions are satisfied at time step $t = 0$.

[***Lemma 5.2***]: *(Upper bound of the Lyapunov Drift)* For all possible values of $\boldsymbol{Q}(t)$ at any time step *t*:

$$\Delta(\boldsymbol{Q}(t)) \leq B - \sum_{w:w\in\boldsymbol{W}} (Q_w(t) - \theta_w)\mathbb{E}\{\mu_w(t) - A_w(t)|\boldsymbol{Q}(t)\}$$

where *B* is a finite constant that satisfies:

$$\frac{1}{2}\sum_{w:w\in\boldsymbol{W}} \mathbb{E}\{(\mu_w(t) - A_w(t))^2|\boldsymbol{Q}(t)\} \leq B \leq \frac{1}{2}\sum_{w:w\in\boldsymbol{W}}(\mu_w^{max})^2 \tag{5.21}$$

Such a constant exists because of the assumption that $\mu_w(t)$ and $A_w(t)$ are bounded by $\mu_w^{max}$.

*Proof*: from the queuing dynamics as a resulted of the SWORD approach in (5.18):

$$\begin{aligned}(Q_w(t+1) - \theta_w)^2 &= \{\max[Q_w(t) - \mu_w(t) + A_w(t), 0] - \theta_w\}^2 \\ &\leq (Q_w(t) - \mu_w(t) + A_w(t) - \theta_w)^2\end{aligned} \tag{5.22}$$

If $Q_w(t) - \mu_w(t) + A_w(t) \geq 0$, the inequality in (5.22) holds with equality; if $Q_w(t) - \mu_w(t) + A_w(t) < 0$, the max[$\cdot$ ,0] operator ensures that

$$(0 - \theta_w)^2 < (Q_w(t) - \mu_w(t) + A_w(t) - \theta_w)^2$$

holds when $\theta_w \geq 0$.

$$\begin{aligned}(Q_w(t) - \mu_w(t) + A_w(t) - \theta_w)^2 &= [(Q_w(t) - \theta_w) - \big(\mu_w(t) - A_w(t)\big)]^2 \\ &= (Q_w(t) - \theta_w)^2 - 2(Q_w(t) - \theta_w)\big(\mu_w(t) - A_w(t)\big) \\ &+ \big(\mu_w(t) - A_w(t)\big)^2\end{aligned} \tag{5.23}$$

Substituting (5.23) into (5.22) and divide both sides of the equation by 2, we have:

$$\begin{aligned}&\frac{1}{2}(Q_w(t+1) - \theta_w)^2 \\ &\quad \leq \frac{1}{2}(Q_w(t) - \theta_w)^2 + \frac{1}{2}\big(\mu_w(t) - A_w(t)\big)^2 \\ &\quad - (Q_w(t) - \theta_w)\big(\mu_w(t) - A_w(t)\big)\end{aligned} \tag{5.24}$$

Summing over all $w \in \boldsymbol{W}$ and taking conditional expectation on both sides of (5.24), we have:

$$\begin{aligned}\Delta\big(\boldsymbol{Q}(t)\big) \leq &\frac{1}{2}\sum_{w:w\in\boldsymbol{W}} \mathbb{E}\left\{\big(\mu_w(t) - A_w(t)\big)^2 \middle| \boldsymbol{Q}(t)\right\} \\ &- \sum_{w:w\in\boldsymbol{W}} (Q_w(t) - \theta_w)\mathbb{E}\{\mu_w(t) - A_w(t)|\boldsymbol{Q}(t)\}\end{aligned}$$

(5.25)

Substituting (5.21) into (5.25), we have:

$$\Delta\big(\boldsymbol{Q}(t)\big) \leq B - \sum_{w:w\in\boldsymbol{W}} (Q_w(t) - \theta_w)\mathbb{E}\{\mu_w(t) - A_w(t)|\boldsymbol{Q}(t)\}$$

which proves *Lemma 5.2*.

[***Theorem 5.1***]: *(Proximity to the Optimal Solution)* For any fixed $V > 0$, if the initial request queues satisfy $0 \leq Q_w(t) \leq (N+1)\mu_w^{max} + Vg_{max}\tau_w^{max}$(proven in *Lemma 5.1*) and the SWORD approach is used over time steps $t \in \{0,1,2,\dots\}$, then:

$$\overline{U} \geq \overline{U}^{opt} - \frac{B}{V} - \frac{\mathbb{E}\{L(\boldsymbol{Q}(0)\}}{Vt} \tag{5.26}$$

where $\overline{U}^{opt}$ is the optimal time averaged social welfare achievable by a request allocation policy $\boldsymbol{A}^*(t)$ that yields for all *t* and all $\boldsymbol{Q}(t)$:

$$\mathbb{E}\{\mu_w(t) - A_w^*(t)|\boldsymbol{Q}(t)\} = 0 \tag{5.27}$$

which means all requests are always served within one time step. The optimal social welfare is thus defined as the situation where all requests are successfully served under $\boldsymbol{A}^*(t)$:

$$U^{opt}(t) \triangleq \sum_{w:w\in\boldsymbol{W}} g_{max}\mu_w(t) - cA_w^*(t)$$

Therefore:

$$\lim_{t\to\infty}\inf \overline{U}(t) \geq U^{opt} - \frac{B}{V}. \tag{5.28}$$

*Proof*: From *Lemma 5.1* and *Lemma 5.2*, the *drift-minus-reward* expression in (5.7) can be re-written as:

$$\begin{aligned}
&\Delta\big(\boldsymbol{Q}(t)\big) - V\mathbb{E}\{U(t)|\boldsymbol{Q}(t)\} \\
&\qquad \leq B - \sum_{w:w\in\boldsymbol{W}} (Q_w(t) - \theta_w)\mathbb{E}\{\mu_w(t) - A_w(t)|\boldsymbol{Q}(t)\} \\
&\qquad - V\sum_{w:w\in\boldsymbol{W}} \mathbb{E}\{g(t)\mu_w(t)\} + V\sum_{w:w\in\boldsymbol{W}} \mathbb{E}\{cA_w(t)\}
\end{aligned} \tag{5.29}$$

Substituting $A_w^*(t)$ into (5.29), we have:

$$\begin{aligned}
&\Delta\big(\boldsymbol{Q}(t)\big) - V\mathbb{E}\{U(t)|\boldsymbol{Q}(t)\} \\
&\qquad \leq B - \sum_{w:w\in\boldsymbol{W}} (Q_w(t) - \theta_w)\mathbb{E}\{\mu_w(t) - A_w^*(t)|\boldsymbol{Q}(t)\} \\
&\qquad - V\sum_{w:w\in\boldsymbol{W}} \mathbb{E}\{g(t)\mu_w(t)\} + V\sum_{w:w\in\boldsymbol{W}} \mathbb{E}\{cA_w^*(t)\}
\end{aligned} \tag{5.30}$$

According to (5.27):

$$\begin{aligned}
\Delta\big(\boldsymbol{Q}(t)\big) - V\mathbb{E}\{U(t)|\boldsymbol{Q}(t)\} &\leq B - V\sum_{w:w\in\boldsymbol{W}} \mathbb{E}\{g(t)\mu_w(t)\} + V\sum_{w:w\in\boldsymbol{W}} \mathbb{E}\{cA_w^*(t)\} \\
&= B - VU^{opt}(t)
\end{aligned} \tag{5.31}$$

Taking expectations of (5.31) over the distribution of $\boldsymbol{Q}(t)$ and according to the law of iterated expectations, we have:

$$\mathbb{E}\{L(\boldsymbol{Q}(t+1)) - L(\boldsymbol{Q}(t)) | \boldsymbol{Q}(t)\} - V\mathbb{E}\{U(t)\} \le B - VU^{opt}. \quad (5.32)$$

Summing (5.32) over time steps $t = \{0,1,2, \dots, T-1\}$, we have:

$$\mathbb{E}\{L(\boldsymbol{Q}(T)) - L(\boldsymbol{Q}(0))\} - V\sum_{t=0}^{T-1}\mathbb{E}\{U(t)\} \le TB - TVU^{opt}. \quad (5.33)$$

Through dividing (5.33) by $TV$ (when $t > 0$ and $V > 0$), we have:

$$\frac{\mathbb{E}\{L(\boldsymbol{Q}(T)) - L(\boldsymbol{Q}(0))\}}{TV} - U(T) \le \frac{B}{V} - U^{opt}$$

Thus,

$$U(T) \ge U^{opt} - \frac{B}{V} + \frac{\mathbb{E}\{L(\boldsymbol{Q}(T)) - L(\boldsymbol{Q}(0))\}}{TV}$$

Since $L(\cdot) \ge 0$ and all $Q_w(0) = 0$,

$$U(T) \ge U^{opt} - \frac{B}{V} - \frac{\mathbb{E}\{L(\boldsymbol{Q}(0))\}}{TV} = U^{opt} - \frac{B}{V} - \frac{\sum_{w:w\in\boldsymbol{W}}(\theta_w)^2}{2TV}$$

When $T \to \infty$, this proves (5.28).

Based on (5.28), by increasing the value of $V$, the term $B/V$ can be reduced and the performance of the SWORD approach in term of social welfare can be made closer to the optimal social welfare. However, by increasing $V$, according to (5.20), the upper limit of $Q_w(t), w \in \boldsymbol{W}$ also increases. This creates a trade-off between the social welfare and the delay in obtaining results for the requesters. According to (5.9), $\theta_w$ is linearly related to the value of $V$. Thus, increasing $V$ will cause the last term in (5.28) to increase. This phenomenon can be understood from another perspective. Due to the requirement on the timeliness of completion of the requests, once the upper limit of $Q_w(t)$ is increased to such an extent that some requests cannot be served within the stipulated deadlines, social welfare will start to decline. Therefore, the social welfare produced by SWORD can only be made closer to the optimal value if the upper limited of $Q_w(t)$ as a result of changing the value of parameter $V$ does not cause delays longer than the deadlines specified by the requesters.

As each requester may specify different deadlines for different requests and the behavior patterns of workers in different MASs may differ, it is difficult to analyze the range of values for parameter $V$ theoretically. In the next chapter, we will perform extensive empirical evaluation of the performance of the SWORD approach in a crowdsourcing scenario to complement the theoretical performance analysis presented in this section.

# Chapter 6

# Evaluating the SWORD Approach in Crowdsourcing

In Chapter 5, we proposed the SWORD approach that introduced the principles of queuing theory into trust-aware task allocation in an MAS with resource constrained trustee agents. Through theoretical analysis, we have proved the existence of various performance guarantees that make the SWORD approach a boundedly-optimal approach. In this chapter, we further evaluate the performance of the SWORD approach against existing state-of-the-art approaches in a crowdsourcing scenario where the trustee agents are human beings with limited cognitive and physical capabilities in serving requests. Through extensive simulation based on behaviorial characteristics observed from the Amazon's Mechanical Turk crowdsourcing system, the SWORD approach is shown to significantly improve the social welfare achieved by given populations of agents with varying behavior patterns while maintaining short waiting time for truster agents to receive high quality results.

## 6.1. Introduction to Crowdsourcing Systems

Crowdsourcing systems divide large problems which are too complex for computers but easy for human beings to solve into smaller tasks, and outsource them to a diverse group of people. They provide a platform where mass collaboration can occur where individual users contribute to solving a small part of the overall problem and be compensated for their effort by people who proposed the problem. Over the years, different types of crowdsourcing systems have emerged. Some of the well-known systems with crowdsourcing features include Amazon's Mechanical Turk (AMT), 99designs, Mob4hire, Youtube, and Wikipedia, etc [Doan *et al.*, 2011].

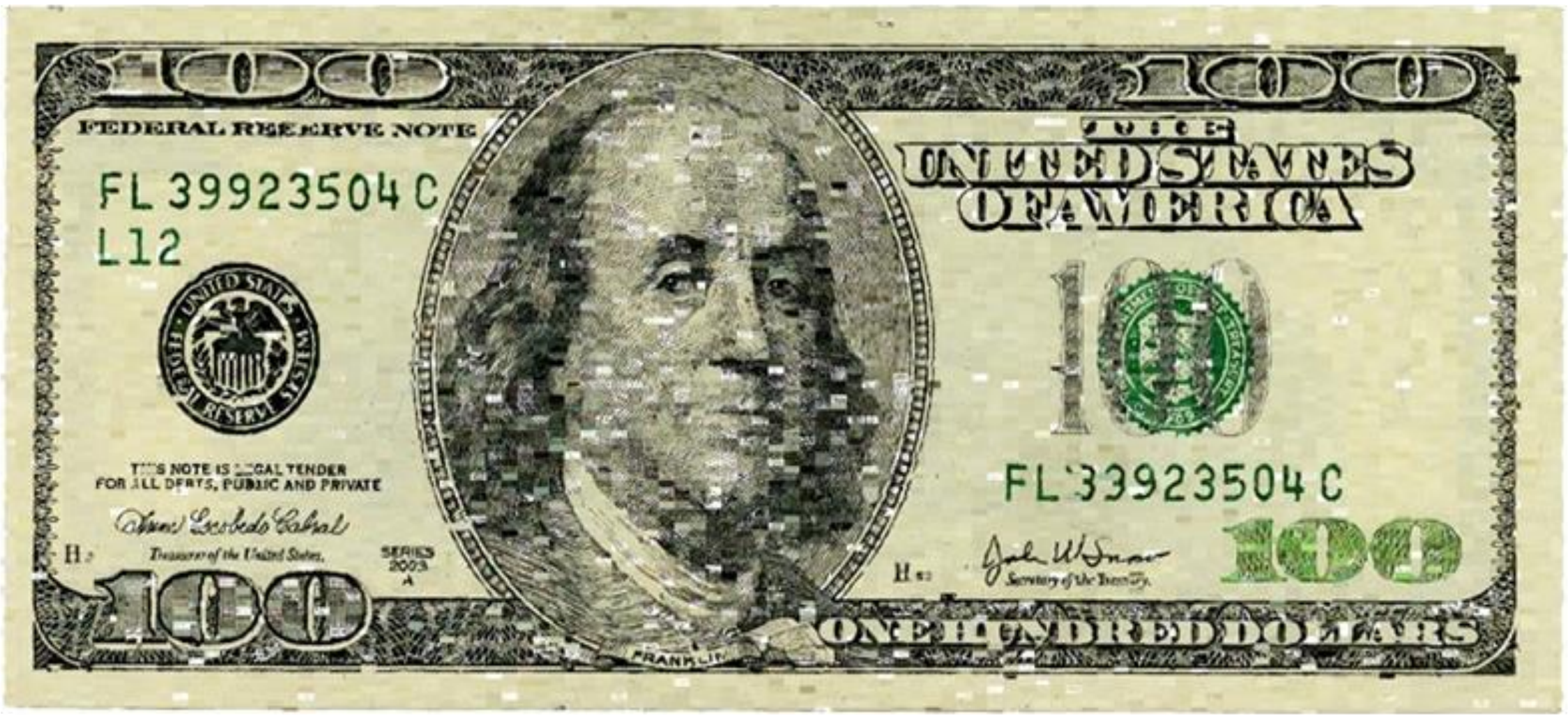

Figure 25. The *Ten Thousand Cents* digital artwork created through crowdsourcing on AMT [Doan *et al.*, 2011].

Among them, AMT, 99designs and Mob4hire employ a business model similar to a job placement agency. *Requesters* break down their tasks into small Human Intelligence Tasks (*HITs*) using proprietary tools provided by the crowdsourcing systems and assign a monetary reward for each HIT. Each task appears as an *HIT group* in the system. Usually, the rewards are the same for HITs in the same HIT group. *Workers* who have registered with the crowdsourcing systems can browse through the available HITs and work on those which they are qualified.

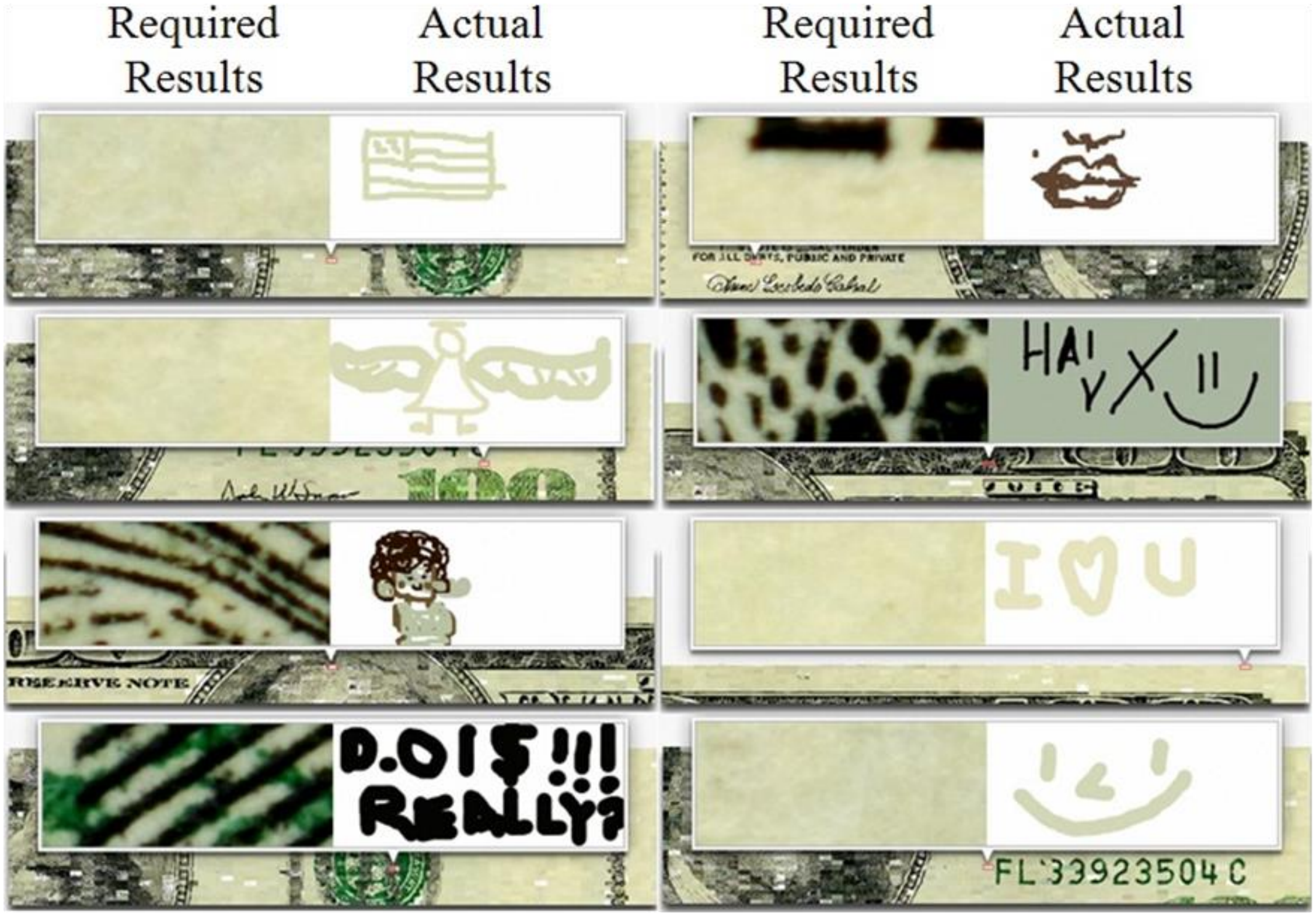

Figure 26. Examples of HITs completed by workers in AMT as compared to what was required by the requester of the *Ten Thousand Cents* project.

Ensuring the quality of the received HIT results is a challenging problem faced by crowdsourcing systems. Workers in crowdsourcing systems behave similarly to those in other types of labor markets – they are self-interested and want to maximize their own benefit. When there is a lack of deterrent, some may resort to behaving dishonestly (e.g., providing low quality results to HITs) to achieve this goal. A series of studies conducted in AMT has discovered that workers indeed lied more when given the opportunity [Paolacci *et al.*, 2010; Horton *et al.*, 2011; Suri and Watts, 2011]. For example, a well-known crowdsourcing experiment called the *Ten Thousand Cents* project was conducted on AMT where a digital artwork was created to represent a US$100 bill as shown in Figure 25. Thousands of individuals working in isolation painted a tiny part of the bill without knowledge of the overall task using a custom drawing tool. Workers were paid one cent each via AMT. It was discovered that many of the returned HITs were not completed according to the requirements. Some examples of low quality HIT results are shown in Figure 26. The figure is organized into two columns. The left column represents the artwork that a worker is required to draw, and the right column represents the corresponding returned result. There are eight pairs of required and actual results shown in Figure 26. It can be seen that low quality results can be drastically different from the required result and they are the causes of the glitches visible in Figure 25.

In popular crowdsourcing systems such as AMT, workers are anonymous in the eyes of the requesters, and it is impossible for the requesters to know who cheated them. The crowdsourcing systems also provide no recourse for requesters when low quality HITs are received. To the best of our knowledge, no crowdsourcing system implements mechanisms for proactive HIT allocation. Workers need to search for HITs and can accept HITs that they qualify on a first-come-first-served basis. In AMT, requesters can specify the minimum HIT approval rate (i.e., the percentage of HITs completed by a worker that has been accepted by the requesters) workers must achieve in order to work on their HIT Groups. However, from studies conducted by [Suri *et al.*, 2011], there is no significant difference in the cheating behavior by workers with different approval rates. There is generally a lack of protection against malicious workers in the current crowdsourcing landscape.

Trust management mechanisms have been suggested by recent research [Doan *et al.,* 2011] to be a viable way to address the problem of malicious workers in crowdsourcing systems. However, crowdsourcing systems present a unique challenge to existing trust management models. Firstly, HITs in a crowdsourcing system are completed by human beings. Naturally, they have limited capacities in terms of skills and time they can contribute to working in the crowdsourcing system each day, which limit the maximum number of HITs they can complete per day. Secondly, requesters in a crowdsourcing system usually associate deadlines with HIT Groups they proposed. In AMT, the deadline is usually two weeks or shorter. Once the deadline is passed, the HIT Group is deactivated and late coming HIT results will not be processed (which are equivalent of being unacceptable results). The characteristics of this application domain are similar to the proposed revised system model and it is suitable for studying the performance of the proposed SWORD approach and existing trust management models.

## 6.2. Key Hypotheses and Evaluation Metrics

Three major types of stakeholders are present in a crowdsourcing system: 1) the requesters who need to rely on the services provided by the workers in order to accomplish their goals (trusters), 2) the workers who provide their services to requesters in exchange of monetary or other rewards, and 3) the crowdsourcing system operator who provides and maintains basic services in the system to facilitate the exchange between the requesters and the workers. In this sense, a crowdsourcing system can also be modeled as an MAS. The requesters' wellbeing depends on them being able to receive as many HIT results within the stipulated deadline as possible with high quality. The workers' wellbeing depends on them receiving as many HITs to work on as they can effectively handle. The system operator’s wellbeing depends on the system being able to attract enough business.

From the above analysis, the requesters are likely to take the average accuracy of the received HIT results as well as the total utility (e.g., the balances in their accounts) derived from the satisfactorily completed on time; the workers are likely to mainly care about the total utility they can derive from completing the HITs assigned to them; the system operator is likely to mainly be concerned with the total amount of utility both the requesters and the workers can derive by collaborating in the system. We refer to the total utility produced by utilizing the capacities of workers in a crowdsourcing system following a given plan as the *social welfare* of the system. In order to sustain the healthy operation of a crowdsourcing system, the interest of these three different groups of stakeholders need to be satisfied.

Since the requesters are the ones who put up the rewards and propose the HITs, their needs are likely to be the central consideration for the system operators. To satisfy the requesters requirements of timeliness, quality and number of HITs completed, the intuitive goal for the crowdsourcing system is to distribute their HIT requests as widely as possible among the workers whose reputations satisfy the requesters’ minimum requirement while minimizing the number of less reputable workers involved. By doing so, the less reputable workers are likely to receive less work than their more reputable counterparts. Nevertheless, this theoretically should not affect the wellbeing of the system as a whole since less trustworthy workers are being marginalized. This complies with the definition of *social equity* from [Adams, 1966] which postulates that individuals should be rewarded in accordance with their contributions to society. It is an integral component to achieving social sustainability.

In this study, we aim to verify the following hypotheses:

- *Hypothesis 1*: the SWORD approach improves the social welfare of a given crowdsourcing system compared to existing approaches.
- *Hypothesis 2*: the SWORD approach improves social equity among workers in a given crowdsourcing system compared to existing approaches.

In the following experiments, we use the time averaged total utility in Equation (5.4) as a measure of the social welfare achieved by a HIT allocation approach. Distributive fairness is measured using a Fairness Index proposed in [Jain *et al.*, 1984] as:

$$F_{Hon} = \frac{[\sum_{w=1}^{N_{Hon}} n_w]^2}{N_{Hon} \cdot \sum_{w=1}^{N_{Hon}} n_w^2} \tag{6.1}$$

where $N_{Hon}$ is the total number of workers considered belonging to the most reputable group (labeled as the *Hon* Group) in a crowdsourcing system; and $n_w \geq 0$ is the total number of HITs which have been assigned to a worker *w* upto the current time step; $F_{Hon}$ denotes the HIT distributive fairness towards the group of the most reputable workers. $F_{Hon} = 0$ represents the most unfair treatment of this group of workers (i.e., all HITs are assigned to one worker and the rest receives to HIT), while $F_{Hon} = 1$ represents the most fair treatment of this group of workers (i.e., everyone of them are assign an equal number of HITs).

## 6.3. Design of the Evaluation Test-bed

In order to study the proposed approach against existing approaches under the crowdsourcing conditions, we design a simulation test-bed based on the characteristics of the AMT system which is one of the most widely studied crowdsourcing systems. In the test-bed, a population of requester agents puts up HITs with associated rewards for worker agents to complete while a population of worker agents seeks HITs to work on in order to earn artificial monetary rewards from the requester agents. Four different groups of worker agents with different behavior patterns are implemented in the test-bed. These agents are designed to simulate the limitations of human workers. Each worker agent can only complete up to a predetermined number of HITs in one time step (assuming all HITs require similar effort levels). The worker agents with different behavior patterns in our experiments are labeled as:

1) *Hon workers*: honest worker agents who return high quality HIT results randomly 90% of the time. Since producing high quality work may require more effort, each honest worker is set to complete at most 5 HITs per time step;
2) *MH workers*: moderately honest worker agents who return high quality HIT results randomly 70% of the time. Each moderately honest worker is set to complete at most 10 HITs per time step;
3) *MM workers*: moderately malicious worker agents who return high quality HIT results randomly 30% of the time. Each moderately malicious worker is set to complete at most 10 HITs per time step;

4) *Mal workers*: malicious worker agents who return high quality HIT results randomly 10% of the time. Each malicious worker is set to complete at most 20 HITs per time step as they do not care about the quality of their work.

The performance characteristics of agents following the four different behavior patterns are illustrated in Figure 27. Intuitively, *Hon* workers fall into the *high-quality-low-capacity* quadrant; *MH* workers fall into the *high-quality-medium-capacity* quadrant; *MM* workers fall into the *low-quality-medium-capacity* quadrant; and *Mal* workers fall into the *low-quality-high-capacity* quadrant.

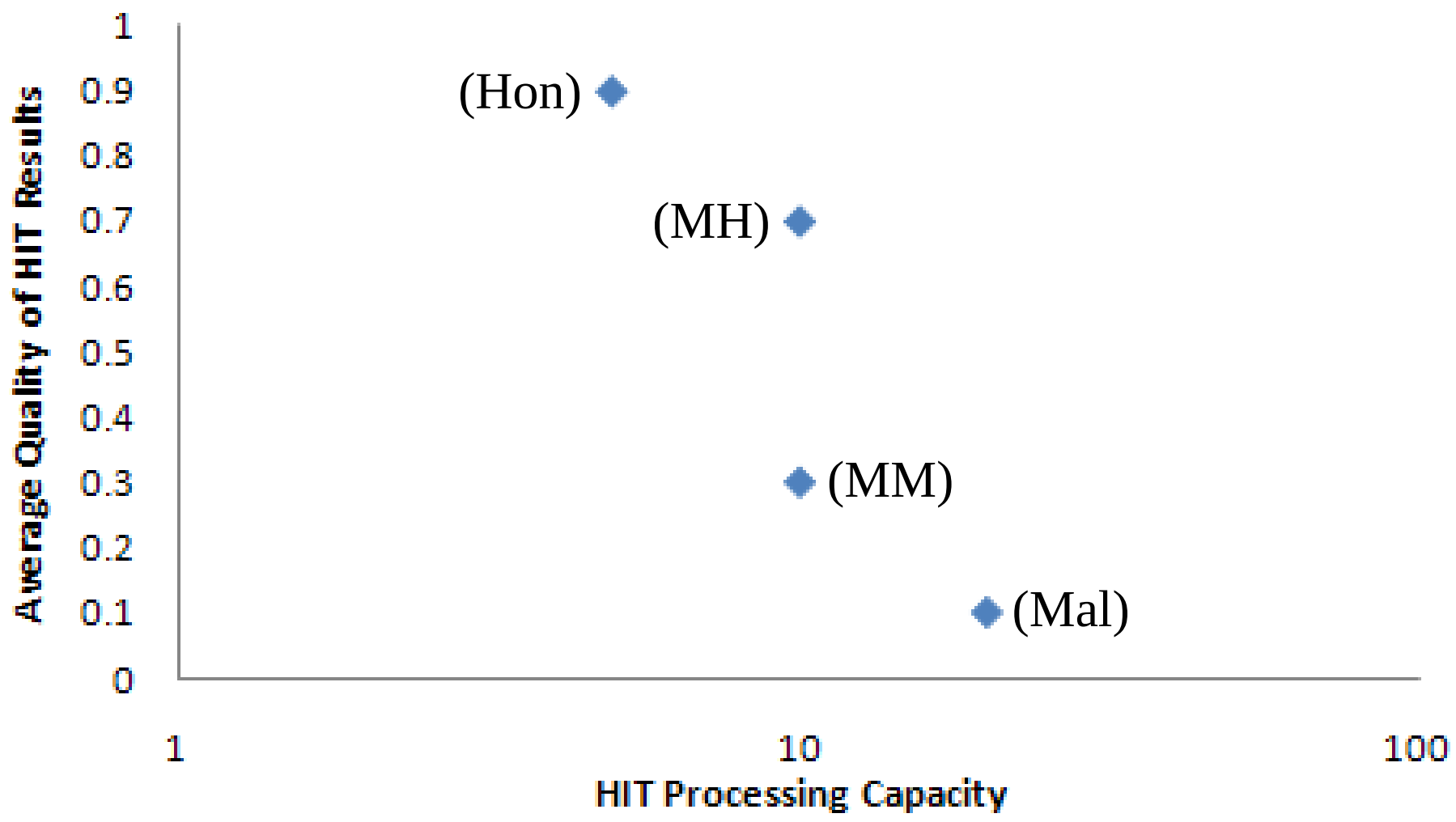

Figure 27. The performance characteristics of different groups of worker agents used in the experiments.

A requester agent $r$ can publish HIT Groups each containing up to $N_r(t)$ HITs. Once an HIT Group is published, it remains open to worker agents until all HITs in it are completed. We assume that there is only one type of task in the test-bed and all workers are qualified to perform this type of tasks. However, since reputation evaluations are usually only valid within a given context (i.e., type of task), the SWORD approach can be trivially extended to handle multiple types of tasks.

According to studies conducted by [Ipeirotis, 2010] and [Ross *et al.*, 2010] on the AMT system from 2009 to 2010, a total of 9,436 requesters and about 200,000 workers have registered with the crowdsourcing system. Over the same period, 165,368 HIT Groups with 6,701,406 HITs have been posted. From these statistics, it can be estimated that the ratio between the worker and the requester populations is about 20-to-1 and the average HITs in an HIT Group is about 40. We use these ratios to select the agent population parameters used our experiments so as to be as close to real crowdsourcing systems as possible.

## 6.4. Benchmark Approaches

In the following experiments, we compare the performance of the proposed SWORD approach against selected existing approaches. Since the currently available trust-aware interaction decision-making approaches are designed for one-to-one interactions, they are not suitable for direct use under crowdsourcing conditions where a requester (i.e., a service consumer) has to disseminate an HIT Group with multiple HITs to substantially more than one worker (e.g., service providers) to take advantage of mass collaboration. Therefore, we extend these approaches in our study to make them suitable for operation in crowdsourcing systems. The extensions are designed to preserve the spirit of the existing approaches so as to reflect the performance of the original approaches under the new system conditions as closely as possible.

### 6.4.1. Extending the Greedy Approach

Currently, trust-aware interaction decision-making approaches can be broadly divided into two categories: 1) *static* and 2) *dynamic*. In the static approach, a truster agent searches for trustee agents either through peer recommendation or random exploration. The reputations of the candidate trustee agents are evaluated using a trust evaluation model of choice and the one with the highest reputation evaluation is selected for interaction. Many existing trust management models adopt this greedy approach [Jøsang and Ismail, 2002; Yu and Singh, 2003; Teacy *et al.*, 2005; Weng *et al.*, 2006; Teacy *et al.*, 2008; Weng *et al.*, 2010; Liu *et al.*, 2011; Shen *et al.*, 2011]. From an individual trustee agent's point of view, in order to maximize its own long term wellbeing, it is rational to always select the best possible option for every interaction.

In our experiments, we choose to extend the Beta Reputation System (BRS) proposed in [Jøsang and Ismail, 2002] which is the most widely used trust model in this area. The extended BRS will use a modified versions of BRS proposed in Chapter 4 – *BRS2012*. The extended BRS be referred to as *BRS2002e* in our experiments. A requester agent equipped with *BRS2002e* explores for potentially more trustworthy workers in a random manner to accumulate direct observations about them. Once enough direct observations are recorded, *BRS2002e* exploits the known workers according to the algorithm listed in Table 6.1. Workers with reputation evaluations higher than a predetermined threshold value $Th$ are selected for interaction and the available new HITs are distributed among them as evenly as possible. Their reputation values are normalized to [0, 1], where 1 represents the most trustworthy and 0 represents the most untrustworthy. *BRS2002e* will spend *10*% of the time allocating HITs to workers with few direct observation records.

### 6.4.2. Extending the Dynamic Approaches

Recently, there are a few dynamic trust-aware interaction decision-making approaches proposed. One approach is denoted as *M2009e*. It is based on [Muñoz *et al.*, 2009] where the authors measure a truster agent's knowledge degrees about trustee agents and use this metric to determine which trustee agent to select for interaction. The knowledge degree depends on the amount of direct past interaction experience with the trustee agent, third-party testimonies about that trustee agent, and the self reported trustworthiness by that trustee agent which the truster agent can gather. The value of the knowledge degree is normalized within the range of [0, 1], with 1 representing "completely known" and 0 representing "no direct interaction experience". In the local record of a truster agent, candidate trustee agents are organized into four groups according to their knowledge degree values:

- Group TK (Totally Known): knowledge degree = 1;
- Group PK (Partially Known): 0 < knowledge degree < 1;
- Group AU (Almost Unknown): knowledge degree = 0;
- Group TU (Totally Unknown): no information available.

If there are enough trustee agents with trustworthiness values higher than a predefined threshold *QT* in Group TK, the truster agent will only select these trustee agents for interaction; otherwise, a number of exploration interacitons will be allocated to trustee agents in groups PK, AU and TU to build up the knowledge degree of trustee agents in these groups and promote them into higher order groups. For the exploitation phase of *M2009e*, the algorithm listed in Table 6.1 is applied only to trustee agents in Group TK.

Another approach is denoted as *H2010e* which is based on [Hoogendoorn *et al.*, 2010]. It measures how much the operating environment has changed and uses this metric to determine the amount of effort a truster agent should devote in exploring for more trustworthy trustee agents. In this approach, each truster agent keeps track of the *long term trust* ($LT_i(t)$) and the *short term trust* ($ST_i(t)$) of candidate trustee agents, where $ST_i(t)$ reflects the changes in an trustee agent $i$'s behavior faster than $LT_i(t)$. The average absolute difference between $LT_i(t)$ and $ST_i(t)$ is used to estimate the changes in the environment $C(t)$. When $C(t)$ is larger than 0, an exploration extent value $E(t)$ is calculated. This, together with the trustworthiness value of each trustee agent, is used to derive a selection probability $RP_i(t)$ for that SP. The candidate SPs are then selected using the Monte Carlo method according to their $RP_i(t)$ values. When $C(t) = 0$, the trustee agent with the highest trustworthiness evaluation is selected for interaction. In *H2010e*, when the changes in the environment necessitate exploration (i.e. $C(t) > 0$,), the number of interactions to be assigned to each known trustee agent is determined according to the weight of its selection probability $RP_i(t)$ among the candidates. When $C(t) = 0$, *H2010e* exploits the most trustworthy trustee agents by using the algorithm listed in Table 6.1. The only difference is that the long term trust $LT_i(t)$ is used to rank the trustee agent in *H2010e*.

Both the static and the dynamic approaches are fundamentally similar in their goals – they aim to increase the overall quality of the results a truster agent can obtain from interactions with trustee agents by minimizing their risk exposure when exploring unfamiliar trustee agents. In our experiments, *BRS2002e*, *M2009e* and *H2010e* will be used as the benchmarks for assessing the performance of the SWORD approach.

**Algorithm 6.1** Extended Exploration Algorithm

1: Rank known workers in descending order of their reputation values $\tau_w(t)$.
2: **if** the number of known wohrkers with $\tau_w(t) \geq Th$ (a.k.a. trustworthy workers) is at least $N_r(t)$ **then**
3: Assign one HIT to each of them.
4: **else**
5: Determine the number of HITs to be assigned to each worker according to the weight of its $\tau_w(\mathrm{t})$ value among selected trustworthy workers.
6: **end-if**

## 6.5. Comparison-based Evaluation

In the first set of experiments, we investigate the performance of the SWORD approach against the benchmark approaches in a comparison-based manner where different approaches operate in separate MASs where agents have no knowledge of the existence of other approaches.

### 6.5.1. Experiment Setup

Five groups of requester agents each equipped with different HIT allocation approaches are compared in the proposed test-bed environment. They are Group *AMT* where requester agents who follow AMT's passive HIT allocation approach (i.e., worker agents to accept HITs proposed in the system on a first-come-first-served basis); Group *BRS2002e*, Group *M2009e*, and Group *H2010e* where requester agents distribute HITs to workers following the approaches extended from [Jøsang and Ismail, 2002; Muñoz *et al.*, 2009], and [Hoogendoorn *et al.*, 2010] respectively; and Group *SWORD* where the HITs proposed by the requester agents are distributed to the worker agents by the crowdsourcing system following the SWORD approach. All groups, except Group *AMT*, adopt the *BRS2012* model as the underlying reputation evaluation model. Each round of experiment runs for 1,000 time steps and is repeated 10 times to reduce the effect of random variation.

The environment of each experiment includes 50 common witness agents who accumulate direct trust evidence about worker agents and provide testimonies to requester agents with HIT assignment approaches that require this information (i.e., Group *M2009e* and Group *H2010e*). These agents are allowed to run for 200 time steps to accumulate some direct trust evidence before agents equipped with various approaches start to operate. A total of $N_w$ worker agents are included in each experiment. The number of them adopting the four different behavior patterns is varied in different experiments to simulate different population configurations. Each worker agent population configuration is denoted as *HonX*. It represents a worker agent population consists of $\frac{1}{2}X\%$ *Hon* worker agents, $\frac{1}{2}X\%$ *MH* worker agents, $\frac{1}{2}(100-X)\%$ *MM* worker agents, and $\frac{1}{2}(100-X)\%$ *Mal* worker agents. Worker agents perform the *Clean Sweep* operation as mentioned in Chapter 4 at every time step to drop any HITs in their own request queues that have become expired.

Table 8. Parameter Values in the Experiments

| Symbol | Description | Value |
|---|---|---|
| $g_{max}$ | Maximum utility for successfully completing an HIT. | 1.0 |
| $c$ | The utility cost to a requester for proposing an HIT (i.e., the reward offered to workers to complete it). | 0.2 |
| $N_r(t)$ | The average number of HITs in a HIT Group. | 40 |
| $N_{req}$ | The total number of requester agents in an experiment. | 50 |
| $N_w$ | The total number of worker agents in an experiment. | 1,000 |
| $N$ | A weight parameter used to calculate the target queue size in the SWORD approach in Equation (5.9). | 1.0 |
| $V$ | The performance tuning parameter in the SWORD approach. | 2.0 |
| $Th_r$ | The minimum reputation value required by the approaches (except Group *AMT*) in the experiment for a worker to quality as a potential candidate for HIT allocation. | 0.6 |
| $\Pr(Exp)$ | The percentage of time spent by the SWORD approach for random exploration. | 10% |
| $T_{dl}$ | The deadline (i.e., maximum number of time steps from the time step an HIT is proposed) before which a HIT must be completed in order to be regarded as completed on time. | 14 |

$N_{req}$ requester agents are involved in each experiment. Requester agents are designed only to propose new HIT Groups when the last proposed HIT Group is completed or becomes expired. In this way, the social welfare produced represents the maximum social welfare a crowdsourcing system made up of the given requester agent and worker agent populations can produce under the given HIT allocation approach. The values for the parameters used in our experiments are listed in Table 8.

### 6.5.2. Analysis of Numerical Results

**Hypothesis 1**

Figure 28 illustrates the performance of various approaches in terms of time averaged social welfare ($\overline{U}$) calculated according to Equation (5.4). Here, the box-plot is used to shown a summary for each approach containing the maximum, the 75-percentile, and median, the 25-percentile and the minimum $\overline{U}$ values achieved by the respective approaches. Comparing the median of the $\overline{U}$ values, Group *SWORD* outperforms all other approaches by wide margins. Group *AMT*, which does not care about the reputation of the workers and adopts a *laissez-faire* approach towards allocating HITs to workers (a.k.a. letting the workers to decide which HITs to accept on a first-come-first-served basis), have achieved the second highest level of $\overline{U}$. In fact, it can be seen that Group *AMT* significantly outperforms Groups *BRS2002e*, *M2009e* and *H2010e* which are trust-aware

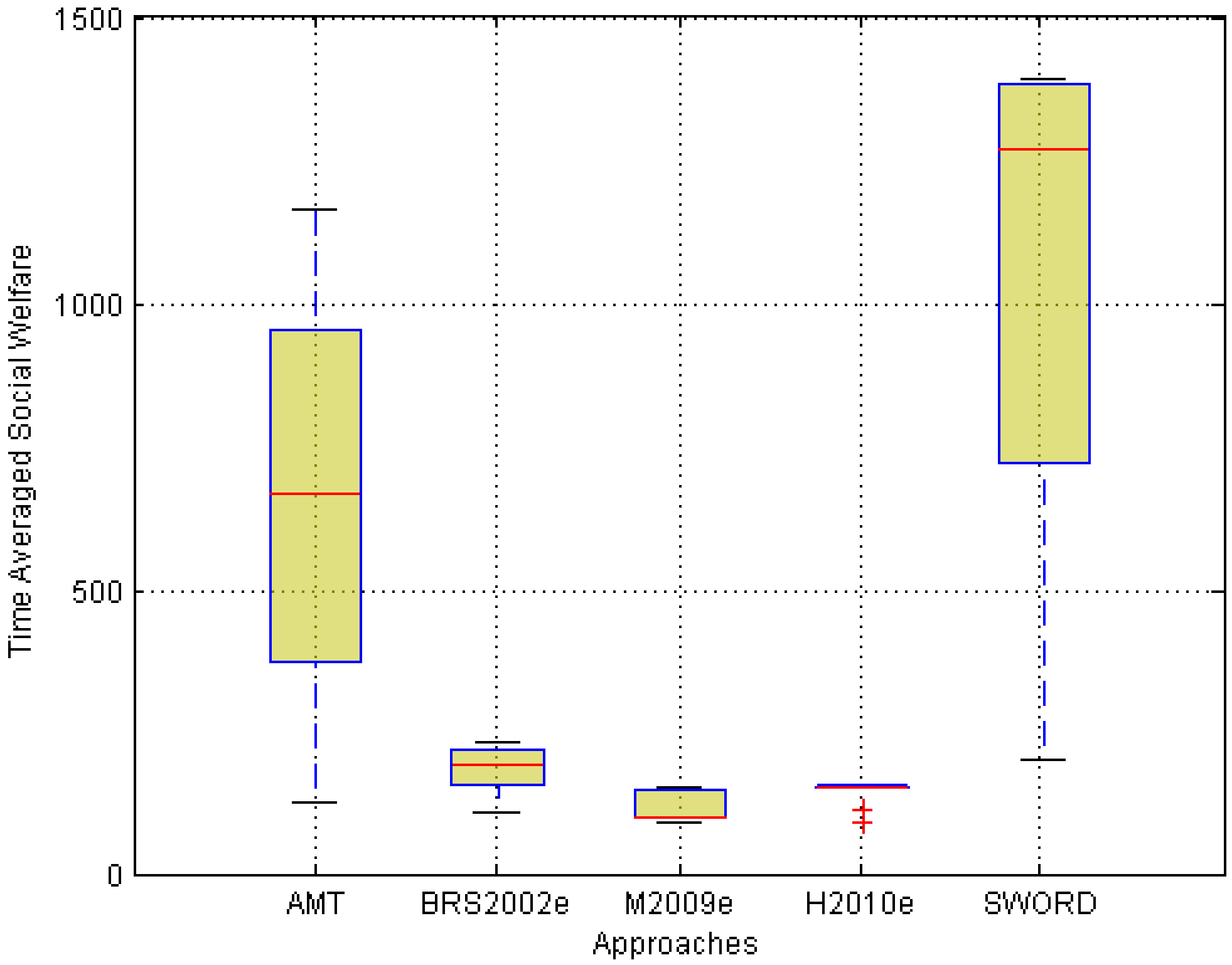

Figure 28. The ranges of the Time Averaged Social Welfare achieved by various HIT allocation approaches under different worker agent population configurations.

HIT allocation. Among these three groups, Group *BRS2002e* which employs a static exploration method has outperformed Groups *M2009e* and *H2010e* which use dynamic exploration methods.

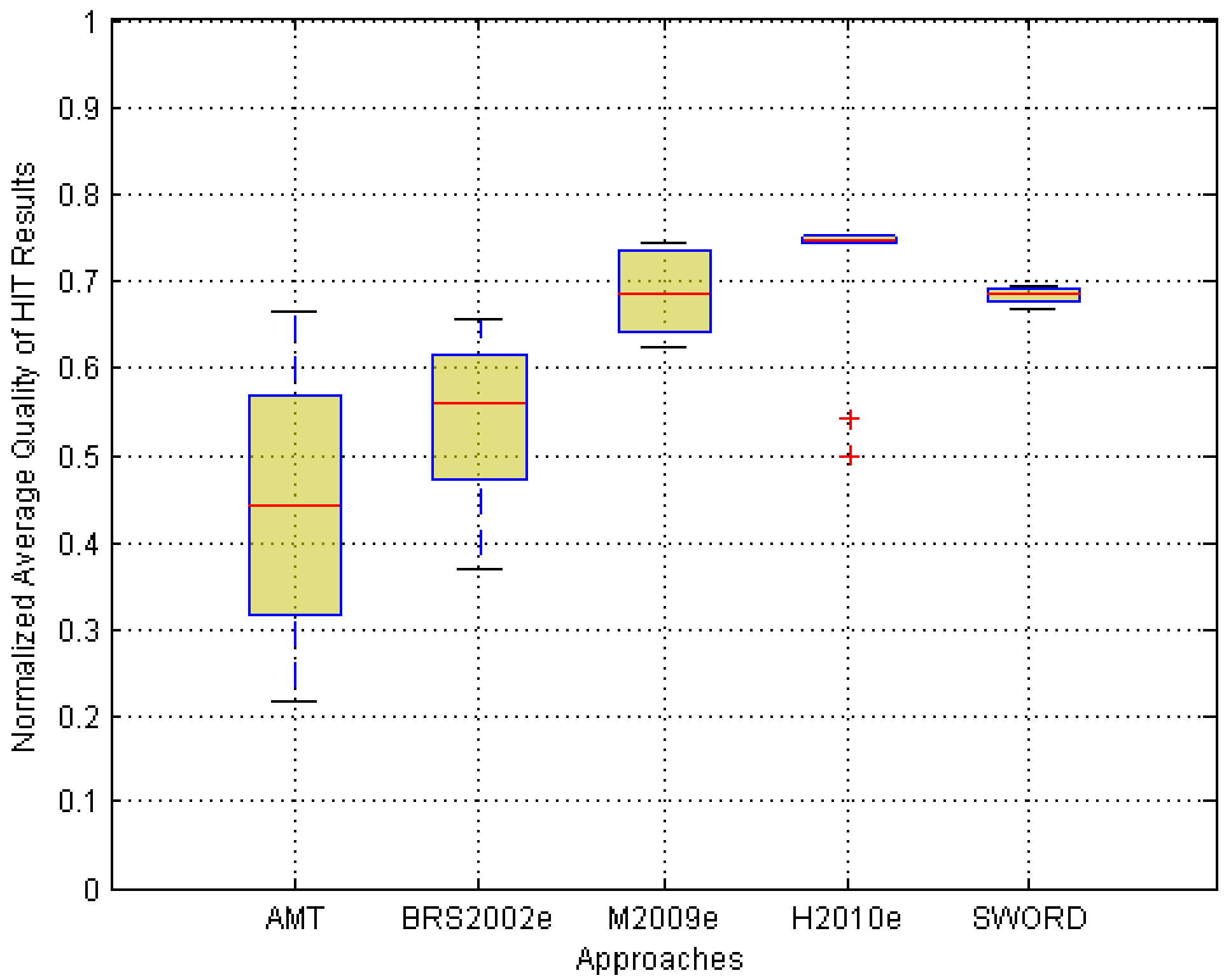

Figure 30. The ranges of the Average Quality of HIT results achieved by various HIT allocation approaches under different worker agent population configurations.

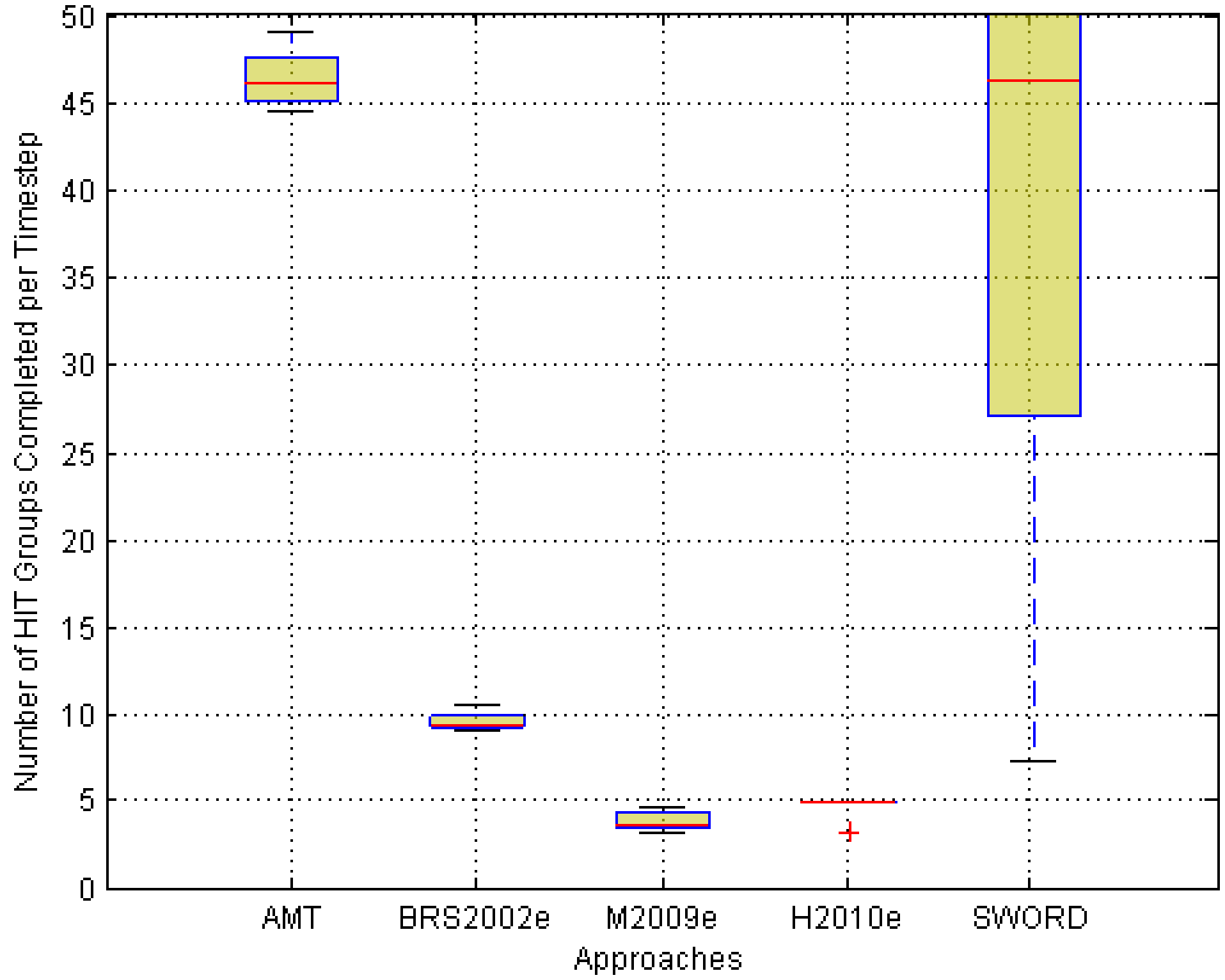

Figure 29. The ranges of the Time Averaged Number of HIT Groups Completed by various HIT allocation approaches under different worker agent population configurations.

It appears that, under crowdsourcing conditions, existing trust-aware interaction decision-making approaches can cause significant deterioration in the overall system wellbeing as measured by $\overline{U}$. To gain deeper understanding of the cause of this outcome, the two factors affecting $\overline{U}$ - the average quality of the HIT results and the total number of HITs completed on time – should be analyzed separately.

Figure 29 shows the box-plot of the average quality of HIT results as reflected by the average normalized utility gain derived from a completed HIT. By comparing the median of this value achieved by the five groups of worker agents, it can be seen that Group *H2010e* outperforms other groups. Groups *M2009e* and *SWORD* have achieved similar levels of average HIT result quality. Group *AMT* has achieved a lower level of average HIT result quality than all trust-aware approaches. Among the four trust-aware approaches, *M2009e* and *H2010e* which use dynamic exploration methods outperform

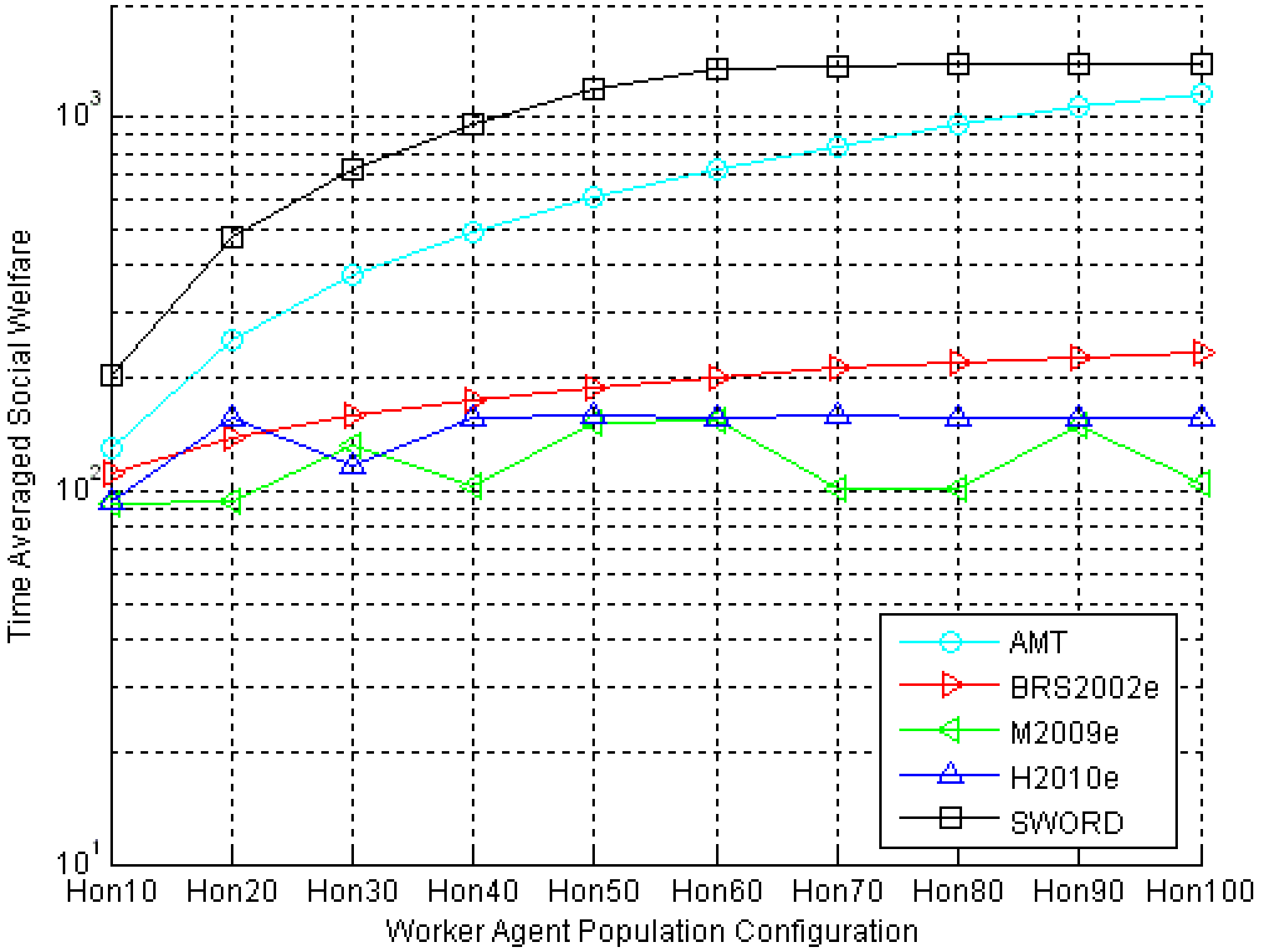

Figure 31. The Time Average Social Welfare achieved by various HIT allocation approaches under different worker agent population configurations.

Groups *BRS2002e* and *SWORD* in terms of average HIT result quality.

Figure 30 shows the box-plot of the time averaged total number of HIT Groups various approaches can complete under different worker agent population configurations. Since, in our experiments, $N_{req} = 50$ and a HIT needs at least one time step to complete, the upper limit for the time averaged total number of HIT Groups completed for each group of requester agents is 50. By comparing the median of this metric achieved by the five approaches, it can be seen that Groups *AMT* and *SWORD* significantly outperform other groups with Group *AMT* doing slight better than Group *SWORD*. Among the three groups adopting existing trust-aware interaction decision-making approaches, Group *BRS2002e* which use a fixed 10% of time exploring unfamiliar workers outperforms Groups *M2009e* and *H2010e* which use dynamic exploration methods.

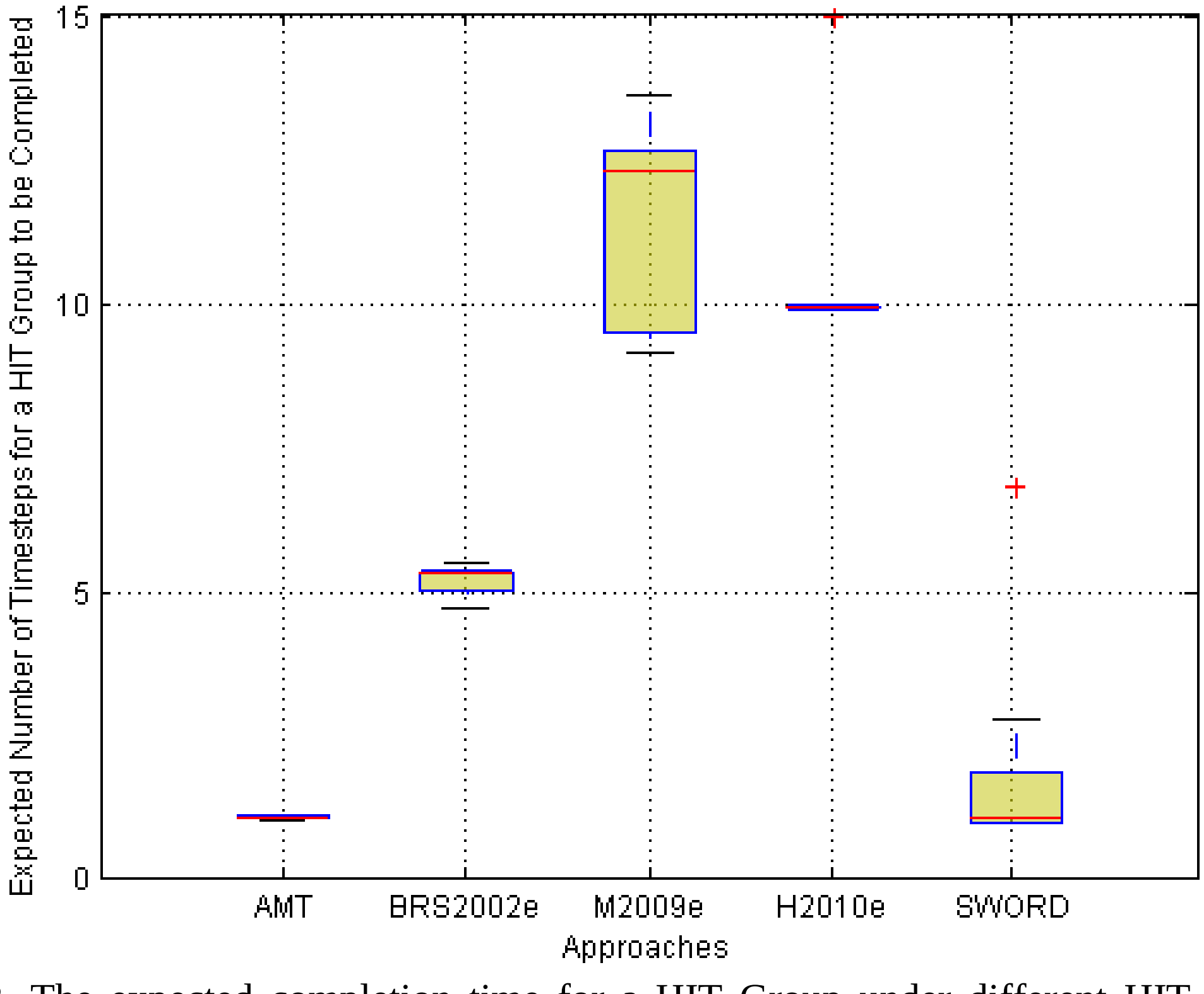

Figure 32. The expected completion time for a HIT Group under different HIT allocation approaches under worker agent population configuration *Hon50*.

The poor performance of Groups *BRS2002e*, *M2009e* and *H2010e* is mainly due to the low number of HIT Groups that can be completed on time based on their worker agent utilization plans. Since they are designed to minimize the risk exposure of the requesters in principle, their plans are not aimed to spread out the HITs among worker agents to utilize mass collaboration, but rather to concentration HIT allocations among a select few highly reputable worker agents. Although Group *AMT* produces the lowest average HIT result quality, the worker agents' capacities utilization rate is high. Even with a large percentage of HITs being completed with poor quality, the social welfare achieved by

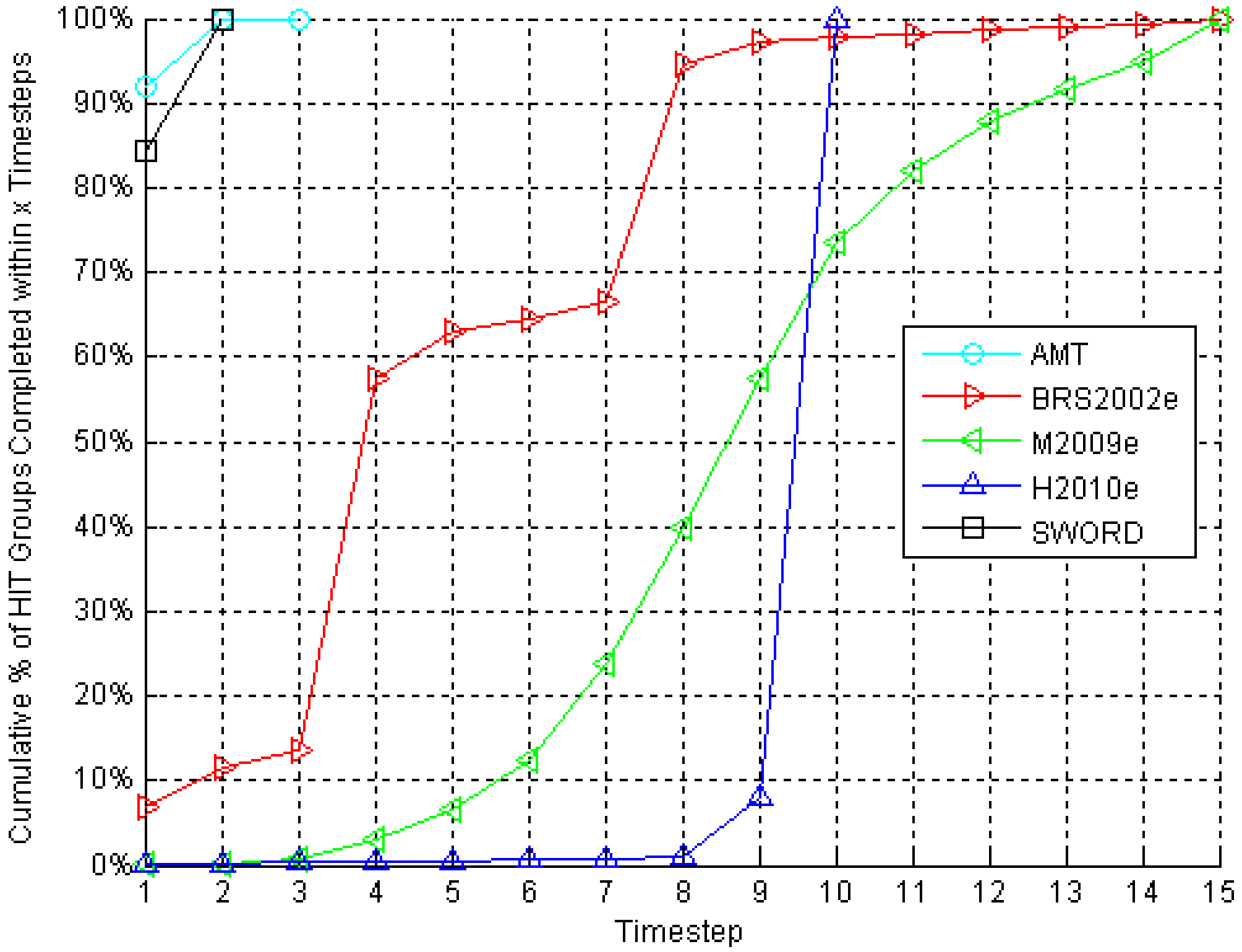

Figure 33. The cumulative percentage of all HIT Groups completed with completion time less than or equal to *x* time steps under worker agent population configuration *Hon50*.

Group AMT is significantly higher than Groups *BRS2002e, M2009e* and *H2010e*.

In Figure 28, the performance of Groups *AMT* and *SWORD* shows significant variations. To investigate this observation, the time averaged social welfare values achieved by the five approaches under each worker population configuration are plotted in Figure 31. The performance of Group *AMT* and *SWORD* are affected more significantly by the changes in worker agent population configurations than the other groups. In the case of Group *AMT*, the changes in worker agent population configuration mainly affect its average HIT result quality. As more *Hon* and *MH* workers are present in the worker agent population, the average HIT result quality achieved by Group *AMT* following what essentially is random HIT allocation to worker agents improves. In the case of Group *SWORD*, changes in worker agent population configuration mainly affect the time averaged total number of HIT Groups completed. In Figure 31, from *Hon10* to *Hon60*, as more reliable worker agents are present, the pool of eligible worker agents satisfying the requirements of the SWORD approach increases. Thus, its $\overline{U}$ value also increases. From *Hon60* to *Hon100*, the aggregate capacity of the pool of eligible worker agents under the management of the SWORD approach surpasses the total number of HITs the requester agents can propose. Therefore, the $\overline{U}$ values achieved by Group *SWORD* stops increasing.

The performances achieved by the SWORD and the AMT approach are partly due to the low HIT turn-around time which is measured by the expected HIT Group completion time shown in Figure 32. Groups *AMT* and *SWORD* achieved the lowest median expected completion time for a HIT Group among the five groups. The maximum expected completion time under the SWORD approach occurred in the worker agent population configuration *Hon10*. Since the clean sweep operation is regularly performed by the five approaches, the maximum completion time for any HIT Group is $(T_{dl} + 1)$, which is 15 time steps in our experiments.

The detailed performance of the five approaches in terms of the HIT Group completion time (under *Hon50*) is illustrated in Figure 33. Around 92% of all HIT Groups are completed in 1 time step under AMT, while around 85% are completed in 1 time step under SWORD. Almost all HIT Groups take 2 time steps to complete under AMT and SWORD. In comparison, 90% of the HIT Groups are completed within 8 time steps and 14 time steps under *BRS2002e* and *M2009e* respectively, and the majority of the HIT Groups take 10 time steps to finish under *H2010e*. Under *BRS2002e* and *M2009e*, there are around 1% and 10% of the HIT Groups not completed within the 14 time step

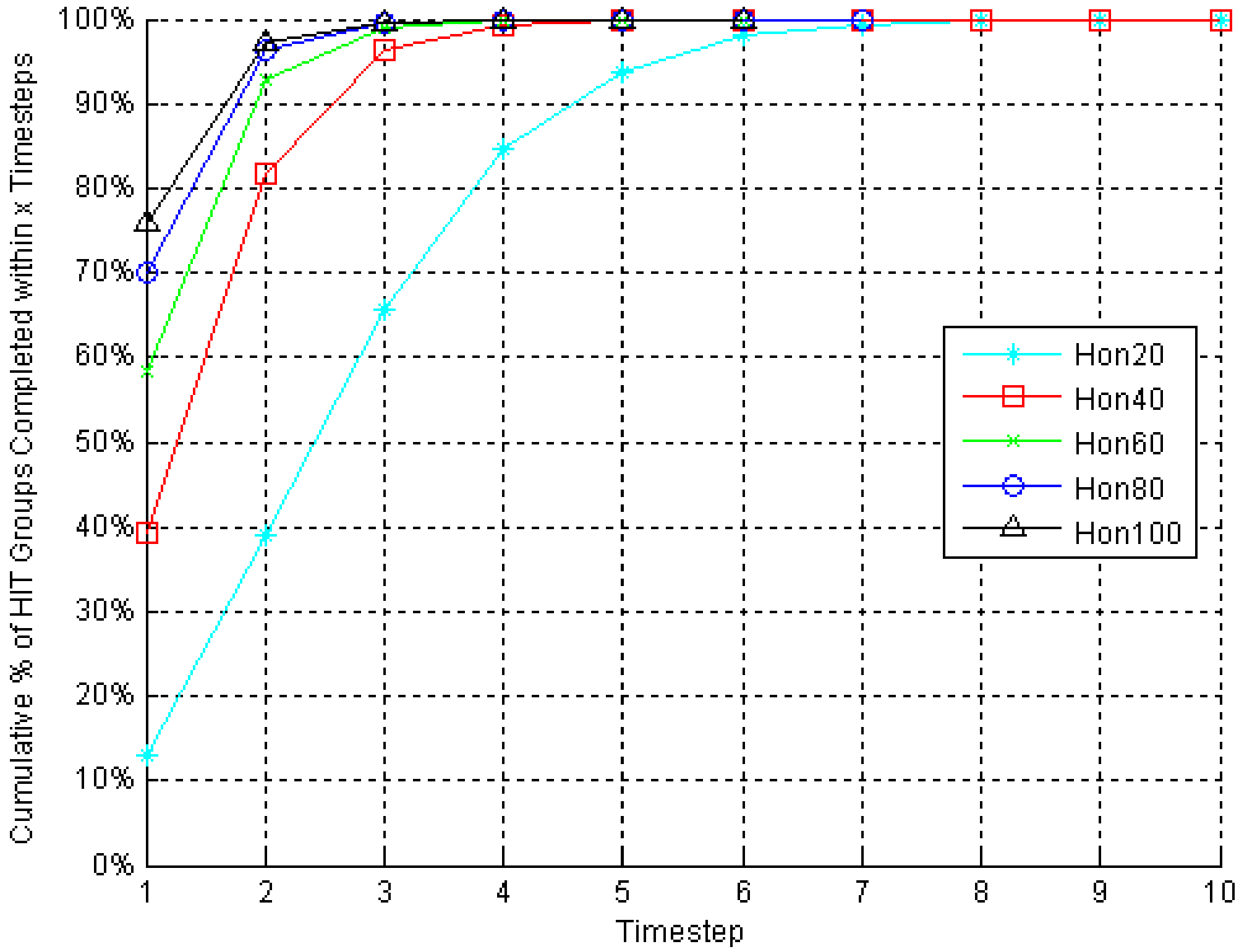

Figure 34. The cumulative percentage of all HIT Groups completed with completion time less than or equal to *x* time steps under the SWORD approach with varying worker agent population configurations.

deadline respectively.

Figure 34 shows the performance of the SWORD approach under different worker agent population configurations. As the percentage of *Hon* worker agents increases, the time taken for HIT Groups to be completed becomes shorter and an increasing percentage of all HIT Groups are completed within fewer time steps. Figures 32~34 not only illustrate the reason for the SWORD approach to achieve significantly higher social welfare than other approaches, but also demonstrate another improvement to the requesters' user experience – shorter waiting time to receive HIT results. Our first hypothesis is considered verified.

**Hypothesis 2**

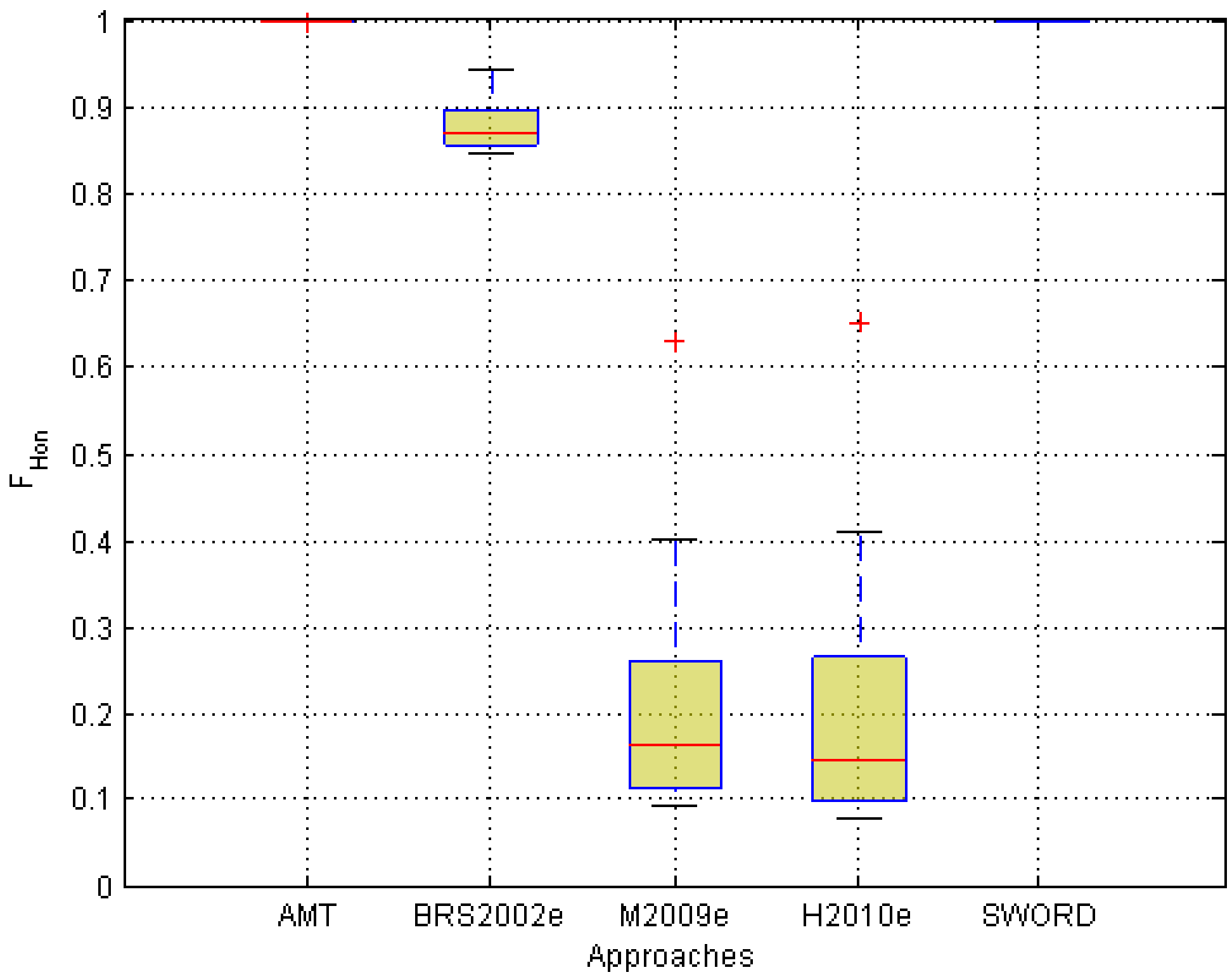

Figure 35. The ranges of the fairness index for *Hon* workers achieved by various HIT allocation approaches under different worker agent population configurations.

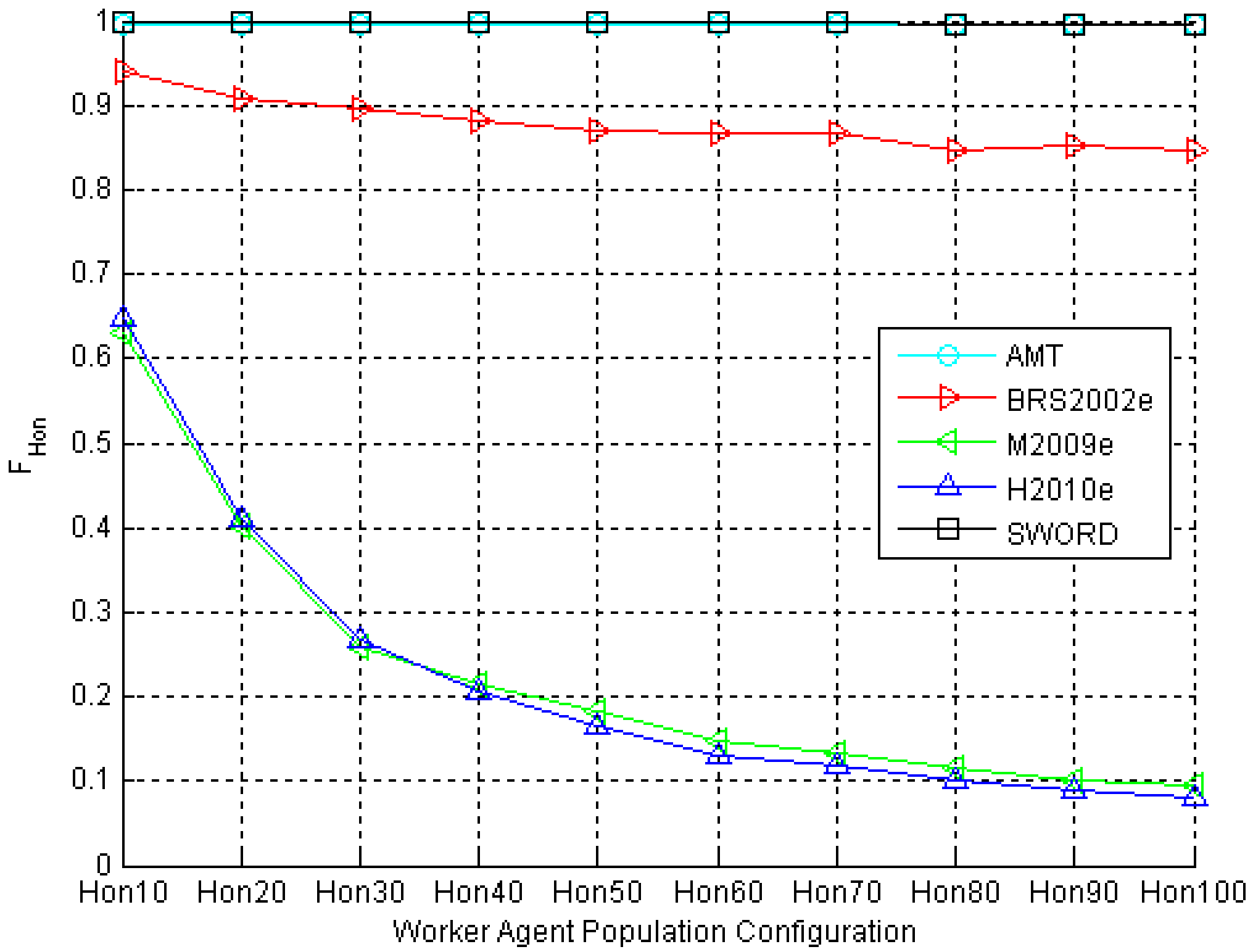

Figure 36. The fairness index for *Hon* workers achieved by various HIT allocation approaches under different worker agent population configurations.

The effect of different approaches on social welfare ultimately depends on how well they build social equity among the workers in a given crowdsourcing system. Measured with the fairness index for *Hon* worker agents ($F_{Hon}$) in Formula (6.1), it can be seen in Figure 35 that the distributive fairness achieved by groups *BRS2002e*, *M2009e* and *H2010e* are significantly lower than that of groups *AMT* and *SWORD*. Groups AMT and SWORD both achieved almost complete distributive fairness among for *Hon* workers over the long run. Figure 36 breaks down the long term $F_{Hon}$ values of the various approaches in correspondence to different worker agent population configurations. The $F_{Hon}$ values for groups *BRS2002e*, *M2009e* and *H2010e* decreases with increasing percentage of *Hon* worker agents in the worker agent population. This trend is also present for Group

*SWORD* albeit the variation in very small and is only visible if we zoom into the scale between $F_{Hon} = 0.995$ and $F_{Hon} = 1$ as shown in Figure 37.

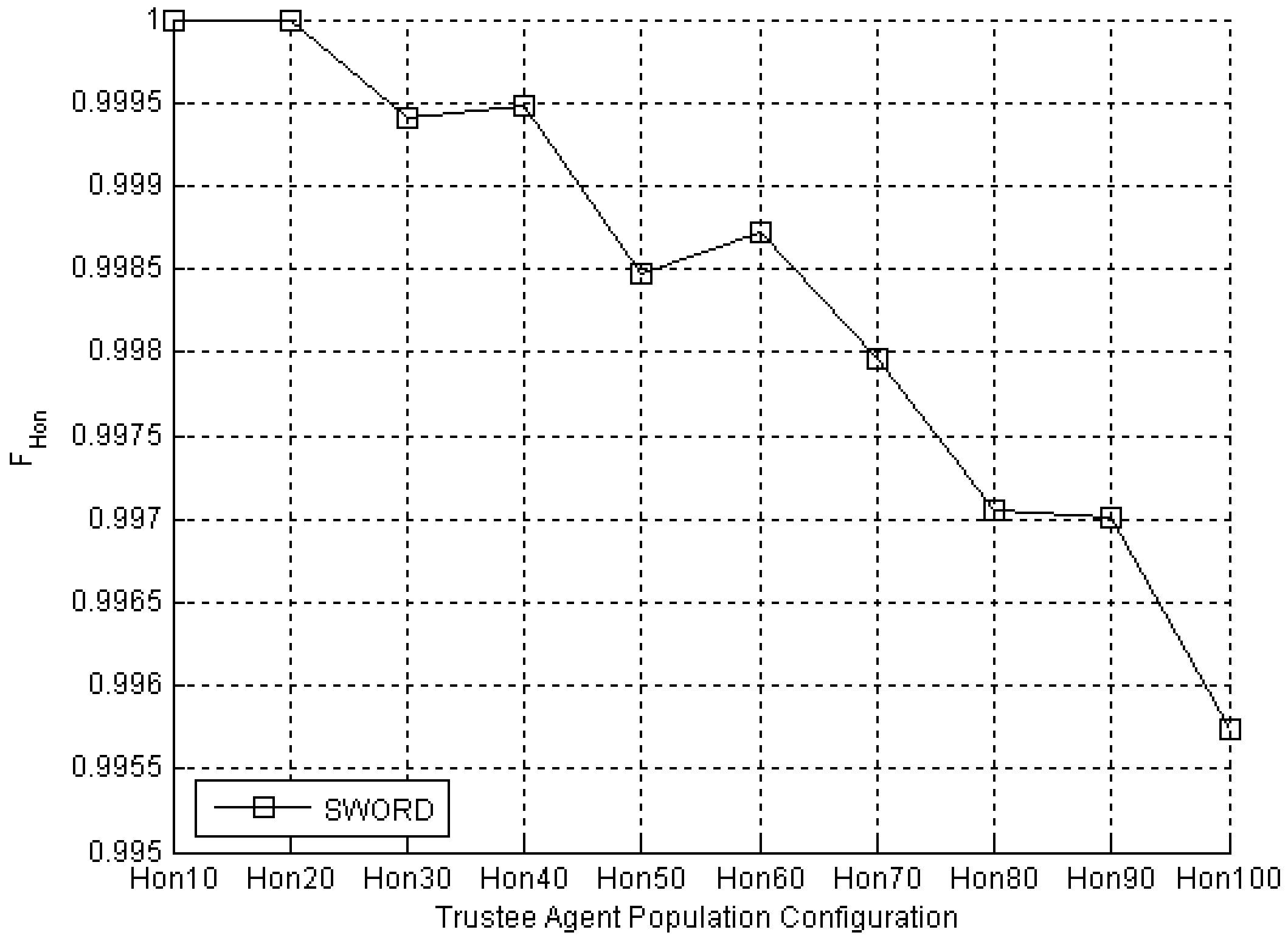

Figure 37. The fairness index for *Hon* workers achieved by Group *SWORD* under different worker agent population configurations.

In the cases of groups *BRS2002e*, *M2009e* and *H2010e*, this phenomenon is caused by the relative size of the highly reputable worker agents recognized by each of these approaches ($N_{Selected}$) v.s. the total number of *Hon* worker agents ($N_{Hon}$). Group *M2009e* and *H2010e* aim to minimize the requesters' risk exposure by minimizing the need for exploration. Thus, their $N_{Selected}$ value remain relatively constant when the worker population configuration changes. Therefore, their $N_{Selected}$:$N_{Hon}$ ratios becomes smaller, causing decreasing distributive fairness with $N_{Hon}$ becomes larger. With a constant predefined exploration probability, *BRS2002e* is able to discover more *Hon* workers as time passes. However, the greedy self-interested HIT allocation decisions made by *BRS2002e* requester agents still cause concentration of requests to a relatively small number of highly reputable workers from their own perspectives. Group *SWORD* is faced with a similar situation. However, the proposed approach is able to decrease the magnitude of the influence of increasing $N_{Hon}$ and achieve a high distributive fairness among *Hon* workers.

As illustrated by Figure 38 which is a snapshot of the total number of HITs allocated to each of the 1,000 worker agents in a simulation run with worker agent population configuration *Hon50*, Group *AMT* does not really social equity in the sense that it treats all worker agents equally (i.e., each worker receives 0.1% of all published HITs) with no regard to their reputations. Groups *M2009e* and *H2010e* achieve good quality HIT results by overly concentration of HIT allocations to a relative small number of highly reputable worker agents determined by their reputation evaluation models. The selected worker agents receive up to 3% of all published HITs which is about 30 times more than the absolute equal workload of 0.1% each, while other worker agents receive no HITs. The variations in the treatment of *Hon* worker agents by requester agent in Group *BRS2002e* are quite significant as shown by the standard deviation compared to its average value in Table 6.2. The SWORD approach is able to efficiently utilize the capacities of the Hon workers over the long run. As illustrated in Table 9, the average total number of HITs completed by a *Hon* worker under *SWORD* is 2.66 times that under *AMT*, 9.49 times that under *BRS2002e*, 6.29 times that under *M2009e*, and 6.11 times that under *H2010e*. Apart from achieving high distributive fairness among *Hon* workers, the SWORD approach also achieves social equity among all worker agents where they receive HITs commensurate to their trustworthiness. Our second hypothesis is considered verified.

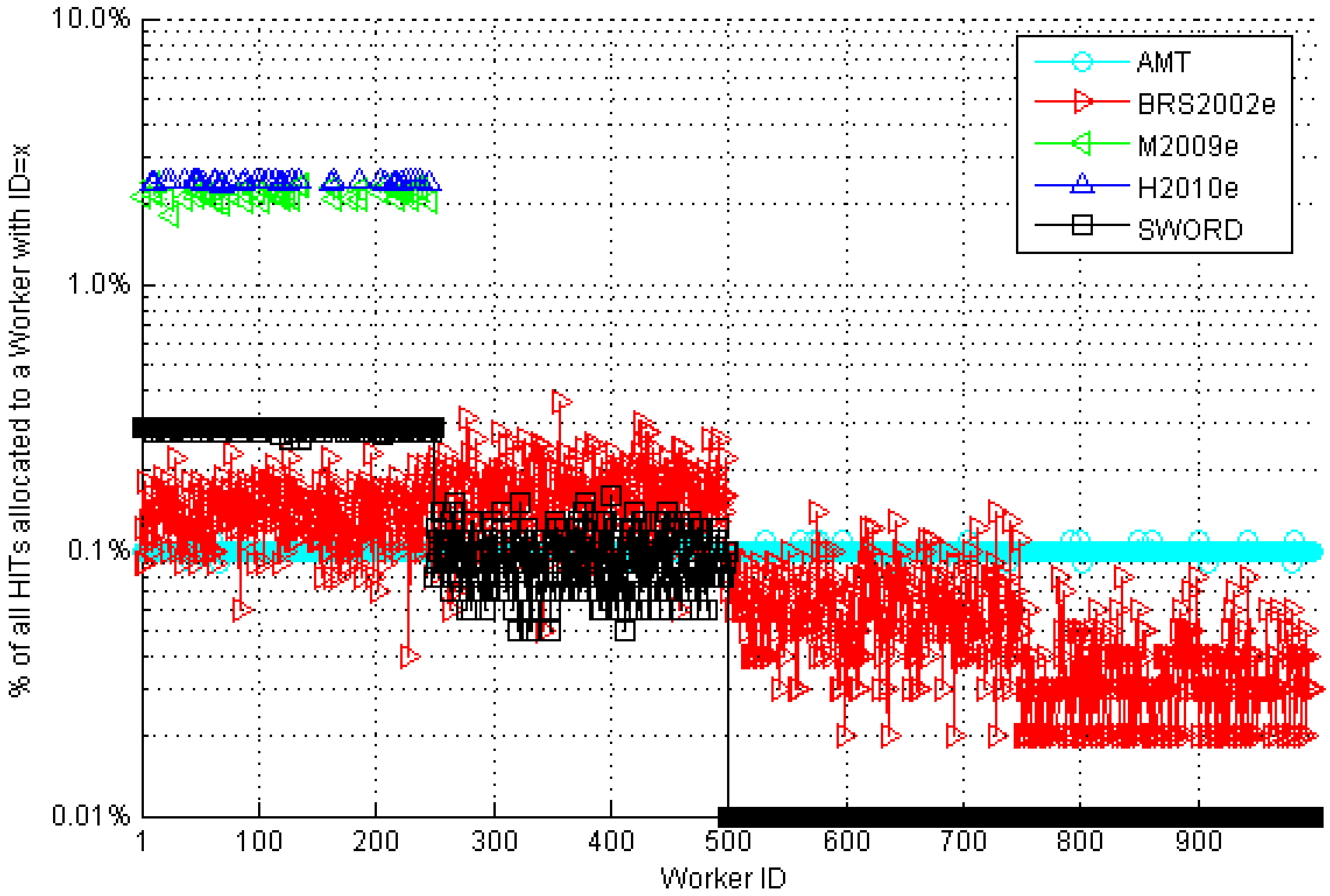

Figure 38. The percentage of all HITs allocated to each worker agent by various approaches over 1,000 time steps of simulation under worker agent population configuration *Hon50*.

Table 9. The distribution of HITs among Hon worker agents from the data in Figure 38.

| | **AMT** | **BRS2002e** | **M2009e** | **H2010e** | **SWORD** |
|---|---|---|---|---|---|
| **Average** | 1851.76 | 520.14 | 784.27 | 807.45 | 4935.45 |
| **Standard Deviation** | 42.34 | 122.97 | 1661.52 | 1831.14 | 227.77 |

## 6.6. Competition-based Evaluation

Comparing the performance of the proposed SWORD approach against benchmark approaches with a fixed number of requesters and workers for each approach may not be a good way to reflect their performances in a competitive crowdsourcing marketplace. In reality, requesters and workers may decide to work in any of the crowdsourcing systems based on how well they are doing under each system. Intuitively, if a requester cannot get high quality results or suffers from long waiting time in a crowdsourcing system, he is not likely to be satisfied with the user experience and may switch to another crowdsourcing system. Similarly, if a worker cannot earn much reward through working in a crowdsourcing system, he may leave the crowdsourcing system and join another one. In this section, the relative performance of the SWORD approach and the benchmark approaches by letting them compete for the preference of self-interested requester and worker agents.

### 6.6.1. Experiment Setup

In this set of experiments, five simulated crowdsourcing systems, each using a different HIT allocation approach, are operating side by side. Similar to Section 6.5.1, the five approaches are *AMT*, *BRS2002e*, *M2009e*, *H2010e* and *SWORD*. All crowdsourcing systems, except *AMT*, adopt the *BRS2012* model as the underlying reputation evaluation model. Each round of experiment runs for 1,000 time steps and is repeated 10 times to reduce the effect of random variation.

The environment of each experiment includes 50 common witness agents who accumulate direct trust evidence about worker agents and provide testimonies to requester agents with HIT assignment approaches that require this information. These agents are allowed to run for 200 time steps to accumulate some direct trust evidence before agents equipped with various approaches start to operate.

A total of $N_w$ worker agents are included in each experiment. The number of them adopting the four different behavior patterns is varied in different experiments to simulate different population configurations. Each worker agent population configuration is denoted as *HonX*. It represents a worker agent population consists of $\frac{1}{2}X\%$ *Hon* worker agents, $\frac{1}{2}X\%$ *MH* worker agents, $\frac{1}{2}(100-X)\%$ *MM* worker agents, and $\frac{1}{2}(100-X)\%$ *Mal* worker agents. Worker agents perform the *Clean Sweep* operation as mentioned in Chapter 4 at every time step to drop any HITs in their own request queues that have become expired.

$N_{req}$ requester agents are involved in each experiment. Requester agents are designed only to propose new HIT Groups when the last proposed HIT Group is completed or becomes expired. In this way, the social welfare produced represents the maximum social welfare a crowdsourcing system made up of the given requester agent and worker agent populations can produce under the given HIT allocation approach.

The worker agents and the requester agents are registered users of all five simulated crowdsourcing system. They are assumed to be self-interested and rational. The agents learn how to divide their time working in each of the five systems based on their experience in the past. Thus, the more reward an agent receives from working in a crowdsourcing system, the more time it is willing to spend working in that crowdsourcing system in the future. All agents adopt an actor-critic learning [Sutton and Barto, 1998] based decision-making method to learn how to allocate their time based on its own wellbeing. For each crowdsourcing system $i$, an agent keeps track of the rewards and penalties it has received as:

$$r_i = \tilde{\mu}(t) \cdot R + (1 - \tilde{\mu}(t)) \cdot P$$

$$\tilde{\mu}(t) = \begin{cases} 1, & if\ a\ reward\ is\ received \\ 0, & if\ a\ penalty\ is\ received \end{cases} \tag{6.2}$$

where $R$ and $P$ are constants representing the reward and penalty values, $T_i$ is the total number of times an agent has performed a task (either publishing a HIT or completing a HIT) in the crowdsourcing system $i$. Once $r_i$ is calculated, it is compared with the baseline wellbeing variable $\widetilde{r_i}$ to update the learning parameter $p_i$ as:

$$p_i \leftarrow p_i + \rho \cdot (r_i - \widetilde{r_i}) \cdot (1 - \pi_i) \tag{6.3}$$

where $\rho$ $(0 < \rho \leq 1)$ is a constant controlling the learning rate. The baseline reward is then updated according to the formula:

$$\widetilde{r_i} \leftarrow \varphi \cdot \widetilde{r_i} + (1 - \varphi) \cdot r_i \tag{6.4}$$

where constant $\varphi$ $(0 < \varphi \leq 1)$ determines the weight given to the previous baseline reward and the new reward received. The probability of an agent choosing to work in any one of the five crowdsourcing systems at time step $t$ is:

$$\pi_i(t) = \frac{e^{p_i}}{\sum_{i=1}^{5} e^{p_i}}. \tag{6.5}$$

Based on this probability, each agent uses a Monte Carlo method to decide the allocation of its working time to the five competing crowdsourcing systems throughout the simulations. Once a choice is made at time step $t$, a worker agent will spend all its capacities working on HITs allocated to him in the chosen crowdsourcing system for that time step and disregard HITs pending its attention in other systems for that time step; while a requester agent will publish a new HIT Group in the chosen crowdsourcing system if its previous HIT Group has been completed. The values for the parameters used in our experiments are listed in Table 10.

Table 10. Parameter Values in the Experiments

| Symbol | Description | Value |
|---|---|---|
| $g_{max}$ | Maximum utility for successfully completing an HIT. | 1.0 |
| $c$ | The utility cost to a requester for proposing an HIT (i.e., the reward offered to workers to complete it). | 0.2 |
| $N_r(t)$ | The average number of HITs in a HIT Group. | 40 |
| $N_{req}$ | The total number of requester agents in an experiment. | 50 |
| $N_w$ | The total number of worker agents in an experiment. | 1,000 |
| $N$ | A weight parameter used to calculate the target queue size in the SWORD approach in Equation (5.9). | 1.0 |
| $V$ | The performance tuning parameter in the SWORD approach. | 2.0 |
| $Th_r$ | The minimum reputation value required by the approaches (except Group *AMT*) in the experiment for a worker to quality as a potential candidate for HIT allocation. | 0.6 |
| $\Pr(Exp)$ | The percentage of time spent by the SWORD approach for random exploration. | 10% |
| $T_{dl}$ | The deadline (i.e., maximum number of time steps from the time step an HIT is proposed) before which a HIT must be completed in order to be regarded as completed on time. | 14 |
| $R$ | The reward received for a successful performance in a crowdsourcing system by an agent. | 1 |
| $P$ | The penalty received for an unsuccessful performance in a crowdsourcing system by an agent. | -1 |
| $\rho$ | The learning rate parameter. | 0.4 |
| $\varphi$ | The weight parameter for updating baseline reward. | 0.6 |

In the next section, we analyze the numerical results about the long term choices made by the agents and the wellbeing of the competing crowdsourcing systems.

### 6.6.2. Analysis of Numerical Results

Figure 39 shows the average percentage of time allocated to working in the five competing crowdsourcing systems by the requester agents after 1,000 time steps of simulation under various worker agent population configurations. In all scenarios, the system using SWORD attracte d the largest share of the requesters' preference. Under worker agent populations *Hon10* to *Hon80*, the requesters spend more than half of their time, on average, using the service provided by SWORD. Although no trust management is used by AMT, it still managed to attract the second largest share of the requesters' time. Crowdsourcing systems using BRS2002e ranked the third according to the requesters' preference while M2009e and H2010e alternate on the fourth and fifth places. As the percentage of *Hon* workers increases, the average quality of HIT results achieved by AMT improves. Thus, its share of the requesters' time also increases.

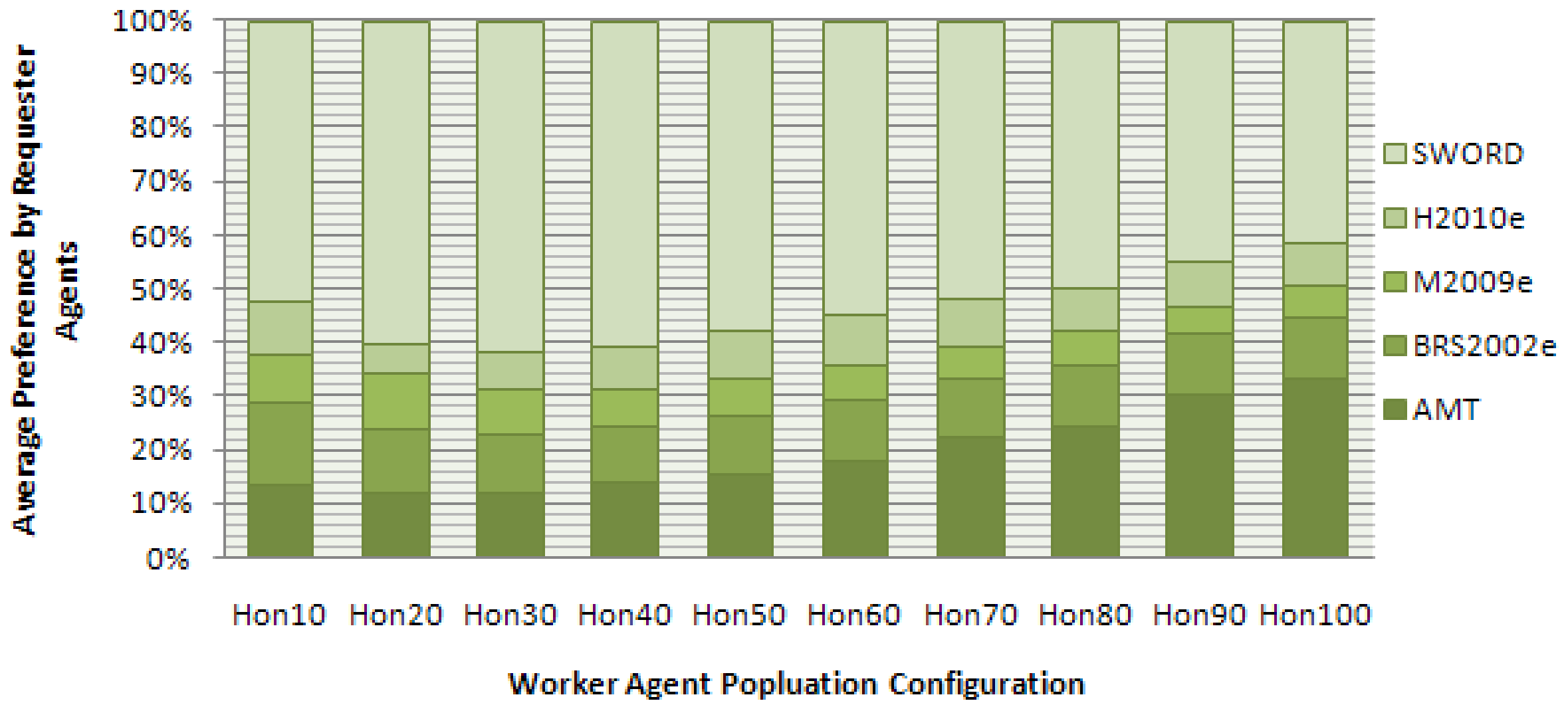

Figure 39. The average preference for the competing crowdsourcing systems by requester agents.

Figure 40 shows the average percentage of time allocated to working in the five competing crowdsourcing systems by the worker agents after 1,000 time steps of simulation under various worker agent population configurations. It appears that, on average, worker agents do not have a clear preference towards any of the five crowdsourcing systems which is against the intuition. However, by plotting individual worker agents' preferences under worker agent population configuration *Hon50* as shown in Figure 41, it can be observed that different types of worker agents have different preferences for the competing crowdsourcing systems. Worker agents with IDs equal to 1~250 (the *Hon* workers) and 251~500 (the *MH* workers) prefer working under SWORD the most, while worker agents with IDs equal to 501~750 (the *MM* workers) and 751~1000 (the *Mal* workers) generally prefer working under M2009e and H2010e the most.

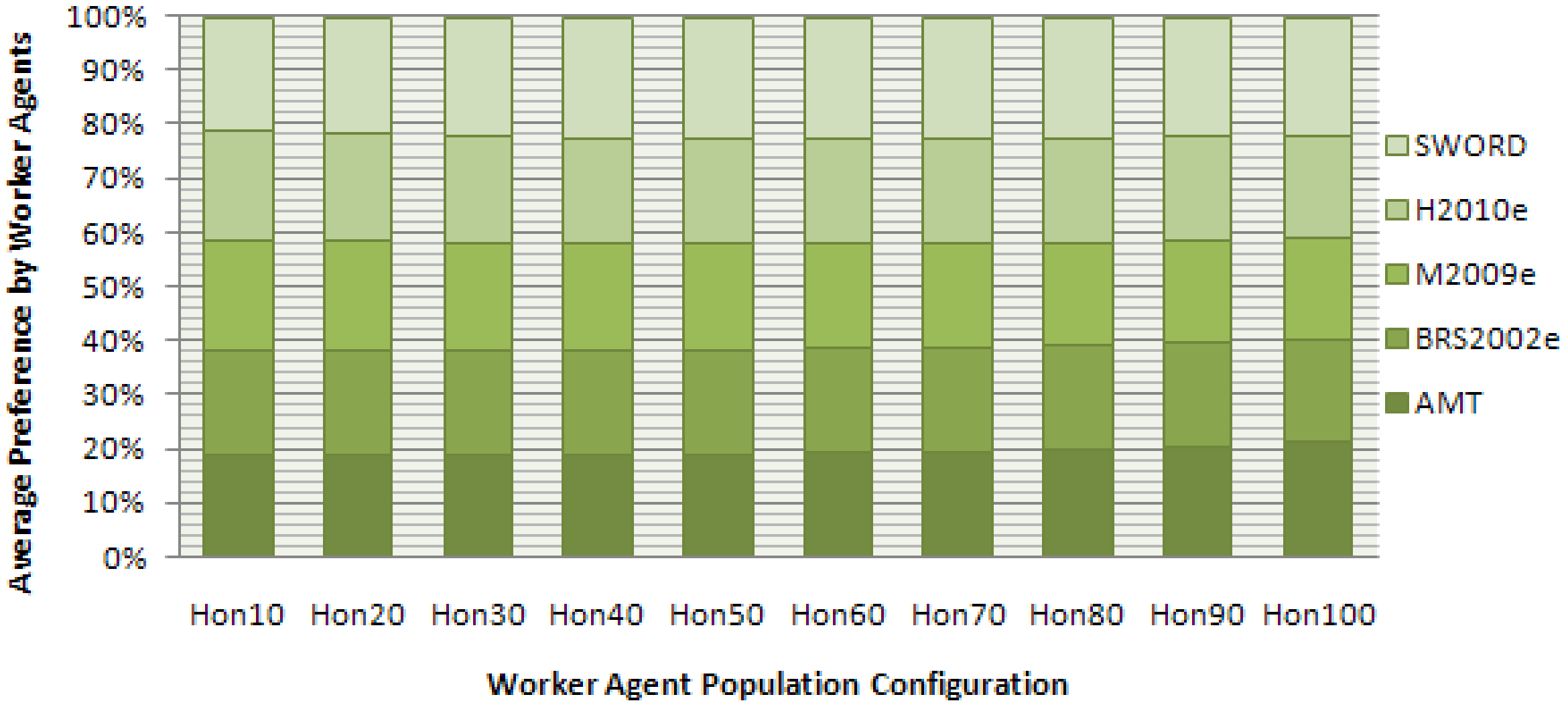

Figure 40. The average preference for the competing crowdsourcing systems by worker agents.

Since the worker agents follow different behavior patterns, the treatment received by different types of worker agents can be very different. Figure 42 summarizes the average reward received by worker agents belonging to different types from the competing crowdsourcing systems under worker agent population configuration *Hon50*. *Hon* workers and *MH* workers received significantly higher net reward (i.e., reward – cost) from the crowdsourcing system adopting the SWORD approach than other competitors. This attracts *Hon* and *MH* workers to spend a large percentage of their time working under the management of SWORD. For *MM* and *Mal* workers, the situation is different. Since the quality of HIT results produced by them is generally low, on average, they did not receive enough payment from crowdsourcing systems using the SWORD, BRS2002e and AMT approaches, which gave them opportunities to service HIT requests either through exploration (in the case of SWORD and BRS2002e) or allowing them to taken on HIT requests on the first-come-first-served basis (in the case of AMT), to compensate the effort they have spent working on the HITs. This appears to the workers as a negative net reward and causes them to reduce their preference towards these three crowdsourcing systems. Under M2009e and H2010e, these two types of worker agents are allocated close to no HITs over the long run. Thus, although they receive no rewards from these two crowdsourcing systems, they incur no cost either. This results in the appearance that *MM* and *Mal* workers prefer M2009e and H2010e the most. In reality, these two types of worker agents are driven towards extinction in all five competing crowdsourcing systems.

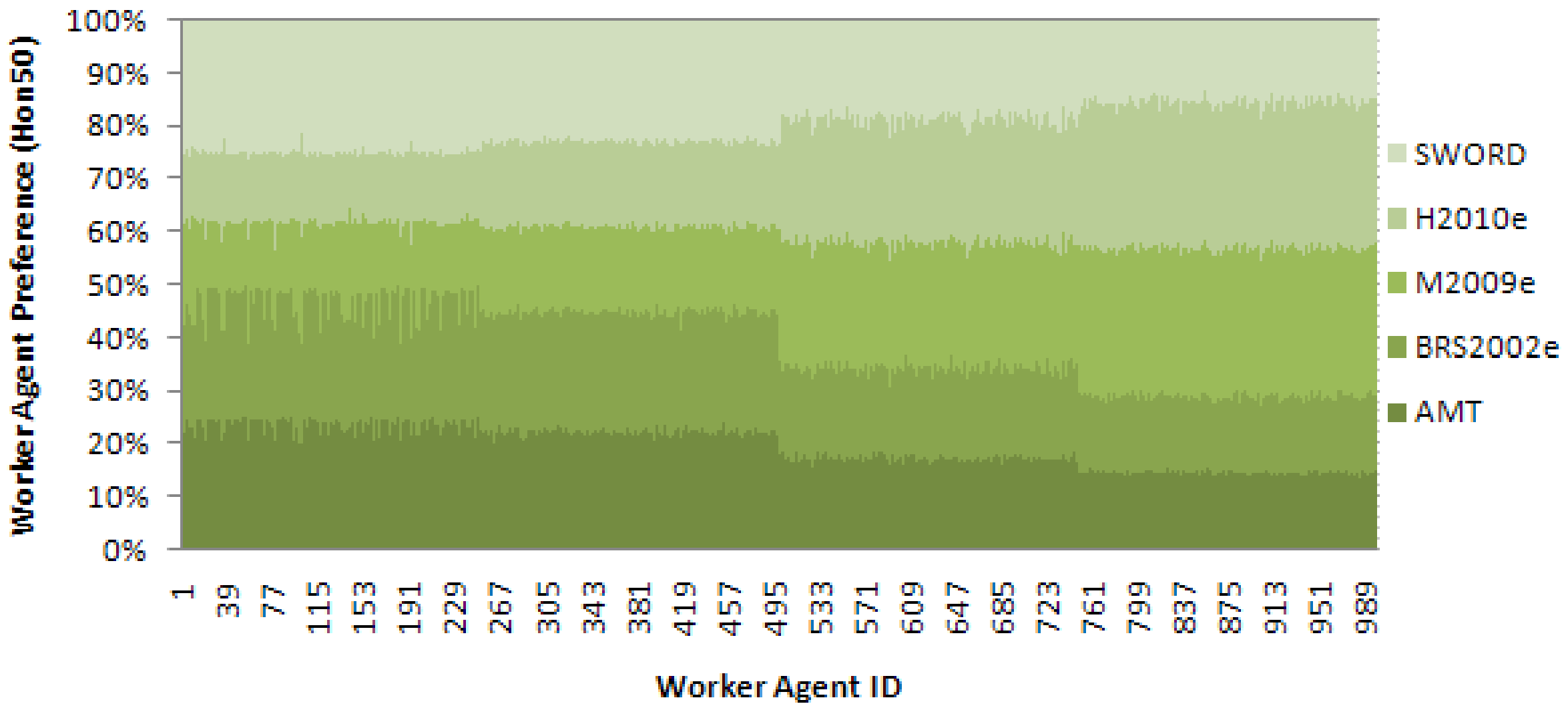

Figure 41. The preference for the competing crowdsourcing systems by individual worker agents under worker agent population configuration *Hon50*.

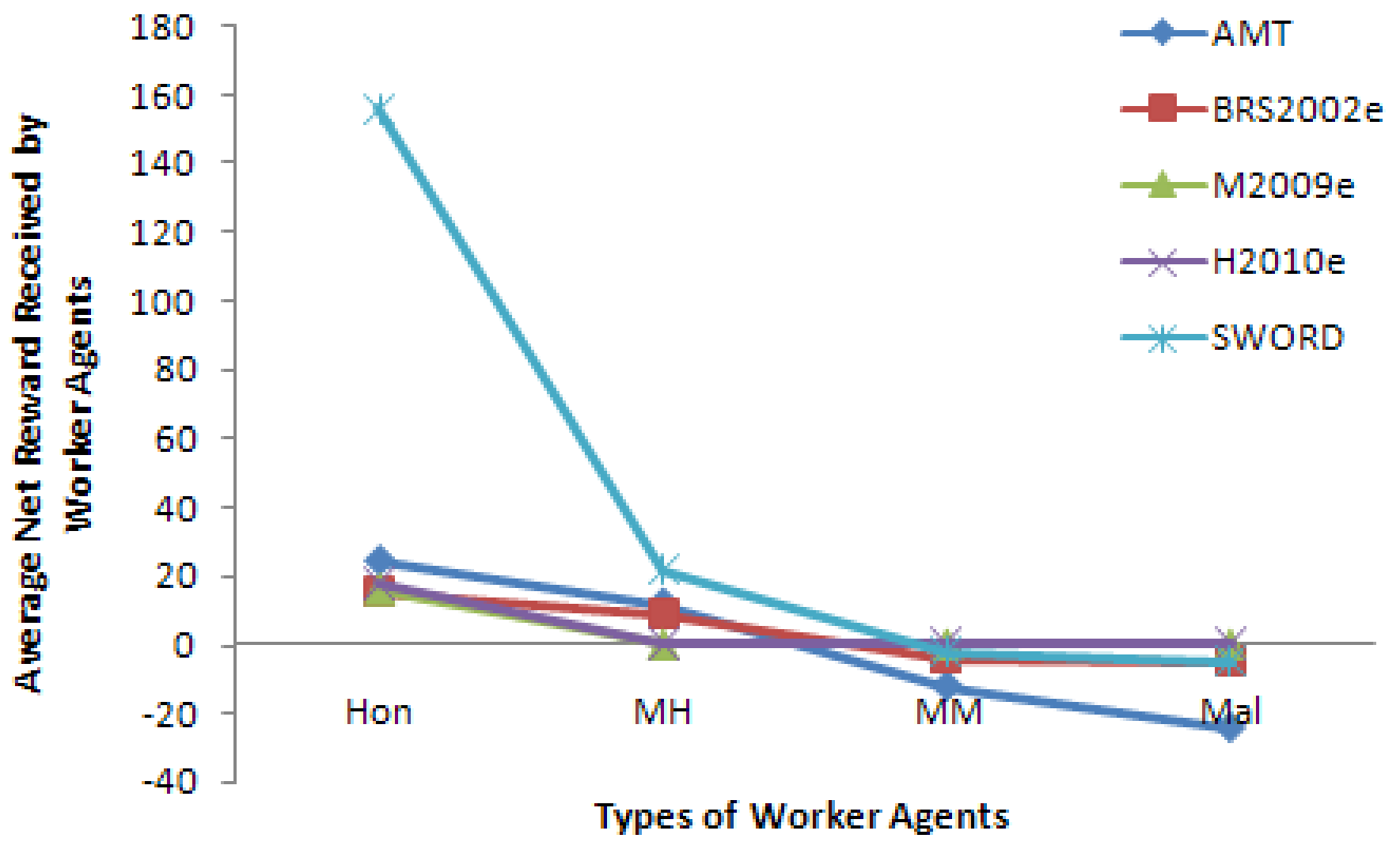

Figure 42. The treatment received from the competing crowdsourcing systems by worker agents following different behavior patterns under worker agent population configuration *Hon50*.

On average, the crowdsourcing system using the SWORD approach is able to attract more *Hon* worker agents over the long run as shown in Figure 43. This is especially the case when the percentage of *Hon* worker agents in the population is low (i.e., when trust management is most needed).

The competitiveness of the fiving competing crowdsourcing systems is illustrated in Figure 44. Under all worker agent population configurations, the proposed SWORD approach achieved the highest share of the total time averaged social welfare produced. Its share of the total social welfare is larger when the general level of trustworthiness in the worker agent population is low (i.e., *Hon50* and below). As general level of worker agent trustworthiness increases, the total time averaged social welfare produced by the five crowdsourcing systems increases. When the worker population has a high level of general trustworthiness (e.g., *Hon100*), the performance of the SWORD approach is similar to that of AMT.

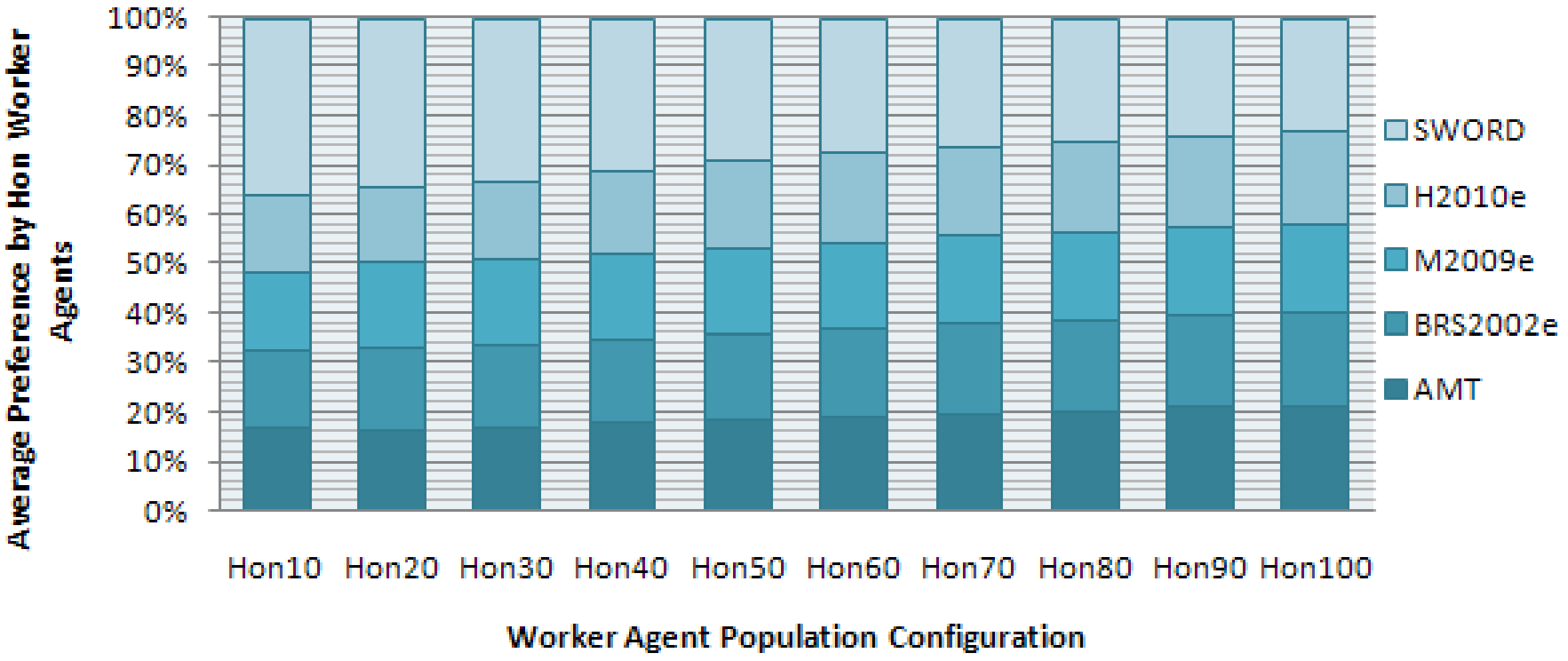

Figure 43. The average preference for the competing crowdsourcing systems by *Hon* worker agents.

Through efficient and trust-aware utilization of the available capacity of worker agents, the SWORD approach is also able to offer the requesters short waiting time to get quality results. Figure 45 shows the performance of various HIT allocation approaches under competition conditions with worker agent population configuration *Hon50*. With more requesters and more trustworthy workers being attracted to SWORD, it is able to complete all published HIT Groups using no more than 7 time steps. The relative performance of other approaches is similar to that under comparison experimental conditions with one important difference. Being unable to retain the interest from *Hon* and *MH* worker agents as effectively as SWORD, BRS2002e, M2009e and H2010e now have less worker agent resources at their disposal to complete HITs while still maintaining the quality of HIT results. This situation results in significant portions of the HITs cannot be completed on time and dropped by overloaded worker agents. The drop rates under M2009e and H2010e are especially high which, in practical operations, will drive these two crowdsourcing systems out of business.

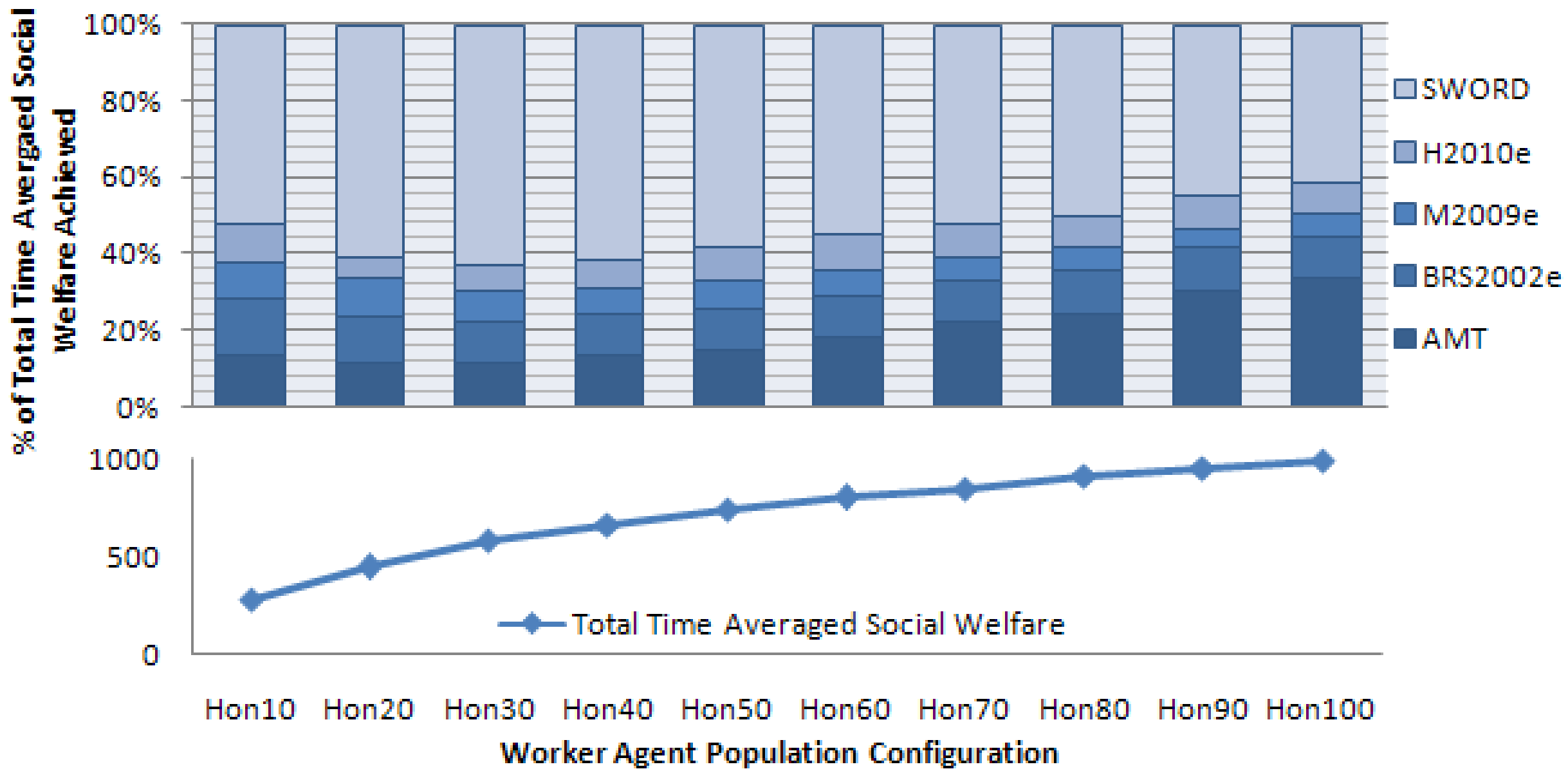

Figure 44. The competitiveness of crowdsourcing systems using different HIT allocation approaches in terms of social welfare.

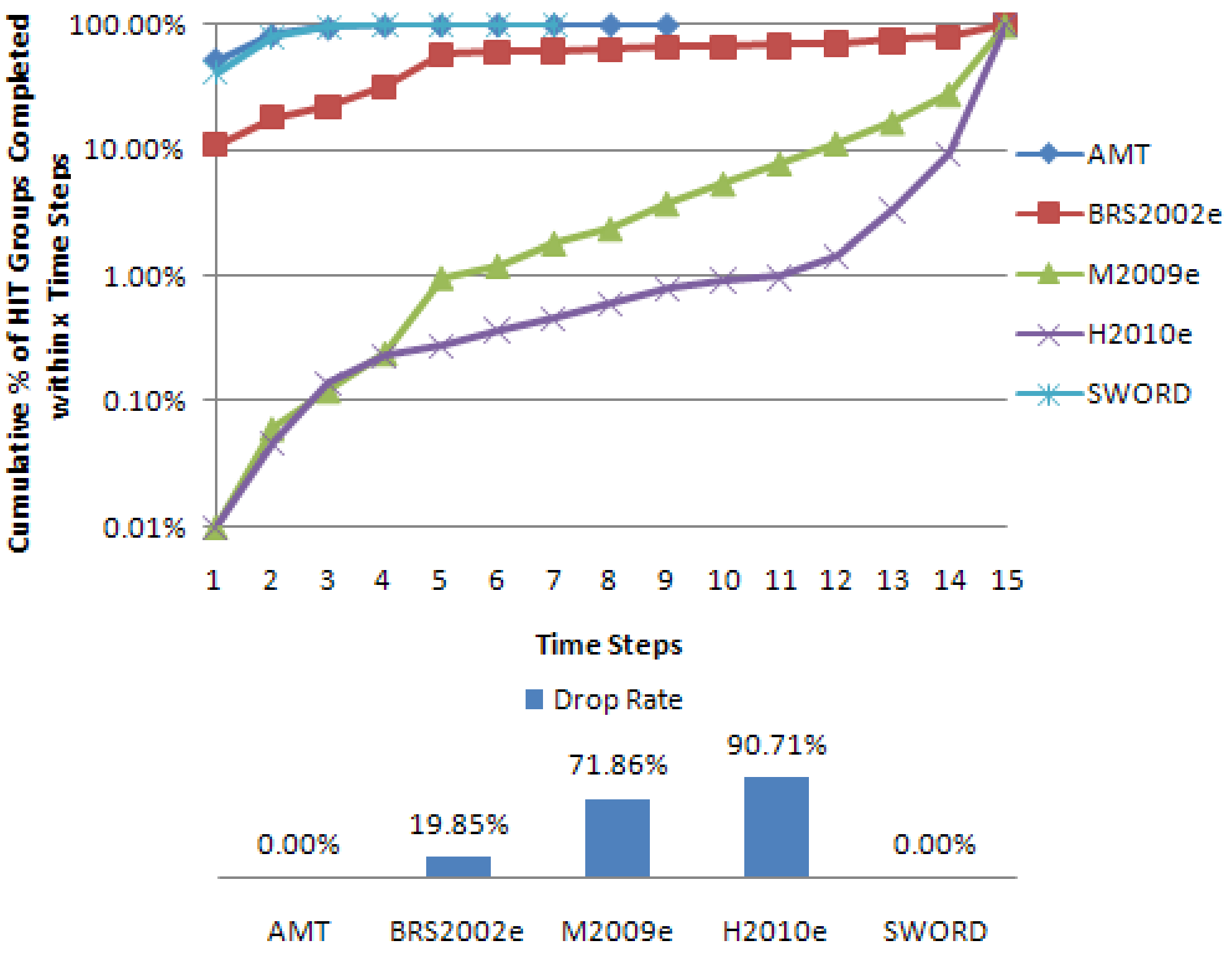

Figure 45. The performance of various approaches in terms of waiting time under competition conditions with worker agent population configuration *Hon50*.

## 6.7. Sensitivity Analysis

In the SWORD approach, the control parameter $V$ offers the system administrator some control over the trade-off between the proximity of the social welfare towards the optimal social welfare of a given crowdsourcing system and the average delay for requesters to obtain HIT results (i.e., their user experience). From Figure 46, it is apparent that by increasing $V$, the average quality of the HIT results increases while the $F_{Hon}$ value decreases. In essence, more HIT requests are being concentrated to highly reputable worker agents. The overall effect of doing so on the proximity of the social welfare in the system to the optimal social welfare is illustrated in Figure 47.

However, giving more importance to quality by increasing *V* cannot push social welfare arbitrarily close to optimal. In fact, as observed in Figure 48, once the *V* value is increased to

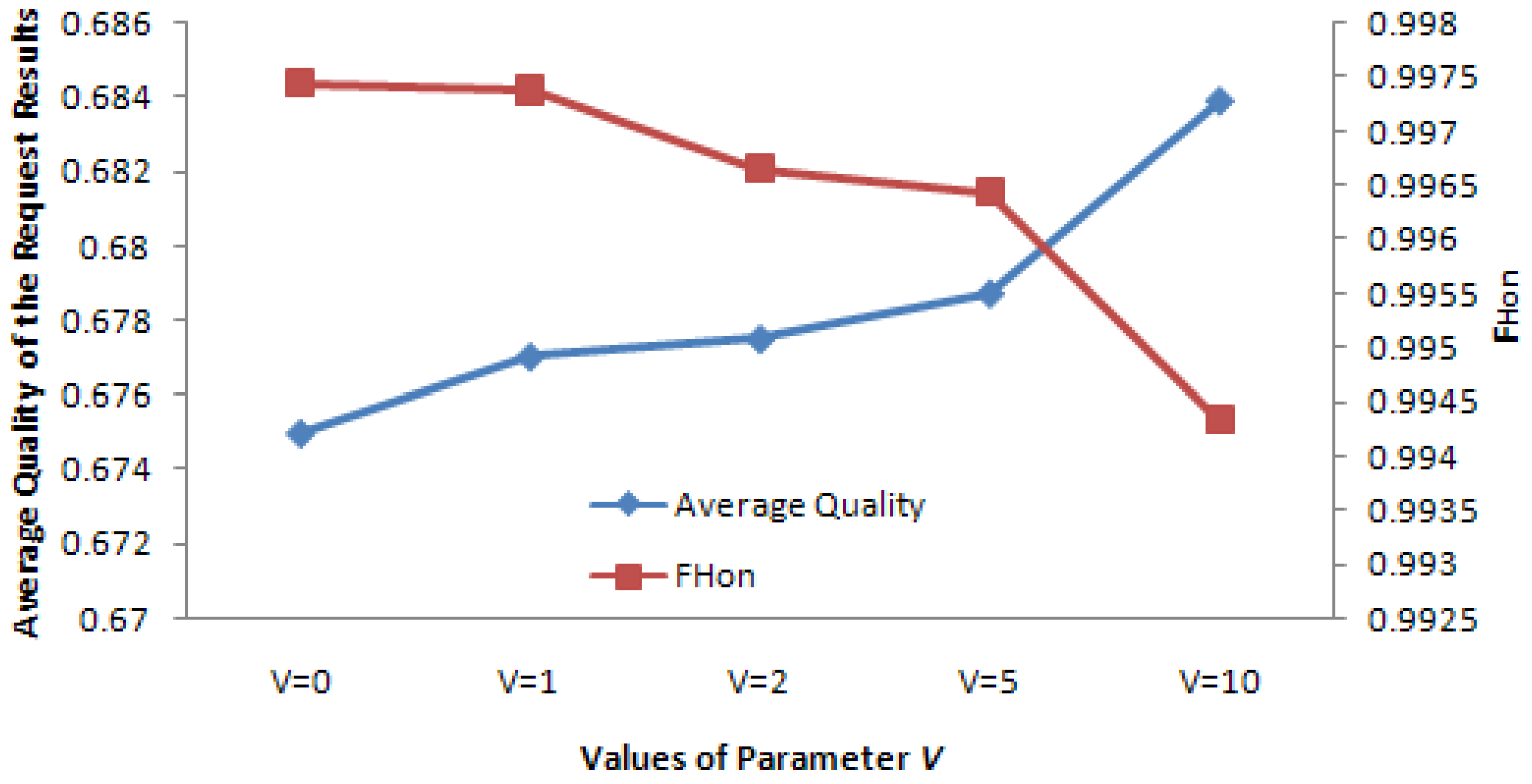

Figure 46. The change in average quality of HIT request results and $\boldsymbol{F_{Hon}}$ values achieved by the SWORD approach with different values for the control parameter *V*.

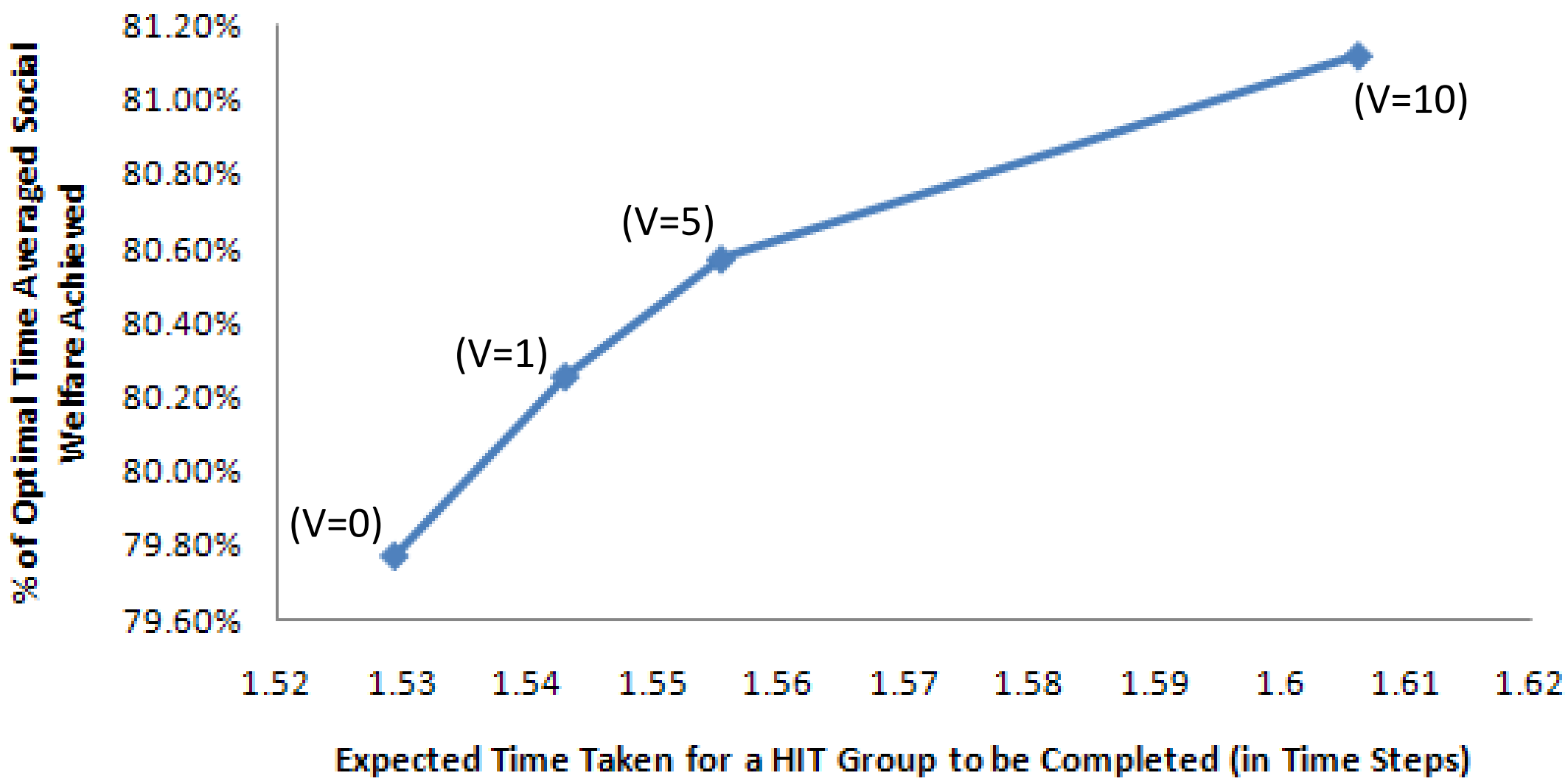

Figure 47. The percentage of optimal time averaged social welfare achieved by the SWORD approach with different values for the control parameter *V*.

such an extent that over-concentration of HIT allocation to reputable workers starts to appear, both the average quality of the HIT results as well as $F_{Hon}$ start to decrease together. In this situation, the reputation damage effect appears and brings down the social welfare achieved by the system as shown in Figure 49. When *V* is sufficiently large, the social welfare achieved by the SWORD approach becomes similar to those achieved by

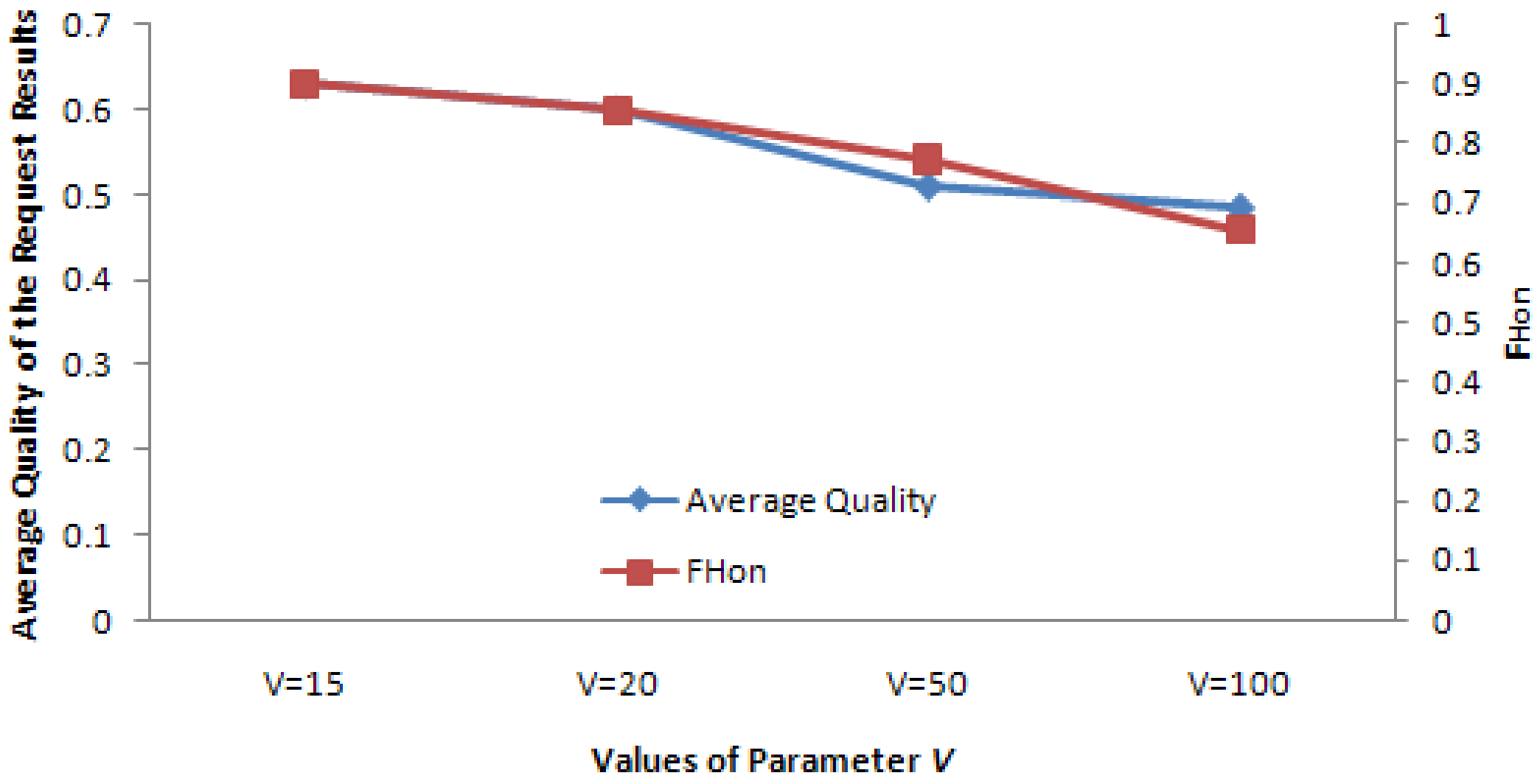

Figure 48. The change in average quality of HIT request results and $\boldsymbol{F_{Hon}}$ values achieved by the SWORD approach with different values for the control parameter *V*.

BRS2002e, M2009e and H2010e as shown in Figure 50 because they suffer from the same reputation damage problem.

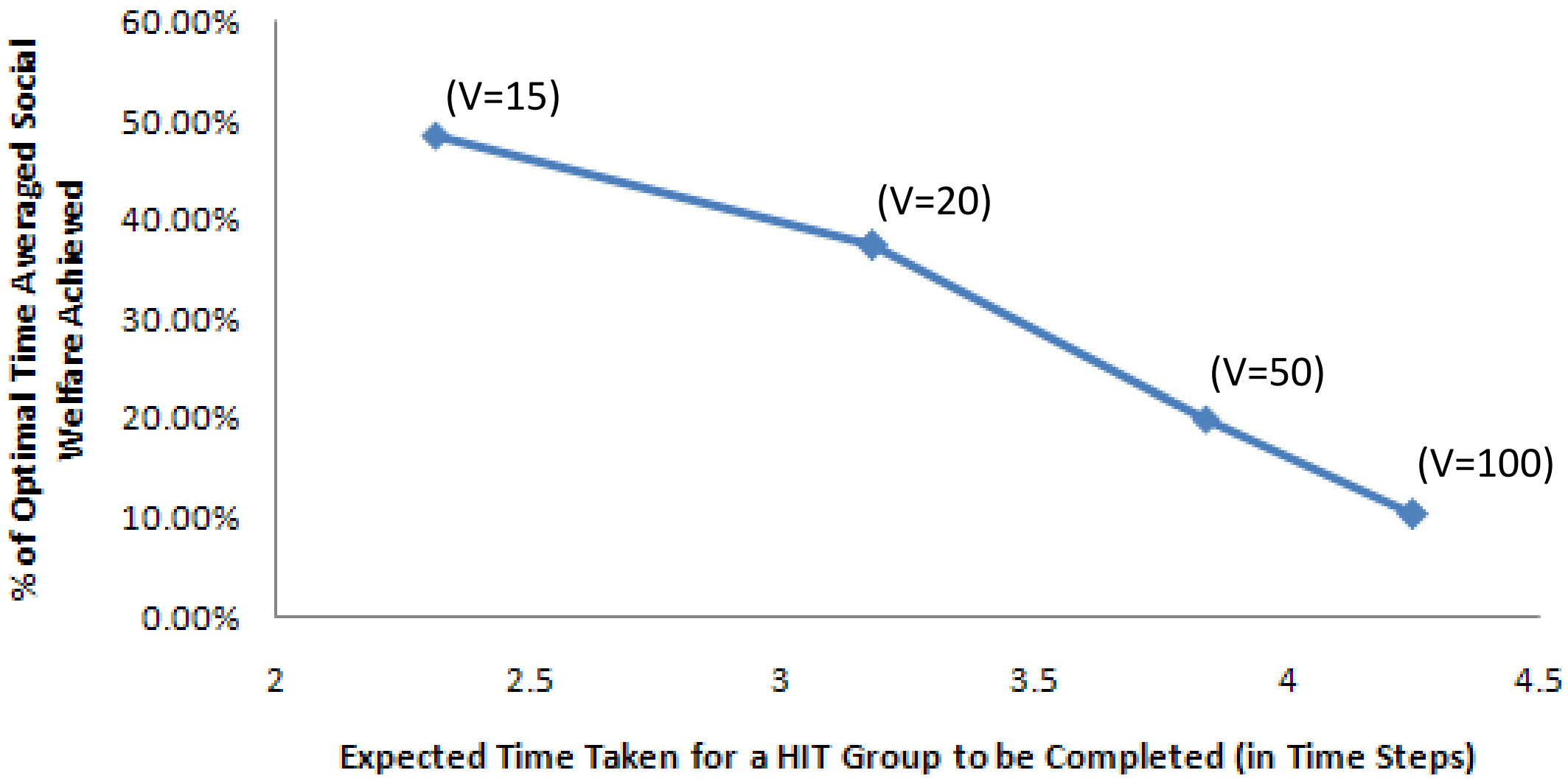

Figure 49. The percentage of optimal time averaged social welfare achieved by the SWORD approach with different values for the control parameter *V*.

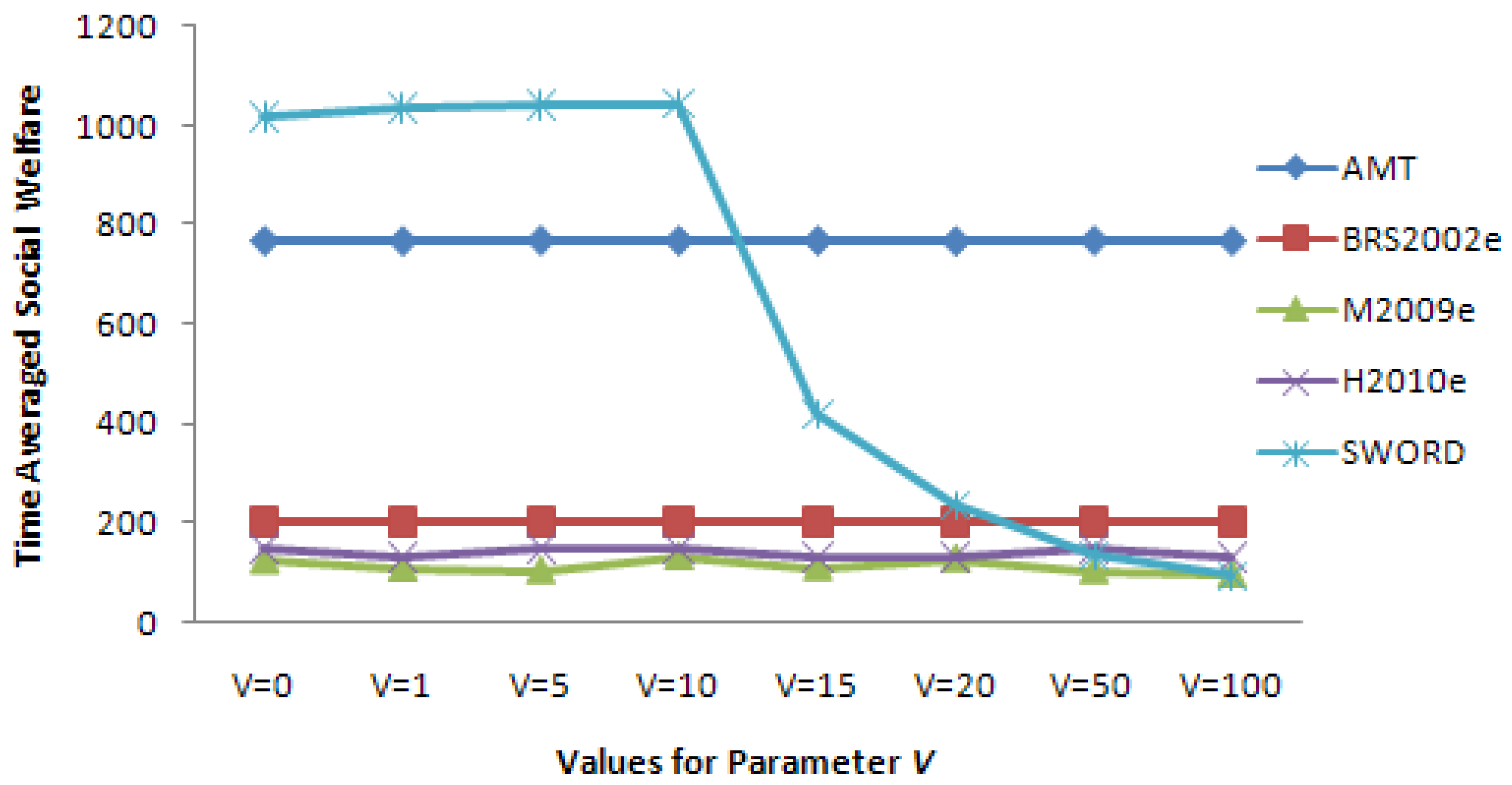

Figure 50. Summary of performance in terms of social welfare against different values for the control parameter *V*.

## 6.8. Summary

In Chapter 4 to 6, we have identified a new research problem mostly overlooked by existing trust-aware decision-making approaches – the reputation damage problem (RDP). The conflict between agents' self-interest and the wellbeing of an MAS is at the centre of the RDP. In order to mitigate the adverse effect of RDP on the social welfare in an MAS, we proposed a centralized trust-aware decision-making approach called SWORD. Theoretical analysis has shown that it provides performance guarantees to the extent limited by the collective capacity of an MAS. Extensive empirical evaluations have shown that the SWORD approach can minimize the impact of the RDP on the social welfare of an MAS by making less greedy and more situation-aware decisions that state-of-the-art approaches. It provides better user experiences for the truster agents through increased utility and shorter waiting time, and better user experiences for the trustee agents through maintaining social equity among them while avoiding overloading top performers with requests. Through competition-based performance evaluations, it has been shown that the SWORD approach can better maintain social sustainability in a crowdsourcing scenario better than the benchmark approaches.

Being a centralized approach, SWORD can access a wide range of information about the trustee agents. This enables it to make efficient, reputation-aware and situation-aware decisions. Nevertheless, this reduces its scalability. For systems without a central entity (e.g., mobile ad hoc networks, distributed sensor networks, etc.), the SWORD approach cannot operate. In view of these limitations, we propose a distributed approach to address the RDP in the next chapter.

# Chapter 7

# DRAFT: A Distributed Decision Approach for Social Sustainability in Multi-agent Systems

In Chapter 5 and 6, we have proposed SWORD which is a centralized decision-making approach to enable existing trust models to make efficient interaction partner selection decisions in multi-agent systems (MASs) to improve their social sustainability. Nevertheless, we are aware that many MAS applications are distributed systems in nature and do not possess a trusted central entity to host the SWORD approach. Therefore, a distributed decision-making approach that can address the reputation damage problem will be able to serve a wider domain of applications.

In this chapter, we propose the Distributed Request Acceptance approach for Fair utilization of Trustee agent services (DRAFT) to achieve this goal. It enables a trustee agent to dynamically determine how many new interaction requests from truster agents should be accepted into its pending request queues at each time step, given its current workload, its assessment of its reputation standing in the MAS based on the truster agents ratings, and the value and cost associated with serving new requests in order for it to achieve and maintain a good reputation. It promotes social equity based on the reputations of the trustee agents (i.e., the contributions of trustee agents to the wellbeing of an MAS) and achieves efficient utilization of their services.

To the best of our knowledge, the DRAFT approach is the first such approach designed for trustee agents in an MAS to use. Previously, most trust models were focused on helping truster agents select trustee agents. They usually assume that trustee agents are always motivated to accept more requests in order to earn more potential rewards. Thus, the prevailing assumption among existing trust models is that trustee agents adopt an *accept-when-requested* approach when their services are being asked for by truster agents. In order to enrich the decision-making approach for trustee agents in an open MAS where reputation is used as a social capital to facilitate interactions among agents with the DRAFT approach, we first propose a more realistic MAS system model to capture the limitation in trustee agents' capacity to serve task requests and the truster agents' tolerance on waiting time.

## 7.1. System Model

In a practical system (e.g., e-commerce), a trustee may only be qualified to perform up to *N* different types of tasks. In computational trust literature, the trust evidence of a trustee agent in serving different types of tasks are often recorded separately (e.g., a trustee agent can have a high reputation for selling T-shirts but a low reputation for selling gloves, although it is qualified to sell both types of items). This is referred to as the *context* of the trust evidence [Weng *et al.*, 2010] and denoted as $c \in (1, \ldots, C)$ in our system model. The maximum utility payoff for successfully completing one task of each type is denoted by the vector $\vec{G} = (G_{max}^{(c)})$. The general effort level required to complete one task of each type is denoted as $\vec{e} = (e^{(c)})$ which should be proportional to its corresponding maximum utility payoff.

For each type of tasks, there is a generally accepted maximum completion time (the deadline) from the perspective of the truster agents represented by the vector $\overrightarrow{T_{dl}} = (T_{dl}^{(c)})$. If a task is completed within the maximum completion time, the trustee agent *n* will receive a rating from the requesting truster agent based on the quality of the result; otherwise, *n* will receive a negative rating regardless of the quality of the result.

Pending requests are stored in trustee agent $n$'s task queues $\vec{Q}(n) = (Q_n^{(c)}(t))$ according to the contexts to which the tasks belong. At each time step, *n* is able to expend up to a maximum of $e_n^{max}(t)$ amount of effort to serve the tasks in its task queues. This value differs from agent to agent due to the agent's own limitations. The tasks in $\vec{Q}(n)$ are served by *n* on a first-come-first-served (FIFO) basis across all queues based on their time of acceptance.

The actual utility payoff gained by *n* for completing a task depends on whether the result is deemed successful by the requesting truster agent:

$$g_n^{(c)}(t) = \begin{cases} G_{max}^{(c)}, & if\ interaction\ is\ successful \\ 0, & Otherwise \end{cases}. \qquad (7.1)$$

As mentioned before, the interaction is deemed successful by the requesting truster agent only if the result produced by the trustee agent $n$ is correct and received by the truster agent completed before the end of the deadline $T_{dl}^{(c)}$.

At each time step, the number of incoming task requests received by trustee agent $n$ are denoted as $\vec{\lambda}(n) = (\lambda_n^{(c)}(t))$. The exact number depends on the interaction decisions made by the truster agents in the MAS which is influenced by $n$'s reputation. The reputation of $n$ in each context is represented by $\tau_n^{(c)}(t)$. As our focus in this study is on decision-making by trustee agents, we do not discuss how $n$'s reputation is evaluated. Instead, we assume the existence of a probabilistic reputation evaluation model (e.g., *BRS2012* as specified in Chapter 5) which produces $\tau_n^{(c)}(t)$ in the range of [0, 1] with 1 denoting completely trustworthy and 0 denoting not trustworthy at all.

## 7.2. Research Objectives

In this chapter, we still define the *social welfare* in an MAS as *the summation of the net utility produced by all individual interactions between truster and trustee agents*. The common goal of multi-agent trust models is to maximize the social welfare (which is the summation of utility derived from all interactions) in the system. We define a utility function $u_n^{(c)}(\mu)$ to represent the utility derived by trustee agent $n$ from completing $\mu$ tasks of type $c$. $u_n^{(c)}(\mu)$ will be defined in later sections of this paper. Then, the social welfare of an ODMAS is the summation of $u_n^{(c)}(\mu)$ over all $n$ and $c$. When trustees are resource constrained, the optimization objective can be expressed as:

$$\text{Maximize:} \quad \sum_{n,c} u_n^{(c)}(\overline{\mu}_n^{(c)}(t)) \tag{7.2}$$

$$\text{Subject to:} \quad \sum_{c}(\overline{\mu}_n^{(c)} \cdot e^{(c)}) \leq e_n^{max}(t) \tag{7.3}$$

$$0 \leq \overline{\mu}_n^{(c)}(t) \leq \lambda_n^{(c)}(t) \quad \text{for all } n, c \text{ and } t \tag{7.4}$$

The vector $(\overline{\mu}_n^{(c)}(t))$ depends partially on the discretion and capability of trustee agent *n*, which are intrinsic properties of *n*. Nevertheless, we propose the DRAFT approach to let *n* dynamically adjust the number of new task requests from truster agents accepted into its own task queues $\overrightarrow{Q_n}(t) = (Q_n^{(c)}(t))$ at each time step which can influence $(\overline{\mu}_n^{(c)}(t))$ through Constraint (7.4).

## 7.3. The DRAFT Approach

In systems where $e_n^{max} \to \infty$, Constraint (7.3) is redundant. Greedy task delegation decisions made by truster agents, combined with the accept-when-requested request acceptance approach by trustee agents can maximize (7.2). However, under the revised system model, this may lead to severe congestion in task queues and inefficient use of trustee agents' capacities. To mitigate the adverse effect of this situation, intuitively, the task backlog in the task queues of all trustee agents should be consistently pushed towards a lower congestion state. We propose the DRAFT approach based on *Lyapunov Optimization* [Neely, 2010] which provides a principled way of managing congestion.

The overall level of task queue congestion in an ODMAS at any time step *t* can be measured by a *Lyapunov function* which, in our case, is defined as

$$L(t) \triangleq \sum_{n,c}(Q_n^{(c)}(t))^2. \tag{7.5}$$

The smaller the value of $L(t)$, the lower the level of congestion in trustee agents' task queues. Given the task queue lengths for all *n* and *c* in an MAS, the conditional *Lyapunov drift* in the Lyapunov function from one time step to the next can be expressed as

$$\Delta(t) \triangleq \mathbb{E}\{L(t+1) - L(t)|\boldsymbol{Q}(t)\} \tag{7.6}$$

where $\boldsymbol{Q}(t) = (\overrightarrow{Q_1}(t), \overrightarrow{Q_2}(t), \dots, \overrightarrow{Q_n}(t))$.

Since the main cause for a trust-aware MAS to achieve low social welfare is shown to be over utilization of reputable trustee agents which results in the reputation damage problem in Chapter 4, in order to improve social welfare, trustee agents should be utilized in a way that promotes social equity in the long run. At each time step, if trustee agents can individually make control decisions about which new task requests to accept to greedily minimize $\Delta(t)$, then the task backlog in their respective task queues should be pushed towards a lower congestion state consistently. Nevertheless, while minimizing the Lyapunov drift, it would be rational for the trustee agents in an MAS to try to maximize their reward (in the form of utility) as well. Combining these two considerations, the Lyapunov *reward-minus-drift* expression ($V \times reward - \Delta(t)$) which can be written as:

$$V \sum_{n,c} \mathbb{E}\{u_n^{(c)}\left(A_n^{(c)}(t)\right) | \boldsymbol{Q}(t)\} - \Delta(t) \tag{7.7}$$

should be maximized (i.e., maximizing *reward* while minimizing *drift*) through the selective acceptance of new task requests based on a trustee agent's current situation. The potential reward is affected by the amount of work the trustee agent can accomplish, the amount of work allocated to it and its reputation standing from the perspective of the truster agents. Therefore, the new objective function in (7.7) can be expressed as:

$$\text{Maximize:} \quad \sum_{n,c}[V \cdot u_n^{(c)}\left(A_n^{(c)}(t)\right) - Q_n^{(c)}(t) \cdot A_n^{(c)}(t)] \tag{7.8}$$

$$\text{Subject to:} \quad \sum_c (A_n^{(c)}(t) \cdot e^{(c)}) \leq e_n^{max}(t) \tag{7.9}$$

$$0 \leq A_n^{(c)}(t) \leq \lambda_n^{(c)}(t) \quad \text{for all } n, c \text{ and } t \tag{7.10}$$

where ($A_n^{(c)}(t)$) are control decisions made by trustee agent *n* at each time step (i.e., the exact number of task requests belonging to each context to be accepted into its task queues), and $V > 0$ is a chosen constant that affects the relative emphasis given to the reward and the drift respectively. In this study, we assume that the potential utility $u_n^{(c)}(A_n^{(c)}(t))$ is linearly related to $A_n^{(c)}(t)$. Thus,

$$u_n^{(c)}\left(A_n^{(c)}(t)\right) = A_n^{(c)}(t) \cdot g_n^{(c)}(t). \tag{7.11}$$

In the long run, the actual utility gain is expected to be related to the probability of trustee agent *n* producing a correct result within the stipulated deadline. Although this probability cannot be definitively known, it can be approximated using *n*'s reputation in serving each type of tasks. Therefore,

$$u_n^{(c)}\left(A_n^{(c)}(t)\right) = A_n^{(c)}(t) \cdot \tau_n^{(c)}(t) \cdot G_{max}^{(c)}. \tag{7.12}$$

By substituting (7.12) into (7.8), we have

$$\sum_{n,c}\left[V \cdot A_n^{(c)}(t) \cdot \tau_n^{(c)}(t) \cdot G_{max}^{(c)} - Q_n^{(c)}(t) \cdot A_n^{(c)}(t)\right]$$

$$= \sum_{n,c}\left[V \cdot \tau_n^{(c)}(t) \cdot G_{max}^{(c)} - Q_n^{(c)}(t)\right] A_n^{(c)}(t) = \sum_{n,c} a_n^{(c)}(t) \cdot A_n^{(c)}(t)$$

(7.13)

**Algorithm 7.1** DRAFT

**Input**: $a_n^{(c)}(t)$ values for all $c$ in trustee agent $n$, the incoming requests $\lambda_n^{(c)}(t)$ for all $c$ at $n$, and $e_n^{max}(t)$.
Rank $Q_n^{(c)}(t)$ in descending order of $\frac{a_n^{(c)}(t)}{e^{(c)}}$
**foreach** $Q_n^{(c)}(t)$ in $n$ **do**
  **if** $\frac{a_n^{(c)}(t)}{e^{(c)}} > 0$ **then**
    **if** $\lambda_n^{(c)}(t) \cdot e^{(c)} \leq e_n^{max}(t)$ **then**
      $A_n^{(c)}(t) = \lambda_n^{(c)}(t)$
    **else**
      $A_n^{(c)}(t) = \left\lfloor \frac{e_n^{max}(t)}{e^{(c)}} \right\rfloor$
    **end-if**
    $e_n^{max}(t) \leftarrow e_n^{max}(t) - A_n^{(c)}(t) \cdot e^{(c)}$
  **else**
    $A_n^{(c)}(t) = 0$
  **end-if**
  *Reject_Requests*($\lambda_n^{(c)}(t) - A_n^{(c)}(t)$, $c$)
**end-for**
Return($A_n^{(c)}(t)$ for all $Q_n^{(c)}(t)$ in $n$)

where $a_n^{(c)}(t) = V \cdot \tau_n^{(c)}(t) \cdot G_{max}^{(c)} - Q_n^{(c)}(t)$ is defined as the *availability score* of each task queue at *n*. The DRAFT approach helps a trustee agent come up with a task request acceptance plan ($A_n^{(c)}(t)$) about how many new tasks of different types it should accept at each time step based on an agent's current situation which is represented by the 3-tuple $\langle \left(a_n^{(c)}(t)\right), \left(\lambda_n^{(c)}(t)\right), e_n^{max}(t) \rangle$. In order to maximize (7.13), the DRAFT approach proceeds as illustrated in Algorithm 7.1. In essence, the higher the payoff per unit effort for a task $t^c$, the higher the reputation of *n* in performing tasks of type *c*, and the more spare capacity *n* currently has in accommodating more requests for performing tasks of type *c*, the more likely $t^c$ will be accepted by DRAFT on behalf of *n*. The *Reject_Requests()* function informs the requesting agent that a number of tasks of a particular type are not accepted by the trustee agent for processing at the current time step so that the truster agent can look for other alternative. The DRAFT approach has a computational complexity of $O(c)$.

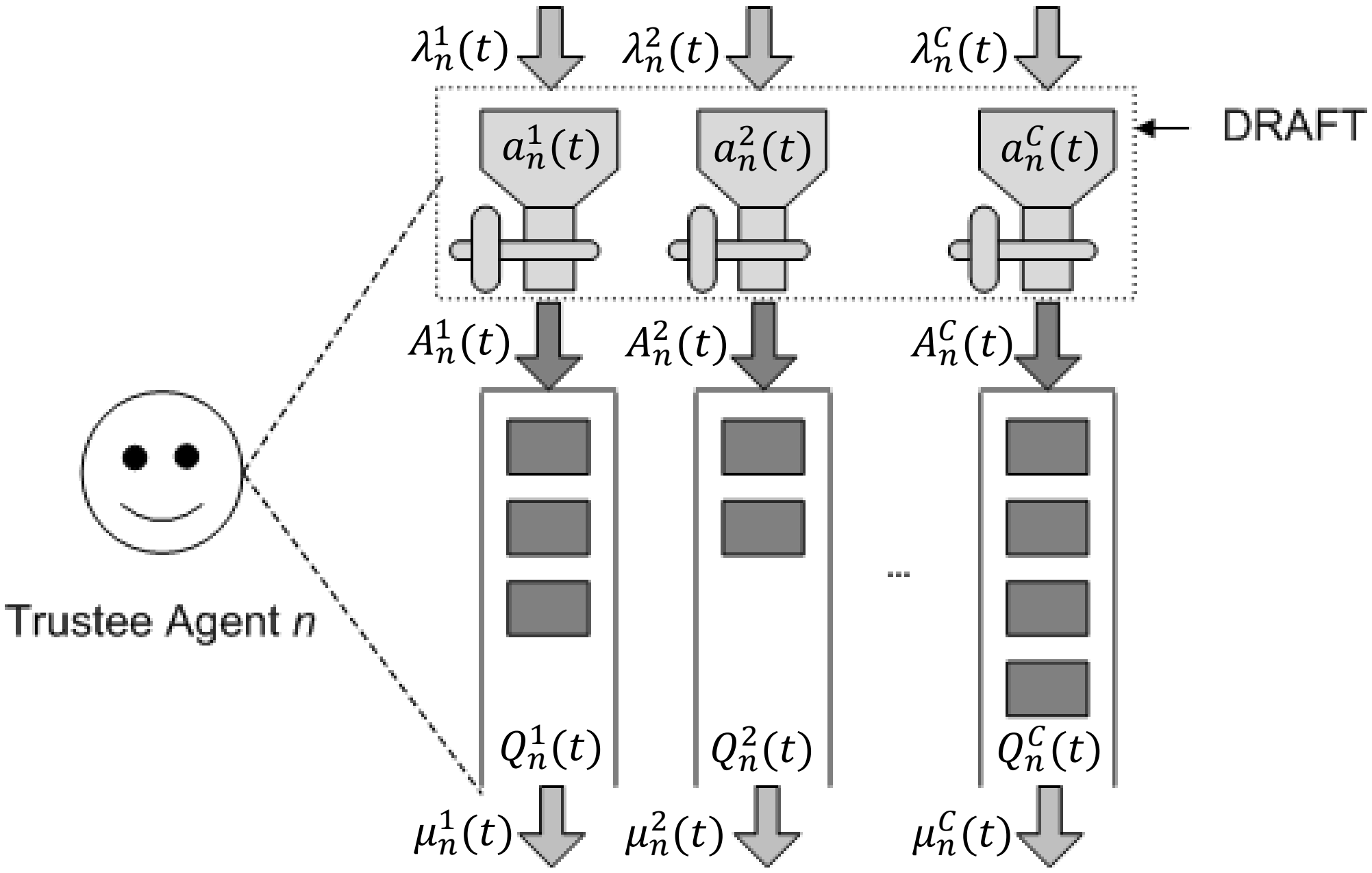

Figure 51. The working principle of the DRAFT approach.

The DRAFT approach enriches the existing trust evaluation models with a principled way for them to achieve more efficient utilization of trustee agent capacities in an MAS. It occupies an overlooked niche in the trust management domain by serving trustee agents instead of directly serving truster agents. With the DRAFT approach, trustee agents are, in effect, fitted with a set of intelligent faucets as illustrated in Figure 51 that dynamically determine how many requests of each type to be let into the trustee agent's pending request queues for the benefit of the trustee agent. Truster agents, on the other hand, will be able to know with more certainty when they are required to widen their scope of exploration for trustee agents when they are informed that the currently chosen trustee agent is not able to accept new requests. In this way, we conjecture that social equity and thus, social sustainability in an MAS will be achieved and maintained through complementing existing trust evaluation models with the use of the DRAFT approach. In the next section, theoretical analysis of the performance of the DRAFT approach is provided.

## 7.4. Theoretical Analysis of the Performance of DRAFT

Assume there are positive constants *B*, *M* and *V* such that the *reward-minus-drift* expression in (7.7) satisfies:

$$V\sum_{n,c}\mathbb{E}\{u_n^{(c)}\left(A_n^{(c)}(t)\right)|\boldsymbol{Q}(t)\} - \Delta(t) \geq VU^{opt} + M\sum_{n,c}Q_n^{(c)}(t) - B \tag{7.14}$$

where $U^{opt}$ is the social welfare produced by the optimal solution for (7.8). Taking the expectations over the distribution of $\boldsymbol{Q}(t)$ on both sides of (7.14), we have:

$$V\sum_{n,c}\mathbb{E}\{u_n^{(c)}\left(A_n^{(c)}(t)\right)\} - \mathbb{E}\{L\big(\boldsymbol{Q}(t+1)\big) - L\big(\boldsymbol{Q}(t)\big)\} \geq VU^{opt} + M\sum_{n,c}\mathbb{E}\{Q_n^{(c)}(t)\} - B$$

which holds for all time steps *t*. Summing both sides over $t \in \{0,1,\dots,T-1\}$, we have:

$$V\sum_{t=0}^{T-1}\sum_{n,c}\mathbb{E}\{u_n^{(c)}\left(A_n^{(c)}(t)\right)\} - \mathbb{E}\{L\big(\boldsymbol{Q}(T)\big) - L\big(\boldsymbol{Q}(0)\big)\}$$
$$\geq VTU^{opt} + M\sum_{t=0}^{T-1}\sum_{n,c}\mathbb{E}\{Q_n^{(c)}(t)\} - BT \tag{7.15}$$

Since $u_n^{(c)}(\cdot) \geq 0$ and $L(\cdot) \geq 0$, and suppose the potential utility reward for completing $A_n^{(c)}(t)$ tasks are bounded by $\sum_{n,c} u_n^{(c)}\left(A_n^{(c)}(t)\right) \leq G_{max}$ for all $t$ for some value $G_{max}$, by re-arranging the terms in (7.15) and dividing both sides by *TM*, we have:

$$\frac{1}{T}\sum_{t=0}^{T-1}\sum_{n,c}\mathbb{E}\{Q_n^{(c)}(t)\} - \frac{\mathbb{E}\{L(\boldsymbol{Q}(0))\}}{TM} \leq \frac{B + VG_{max} - VU^{opt}}{M} \leq \frac{B + VG_{max}}{M}$$

(7.18)

By taking lim sup as $T \to \infty$ on both sides of (7.18):

$$\lim_{T\to\infty} \sup\left\{\frac{1}{T}\sum_{t=0}^{T-1}\sum_{n,c}\mathbb{E}\{Q_n^{(c)}(t)\}\right\} \leq \frac{B + VG_{max}}{M}$$

(7.19)

Similarly, by re-arranging (7.15) and dividing both sides by *TV*, we can also have:

$$\frac{1}{T}\sum_{t=0}^{T-1}\sum_{n,c}\mathbb{E}\{u_n^{(c)}\left(A_n^{(c)}(t)\right)\} \geq U^{opt} + \frac{M}{TV}\sum_{t=0}^{T-1}\sum_{n,c}\mathbb{E}\{Q_n^{(c)}(t)\} - \frac{B}{V} - \frac{\mathbb{E}\{L(\boldsymbol{Q}(0))\}}{TV}$$

$$\geq U^{opt} - \frac{B}{V} - \frac{\mathbb{E}\{L(\boldsymbol{Q}(0))\}}{TV}$$

(7.18)

By taking lim inf as $T \to \infty$ on both sides of (7.18):

$$\lim_{T\to\infty} \inf\left\{\frac{1}{T}\sum_{t=0}^{T-1}\sum_{n,c}\mathbb{E}\{u_n^{(c)}\left(A_n^{(c)}(t)\right)\}\right\} \geq U^{opt} - \frac{B}{V}$$

(7.19)

From the above analysis, it can be deduced that if the condition in (7.14) can be fulfilled, which can be done through careful choice of the values for *B*, *M* and *V*, then, based on (7.17), it can be concluded that a theoretical upper bound exists for all pending request queues in all trustee agents over the long run under the management of the DRAFT approach. This ensures that the request queue lengths will not keep on increasing and the trustee agents always can stop the growth of their request queues so that they will not be overwhelmed by incoming requests.

In addition, based on (7.19), it can be deduced that the time averaged social welfare achieved through following the DRAFT approach can approach that achieved by the optimal solution within *B/V* in the long run. By increasing *V*, the social welfare produced by the DRAFT approach can be made closer to the optimal social welfare. However, increasing *V* also causes the upper bound to the pending request queue lengths to rise according to (7.17), thereby increasing the expected time taken to complete a request. Due to the physical limitations of the trustee agents in a given MAS, if the increase in the value of *V* causes the expected completion time of tasks to exceed the stipulated deadlines, social welfare will start to decrease as the level of satisfaction from truster agents will be reduced. Setting the value of *V* arbitrarily high will not make the social welfare produced by the DRAFT approach be indefinitely close the optimal since doing so will also require the value of *B* to be increased in order to satisfy the (7.14). Thus, the trade-off between social welfare and the delay in receiving services only exist within a certain range of the value of *V* which may be different based on the physical limitations of the agent populations in each given MAS.

## 7.5. Evaluation

In order to further evaluate the performance of the DRAFT approach, it is implemented within a simulated multi-agent environment which is designed based on the revised system model. The objectives of the experiments are to investigate whether enabling trustee agents to make control decisions on which new task requests to accept through the use of the DRAFT approach can be beneficial to themselves as well as the agent society in an MAS. Our hypotheses waiting to be verified through these experiments are:

- *Hypothesis 1*: A trustworthy trustee agent can better mitigate the adverse effect of reputation damage by using the DRAFT approach than using the traditional accept-when-requested approach.
- *Hypothesis 2*: The DRAFT approach achieves equitable distribution of task requests when applied under an existing trust evaluation model.
- *Hypothesis 3*: The social welfare of an MAS can be improved through the use of the DRAFT approach compared to the traditional accept-when-requested approach.

We now discuss the design of the experiments used to investigate our hypotheses.

### 7.5.1. Design of the Simulation Test-bed

In the simulated MAS, the trustee agent population consists of 100 agents belonging to four groups with different behavior patterns. They are labeled as:

1) *Hon agents*: honest trustee agents who return high quality task results randomly 90% of the time;

2) *MH agents*: moderately honest trustee agents who return high quality task results randomly 70% of the time;

3) *MM agents*: moderately malicious trustee agents who return high quality task results randomly 30% of the time;

4) *Mal agents*: malicious trustee agents who return high quality task results randomly 10% of the time.

The number of them adopting each of the four different behavior patterns is varied in each experiment to simulate different trustee agent population configurations. A trustee agent population configuration is denoted as *HonX*. It represents a trustee agent population consisting of $0.5X\%$ *Hon* trustee agents, $0.5X\%$ *MH* trustee agents, $0.5 \times (100 - X)\%$ *MM* trustee agents, and $0.5 \times (100 - X)\%$ *Mal* trustee agents. The maximum effort each type of trustee agents can expend per time step is shown in Table 11. More trustworthy trustee agents can complete less number of tasks in each time step than less trustworthy ones in order to maintain the quality of their work.

Table 11. The $\boldsymbol{e_n^{max}}$ values for trustee agent in each group.

| | *Hon* | *MH* | *MM* | *Mal* |
|---|---|---|---|---|
| $e_n^{max}$ | 25 | 30 | 35 | 40 |

Five categories of tasks are available for trustee agents to serve. Their properties used in the experiments are listed in Table 12.

Table 12. The properties of the types of tasks in the study.

| *c* | *1* | *2* | *3* | *4* | *5* |
|---|---|---|---|---|---|

| $G_{max}^{(c)}$ | 5 | 4 | 3 | 2 | 1 |
|---|---|---|---|---|---|
| $e^{(c)}$ | 5 | 4 | 3 | 2 | 1 |
| $T^{(c)}$ | 1 | 2 | 2 | 3 | 3 |

1,000 truster agents equipped with BRS2012 as presented in Chapter 4 for evaluating the reputations of the trustee agents are included in the simulation. Each truster agent randomly use 15% of its time for exploration. During exploitation rounds, truster agents require a trustee agent to have a reputation of over 2/3 in order to consider it as a candidate. At each time step, an equal number of requests for each type of task are sent out to the trustee agents based on their reputations by the truster agents (assuming truster agents do not intentionally distort their reputation ratings about the trustee agents).

### 7.5.2. Experiment Setup for DRAFT

Two MASs with the same environment conditions are run in parallel. In one MAS, the trustee agents adopt the traditional *accept-when-requested* approach for handling incoming task requests. In the other, the trustee agents adopt the proposed DRAFT approach for handling incoming task requests. The results from these two sets of experiments are labeled as *TRD* and *DRAFT* respectively in the figures in this chapter. If no trustee agent is willing to accept a task request under *DRAFT*, the truster agent will attempt to propose the same task in the following time steps. Each simulation is run for 1,000 time steps and repeated 10 times to reduce the effect of random variations.

### 7.5.3. Analysis of Results

#### Hypothesis 1

Figure 52 shows a snapshot of the change in the reputation evaluation of a trustee agent *n* which belongs to the *Hon* group (trustworthiness is 0.9) over 500 time steps under both *TRD* and *DRAFT*. It can be seen that under *TRD*, the agent's reputation fluctuates significantly. The sequence of event is: during time steps 1~15, *n* was discovered by a few truster agents through exploration and its good performance had gained it a high reputation (about 0.8) compared to other trustee agents. This attracted many other truster agents to request *n*'s services. The sudden influx of requests to *n* resulted in long backlog in its task queues. During time steps 15~270, *n* kept working on these tasks and producing high quality results. However, as most of them were completed with long delays, it received a large number of negative ratings from unsatisfied truster agents and its reputation dropped to a very low level (about 0.1). As its reputation dropped, the number of requests received by *n* also decreased. Over time, the backlog in *n*'s task queues had gradually been worked off. From time step 270 to 300, *n*'s high quality service was rediscovered by a few truster agents through exploration. Then, a similar sequence of events occurred, causing another round of severe fluctuation in its reputation.

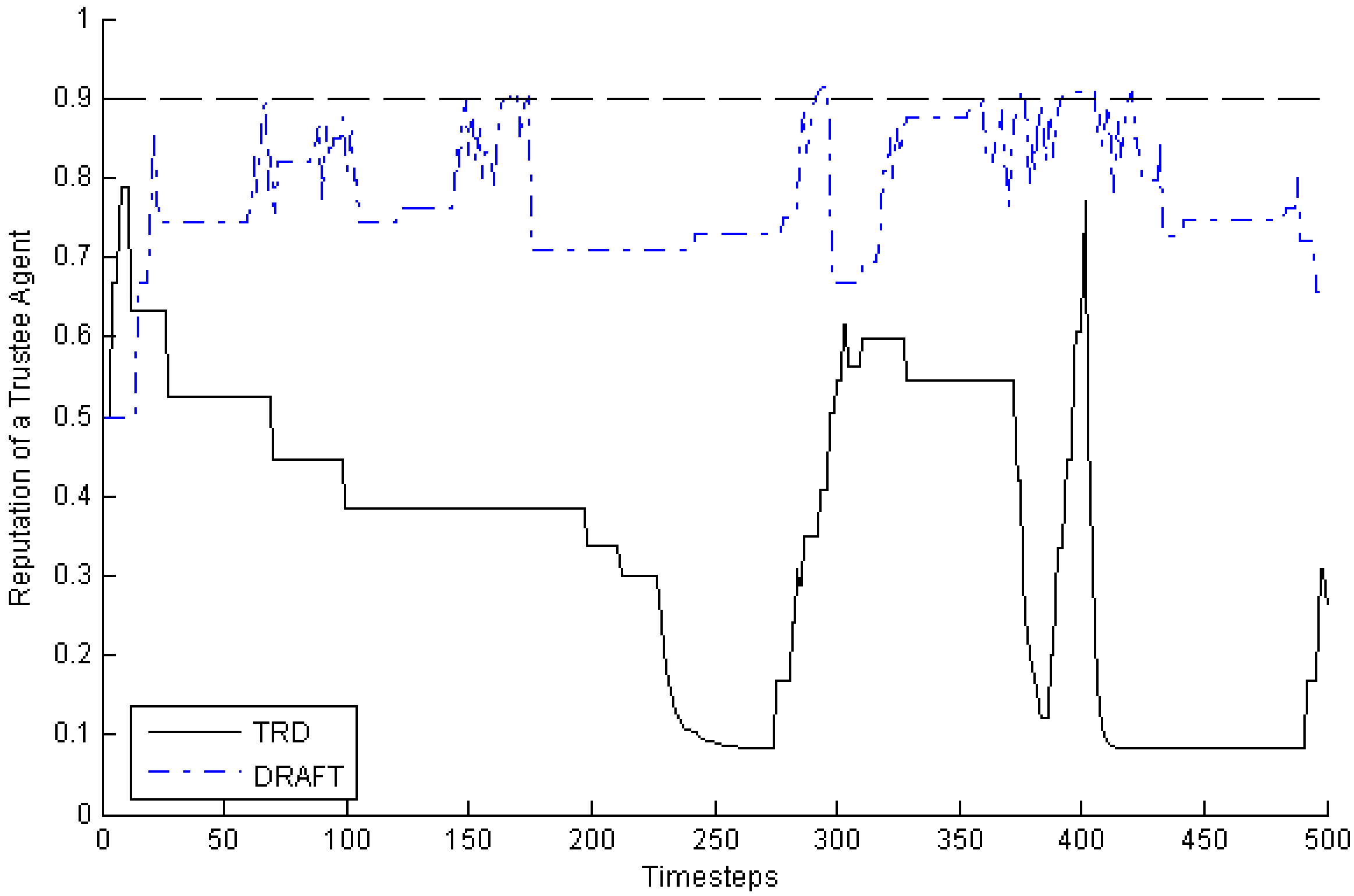

Figure 52. Evolution of reputation evaluation of a *Hon* trustee agent under different approaches over 500 time steps.

On the other hand, the reputation fluctuation of *n* was significantly smaller under *DRAFT* than under *TRD*. By dynamically deciding when to accept and decline task requests based on its current situation, *n* was protected from the adverse effect of reputation damage caused by truster agents seeking higher utility in an uncoordinated fashion. The fluctuations in *n*'s reputation under *DRAFT* were due to the 10% of time when it produced low quality results.

**Hypothesis 2**

Overall, the social equity among the group of the most trustworthy trustee agents (*Hon*) can be gauged through the fairness index ($F_{Hon}$) as defined in Chapter 6.

From Figure 53, it can be shown that $F_{HON}$ under *DRAFT* is consistently close to 1 (its value varies from 0.996 to 0.999), whereas $F_{HON}$ under *TRD* fluctuates erratically over different trustee agent population configurations. Although in some cases, *TRD* can also lead to highly fair distribution of tasks among trustee agents, when left to chance, such performance cannot be consistently achieved.

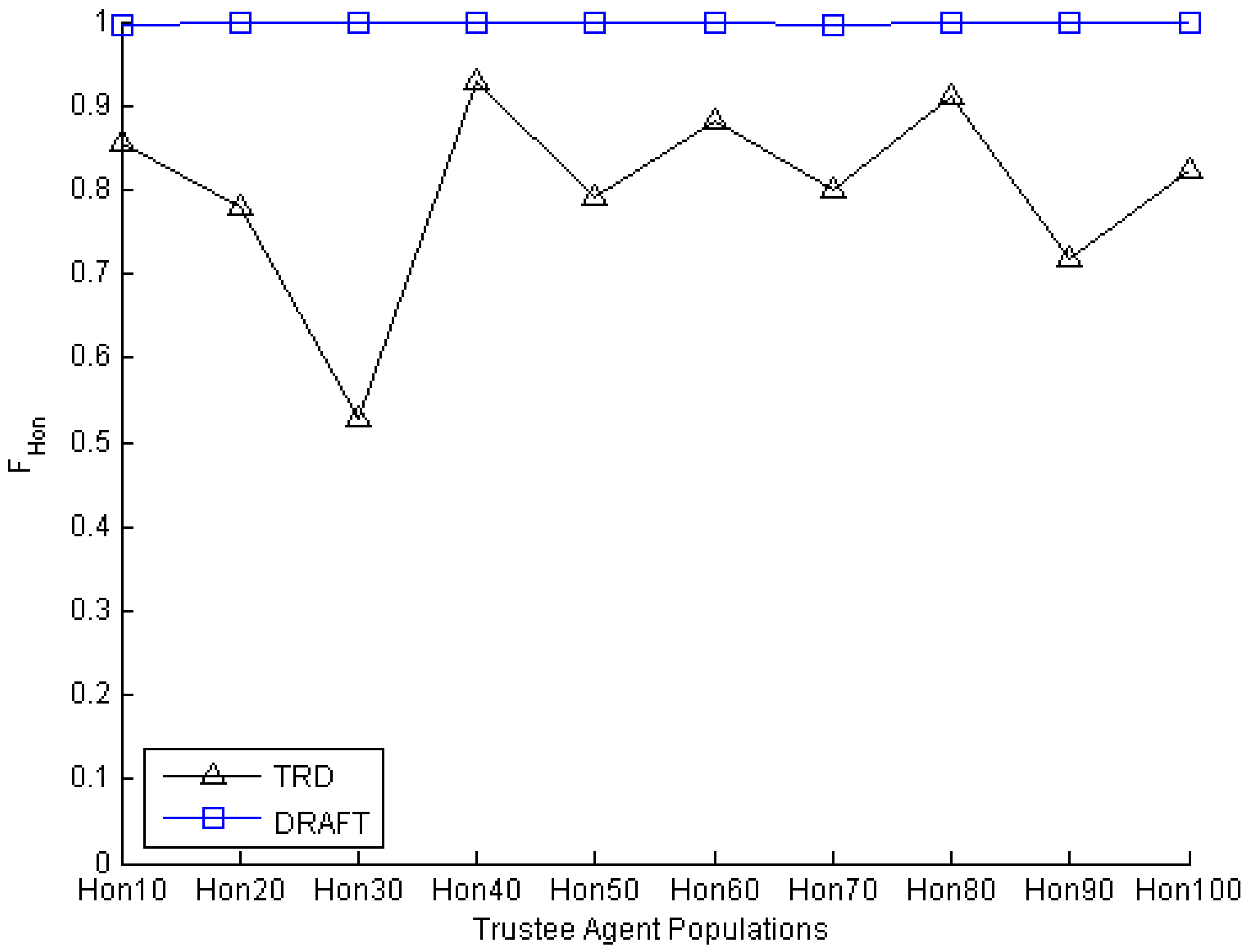

Figure 54. The fairness index for trustee agents belonging to Group *Hon* in various trustee agent population configurations.

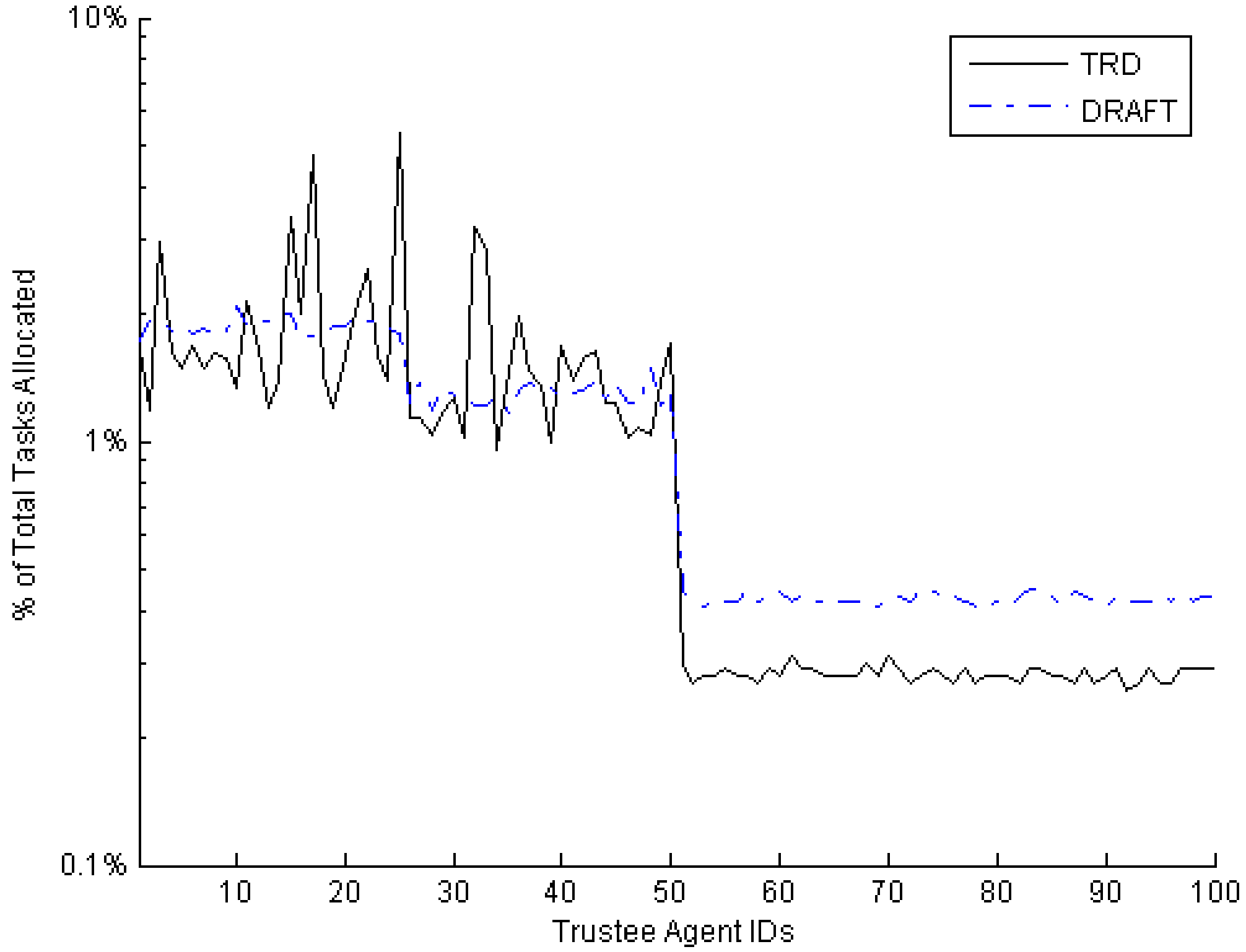

Figure 53. A snapshot of task distribution among trustee agents under agent population *Hon50*.

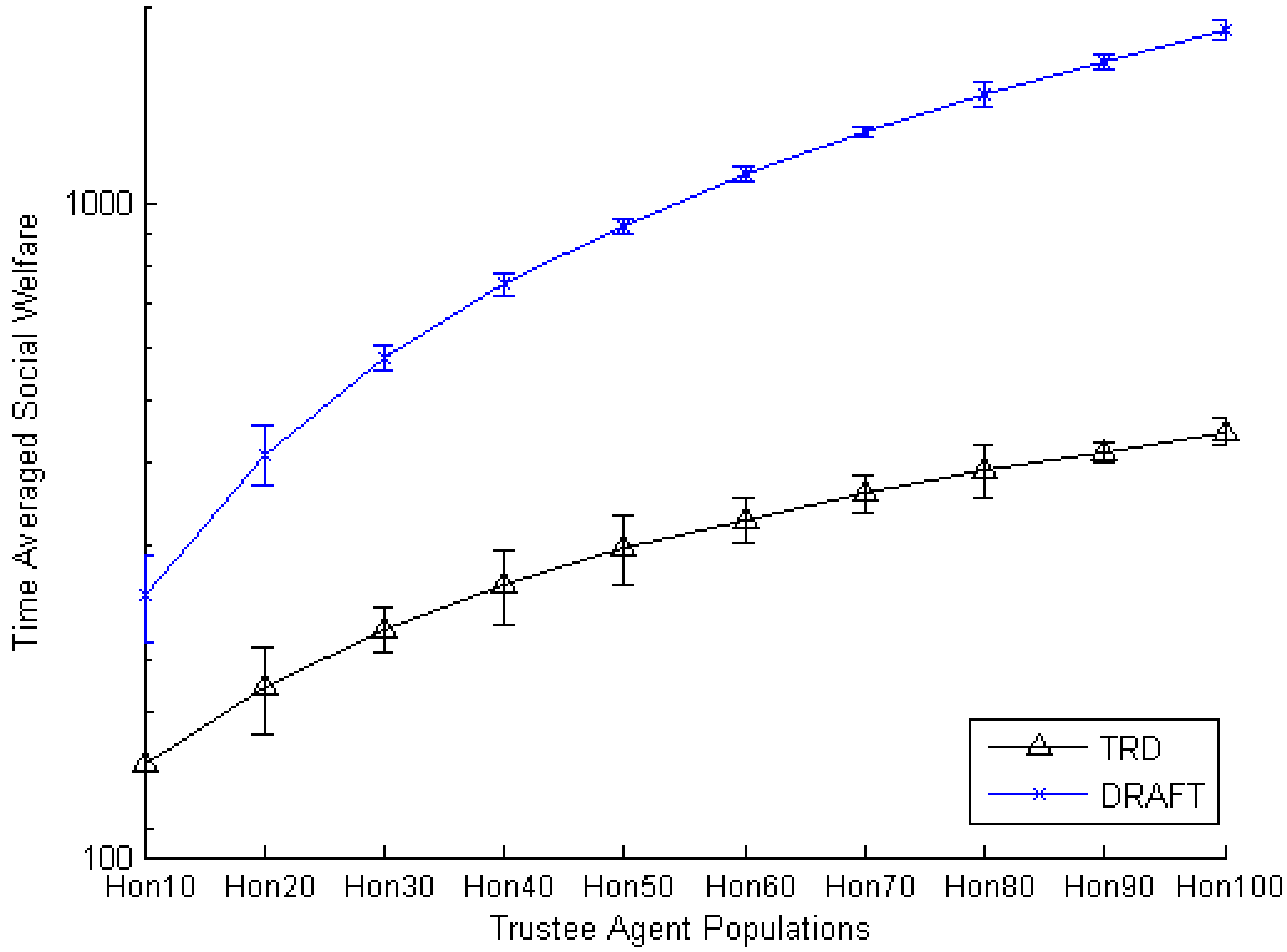

Figure 55. Time averaged social welfare achieved under different trustee population configurations.

A snapshot of the task distribution among trustee agents under agent population configuration *Hon50* is shown in Figure 54. Under *TRD*, clear peaks representing high concentration of tasks on a few trustee agents can be seen; whereas the distribution of tasks is smoother under the management of the DRAFT approach which represents better utilization of the trustee agents' capacities. In addition, the DRAFT approach helps existing reputation-aware interaction decision-making approaches to ensure that: 1) more trustworthy trustee agents receive more tasks than less trustworthy ones, and 2) similarly trustworthy trustee agents receive equitable amounts of tasks.

**Hypothesis 3**

Figure 55 shows the time averaged social welfare achieved by given trustee agent populations under *TRD* and *DRAFT*. By reducing the adverse effect of reputation damage through efficient utilization of trustee agents' capacities, the MAS equipped the DRAFT approach consistently achieved significantly higher social welfare than the MAS where trustee agents use the traditional *accept-when-requested* approach. The DRAFT approach is able to help a given trustee agent population to achieve close to its full potential in terms of social welfare which depends on the characteristics of the trustee agents.

From Figure 56, it can be seen that under *DRAFT*, all tasks accepted by the trustee agents can be completed on time (within a maximum of 3 time steps in this study). However, only 23.73% of the tasks completed by trustee agents under *TRD* are on time. This is the reason behind the low percentage of optimal social welfare achieved by *TRD*. The task backlog of trustee agents under *TRD* can become very long. A small percentage of the tasks (about 0.01%) even took almost the entire duration of a simulation to complete.

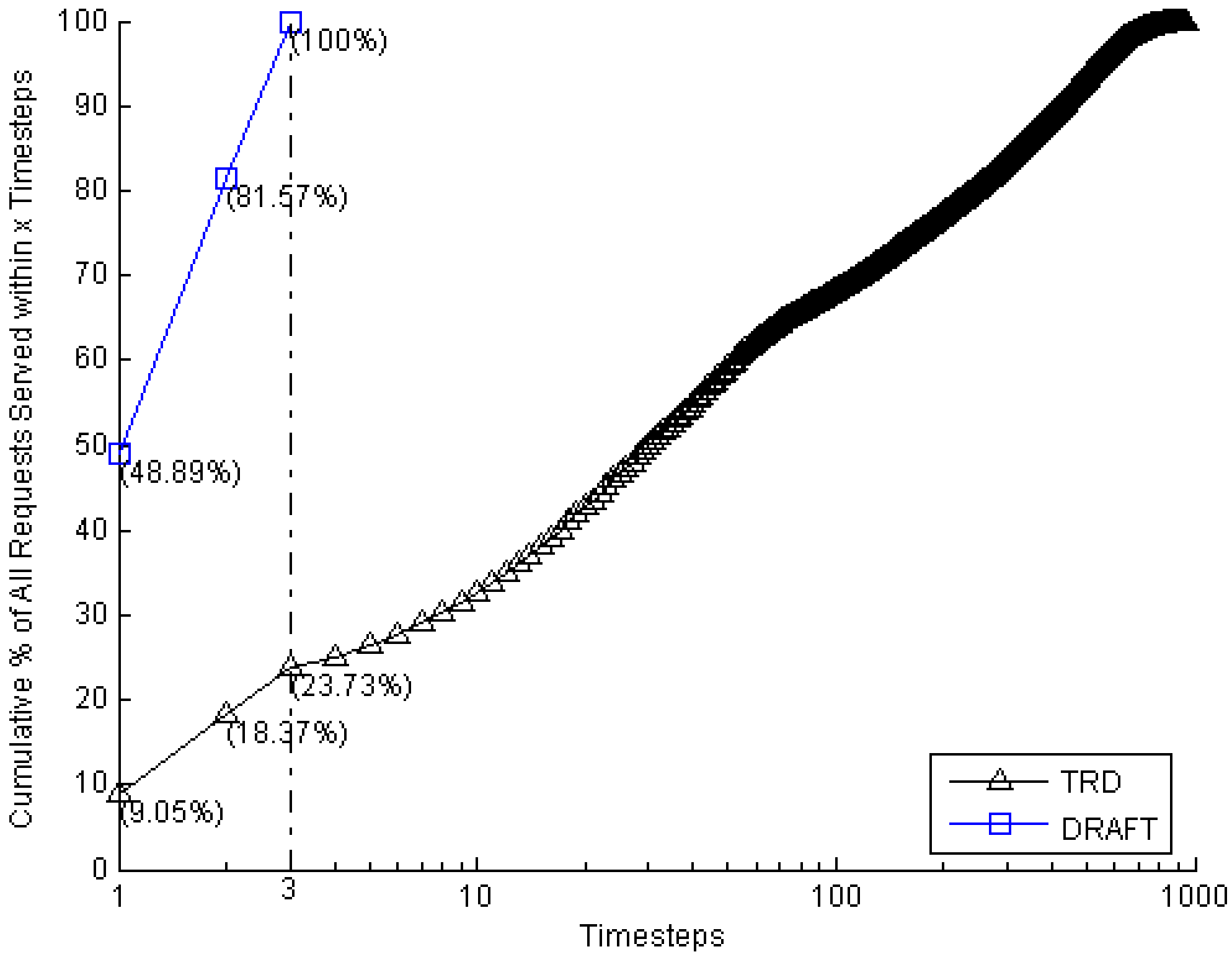

Figure 56. Cumulative percentage of requests served within *x* time steps.

## 7.6. Sensitivity Analysis

The parameter *V* control the trade-off between social welfare and average waiting time (which can be gauged by the average task backlog for every trustee agent across different agent population configurations at any given time step). When $V = 0$, only the Lyapunov drift will be minimized; when *V* is sufficiently large, only the reward will be maximized (in this case, according to (7.13), the DRAFT approach is equivalent to the traditional *accept-when-requested* approach). If we vary the value of *V*, the trade-off between social welfare and waiting time is illustrated in Figure 57. Increasing the value of *V* results in increases in social welfare and results longer waiting time. However, the rate of increase in social welfare drops rapidly as the value of *V* increases. Beyond the point of $V = 10^2$, further increasing the value of *V* starts to cause congestions in reputable trustee agents again and results in reductions in social welfare. These two parameters provide the owners of the trustee agents with some flexibility in customizing their trustee agents' behaviors according to their own personal preference.

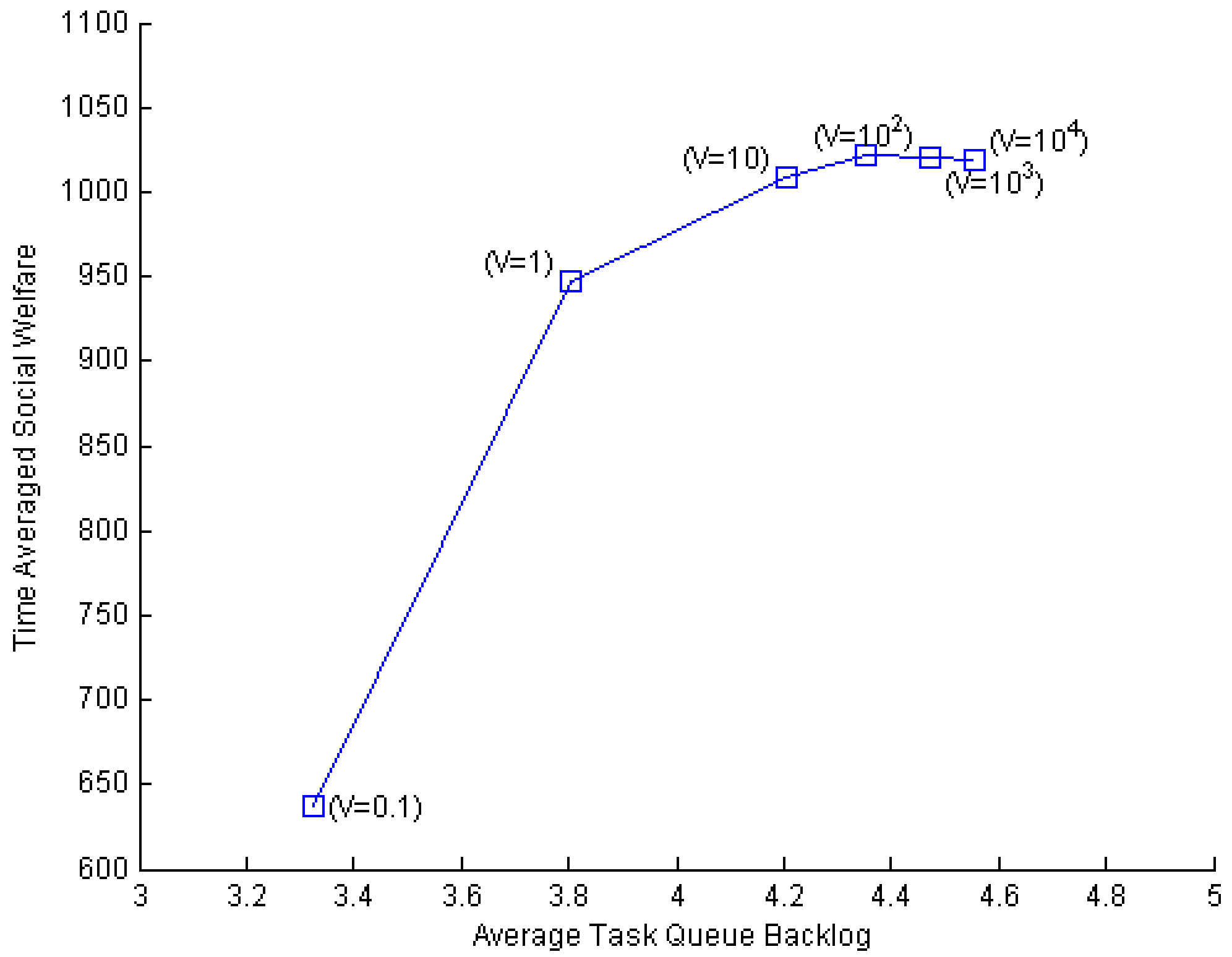

Figure 57. The effect of varying the value of parameter *V* on the performance of the DRAFT approach.

## 7.7. Summary

Trust is necessary for protecting individual truster agents in open MASs characterized by uncertainty. However, our study has shown that, under a more realistic system model, existing trust management approaches may result in highly unfair treatment of trustee agents even when their underlying trustworthiness is similar.

In this chapter, we proposed the DRAFT approach which is a distributed decision-making approach for trustee agents to manage their own reputations through the dynamic management of their workloads. The DRAFT approach is simple and requires only local knowledge to operate. It can complement existing trust evaluation models and trust-aware interaction decision-making models by enabling trustee agents to dynamically determine how many new interaction requests from truster agents should be accepted at each time step, given its current workload, its assessment of its reputation standing in the MAS based on the truster agents' ratings, and the value and cost associated with serving new requests.

By doing so, individual trustee agents can benefit from less fluctuation of their reputations and more equitable access to task requests from truster agents. Opportunities are passed up by trustee agents who are busy and picked up by agents who have more spare capacities while taking into account of their reputations. The collective capacity of the trustee agents can, thus, be efficiently utilized, which increases the social welfare in the MAS and improves social sustainability in the long run.

# Chapter 8

# Conclusions and Future Research

This thesis sets out to investigate how to make socially sustainable interaction decisions in multi-agent systems based on accurate assessment of agents' trustworthiness in the face of possible collusion among them. In this final chapter, we will summarize the research contributions of this work and discuss potential directions for future research.

## 8.1. Contributions

In an open MAS where agents come from various background and have potentially conflicting objectives, establishing trust can be regarded as a multi-step decision process where a truster agent's utility is a function of other agent's behavior. During the trust building stage, behaviors of other truster agents who act as witnesses can affect the accuracy of the information a truster agent uses to evaluate the reputation of a trustee agent. During the trust-aware task delegation stage, the number of truster agents who selected the same trustee agent can affect the expected quality of service received by each one of them, especially in cases where the trustee agent has limited capacity.

The work described in this thesis makes a number of important contributions to the state of the art in the area of trust management in multi-agent systems by extending the theoretical framework of trust-aware interaction decision-making to more realistic settings involving considerations on the limitations in the capabilities of the trustee agents as well as the real-time situation facing them. The contributions of this work can be summarized as follows:

(i) *The ACT Model*: Although there are a large number of research works over the years focused on addressing the problem of unfair testimonies from public reports of direct trust experience by witness agents. Many of these approaches generally suffer from three main types of shortcomings: 1) *relying on assumptions about the characteristics of the witness agent population*: They are usually majority voting based and perform poorly in situations where the majority of the witness agent population are compromised; 2) *tightly coupled with specific operating environments*: They often require additional infrastructural support (e.g., payment systems, knowledge of social relationships among agents, etc.) in order to work; or 3) *involving manual tuning of parameters crucial to the performance of the model*: This reduces the adaptability of the models in the face of a dynamically changing environment and makes regular human intervention necessary. We propose a trust evidence aggregation model based on the principles of reinforcement learning called the Actor-Critic Trust (ACT) model (*Chapter 3, Section 3.1 to 3.7*). It enables truster agents to dynamically learn the appropriate values of a large number of parameters based on their interaction experience. Compared to existing work, the ACT model does not require additional information or infrastructure support other than the third-party testimonies received by a truster agent. Experimental results show that it outperforms related work. The ACT model was applied to solve the collaborative spectrum sensing problem in Cognitive

Radio Networks (CRNs) and demonstrated effectiveness in preserving the wellbeing of the network under various attack scenarios (*Chapter 3, Section 3.8*).

(ii) *Multi-agent Trust Game*: In Chapter 4, the assumption made by most existing multi-agent trust research that trustee agents' perceived performance is not affected by the amount of workload assigned to them is removed. The multi-agent trust management problem is reframed into a new framework of thinking that is capable of taking the limitations in trustee agents' capacities into account. We redefine the trust-aware interaction decision-making (TID) problem as a Multi-agent Trust Game (MTG) based on the concept of Congestion Games [Monderer and Shapley, 1996] (*Chapter 4, Section 4.2*). The MTG complements existing research [Mikulski *et al.*, 2011] by providing a theoretical framework for analyzing TID approaches under more realistic conditions where trustee agents have limited resources and capabilities to handle workload and the delay experienced by truster agents is a function that is partially affected by the choices of interaction partners made collectively by them. By explicitly including these limitations into the analysis of trust in MASs, the MTG can facilitate the design of TID approaches that can produce higher system-wide social welfare.

(iii) *The SWORD Approach*: Based on a system model under MTG, we propose the Social Welfare Optimizing Reputation-aware Decision-making (SWORD) approach (*Chapter 5*). Designed for a system manager such as an e-commerce platform operator to use, the SWORD approach observes agents' real-time situations (such as their reputations, current workload, historical performance in handling assigned tasks per unit time, etc.) and helps truster agents who want to delegate tasks at the current time select interaction partners subject to multiple constraints. The SWORD approach is the first TID approach designed with the objective to mitigate the adverse effect of the RDP. The objective is to compute a decision policy for the truster agents that minimize their collective waiting time and maximize the sum of their utility over a potentially infinite horizon of interactions. Based on the principles of *Lyapunov Drift* analysis [Neely, 2010], it produces solutions to the MTG which can be proven to achieve social welfare values in an MAS close to the optimal value. Solutions are produced in polynomial time. Experiments conducted under crowdsourcing system environments have shown that the SWORD approach significantly outperforms related work in terms of promoting social equity and enhancing social welfare (*Chapter 6*). The SWORD approach protects trustee agents from being overloaded with requests, makes efficient use of the overall trustee agent resources in an MAS, reduces truster agents' waiting time for delegated tasks to be completed, and increases the throughput of interactions among agents in an MAS.

(iv) *The DRAFT Approach*: being a centralized approach, SWORD depends on the existence of a central trusted entity and may face difficulty scaling up. In order to address these issues, we propose the Distributed Request Acceptance approach for Fair utilization of Trustee agent services (DRAFT) in Chapter 7. The key to solving the RDP is to coordinate decisions made by truster agents. In a distributed MAS, this is difficult to achieve and incurs significant communication overhead. To address these challenges, we take a novel approach. Instead of designing a distributed TID model for truster agents, we propose the DRAFT approach to help trustee agents determine how to react to potentially uncoordinated truster agent decisions. DRAFT is the first approach in the state of the art to enable trustee agents to make situation-aware decisions about whether to accept incoming task delegation requests. Through an analysis of its current situation taking into account its current reputation standing in the MAS, its current workload, and the anticipated effort level it can spend on completing tasks over the current time step, the trustee agent can adjust the degree of greediness for accepting incoming requests with the DRAFT approach. Based on the same design principle as the SWORD approach, the DRAFT approach can also be proven to produce solutions for the MTG achieving close to optimal social welfare over the long run. The solutions can also be produced in polynomial time. Experiments conducted in open dynamic environments show that the DRAFT approach achieves significantly higher social welfare than related work and promotes social equity in the community.

The impact of the proposed models and approaches on the state of the art in trust management process is shown in Figure 58. The ACT model enable truster agents to dynamically adapt to changing behaviors of witness agents when they report testimonies about other trustee agents, and maintain the quality of the reputation values produced from these potentially biased opinions. In MASs with a central trusted entity, the SWORD approach can help truster agents coordinate their interaction partner selections in real time to achieve near optimal utilization of system resources. In fully distributed MASs, the DRAFT approach can still achieve this goal without incurring significant communication overhead needed for coordinating truster agent decisions in a distributed manner.

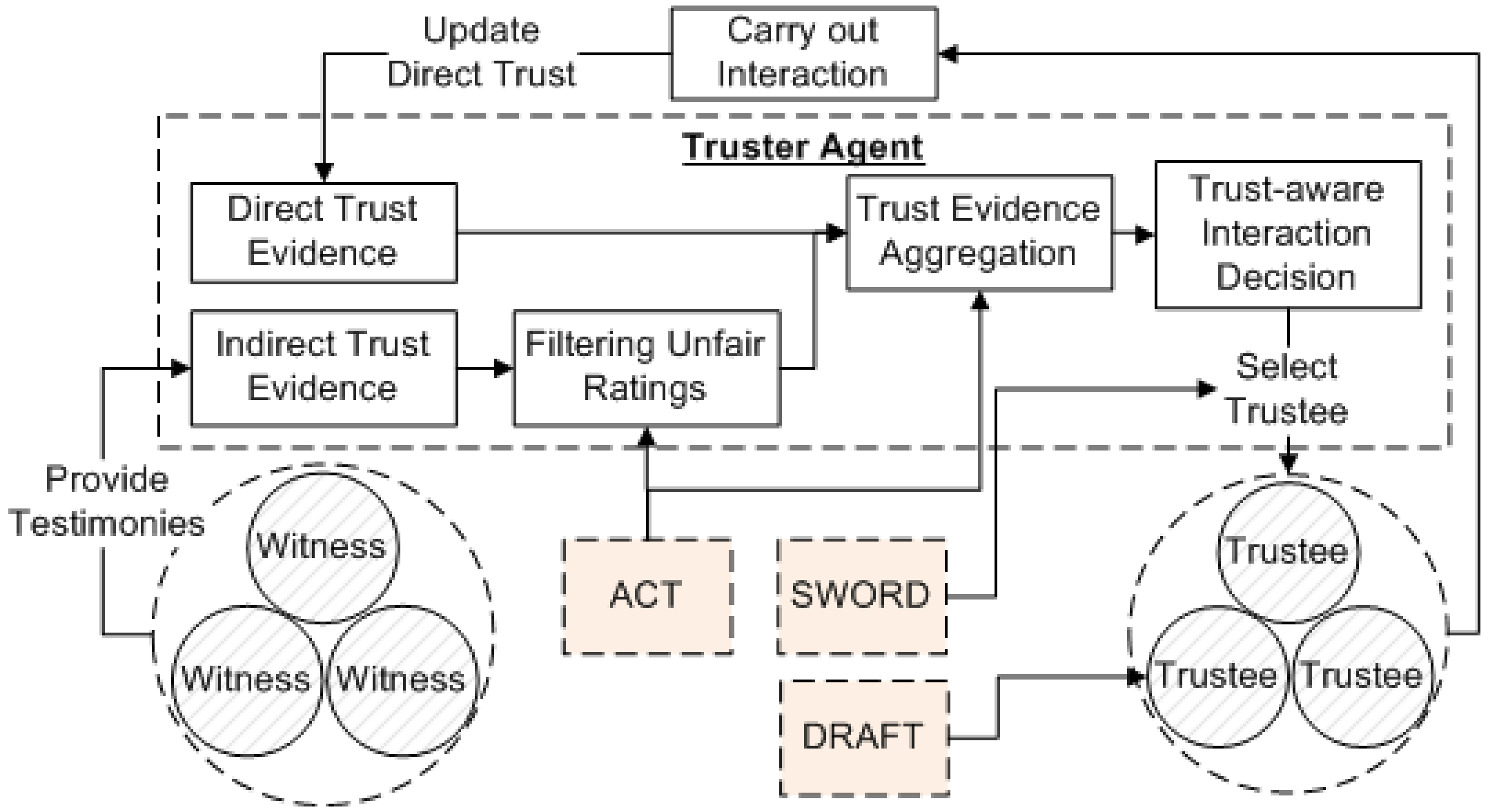

Figure 58. The areas of application of the proposed models and approaches in the trust management process.

With the proposed models and approaches, this work provides various stakeholders in MASs (such as truster agents, trustee agents and supervisory agents if any) with means to combat collusion and make near optimal use of the collective capabilities in the MASs. In this way social equity among agents can be achieved through the proposed approaches which will help sustain trustworthy interactions in open MASs over the long term. These contributions, especially SWORD and DRAFT, bridged important gaps in existing multi-agent trust management research, and have the potential to enable trust models to operate more effectively in MASs where agents and human beings co-exist.

## 8.2. Future Research Directions

There are several areas in which current research can be extended:

(i) Reputation rating distortions have been reported in some of the world's largest online e-commerce systems as one of the goals of people participating in such schemes is to quickly build up their reputations through illegitimate transactions [Hexun.com, 2012]. Although their subsequent behaviors may not necessarily be malicious, such a practice is very disruptive to the e-commerce community. In general, reputable sellers whose reputations are gained through illegitimate means often appear to be building up their reputations much faster than peer members in the community. Integrating the analysis of the temporal aspect of the reputation building process may be a research direction which has the potential to yield more effective trust evidence filtering and aggregation models. In the largest e-commerce platform in China – Taobao.com, the buyer communities are starting to respond to this problem by coming up with some rudimentary guidelines on helping buyers spot collusive sellers through looking at their historical reputation scores. Nevertheless, these collusive sellers are adapting to these self-protection mechanisms by slowing down the rate at which their reputation scores grow through less greedy behavior. Future research attempts in incorporating temporal analysis into trust evidence filtering and aggregation models should therefore, be rooted in analyzing real world data to identify and respond to these strategic behavior patterns.

(ii) In some applications, what really matters to a truster is not always the immediate outcome of trusting a trustee, but rather the long term impact of such an act. An example is in investment planning. By taking portfolio adjustment actions at a particular point in time following the advice of an investment planner (who is the trustee in this case), the truster expects to achieve long term financial wellbeing. If the portfolio value drops immediately after taking the actions recommended by the financial planner but rises to much higher values over a longer period of time (e.g., after a few months), should the interaction be considered a failure? What should be the case if the portfolio value first rises then drops?

In this scenario, where the truster is an investor, depending on his investment horizon, either case could be considered a success. Current interaction outcome evaluation models in trust management are designed more for situations where the effect of an interaction can be fully known relatively quickly after the completion of the interaction. Directly applying them into application domains where the effect of an interaction differs with changing time and situation may mislead truster agents into trusting the wrong trustee.

To handle such scenarios effectively, existing interaction outcome evaluation models should be enriched with the following considerations:

1) *Feedback may be given not necessarily immediately after the conclusion of an interaction*: truster agents should be given more time in the interaction protocol to assess the effect of the last interaction until he has reached a level of uncertainty about the result which he is comfortable with to give a rating to the trustee. This requires the incorporation of the factor of time into the current uncertainty evaluation models for trust management (which are mainly concerned with aggregating trust evidence from multiple sources [Wang and Singh, 2007]). Alternatively, new feedback reporting protocols could be designed to allow the truster agents to withdraw or modify a previous feedback rating based on newly available evidence with regard to its impact.
2) *Improving the accountability of trustee agents through more sophisticated feedback information*: if the time between the completion of an interaction and the time when the feedback for it is given is extended, there may be cases where other activities that may affect the outcome of the previous interaction taking place during this interval. Therefore, the impact of each interaction should be clearly identified. This may require incorporating models used for describing an interaction with features to distinguish its own sphere of influence on the wellbeing of the truster agent. Such an approach may facilitate the analysis of accountability in cases where an eventual impact is due to the outcome of several interactions between a truster agent and possibly more than one trustee agents.

(iii) As human beings with limited resources (in terms of task processing capacity, time, health, etc.) are starting to play the role of trustees in many online communities (e.g., e-commerce systems, crowdsourcing systems, etc.), research on modeling their utility functions using a human centric approach is necessary for trust-aware interaction decision-making mechanisms. For example, a decision model that takes the holistic wellbeing of a human trustee into account may not necessarily adopt a utility function which is always linear to the amount of work allocated to him. Instead, a more complex utility function that allows the decision model to vary the delegation of tasks to trustees in such a way as to achieve work-life-balance for the trustees, while satisfying the overall goal of the community, will be desirable in future human-agent collectives.

The above-mentioned issues are important for constructing and sustaining a trust-based digital ecosystem in which human beings and artificial intelligence agents coexist. In order to pursue these research directions, the field of computational trust and reputation management needs to deepen the collective understanding in human strategic behavior by involving more evidence-based research into the current predominantly model-based research practice. When these research issues are successfully tackled, the resulting trust and reputation management system will be of greater relevance to practical applications, such as e-commerce and crowdsourcing, than existing systems.

# Appendix A

# List of Publications

The research work reported in this thesis has contributed to a total 31 publications including rigorously peer reviewed international journal articles (e.g., Proceedings of the IEEE, IEEE Access), high quality international conferences (e.g., IJCAI, AAMAS, IUI, IAT), workshop presentations, and pending intellectual property disclosures. The list is given below.

## Journal Papers

[1] **H. Yu**, Z. Shen, C. Miao, B. An, and C. Leung, "Filtering Trust Opinions through Reinforcement Learning," *Decision Support Systems*, vol. 66, pp. 102-113, 2014.

[2] **H. Yu**, Z. Shen, C. Leung, C. Miao, and V.R. Lesser, "A Survey of Multi-agent Trust Management Systems," *IEEE Access*, vol. 1, no. 1, pp. 35-50, 2013.

[3] Z. Shen, **H. Yu**, C. Miao, and J. Weng, "Trust-based Web-service Selection in Virtual Communities," *Journal for Web Intelligence and Agent Systems (WIAS)*, vol. 9, pp. 227-238, 2011.

[4] **H. Yu**, Z. Shen, C. Miao, C. Leung, and D. Niyato, "A Survey of Trust and Reputation Management Systems in Wireless Communications," *Proceedings of the IEEE*, vol. 98, pp. 1755-1772, 2010.

## Conference Papers

## Workshop Presentations

[2] **H. Yu**, Z. Shen, Q. Wu, and C. Miao, "Designing Socially Intelligent Virtual Companions," presented at the 25th International Conference on Computer Animation and Social Agents (CASA'12), 2012.

[3] **H. Yu**, Z. Shen, X. Li, C. Leung, and C. Miao, "Whose Opinions to Trust more, your own or others'?," presented at the Workshop on Incentives and Trust in E-commerce - the 13th ACM Conference on Electronic Commerce (EC'12), 2012.

[4] **H. Yu**, Y. Cai, Z. Shen, X. Tao, and C. Miao, "Intelligent Learning Companions for Virtual Learning Environments," presented at the Multi-agent for Education and Interactive Entertainment (MEIE) Workshop - the 9th International Conference on Autonomous Agents and Multiagent Systems (AAMAS'10), 2010.

## Pending Patents

[1] **H. Yu**, C. Miao, and Z. Shen, "An Apparatus for Efficient Task Allocation in Crowdsourcing," U.S. Provisional Patent US 61/951,785, 2014.

[2] **H. Yu**, C. Miao, Z. Shen, and C. Leung "An Apparatus for Protecting E-Commerce Business Owners from Reputation Damage," U.S. Provisional Patent US 61/951,767, 2014.